\documentclass{aa}
\usepackage{natbib}
\bibpunct{(}{)}{;}{a}{}{,} % to follow the A&A style
\usepackage{graphicx}

\usepackage{txfonts}
\usepackage[allcolors=blue]{hyperref}

\begin{document} 

   \title{Protostellar Interferometric Line Survey of the Cygnus X region (PILS-Cygnus)%\thanks{Table 8 is only available in electronic form at the CDS via anonymous ftp to http://cdsarc.u-strasbg.fr (130.79.128.5) or via http://cdsarc.u-strasbg.fr/viz-bin/qcat?J/A+A/}
}

   \subtitle{First results: observations of CygX-N30}

   \author{S. J. van der Walt\inst{1} \and L. E. Kristensen\inst{1} \and J. K. J\o rgensen\inst{1} \and H. Calcutt\inst{2,}\inst{3} \and S. Manigand\inst{1} \and M. el Akel\inst{4,1} \and R. T. Garrod\inst{5} \and K. Qiu\inst{6,7} 
   }

   \institute{Niels Bohr Institute \& Centre for Star and Planet Formation, University of Copenhagen, \O ster Voldgade 5-7, 1350 Copenhagen K., Denmark\\
              \email{sarel.vanderwalt@nbi.ku.dk}
    \and
    Department of Space, Earth and Environment, Chalmers University of Technology, SE-412 96 Gothenburg, Sweden
    \and
    Institute of Astronomy, Faculty of Physics, Astronomy and Informatics, Nicolaus Copernicus University, Grudziadzka 5, 87-100 Torun, Poland
    \and
    CY University, Observatoire de Paris, CNRS, LERMA, F-95000 Cergy, France
    \and
    Departments of Astronomy and Chemistry, University of Virginia, Charlottesville, VA 22904 USA
    \and
    School of Astronomy and Space Science, Nanjing University, 163 Xianlin Avenue, Nanjing 210023, People’s Republic of China
    \and
    Key Laboratory of Modern Astronomy and Astrophysics (Nanjing University), Ministry of Education, Nanjing 210023, People’s Republic of China
             }

   \date{Received \ldots
   ; accepted \ldots}

  \abstract
   {Complex organic molecules (COMs) are commonly detected in and near star-forming regions. However, the dominant process in the release of these COMs from the icy grains ---where they predominately form--- to the gas phase is still an open question.}
   {We investigate the origin of COM emission in a high-mass protostellar source, CygX-N30 MM1, through high-angular-resolution interferometric observations over a continuous broad frequency range. }
  % methods heading (mandatory)
   {We used  32 GHz  Submillimeter Array (SMA) observations with continuous frequency coverage from 329 to 361 GHz at an angular resolution of $\sim 1''$ to do a line survey and obtain a chemical inventory of the source. The line emission in the frequency range was used to determine column densities and excitation temperatures for the COMs. We also mapped out the intensity distribution of the different species. }
  % results heading (mandatory)
   {We identified approximately $400$ lines that can be attributed to 29 different molecular species and their isotopologues. We find that the molecular peak emission is along a linear gradient, and coincides with the axis of red- and blueshifted H$_2$CO and CS emission. Chemical differentiation is detected along this gradient, with the O-bearing molecular species peaking towards one component of the system and the N- and S-bearing species peaking towards the other. The chemical gradient is offset from but parallel to the axis through the two continuum sources. The inferred column densities and excitation temperatures are compared to other sources where COMs are abundant. Only one deuterated molecule is detected, HDO, while an upper limit for CH$_2$DOH is derived, leading to a D/H ratio of <0.1\%.}
  % conclusions heading (optional), leave it empty if necessary 
   {We conclude that the origin of the observed COM emission is probably a combination of the young stellar sources along with accretion of infalling material onto a disc-like structure surrounding a young protostar and located close to one of the continuum sources. This disc and protostar are associated with the O-bearing molecular species, while the S- and N- bearing species on the other hand are associated with the other continuum core, which is probably a protostar that is slightly more evolved than the other component of the system. The low D/H ratio likely reflects a pre-stellar phase where the COMs formed on the ices at warm temperatures ($\sim$ 30 K), where the deuterium fractionation would have been inefficient. The observations and results presented here demonstrate the importance of good frequency coverage and high angular resolution when disentangling the origin of COM emission.}

   \keywords{Astrochemistry; Stars: formation; 
   Stars: protostars; 
   ISM: molecules; 
   ISM: individual objects: \object{W75N(B); 
   Submillimeter: ISM}.}

   \maketitle
%
%________________________________________________________________

\section{Introduction}

Complex organic molecules \citep[COMs; molecules with at least six atoms, of which at least one is carbon;][]{Herbst-2009} are found everywhere in and near star-forming regions \citep[e.g.][]{Caselli-2012,Ceccarelli-2014,Joergensen-2020}. Our current understanding is that most COMs form on the surfaces of ice-covered dust grains, either from simple volatile gas species frozen out onto the grains, or from radicals in the ice \citep{Oberg-2016}. Near the forming stars, where $T_{\rm dust} \gtrsim 100$~K, the water-rich ice mantels sublimate off the dust grains, releasing COMs into the gas phase, where they can be directly observed at submillimetre (submm) wavelengths \citep[the so-called hot cores; e.g.][]{Kurtz-2000, Cesaroni-2010, joergensen-2016}. However, other non-thermal mechanisms can be efficient at desorbing COMs into the gas phase, and have been proposed in the literature. These include sputtering of the ices in shocks, where the shocks are generated by jets and outflows \citep[e.g.][]{Avery_and_Chiao-1996,Joergensen-2004,Arce-2008,Sugimura-2011,Lefloch-2017}, accretion shocks from the envelope onto the disc \citep[e.g.][]{Podio-2015,ArturdelaVillarmois-2018,Csengeri-2018,Csengeri-2019}, explosive events near forming stars \citep[e.g. Orion KL;][]{2011A&A...529A..24Zapata,Orozco-Aguilera-2017}, and UV irradiation of the outflow cavity walls \citep[e.g.][]{Drozdovskaya-2015}. Which of these processes dominate in releasing the molecules from the ice to the gas phase is in many cases unclear, as is the impact on the observed chemistry. 

Recently, \cite{Belloche-2020} presented observations of a sample of 26 solar-type protostars as part of the CALYPSO survey performed with the NOrthern Extended Millimeter Array (NOEMA). COM emission from methanol, CH$_3$OH, the `simplest COM' was detected towards 12 of the sources, with 8 of these sources having detections of at least three COMs. These latter authors found that a canonical hot-corino origin may account for only four of the sources with COM emission, while an accretion-shock origin fits best with two or possibly three sources, and an outflow origin fits three others. It is therefore becoming increasingly clear that the origin of COMs in scenarios other than the canonical hot-corino (and hot cores) are relatively common, and through large surveys like CALYPSO we may get a better understanding of how these COMs form.

%3. PILS Cygnus survey
The work presented here is part of a large-frequency-range line survey of ten intermediate- to high-mass protostellar sources in the Cygnus X molecular cloud, which forms part of the Great Cygnus Rift in the Galactic plane. Cygnus X is a prolific star-forming region, with many newly formed stars of various masses, and has been the focus of numerous studies \citep[for a full review of the region, see][and references therein]{Reipurth-2008}. Given its location in the Perseus Galactic spiral arm, the distance to Cygnus X is uncertain, with values in the literature ranging between 1.3 and 3 kpc \citep[e.g.][]{Campbell-1982,Odenwald-1993,Rygl-2012}. We assume the distance of $\sim$1.3 kpc obtained by \cite{Rygl-2012} from maser parallax measurements. It is home to one of the most massive OB associations known in the Galaxy, Cyg OB2, with hundreds of young O-type stars and perhaps thousands of B-type stars \citep[][]{Knodlseder-2000}. These massive stars provide a huge amount of ionising radiation to the region, making it an ideal target for the study of how the external environment created by these massive stars affects the chemistry of the surrounding star forming regions. To this end, full chemical inventories of a large sample of surrounding protostars are required. 

This kind of survey has historically been observationally very expensive, because it requires high angular resolution and many hours of observation time to obtain the wide frequency range needed. However, in recent years it has become much cheaper thanks to the upgraded receivers on the Submillimeter Array (SMA), and its new SWARM\footnote{SWARM is an acronym for SMA Wideband Astronomical ROACH2 (second generation Reconfigurable Open Architecture Computing Hardware) Machine.} correlator. It is now possible to obtain 32 GHz of continuous frequency coverage in a fraction of the time required in the past. Taking advantage of these new capabilities of the SMA, we obtained data from ten newly formed sources surrounding the Cyg OB2 association in order to study how the chemistry of these young stars is affected by the environment in which they form. Together, these observations form the Protostellar Interferometric Line Survey of Cygnus-X (PILS-Cygnus). 

%4. Source - this paper N30 - other authors

The focus of this paper is one of the most studied sources in the Cygnus X cloud, W75N (B) \citep[e.g.][]{Haschick-1981,Persi-2006}, or CygX-N30 \citep[N30 hereafter, using the designation by][]{Motte-2007}. This source contains three mm continuum cores, MM1, MM2, and MM3, where MM1 is the brightest source and is composed of MM1a and MM1b \citep[e.g.][]{shepherd-2001, minh-2010}. Three radio continuum sources were also detected towards MM1 (VLA1, VLA2 and VLA3), which have been identified as ultra-compact HII (UC HII) regions or thermal jets \citep[][]{hunter-1994,1997ApJ...489..744T,shepherd-2003,Carrasco-Gonzalez-2010,Rodriguez-Kamenetzky-2020}. \cite{shepherd-2004} found that the VLA sources drive high-velocity molecular outflows traced by SiO (2$-$1 and 1$-$0) emission. VLA1 and VLA2 were associated with OH, H$_2$O, and CH$_3$OH masers \citep[][]{1997ApJ...489..744T,Minier-2001,Hutawarakorn-2002,Fish-2005,Surcis-2009}, while VLA3 was found to be associated with one H$_2$O maser \citep{shepherd-2001} with a radio jet and SiO (1$-$0) emission \citep{Carrasco-Gonzalez-2010}. A large-scale CO outflow has been observed and found to originate from MM1 and extending $\sim$1 $\mathrm{pc}$ in both directions \citep[see][]{Fischer-1985,Hutawarakorn-2002,shepherd-2003,Gibb-2003,Birks-2006,Surcis-2009,Surcis-2011}.  Through MERLIN\footnote{Multi-Element Radio Linked Interferometer Network} measurements of OH masers, \cite{Hutawarakorn-2002} found that N30 hosts a rotating molecular disc with a radius of 3000 AU, a rotation velocity of 6 km.s$^{-1}$, and a mass of 120 $M_\odot$. These latter authors found that this disc has a position angle of 155$^\circ$ (measured from north), which is orthogonal to the large-scale molecular outflow seen in CO and H$_2$ ($PA \sim 65^\circ$). The authors note that the compact cluster of high-velocity OH and H$_2$O masers coinciding with VLA2 appears to mark the centre of the outflow. 

\cite{minh-2010} observed N30 with the SMA in the 215 and 345 GHz spectral windows with 8 GHz frequency coverage, which indicated that a hot core is located at MM1b and is associated with a thermal jet from VLA1. These authors also found a chemical difference between the continuum cores MM1a and MM1b, which they suggest is the result of the evolution of a massive star-forming core. They argue that VLA1 is the heating source of the hot core, which suggests that the region associated with VLA1 is the site of recent star formation. \cite{minh-2010} observed SiO emission at the position of VLA2, which they interpret as a spherical shock driven by the recent star formation at the site of VLA1.

In this paper we present new SMA observations of N30, providing 32 GHz of continuous frequency coverage in the 345 GHz atmospheric window. This large frequency coverage makes it possible to perform an unbiased survey of the spectroscopic signatures of the different components in the region and establish their molecular inventories. The paper is laid out as follows: a detailed description of the observational setup and data reduction is given in Section \ref{Sec:Observations}. In Section \ref{Sec:Results} we present the results of our analysis and then discuss them in Section \ref{Sec:Discussion}. We end with a summary and conclusions in Section \ref{Sec:Conclusion}.

%________________________________________________________________

\section{Observations}
\label{Sec:Observations}
The SMA observations of N30 form part of the PILS-Cygnus programme (PI: Kristensen, project ID 2017A-S028), which in turn is an extension of the PILS programme \citep[][]{joergensen-2016}, in which 34 GHz continuous frequency coverage observations of the low-mass protostar, IRAS 16293--2422, obtained from the Atacama Large Millimeter/Submillimeter Array (ALMA) were studied. The ten intermediate- to high-mass protostars of PILS-Cygnus were selected because they are all located in the same molecular cloud structure, and because of their proximity to the Cyg-OB2 association, which provides an opportunity to study the role of the external environment in setting the chemistry of newly formed stars. This paper presents the first results of the PILS-Cygnus programme.

\subsection{Calibration}
The observations were performed using the SMA in a combination of the compact and extended configurations, that is, with projected baselines ranging from 7 to 211 m. The full survey covered ten sources and every source was observed for approximately an equal amount of time over ten tracks: five tracks in the compact configuration and five tracks in the extended configuration. This was done to ensure approximately equal sensitivity for all observed sources. The compact-configuration tracks were executed between 27 June 2017 and 7 Aug 2017, while the extended-configuration tracks were observed between 20 Oct 2017 and 10 Nov 2017. The number of antennas available in the array was between six and eight. 

MWC349A was used as the complex gain calibrator for all observations, while the quasars 3c273, 3c454.3, and 3c84 were used for bandpass calibration depending on the time of the observations. Neptune, Titan, Callisto, and Uranus were used for flux calibration, again depending on the time of the observations. A detailed observing log is presented in Appendix \ref{sect:obslog}, which also includes information on the weather at the time of the observations.

The receivers at the SMA were tuned such that they covered the entire frequency range from 329 to 361 GHz continuously. Specifically, the 345 GHz receivers were tuned to cover the spectral range from 329.2 to 337.2 GHz  in the lower sideband and the range from 345.2 to 353.2 GHz in the upper sideband. The 400 GHz receivers filled in the 8 GHz gaps, and covered the ranges of 337.2--345.2 GHz and 353.2--361.2 GHz in the lower and upper sidebands, respectively. Each sideband consists of four chunks, and each chunk covers 2 GHz. The SWARM correlator provides a uniform spectral channel size of 140 kHz across the spectrum, corresponding to 0.12 km s$^{-1}$ at 345 GHz. Prior to calibration, all spectral data were rebinned by a factor of four to a spectral resolution of 560 kHz, or 0.48 km s$^{-1}$ at 345 GHz to improve the noise level. 

All data were calibrated in CASA 4.7 \citep[Common Astronomy Software Applications,][]{McMullin-2007}. Calibration consisted first of flagging the channels at the edges and any anomalously high intensity spikes. The bandpass calibrator was then phase-calibrated with a 30 sec solution interval before the complex gains and absolute flux were calibrated.

\subsection{Imaging}

Self-calibration and imaging were performed using CASA 4.7. Two rounds of phase self-calibration were performed on the continuum, the first with a solution interval of 240 sec, followed by a second round with a smaller solution interval of 60 sec. This resulted in a reduction in rms level of more than 30$\%$. Only phase self-calibration was performed.

Cleaning and imaging of the data set proceeded along two axes. First, the continuum data were cleaned in order to achieve high angular resolution. This was done by setting the `robust' parameter to $-0.5$, where robust$=2$ gives natural weighting, and robust$=-2$ gives uniform weighting. This was an affordable solution, because the continuum data had very high signal-to-noise ratio. The resulting beam size of the continuum data is $0$\farcs$85 \times 0$\farcs$66$ (PA = $12.1^\circ$). The continuum rms sensitivity of the final line-free and self-calibrated image (Fig. \ref{Fig:N30_continuum}) is 0.008 Jy beam$^{-1}$. 

In a second step, the line data were first continuum-subtracted using line-free channels. This was done by taking a spectrum towards MM1a in the image plane and going through the spectrum by hand to identify all the line-free channels. These frequency ranges were then used throughout the cube, with the continuum subtraction done in the uv-plane. This exercise was then repeated to ensure that all the line-free channels remained line-free. The obtained self-calibration solution was then applied to the line data. To clean the line data, natural weighting was used in order to get the highest possible line sensitivity. For the full data cube, non-interactive cleaning was used with a circular cleaning mask centred at coordinates R. A. =  20$^\mathrm{h}$38$^\mathrm{m}$ 36$\fs$62 and Dec. = 42$^\circ$37$'$31$\farcs$81, with a radius of 24$''$, and including the MM1, MM2, and MM3 cores. The resulting beam size is $\sim 1\farcs65 \times 1\farcs55$ (PA = 11$^\circ$). The noise level varies across the spectrum from $\sim 0.2 - 0.7 $ Jy beam$^{-1}$ as a function of receiver, as is clearly visible in Fig. \ref{Fig:N30_spectrum}, particularly around a frequency of 353.2 GHz. The noise level also varies as a function of the Intermediate Frequency (IF) performance of the receiver, as can be seen from the edge effects. A typical value of the rms is $\sim$0.5 Jy beam$^{-1}$ in 0.48 km s$^{-1}$ channels. The blanked channels in Fig. \ref{Fig:N30_spectrum} are high-amplitude edge channels.

\section{Results}
\label{Sec:Results}
\subsection{Continuum emission}
\begin{figure}
\centering
\includegraphics[width=9cm]{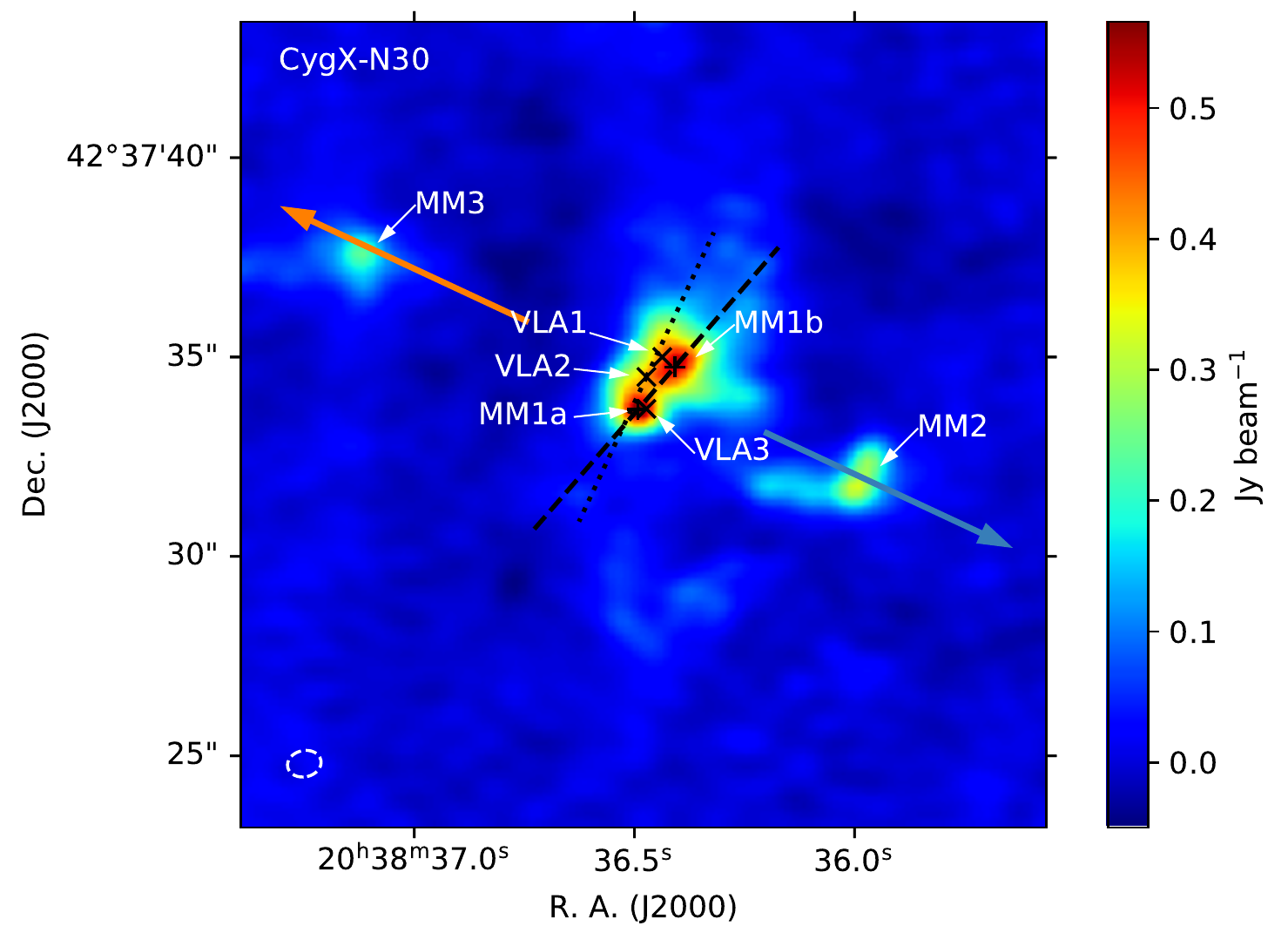}
  \caption{Observed continuum emission at a frequency of 345 GHz, showing the four continuum sources MM1a, MM1b, MM2, and MM3. The beam is shown in the bottom-left corner, and has dimensions $0$\farcs$85 \times 0$\farcs$66$, with position angle $12.1^\circ$. The VLA radio sources VLA1, VLA2, and VLA3 are labelled and marked with black crosses, with the emission peak positions of MM1a and MM1b shown by plus symbols, and the dashed line representing the axis through these positions. Also shown are the large-scale CO emission (red and blue arrows) and the disc-like structure (dotted line) identified by \cite{Hutawarakorn-2002}}.
  \label{Fig:N30_continuum}
\end{figure}

The 345 GHz ($850 \mathrm{ \ \mu m}$) continuum image is shown in Fig. \ref{Fig:N30_continuum}. The four previously identified sources are marked as MM1a, MM1b, MM2, and MM3 \citep[e.g.][]{shepherd-2001, minh-2010}. MM1a and MM1b are considerably brighter than MM2 and MM3, peaking at 0.53 and 0.48 $\mathrm{Jy \ beam^{-1}}$, respectively. All sources show extended emission, with large-scale structure visible to the west of the MM1 region, as well as northwest of MM1b. There is some extended emission to the east of MM2, and some faint structure just visible to the south of MM1. As with previous studies \citep[e.g.][]{minh-2010}, we detect no COM emission from the continuum cores MM2 and MM3, and this paper therefore focuses on emission towards MM1a and MM1b. The positions of the peak continuum emission for MM1a and MM1b were derived using 2D Gaussian fits, and are shown in Table \ref{table:N30_cores} and marked with plus symbols in Fig. \ref{Fig:N30_continuum}, with the dashed line representing the axis going through these two positions. MM1a and MM1b are separated by $\sim$$1 ''$, or $\sim$1300 AU, at a distance of 1.3 kpc.

%______________________________________________________________
%                                  One column figure
%----------------------------------------------------------- 

%______________________________________________________________
%         Two column figure (place early!)
%___________________________________________________________________
\begin{figure*}
    \centering
    \includegraphics[width=17.6cm]{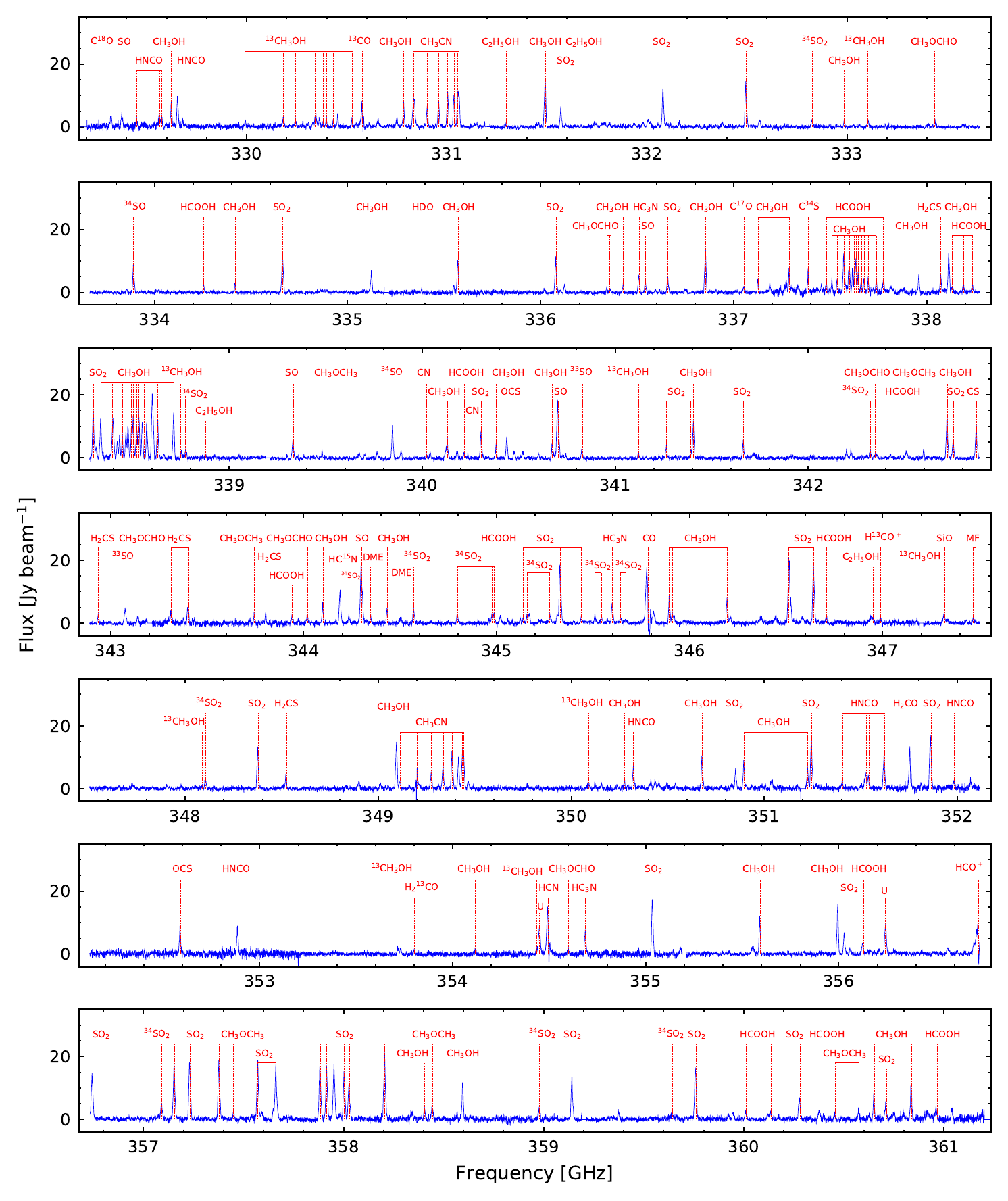}
    \caption{Spectrum towards N30 at the continuum peak of MM1a. The spectrum shows many molecular lines, with the identified lines marked and labelled. Some abbreviations are used where space is limited; `MF' and `DME' refer to methyl formate (CH$_3$OCHO) and dimethyl ether (CH$_3$OCH$_3$), respectively, while `U' stands for `unidentified'.}
    \label{Fig:N30_spectrum}%
\end{figure*}

%_____________________________________________________________
%                                             One column Table 
%_____________________________________________________________
%
\begin{table}
\caption{Properties of the N30-MM1 continuum cores.}             
\label{table:N30_cores}      
\centering          
\begin{tabular}{l c c r}     % 2 columns 
\hline
\hline       
    Name & Right Ascension & Declination & $S_{850 \mathrm{\ \mu m}}$\\ 
     &  &  & [$\mathrm{Jy \ beam^{-1}}$]\\ 
\hline                    
    MM1a & 20$^{\text{h}}$ 38$^{\text{m}}$ 36$\fs$51 & 42$^{\circ}$ 37$'$ 33$\farcs$48 &  0.53\\  
    MM1b & 20$^{\text{h}}$ 38$^{\text{m}}$ 36$\fs$42 & 42$^{\circ}$ 37$'$ 34$\farcs$58  & 0.48\\
\hline                  
\end{tabular}
\end{table}

In addition to the 345 GHz continuum cores, Fig. \ref{Fig:N30_continuum} also shows the positions of the three radio continuum sources VLA1, VLA2, and VLA3, which were identified previously  \citep{hunter-1994,1997ApJ...489..744T,shepherd-2003} using mm and centimetre (cm) observations from the Very Large Array (VLA) telescope. VLA1 is located to the west and slightly to the north of the MM1b submm peak, while VLA2 is to the north of the MM1a peak (southwest of MM1b). VLA3 is located very close to but slightly to the east of MM1a \citep[see also ][for recent high-resolution VLA observations of the region]{Rodriguez-Kamenetzky-2020}. The red and blue arrows represent the large-scale CO emission ($PA\sim$65$^{\circ}$ from north) centred at VLA2, with the dotted line representing the disc-like structure ($PA\sim$155$^{\circ}$) traced by OH masers, as identified by \cite{Hutawarakorn-2002}. 

To check whether the dust is optically thick or not, the first step is to calculate the dust mass. Here, this is only done towards the MM1a source, which is the brightest continuum source in the field. The total mass (gas + dust) can be calculated using the following relation \citep[e.g. ][]{ArturdelaVillarmois-2018}:
\begin{equation}
    M = \frac{S_\nu d^2}{\kappa_\mathrm{\nu}B_\nu(T)},
\end{equation} 
where $S_\nu$ = 0.53 Jy beam$^{-1}$ is the peak intensity of MM1a, $d = 1.3$ kpc, and $B_\nu(T)$ is the Planck function at the specific frequency and temperature. $\kappa_\mathrm{\nu}$ is the dust opacity \citep[at $\nu = 345$ GHz, $\kappa_{\nu}$ = 0.0175 cm$^2$ per gram of gas for a gas-to-dust ratio of 100;][]{Ossenkopf-1994}. For a beam size of $0$\farcs$85 \times 0$\farcs$66$, and taking $T$ = 30 K, we get $M = 3.0 M_\sun$ of gas and dust. If this mass is spread evenly over the beam, this corresponds to an H$_2$ column density of 5.3 $\times$ 10$^{24}$ cm$^{-2}$, where a mean molecular weight of $\mu$ = 2.8 was used to account for Helium \citep{kauffmann08}. This may be converted to optical depth using the following expression \citep[e.g.][]{Schoier-2002}:
\begin{equation}
    \tau_\mathrm{\nu} = \kappa _\mathrm{\nu} \mu m_{\rm H} N_\mathrm{H_2} \ .
\end{equation} 
We note that \citet{Schoier-2002} includes a dust-to-gas ratio, $\delta=0.01$, and $\kappa$ is their dust opacity (dust only). As we have folded the dust-to-gas ratio into $\kappa_\nu$ above (gas + dust), this parameter is not needed in the expression for the dust optical depth. Furthermore, we correct a typo in \citet{Schoier-2002} and use $\mu m_{\rm H}$ instead of $m_{\rm H_2}$, again with $\mu=2.8$. This yields a dust opacity of 0.43. 
We conclude that the dust emission is at least marginally optically thin for a gas-to-dust ratio of 100.

\subsection{Molecular line emission}

Figure \ref{Fig:N30_spectrum} shows a spectrum from the data taken at the position of peak continuum emission of MM1a. The 32 GHz continuous frequency coverage of the data (329 -- 361 GHz) included approximately 400 molecular lines detected above 3$\sigma$ towards the MM1 source, or a line temperature sensitivity of $\sim10$ K. Line identification was done by covering the points as described in \cite{Snyder-2005} and using the software package CASSIS\footnote{http://cassis.irap.omp.eu} \citep[Centre d’Analyse Scientifique de Spectres Instrumentaux et Synth$\Acute{\text{e}}$tiques;][]{Vastel_CASSIS-2015}, in which the CDMS\footnote{https://cdms.astro.uni-koeln.de/} \citep{Muller-2001-CDMS,Muller-2005-CDMS,Endres-2016-CDMS} and JPL\footnote{http://spec.jpl.nasa.gov/} \citep{Pickett-1998-JPL} molecular spectroscopy databases were used to construct a synthetic spectrum. Previously identified molecules were added to the synthetic spectrum first, which was then compared to the observed spectrum in order to find and identify the remaining lines above 3$\sigma$. From the observed lines, we identified 29 different molecules and their isotopologues, including five COMs, (CH$_3$OH, C$_2$H$_5$OH, CH$_3$OCH$_3$, CH$_3$OCHO, and CH$_3$CN).

The molecular transitions detected at $> 3\sigma$ are listed below, with the oxygen-bearing species in Table \ref{table:Molecules_COMs}, sulphur-bearing species in Table \ref{table:Molecules_S_species}, and nitrogen-bearing species shown in Table \ref{table:Molecules_N_SiO_HDO}, together with SiO, HDO, CO, and HCO$^+$. There are approximately 40 lines that remain unidentified, including two bright lines (at frequencies not corrected for the systemic velocity of 354.44896 GHz and 356.24131 GHz, and labelled `U' in Fig. \ref{Fig:N30_spectrum}). A full list of detected molecules above 3$\sigma$ are listed in Table \ref{table:Molecules_all}.

\subsection{Integrated molecular intensity maps}
The molecular peak positions were found by making integrated intensity emission maps of each molecule. Where a molecule had more than one line, the lines were stacked using a 1/$\sigma^2$ weighting scheme ---where $\sigma$ is the RMS measured on an emission-free region for each channel map--- in order to increase the intensity and determine an average peak position of the molecule. Only unblended lines were used.

\begin{figure}
\centering
\includegraphics[width=9cm]{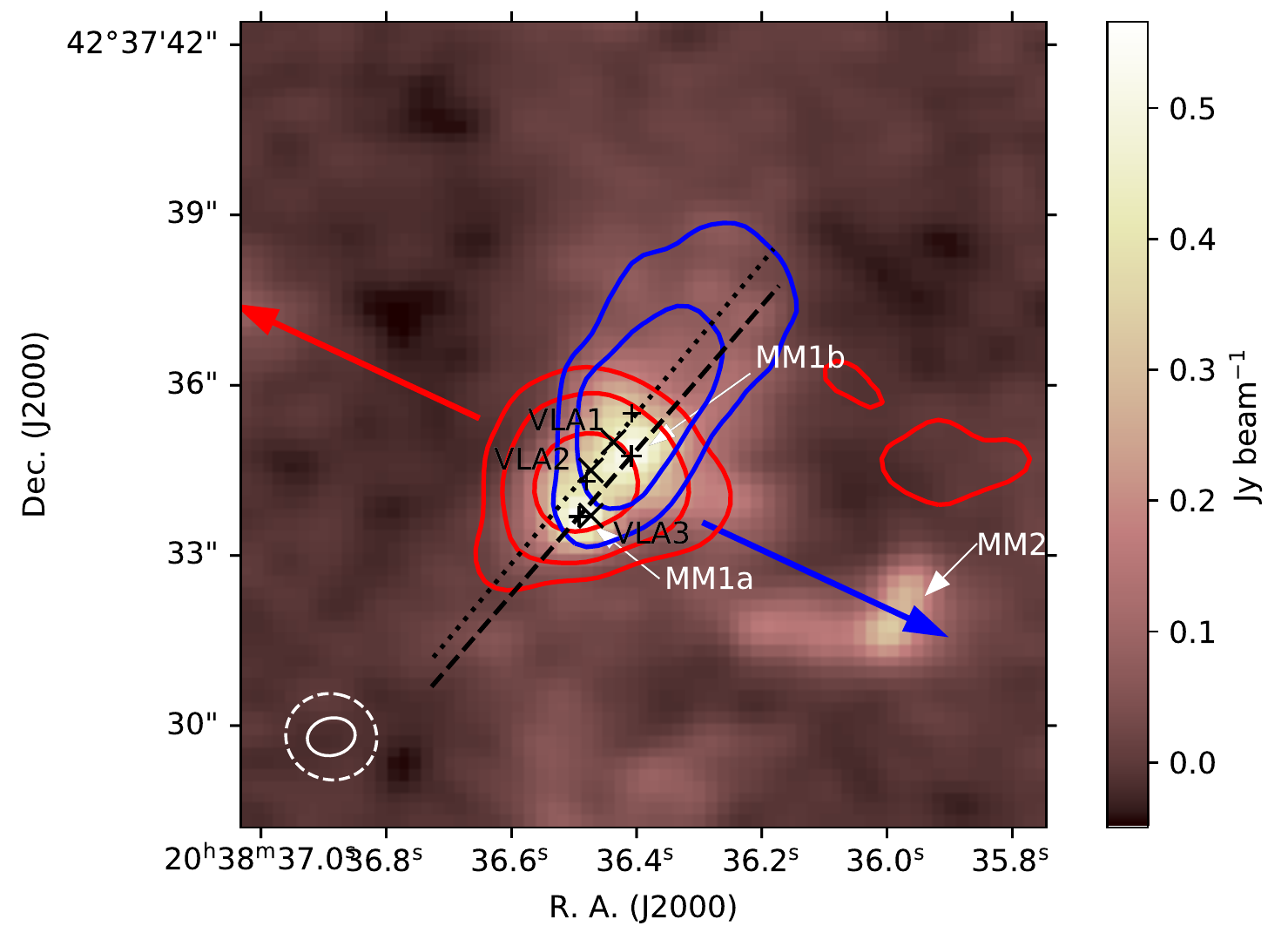}
  \caption{Continuum image of N30, with the red and blueshifted 351.77 GHz H$_2$CO, $ J=5_{1, 5} - 4_{1, 4}$, transition shown in contours. The blueshifted component shows more concentrated emission along the outflow axis (represented with a dotted line, running nearly parallel to the axis through the continuum peaks, represented with a dashed line), while the redshifted emission is slightly more extended perpendicular to the outflow axis, with some emission to the north of the MM2 continuum core. The velocity ranges extend to $\pm$ 14.25 km s$^{-1}$ from the systemic velocity of 9.5 km s$^{-1}$. Beam sizes are shown in the bottom left corner, and are $0$\farcs$85 \times 0$\farcs$66$, with $PA = 12.1^\circ$, for the continuum, and $1$\farcs$61\times 1$\farcs$51$, $PA=162.24^\circ$, for the molecular emission. Contour levels are at 3$\sigma$, 6$\sigma$, and 12$\sigma$. The blue and red arrows represent the large-scale CO emission.}
  \label{Fig:N30_H2CO_contours}
\end{figure}

In the cases where red- and blueshifted components are present, the components were separated with integrated maps made for each component. A map of blue- and redshifted H$_2$CO emission is shown in Fig \ref{Fig:N30_H2CO_contours} as an example (frequency = 351.77 GHz, $J = 5_{1, 5} - 4_{1, 4}$, and $E_{\mathrm{up}} = 62.45$ K). In this figure, the continuum emission is depicted in colour, with the red- and blueshifted integrated intensity maps shown in red and blue contours, respectively.  

\begin{figure}
   \centering
   \includegraphics[width=9cm]{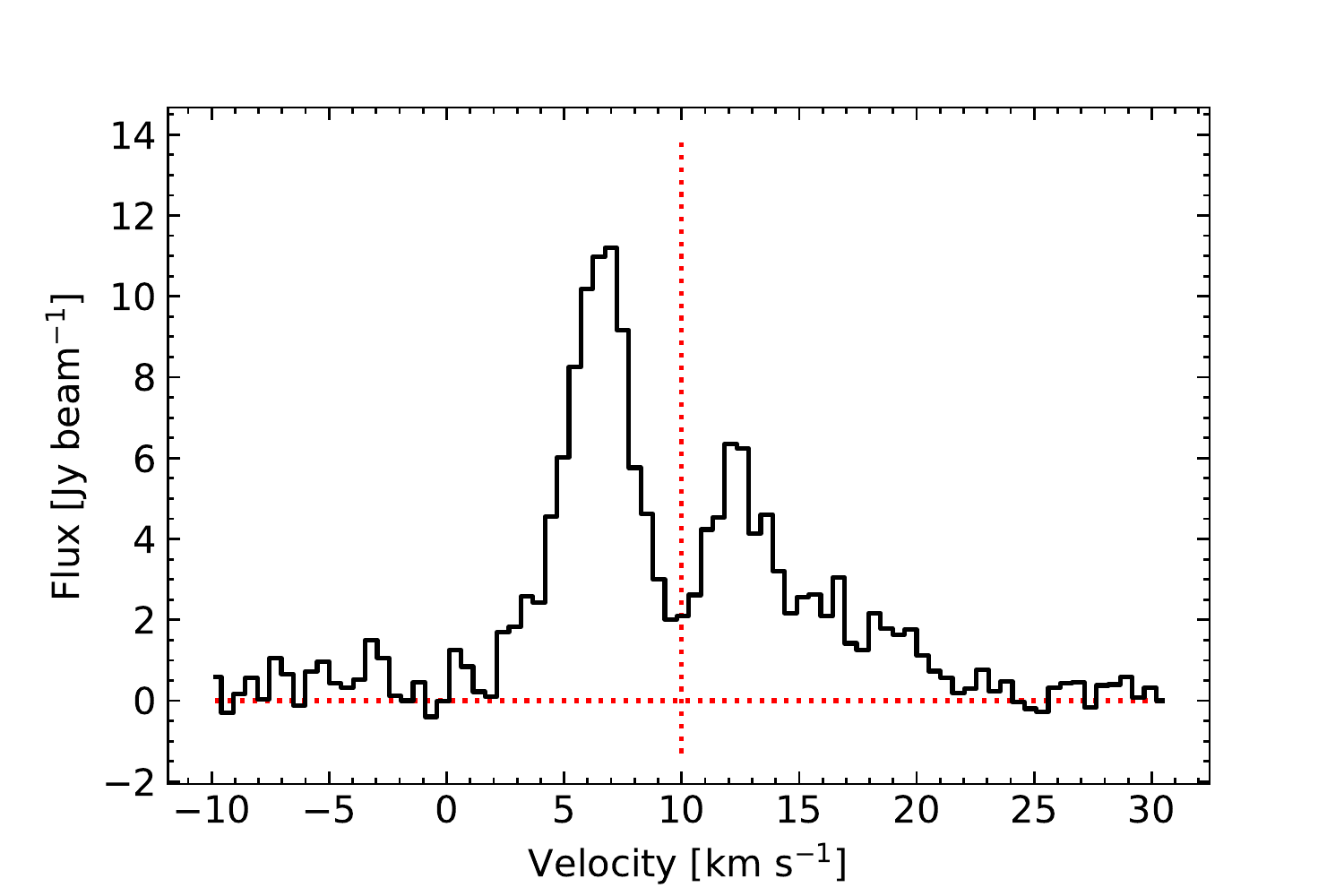}
      \caption{Blueshifted peak of the 351.77 GHz H$_2$CO, $ J=5_{1, 5} - 4_{1, 4}$ line, which peaks at an offset of $(-0.075,1.72)$ arcseconds from the MM1a continuum peak (see Table \ref{table:Molecules_COMs}).}
      \label{Fig:N30_H2CO_line_b}
   \end{figure}

\begin{figure}
   \centering
   \includegraphics[width=9cm]{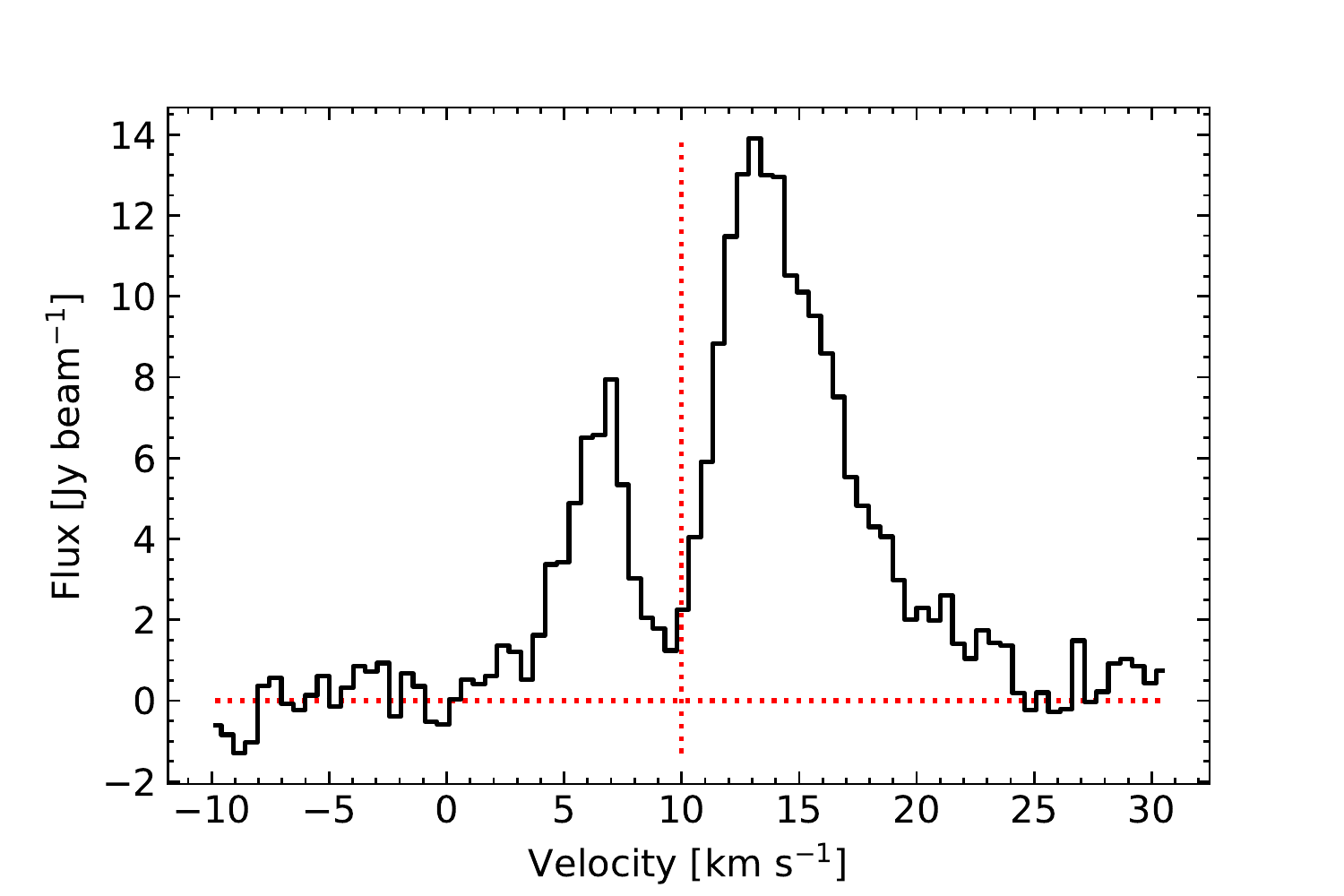}
      \caption{Redshifted peak of the 351.77 GHz H$_2$CO, $ J=5_{1, 5} - 4_{1, 4}$ line, which peaks at an offset of $(-0.003,0.52)$ arcseconds from the MM1a continuum peak (see Table \ref{table:Molecules_COMs})}
      \label{Fig:N30_H2CO_line_r}
   \end{figure}

The maximum intensity for the components were found at positions offset from the MM1a continuum peak: $(-0 \farcs 83,1 \farcs 72)$ for the blueshifted components and $(-0 \farcs 03,0 \farcs 52)$ for the redshifted ones. The line plots for the blue- and redshifted components are shown in Figs. \ref{Fig:N30_H2CO_line_b} and \ref{Fig:N30_H2CO_line_r}, respectively.

%______________________________________________________________
%                                  One column figure
%----------------------------------------------------------- 

%______________________________________________________________
%                                  One column figure
%-------------------------------------------------------------- 

   \begin{figure}
   \centering
   \includegraphics[width=9cm]{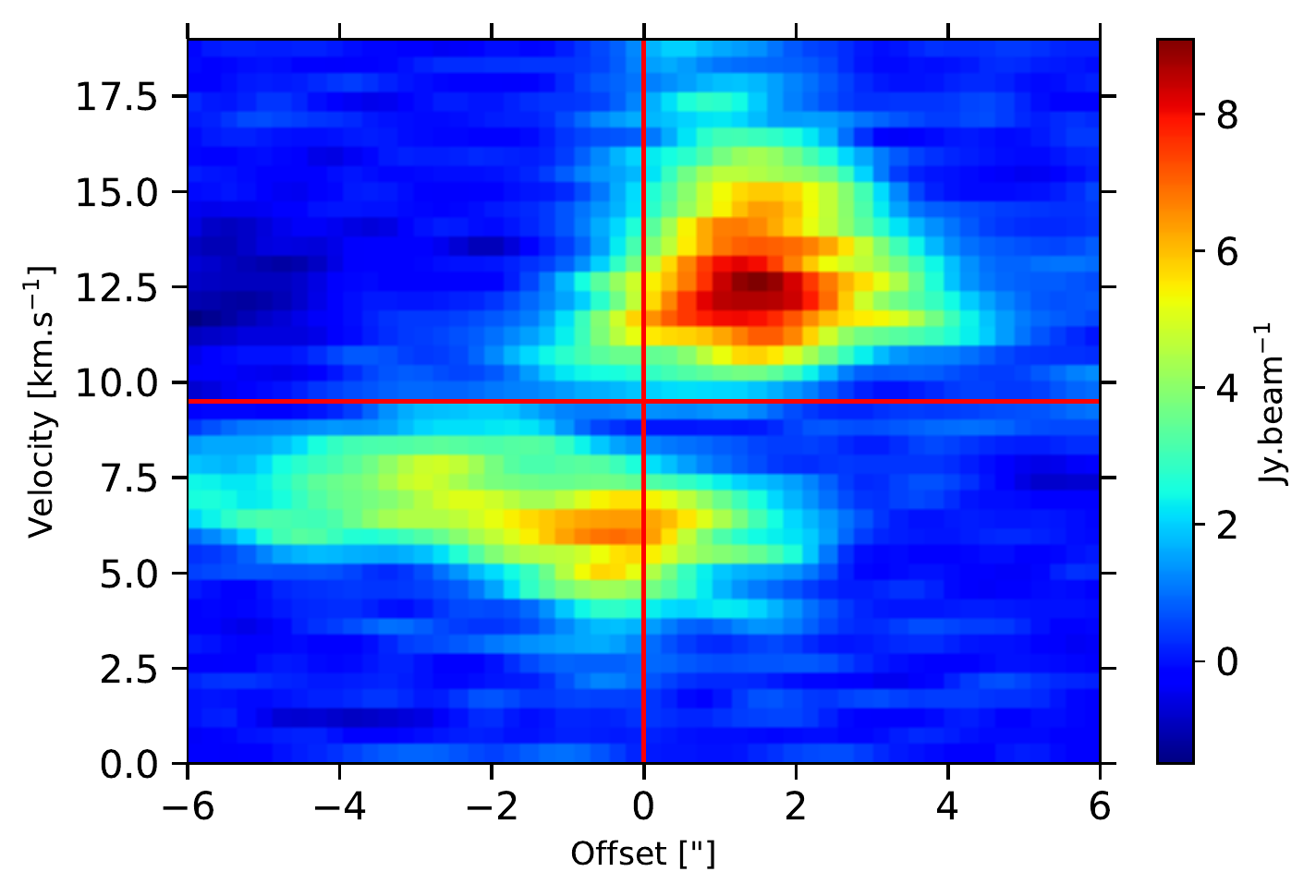}
      \caption{Position--velocity map of the 351.77 GHz H$_2$CO, $ J=5_{1, 5} - 4_{1, 4}$ line. The high-velocity  emission  is located close to the position of zero offset (located between the red- and blueshifted emission peaks), whereas the emission at larger offsets from the centre has velocity closer to the systemic velocity.}
      \label{Fig:N30_PV_plot_H2CO}
   \end{figure}
   
\begin{figure}
\centering
\includegraphics[width=9cm]{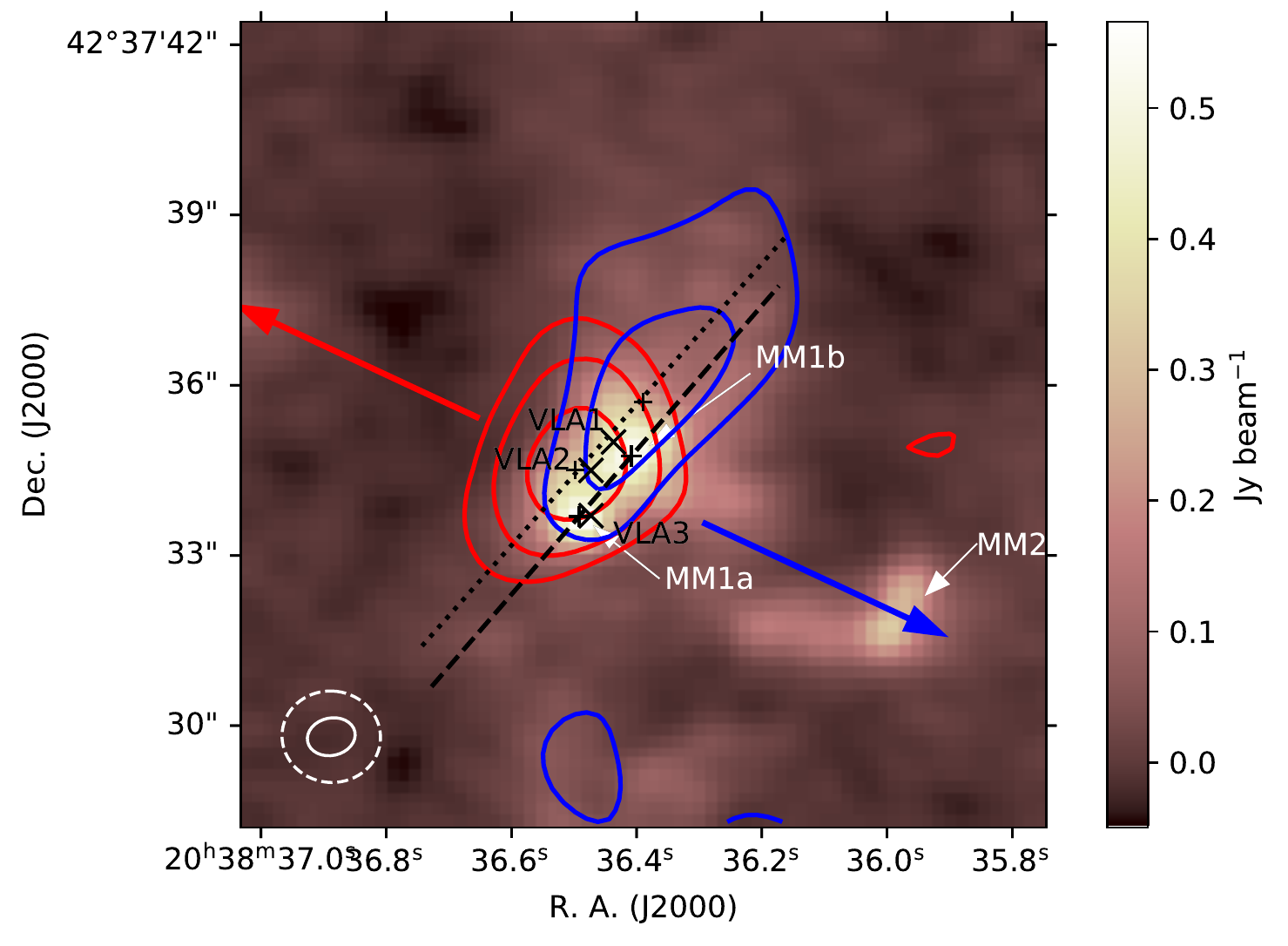}
  \caption{Continuum image of N30, with the red- and blueshifted 342.88 GHz CS line transition $J=7 - 6$ shown in contours. As with the H$_2$CO line, CS shows red- and blueshifted emission along the same outflow axis. The blueshifted emission shows some extended emission further out to the northeast, while the redshifted emission is slightly more concentrated than for H$_2$CO. Here, the velocity ranges extend to $\pm$ 15.6 km s$^{-1}$ from the systemic velocity of 8.75 km s$^{-1}$. Beam sizes and contour levels are as in Fig. \ref{Fig:N30_H2CO_contours}.}
  \label{Fig:N30_CS_contours}
\end{figure}

As can be seen from Fig. \ref{Fig:N30_H2CO_contours}, the blue- and redshifted components of H$_2$CO emission falls on an axis ---marked with a dotted line--- that runs parallel to the axis drawn between the positions of the MM1a and MM1b peak continuum emission marked with plus signs and represented with a dashed line. The blue contours are more collimated along the axis, but extend further out than the red contours, while the red contours have some extended emission to the west of MM1b, and north of the MM2 core. Similarly, the red- and blueshifted 342.88 GHz CS $ J=7 - 6$ transition is shown in Fig. \ref{Fig:N30_CS_contours}. Other molecules that exhibit red- and blueshifted components are CO, $^{13}$CO, HCN, and HCO$^+$.

\begin{figure}
   \centering
   \includegraphics[width=9cm]{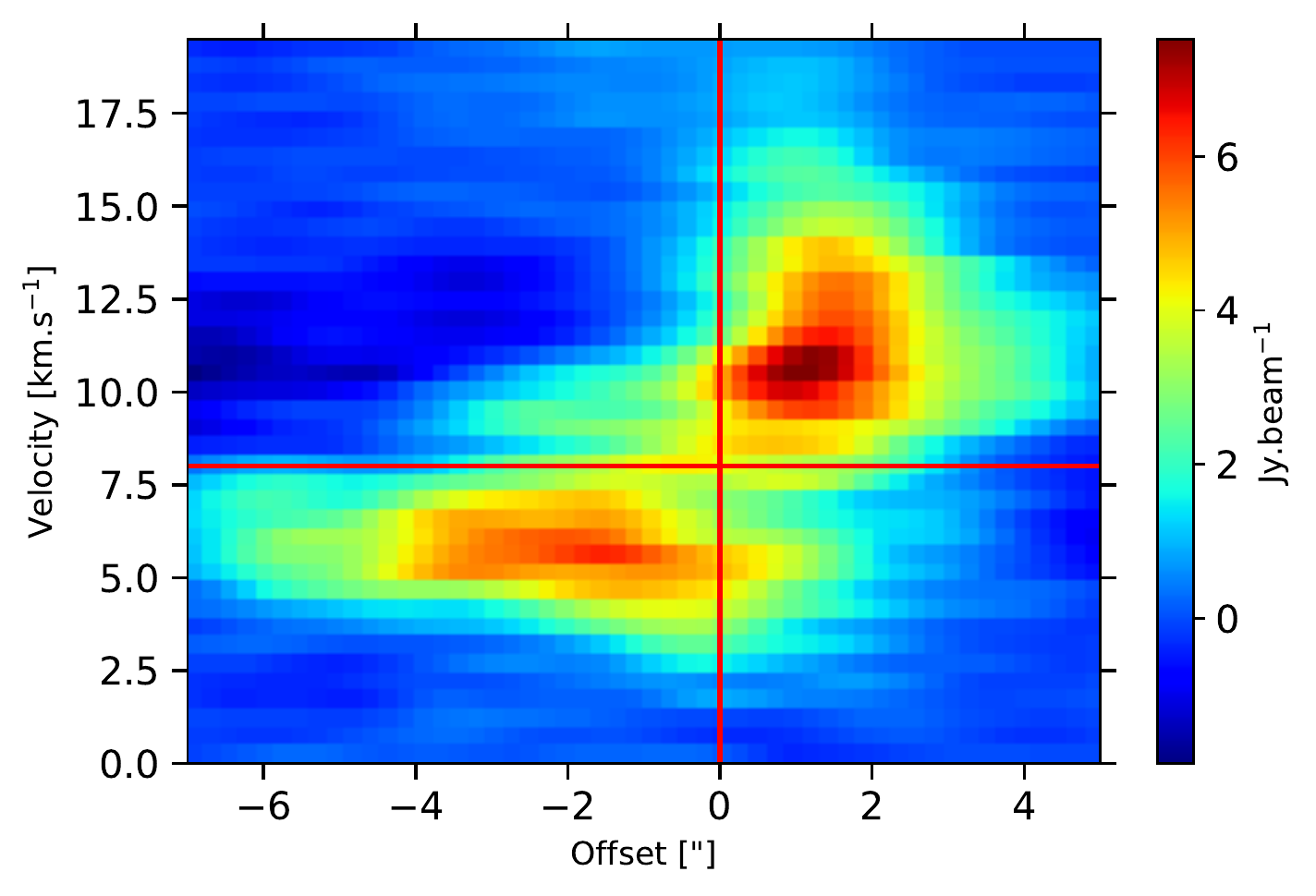}
      \caption{Position--velocity map of the 342.88 GHz CS $ J=7 - 6$ line. As with the H$_2$CO map, the emission at larger velocity is concentrated close to the centre of the emission (between the red- and blueshifted peaks), whereas the emission close to the systemic velocity extends out to larger spacial scales.}
      \label{Fig:N30_PV_plot_CS}
   \end{figure}

In order to investigate this large-scale velocity gradient, Figs. \ref{Fig:N30_PV_plot_H2CO} and \ref{Fig:N30_PV_plot_CS} show position--velocity maps of the H$_2$CO and CS emission, respectively. The high-velocity emission of H$_2$CO and CS is close to the centre (at 0$''$ offset, between the red- and blueshifted emission peaks), whereas the emission at a larger offset from the centre shows a velocity closer to the systemic velocity of the protostar. This could trace an infall motion coupled to a rotation pattern, which might be due to a disc-like structure \citep[e.g.][]{Zhu-2011}. A comprehensive study of the kinematics of the system is beyond the scope of this paper, but position--velocity maps of a selection of other transitions are shown in Appendix \ref{App:PV_plots}.

%______________________________________________________________
%                                  One column figure
%----------------------------------------------------------- 

%______________________________________________________________
%                                  One column figure
%-------------------------------------------------------------- 
   \begin{figure}
   \centering
   \includegraphics[width=9cm]{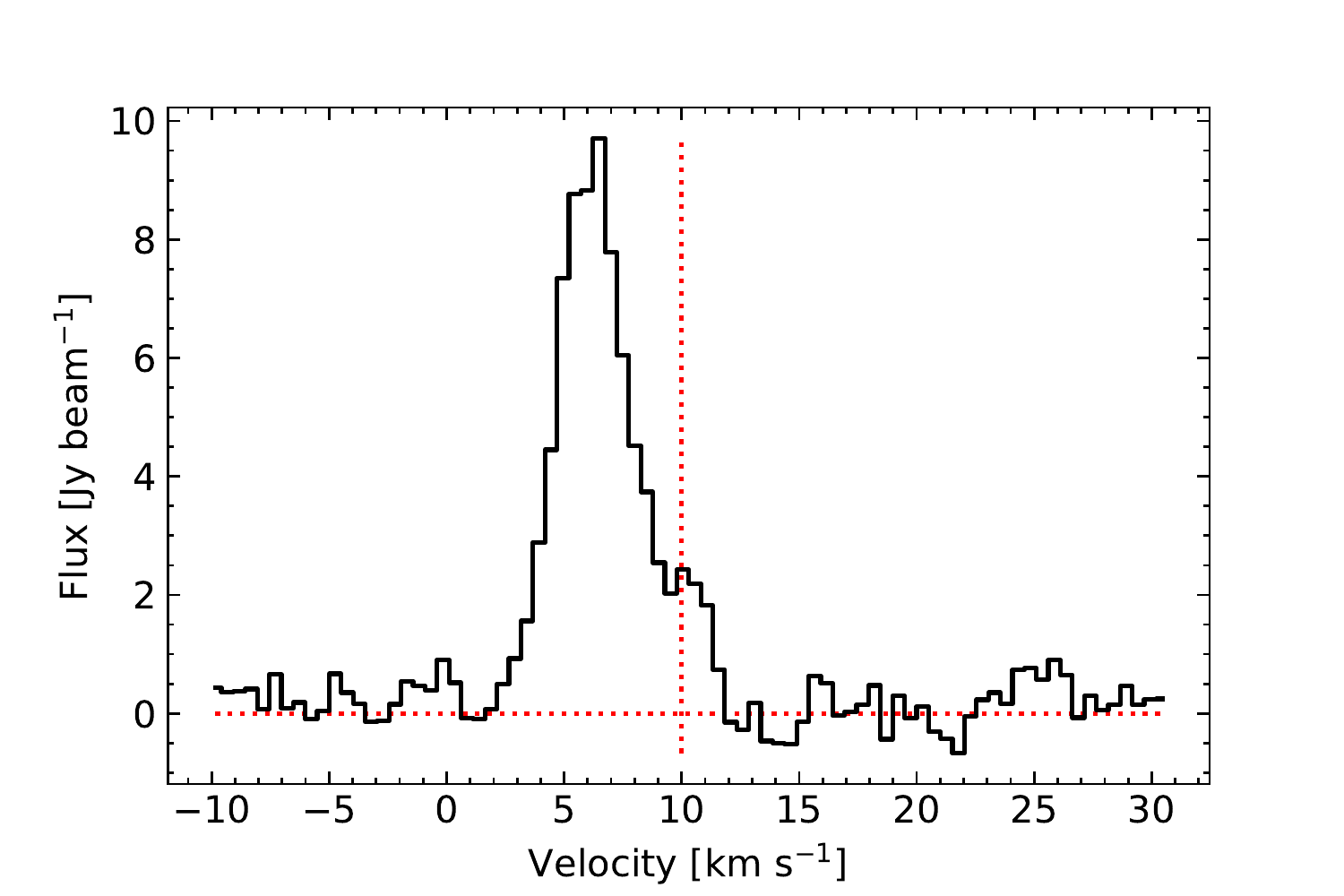}
      \caption{Blueshifted peak of the 342.88 GHz CS $ J=7 - 6$ line, which peaks at an offset of $(-0.094,1.92)$ arcseconds from the MM1a continuum peak (see Table \ref{table:Molecules_S_species}).}
      \label{Fig:N30_CS_line_b}
   \end{figure}
   
   \begin{figure}
   \centering
   \includegraphics[width=9cm]{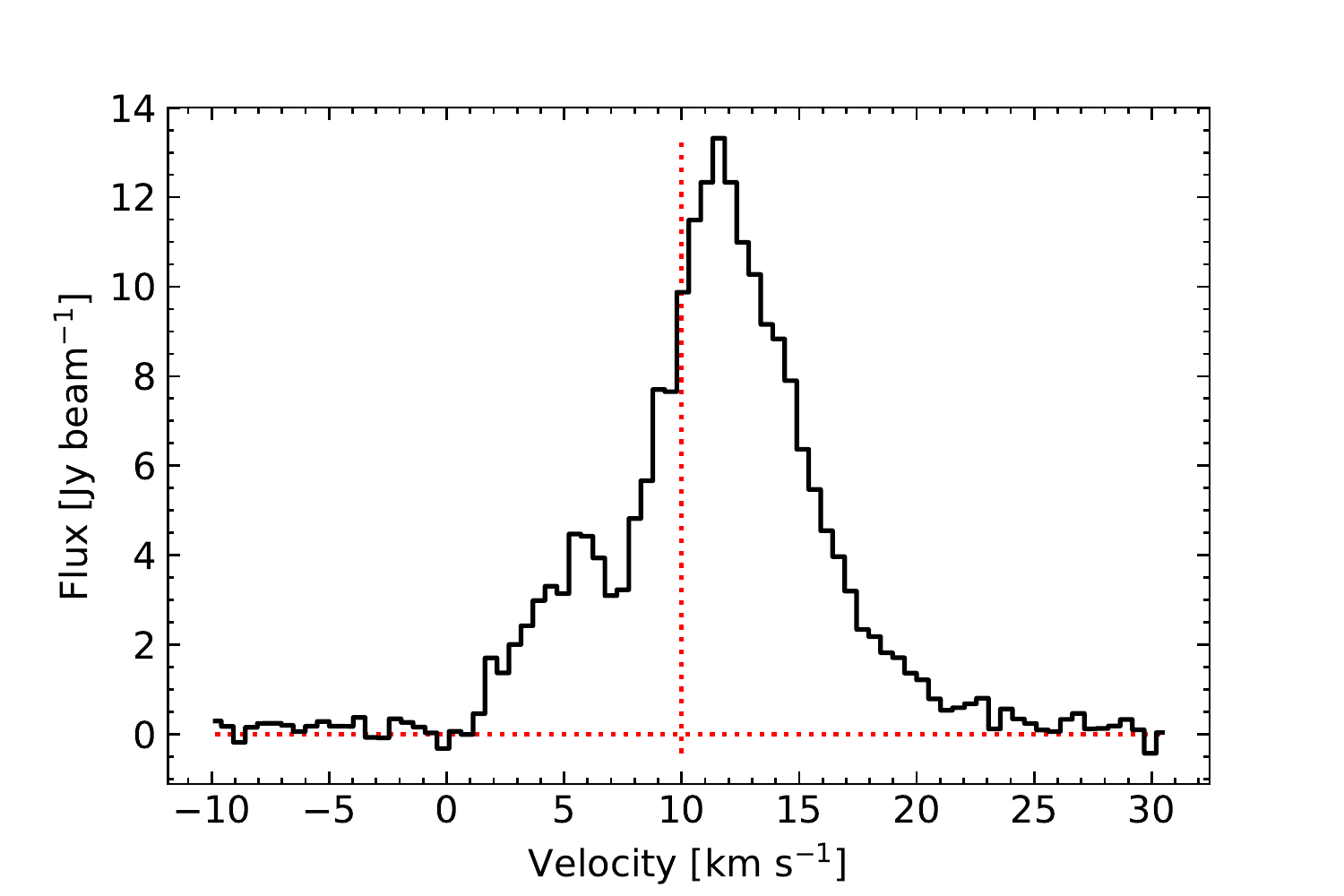}
      \caption{Redshifted peak of the 342.88 GHz CS $ J=7 - 6$ line, which peaks at an offset of $(0.015,0.72)$ arcseconds from the MM1a continuum peak (see Table \ref{table:Molecules_S_species})}
      \label{Fig:N30_CS_line_r}
   \end{figure}
   
%______________________________________________________________
%                                  One column figure
%----------------------------------------------------------- 

\begin{figure*}
    \centering
    \includegraphics[width=15.0cm]{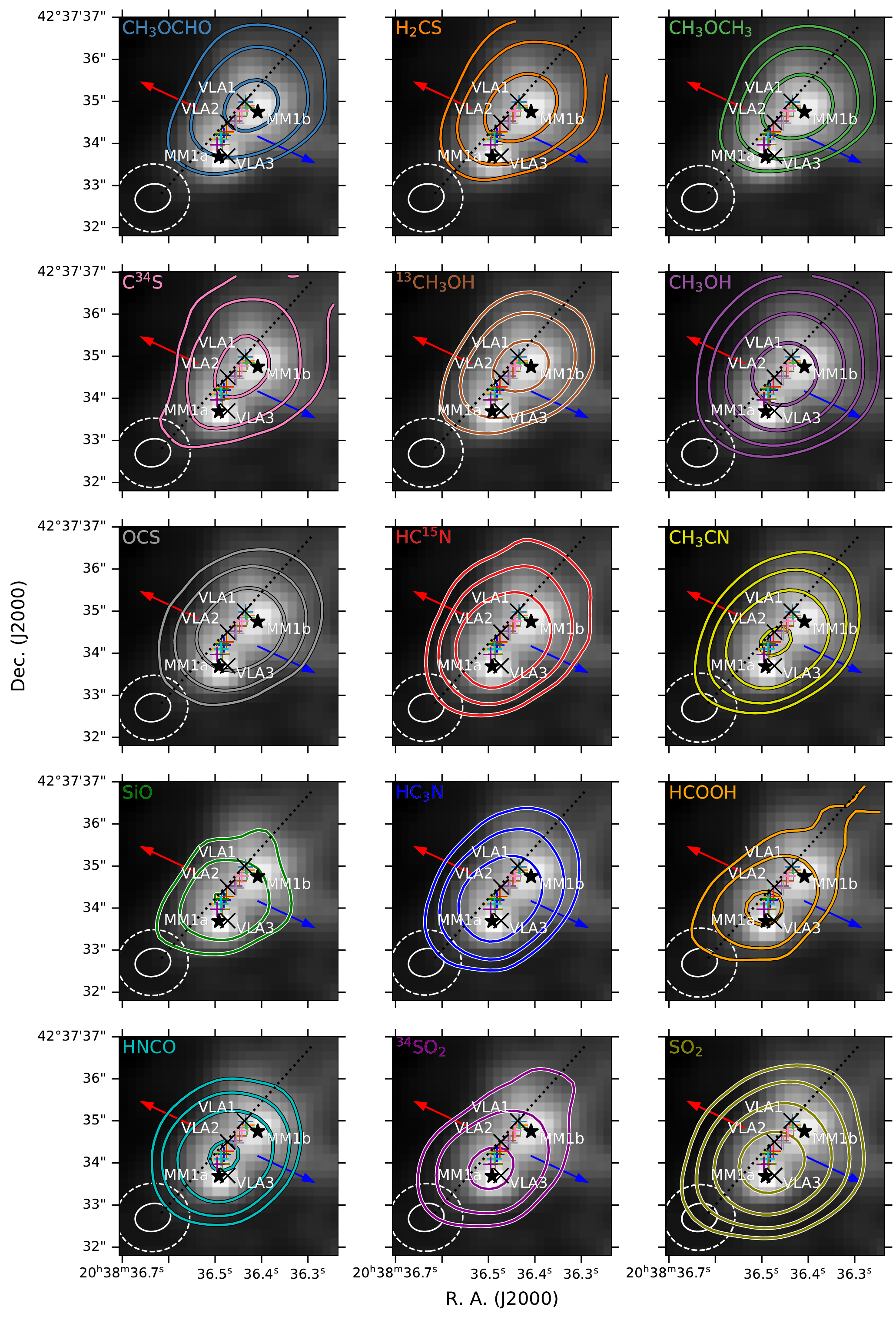}
    \caption{N30 molecules in contours over continuum in grey-scale. The peak positions of the 2D Gaussian fits to the continuum cores, MM1a and MM1b, are represented with black stars, while the VLA sources are shown with crosses. The molecular peak positions are marked with coloured plus symbols of the same colour as the respective molecule contour plot. The contour levels are 3$\sigma$, 6$\sigma$, 12$\sigma$, and 24$\sigma$. The beam sizes are represented by the inner and outer ellipses, with inner representing the continuum, with dimensions as in Fig. \ref{Fig:N30_continuum}, and the outer, the beam size of the molecular emission map, which have dimensions of $\sim$$1\farcs65 \times 1\farcs55$, and position angle of $\sim$$11^\circ$ for all molecules. The dotted line represents the red- and blueshifted H$_2$CO emission, while the arrows represent large-scale CO emission. The panels are sorted according to position of the peak emission, from top to bottom, and left to right.}
    \label{Fig:N30_mols}%
\end{figure*}

The panels in Fig. \ref{Fig:N30_mols} show the molecular peaks of each molecule; these peaks are concentrated along the axis of the red- and blueshifted components of H$_2$CO and CS emission. Each panel shows a contour map of the integrated molecular line emission of a different molecule. In the case of molecules with more than one unblended line, such as for example CH$_3$OH, the images show stacked emission for better S/N. The peak for each molecule was found using 2D Gaussian fits to the integrated (and again stacked where more than one unblended line was detected) molecular line emission, and is indicated with a plus symbol of the same colour as the contour for that molecule. To verify these fit results for the peak positions, an independent fit of the strongest unblended CH$_3$OH and SO$_2$ lines was performed in $uv$-space prior to cleaning the data. These fits confirm the results from the image plane, and typical differences are less than 0\farcs2, the adopted pixel size, and one-fifth of the beam. We therefore adopt a typical uncertainty on the peak locations as 0\farcs2 throughout. However, we note  that the formal uncertainties on the peak locations from the fits in the $uv$-plane and image plane, respectively, yield uncertainties of typically 0\farcs01--0\farcs02.

The full width at half maximum (FWHM) and source velocities ($\varv_{\text{source}}$) of the lines were determined by making 1D Gaussian fits to the respective lines. The detailed values obtained for each line are shown in Table \ref{table:Molecules_COMs} for the oxygen-, Table \ref{table:Molecules_S_species} for the sulphur-, and Table \ref{table:Molecules_N_SiO_HDO} for the nitrogen-bearing species, with this table also showing SiO, HDO, CO, and HCO$^+$.

%______________________________________________________________
%         Two column figure (place early!)
%___________________________________________________________________

%_____________________________________________________________
%                                             Two column Table 
%_____________________________________________________________

\begin{table*}
\caption{Gaussian fits for molecular transitions of COMs, including t-HCOOH and H$_2$CO.\tablefootmark{a}}
\label{table:Molecules_COMs}      
\centering          
\begin{tabular}{l c c c c c c r}     % 8 columns 
\hline
\hline 
    Molecule & Frequency & Transition & Position & $I_{\mathrm{peak}}$\tablefootmark{b} & $I_{\mathrm{int}}$\tablefootmark{b} & FWHM\tablefootmark{b}  & $\varv_{\text{source}}$\tablefootmark{b} \\ 
     &  [GHz] & &  [$'',''$] & [$\mathrm{Jy \ beam^{-1}}$] &  [$\mathrm{Jy \ beam^{-1} \ km \ s^{-1}}$] & [$\mathrm{km \ s^{-1}}$] & [$\mathrm{km \ s^{-1}}$]\\ 
\hline
    A-CH$_3$OH & 331.5023 & $11_{1, -0} - 11_{0, +0}$ &  $(     -0.31,      0.70)$ &       16.0 &      123.0 &        8.5 &        9.9 \\
     & 336.8651 & $12_{1, -0} - 12_{0, +0}$ &  $(     -0.33,      0.82)$ &       14.9 &      112.4 &        8.2 &       10.2 \\
     & 338.4087 & $7_{0, +0} - 6_{0, +0}$ & $(     -0.48,      0.85)$ &       14.7 &      124.6 &       10.4 &        9.9 \\
     & 341.4156 & $7_{1, -0} - 6_{1, -0}$ &  $(     -0.29,      0.86)$ &       12.6 &       87.7 &        8.0 &        8.7 \\
     & 342.7298 & $13_{1, -0} - 13_{0, +0}$ &  $(     -0.27,      0.75)$ &       14.9 &      109.5 &        8.3 &        9.1 \\
     & 349.1070 & $14_{1, -0} - 11_{0, +0}$ &  $(     -0.40,      0.76)$ &       15.7 &      120.4 &        8.3 &       11.0 \\
     & 355.6029 & $13_{0, +0} - 11_{1, +0}$ &  $(     -0.41,      1.03)$ &       13.9 &       99.5 &        7.5 &       10.4 \\
     & 356.0072 & $15_{1, -0} - 15_{0, +0}$ &  $(     -0.37,      0.82)$ &       17.1 &      134.4 &        8.5 &       10.8 \\
     & 360.6616 & $3_{1, +1} - 4_{2, +1}$ &  $(     -0.45,      0.98)$ &       10.6 &       76.1 &        7.4 &        9.5 \\
\hline 
    E-CH$_3$OH & 338.1244 & $7_{0, 0} - 6_{0, 0}$ & $(     -0.45,      0.87)$ &       14.8 &      102.8 &        7.4 &        9.8 \\
     & 338.3446 & $7_{-1, 0} - 6_{-1, 0}$ & $(     -0.50,      0.90)$ &       14.5 &      104.9 &        8.2 &        9.8 \\
     & 360.8489 & $11_{0, 0} - 10_{1, 0}$ & $(     -0.41,      0.93)$ &       13.6 &      100.9 &        7.8 &       10.0 \\
\hline    
    $^{13}$CH$_3$OH & 330.0018 & $7_{0, 7, 0} - 6_{0, 6, 0}$ & $(     -0.59,      0.99)$ &            3.5 &       18.8 &        5.8 &        7.9 \\
     & 330.2528 & $7_{0, 7, +0} - 6_{0, 6, +0}$ & $(     -0.55,      1.07)$ &        3.8 &          24.6 &        8.0 &        8.3 \\
     & 330.4424 & $7_{1, 6, 0} - 6_{1, 5, 0}$ & $(     -0.51,      0.93)$ &        3.5 &            16.7 &        5.5 &        8.0 \\
     & 330.4639 & $11_{1, 10, -0} - 11_{0, 11, +0}$ & $(     -0.52,      0.94)$ &        6.1 &       32.1 &        6.1 &        7.7 \\
     & 330.5352 & $7_{2, 5, 0} - 6_{2, 4, 0}$ &  $(     -0.53,      1.02)$ &        4.9 &           26.6 &        6.1 &        7.6 \\
     & 333.1148 & $7_{1, 6, -0} - 6_{1, 5, -0}$ & $(     -0.61,      0.98)$ &        3.5 &           22.6 &        6.6 &        8.8 \\
     & 335.5602 & $12_{1, 11, -0} - 12_{0, 12, +0}$ & $(     -0.57,      1.05)$ &        4.0 &       22.8 &        5.8 &        9.2 \\
\hline    
    C$_2$H$_5$OH & 338.8862 & $15_{7, 8, 2} - 15_{6, 9, 2}$ & $(     -0.94,      1.45)$ &            1.9 &       11.1 &        8.4 &        7.6 \\
    & 339.3126 & $12_{7, 5, 2} - 12_{6, 6, 2}$ & $(     -1.09,      1.67)$ &        1.1 &            7.8 &        7.6 &        9.1 \\
    & 339.3984 & $11_{7, 4, 2} - 11_{6, 5, 2}$ & $(     -1.12,      1.63)$ &        1.0 &            6.4 &        7.8 &        8.4 \\
\hline    
    CH$_3$OCH$_3$ & 339.4916 & $19_{1, 18, 5} - 18_{2, 17, 5}$ & $(     -0.54,      1.04)$ &        3.5 &       18.6 &        5.2 &        9.4 \\
     & 340.6126 & $10_{3, 7, 1} - 9_{2, 8, 1}$ & $(     -0.89,      1.21)$ &        2.9 &           16.2 &        6.6 &        7.7 \\
     & 342.6080 & $19_{0, 19, 0} - 18_{1, 18, 0}$ & $(     -0.57,      1.22)$ &        4.7 &       27.7 &        5.7 &        9.4 \\
     & 344.3578 & $19_{1, 19, 3} - 18_{0, 18, 3}$ & $(     -0.64,      1.15)$ &        4.4 &       27.3 &        6.6 &        8.8 \\
     & 357.4602 & $18_{2, 17, 1} - 17_{1, 16, 1}$ & $(     -0.69,      1.08)$ &        4.8 &       28.3 &        7.0 &        8.0 \\
     & 358.4541 & $5_{5, 1, 1} - 4_{4, 1, 1}$ & $(     -0.76,      1.17)$ &        7.3 &       63.0      &       10.6 &       10.2 \\
     & 360.4660 & $20_{1, 19, 0} - 19_{2, 18, 0}$ & $(     -0.79,      1.22)$ &        4.0 &         25.9 &        6.1 &       10.9 \\
     & 360.5847 & $20_{0, 20, 0} - 19_{1, 19, 0}$ & $(     -0.80,      1.23)$ &        5.3 &       35.8 &        6.9 &       10.1 \\
\hline    
    CH$_3$OCHO & 333.4494 & $31_{1, 31, 0} - 30_{0, 30, 0}$ & $(            -0.69,1.14)$ & 4.0 & 23.5 & 6.0 & 8.9 \\
      & 336.3514 & $27_{6, 22, 1} - 26_{6, 21, 1}$ & $(     -0.82,      1.24)$ &   2.2 & 15.6 &  7.0 &  10.3 \\
      & 344.0296 & $19_{1, 18, 5} - 18_{2, 17, 5}$ & $(     -0.52,      1.08)$ &  4.4 & 29.4 &   7.1 & 10.1 \\
      & 354.6081 & $19_{1, 18, 5} - 18_{2, 17, 5}$ & $(     -0.65,      1.23)$ &   4.0 & 24.4 & 6.7 &  9.5 \\
\hline 
    t-HCOOH & 334.2658 & $15_{2, 14} - 14_{2, 13}$ & $(     -0.03,      0.46)$ &        2.5 &        15.8 &        6.9 &       10.2 \\
    & 343.9523 & $15_{1, 14} - 14_{1, 13}$ & $(      0.03,      0.37)$ &        2.4 &       17.1 &    7.8 &       11.2 \\
\hline
    H$_2$CO\tablefootmark{c} & 351.7686 & $5_{1, 5} - 4_{1, 4}$ & $(     -0.03,      0.52)$ &       13.1 &           88.0 &        6.3 &       13.9 \\
     &  & $5_{1, 5} - 4_{1, 4}$ & $(     -0.83,      1.72)$ &       11.0 &       47.1 &        3.7 &        6.5 \\

\hline

\end{tabular}
\tablefoot{\tablefoottext{a}{The transitions listed here, as well as the transitions listed in Tables \ref{table:Molecules_S_species} and \ref{table:Molecules_N_SiO_HDO}, were relatively isolated and unblended, and were used to make the emission maps shown in Fig. \ref{Fig:N30_mols}.}
\tablefoottext{b}{Typical uncertainties on the Gaussian centroid fits are $(0\farcs2,0\farcs2)$ for the position, 0.3 $\mathrm{Jy \ beam^{-1}}$ for $I_{\mathrm{peak}}$,  $\mathrm{Jy \ beam^{-1} \ km \ s^{-1}}$ for $I_{\mathrm{int}}$, and 0.4 $\mathrm{km \ s^{-1}}$ for the FWHM and $\varv_{\text{source}}$.} 
 \tablefoottext{c}{The two entries for H$_2$CO are red- and blueshifted lines, respectively. See also entries for CS.} }
\end{table*}

\begin{table*}
\caption{Gaussian fits for molecular transitions of S-bearing species.} 
\label{table:Molecules_S_species}      
\centering          
\begin{tabular}{l c c c c c c r}     % 8 columns 
\hline
\hline 
    Molecule & Frequency & Transition & Position & $I_{\mathrm{peak}}$ & $I_{\mathrm{int}}$  & FWHM  & $\varv_{\text{source}}$ \\ 
     &  [GHz] & &  [$'',''$] & [$\mathrm{Jy \ beam^{-1}}$] &  [$\mathrm{Jy \ beam^{-1} \ km \ s^{-1}}$] & [$\mathrm{km \ s^{-1}}$] & [$\mathrm{km \ s^{-1}}$]\\ 
\hline
\hline 
    CS & 342.8828 & $7_{0} - 6_{0}$ & $(      0.17,      0.72)$ & 11.1 &       69.7 &           8.5 &  9.7 \\
     & 342.8828 & $7_{0} - 6_{0}$ & $(     -1.03,      1.92)$ & 9.2 &       47.3 &   4.2 &  7.9 \\
\hline
    C$^{34}$S & 337.3965 & $7_{0} - 6_{0}$ &  $(     -0.54,      1.08)$ &        8.3 &       49.1  &        6.0 &        9.0 \\
\hline
    SO & 340.7141 & $8_{7} - 6_{ 6} $ & $(      0.13,      0.24)$ &       16.3 &       84.9 &        9.5 &        5.4 \\
     & 334.4311 & $8_{8} - 7_{7} $ & $(      0.14,      0.25)$ &       17.7 &       98.5 &          10.6 &        5.2 \\
     & 346.5281 & $8_{9} - 7_{8} $ & $(      0.06,      0.25)$ &       19.4 &      140.0 &          11.1 &        6.7 \\
\hline
    $^{34}$SO & 333.9010 & $8_{7} - 7_{ 6} $ & $(     -0.07,      0.22)$ &        8.1 &       57.4     &        7.5 &        9.9 \\
     & 337.5801 & $8_{8} - 7_{7} $ & $(     -0.22,      0.29)$ &       10.7 &       80.9 &           9.0 &        8.7 \\
     & 339.8573 & $8_{9} - 7_{8} $ & $(     -0.16,      0.18)$ &        9.0 &       64.1 &           7.5 &       10.5 \\
\hline
    SO$_2$ & 332.0914 & $21_{2, 20} - 21_{ 1, 21} $ & $(     -0.14,      0.17)$ &       11.4 &       86.6 &        8.0 &       10.8 \\
     &  332.5052 & $4_{3, 1} -  3_{ 2, 2} $ & $(     -0.06,      0.30)$ &       12.9 &       95.7    &        8.3 &        9.9 \\
     &  334.6733 & $8_{2, 6 } -  7_{ 1, 7}  $ & $(     -0.09,      0.23)$ &       12.0 &             85.6     &        7.8 &        9.5 \\
     & 336.0892 & $23_{ 3, 21 } -  23_{ 2, 22}$ & $(     -0.15,      0.29)$ &       10.1 &          76.5 &        8.2 &       10.9 \\
     & 336.6695 & $16_{ 7, 9 } -  17_{ 6, 12  }$ & $(      0.03,      0.14)$ &        4.8 &           35.2 &        7.6 &       10.8 \\
     & 338.3060 & $18_{ 4, 14 } - 18_{ 3, 15}$ & $(     -0.06,      0.23)$ &       14.3 &            107.0 &        8.3 &       10.0 \\
     & 340.3164& $28_{ 2, 26 } - 28_{ 1, 27}$ & $(     -0.05,      0.27)$ &        8.0 &            58.5 &        7.9 &       10.6 \\
     & 345.3385 & $13_{ 2, 12} -  12_{ 1, 11} $ & $(     -0.15,      0.33)$ &       16.7 &          137.2 &       10.6 &       10.4 \\
     &  346.6522 & $19_{ 1, 19 } -  18_{ 0, 18}  $ & $(     -0.09,      0.27)$ &       17.4 &       133.7 &        9.1 &        9.4 \\
     & 350.8628 & $10_{ 6, 4 } -  11_{ 5, 7  } $ & $(      0.01,      0.07)$ &        5.6 &          41.7 &        8.1 &        9.8 \\
     & 351.8739 & $14_{ 4, 10 } -  14_{ 3, 11}  $ & $(     -0.11,      0.19)$ &       15.2 &          116.2 &        8.7 &        9.9 \\
     & 355.0455 & $12_{ 4, 8 } - 12_{ 3, 9  }$ &  $(     -0.09,      0.27)$ &       16.2 &         122.4 &        8.5 &        9.7 \\
     & 357.1654 & $13_{ 4, 10 } -  13_{ 3, 11} $ & $(     -0.21,      0.37)$ &       16.1 &         124.4 &        9.2 &       11.6 \\
     & 357.2412 & $15_{ 4, 12 } -  15_{ 3, 13}  $ & $(     -0.19,      0.29)$ &       17.0 &         129.7 &        8.7 &       10.3 \\
     & 357.3876 & $11_{ 4, 8 } -  11_{ 3, 9}$  & $(     -0.15,      0.27)$ &       16.4 &            129.2 &        9.1 &       10.2 \\
     & 357.5814 & $8_{ 4 ,4 } -  8_{ 3, 5 }$ & $(     -0.14,      0.22)$ &       17.0 &      131.8    &        9.0 &       10.2 \\
     & 357.8924 & $7_{ 4, 4 } - 7_{ 3, 5  }$ & $(     -0.11,      0.27)$ &       15.7 &      123.6    &        9.2 &       10.2 \\
     &  357.9258 & $6_{ 4, 2 } -  6_{ 3, 3 }$ & $(     -0.15,      0.29)$ &       15.6 &            121.3 &        9.1 &       10.5 \\
     & 357.9629 & $17_{ 4, 14 } -  17_{ 3, 15} $ & $(     -0.09,      0.26)$ &       15.7 &          126.6 &        9.4 &       10.0 \\
     &  358.0131 & $5_{ 4, 2 } -  5_{ 3, 3}$ & $(     -0.27,      0.37)$ &       13.6 &      105.0    &        8.9 &       10.1 \\
     & 358.2156 & $20_{ 0, 20 } -  19_{ 1, 19} $ & $(     -0.17,      0.25)$ &       19.0 &         150.8 &        9.5 &       10.4 \\
     & 359.7707 & $19_{ 4, 16 } -  19_{ 3, 17}$ & $(     -0.16,      0.29)$ &       15.5 &         113.7 &        8.9 &       12.0 \\

\hline
    $^{34}$SO$_2$ & 342.2089  & $5_{3, 3} - 4_{2, 2}$ & $(     -0.03,      0.37)$ &        2.4 &       15.4 &        7.2 &        9.7 \\
     & 342.3320 & $12_{4, 8} - 12_{3, 9}$ & $(      0.08,      0.29)$ &        2.8 &       19.7 &        7.2 &       10.5 \\
     & 344.2453 & $10_{4, 6} - 10_{3, 7}$ & $(     -0.02,      0.25)$ &        2.3 &       17.7 &        9.1 &       10.5 \\
     & 344.5810 & $19_{1, 19} - 18_{0, 18}$ &  $(     -0.06,      0.20)$ &        4.5 &       35.3   &        8.3 &       11.0 \\
     & 344.8079 & $13_{4, 10} - 13_{3, 11}$ &  $(     -0.05,      0.21)$ &        2.7 &       20.0     &        8.3 &       11.0 \\

\hline
    OCS & 340.4493 & $ 28 - 27 $ & $(     -0.45,      0.79)$ &        8.2 &       55.4 &            6.8 &        9.8 \\
     & 352.5996 & $ 29 - 28 $ & $(     -0.48,      0.70)$ &        9.4 &       70.8 &        7.7 &       10.2 \\
\hline
    OC$^{34}$S & 332.1297 & $ 28 - 27 $ & $(     -1.04,      1.44)$ &        1.1 &        6.0 &        7.8 &       10.9 \\
     & 343.9833 & $ 29 - 28 $ & $(     -1.04,      1.49)$ &        2.0 &       11.7 &        5.4 &        9.6 \\
\hline     
    H$_2$CS & 338.0832 & $ 10_{ 1, 10} - 9_{ 1, 9} $  & $(     -0.67,      1.16)$ &        7.7 &       47.4 &        6.8 &        8.4 \\
     & 342.9464 & $ 10_{ 0, 10} -  9_{ 0, 9} $ & $(     -0.72,      1.22)$ &        4.4 &          26.6 &        6.6 &        8.3 \\
     & 343.4141 & $ 10_{ 3, 7} - 9_{ 3, 6} $ & $(     -0.64,      1.00)$ &        4.3 &       23.0    &        5.1 &        9.0 \\
     & 343.8132 & $ 10_{ 2, 8} - 9_{ 2, 7} $ &  $(     -0.73,      1.23)$ &        7.1 &             46.1 &        6.8 &        9.7 \\
\hline    
\end{tabular}
\tablefoot{Uncertainties are as in Table \ref{table:Molecules_COMs}.}
\end{table*}

\begin{table*}
\caption{Gaussian fits for molecular transitions of N-bearing species, CO, HCO$^+$, SiO and HDO.}             
\label{table:Molecules_N_SiO_HDO}      
\centering          
\begin{tabular}{l c c c c c c r}     % 8 columns 
\hline
\hline 
    Molecule & Frequency & Transition & Position & $I_{\mathrm{peak}}$ & $I_{\mathrm{int}}$  & FWHM  & $\varv_{\text{source}}$ \\ 
     &  [GHz] & &  [$'',''$] & [$\mathrm{Jy \ beam^{-1}}$] &  [$\mathrm{Jy \ beam^{-1} \ km \ s^{-1}}$] & [$\mathrm{km \ s^{-1}}$] & [$\mathrm{km \ s^{-1}}$]\\ 
\hline
   \multicolumn{8}{c}{N-bearing species} \\
\hline 
    CH$_3$CN & 330.7603 & $18_{7} - 17_{7}$    &  $(     -0.22,      0.50)$ &        6.1 &            46.2 &        7.4 &       10.0 \\
     & 330.9126 &  $18_{5} - 17_{5}$   &  $(     -0.18,      0.47)$ &        7.8 &       56.2 &        7.1 &        9.5 \\
     & 330.9698 &  $18_{4} - 17_{4}$   &  $(     -0.22,      0.51)$ &       11.4 &       82.3 &        7.5 &        9.2 \\
     & 331.0143 & $18_{3} - 17_{3}$    &  $(     -0.24,      0.55)$ &       10.0 &       72.6 &        7.2 &        9.2 \\
     & 331.0461 &  $18_{2} - 17_{2}$   &  $(     -0.35,      0.57)$ &        7.5 &       55.3 &        7.4 &        9.5 \\
     & 349.3933 & $19_{3} - 18_{3}$    &  $(     -0.20,      0.56)$ &       11.7 &       88.8 &        8.1 &        9.7 \\
     & 349.4268 &  $19_{2} - 18_{2}$   &  $(     -0.20,      0.60)$ &        9.8 &       73.0 &        7.7 &       10.1 \\
\hline
    HNCO  & 329.6644 & $15_{0,15} - 14_{0,14} $ &  $(     -0.14,      0.42)$ &        8.8 &           66.9 &        8.3 &        9.2 \\
     &  350.3330 & $16_{1,16} - 15_{1,15} $ &  $(     -0.05,      0.35)$ &        7.6 &       61.4    &        8.4 &       10.2 \\
\hline
    HCN\tablefootmark{a} & 354.5055 & 4 - 3 & $(      0.17,      0.52)$ &       17.8 &      127.2 &        6.3 &       14.2 \\

\hline
    HC$^{15}$N & 344.2001 & $4 - 3$ & $(     -0.19,      0.59)$ &       10.2 &       78.7 &        8.6 &        9.9 \\
\hline
    HC$_3$N  & 336.5201 & $37 - 36 $ &  $(     -0.14,      0.53)$ &        5.6 &       44.2 &        7.9 &       10.7 \\
     & 345.6090  & $38 - 37 $ &  $(     -0.15,      0.47)$ &        6.5 &       51.8 &        8.6    &       10.1 \\
     & 354.6975 & $39 - 38 $ &  $(     -0.18,      0.39)$ &        6.7 &       53.8 &        8.7 &       10.6 \\
\hline 

    CN & 340.2478 & $3_{0,3.5}$ - 2$_{0,2.5}$ & $(     -0.49,      1.41)$ &        1.4 &           11.3 &        9.8 &        9.4 \\
\hline    

    \multicolumn{8}{c}{SiO and HDO} \\
\hline 
    SiO & 347.3306 & $8_{0} - 7_{0}$ &  $(     -0.07,      0.47)$ &        2.8 &       33.6 &       11.7 &       11.2 \\

\hline 
    HDO\tablefootmark{b} & 335.39550 & $3_{3, 1} - 4_{2, 2}$ &  $(     -0.43,      0.92)$ &        1.6 &       11.5     &        2.1 &       10.8 \\
\hline  
   \multicolumn{8}{c}{CO and HCO$^+$ } \\
\hline 
    CO\tablefootmark{a} & 345.7960 & $3 - 2$ & $(      0.35,      0.68)$ &       18.2 &      156.0 &       10.1 &       11.6 \\

\hline  
    $^{13}$CO & 330.5880 & $3 - 2$ & $(      0.57,     -0.12)$ &       10.7 &       51.9 &        4.5 &       10.6 \\
\hline 
    C$^{18}$O & 329.3305 & $3 - 2$ & $(     -0.64,      0.37)$ &        3.5 &       25.2 &        8.2 &        9.1 \\
\hline 
    H$^{13}$CO$^+$ & 346.9983 & 4 - 3 &  $( -1.45, 0.50)$ &  3.74 &      -- &        4.31 &        9.20 \\
\hline 

\hline  
\end{tabular}
\tablefoot{Uncertainties are as in Table \ref{table:Molecules_COMs}.
\tablefoottext{a}{For the HCN and CO lines, only the brightest, redshifted components are listed.} 
\tablefoottext{b}{HDO emission is extended, so the peak position is of the maximum flux position, not derived from a 2D Gaussian fit.}}

\end{table*}

\subsection{Molecular gradient}
%______________________________________________________________
%                                  One column figure
%----------------------------------------------------------- 
\begin{figure}
\centering
\includegraphics[width=9cm]{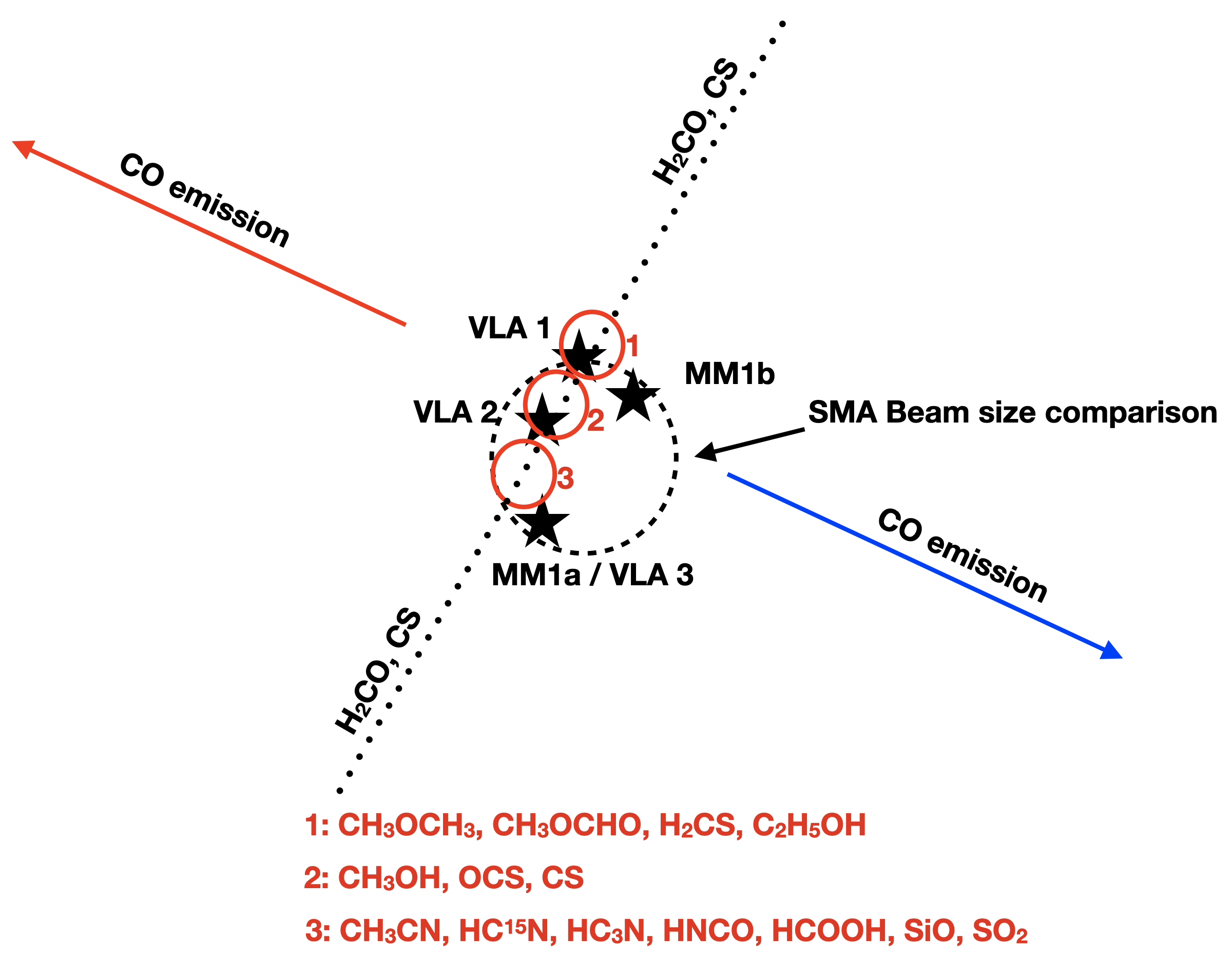}
  \caption{Representation of the observed molecular gradient towards the MM1 source. The molecules are separated into three groups, with the O-bearing COMs in the first and second groups, with group 1 including H$_2$CS, and group 2 including CS and OCS. Group 3 includes all the sulphur- and nitrogen-bearing species, as well as SiO and HCOOH.}
  \label{Fig:N30_Mol_Grad}
\end{figure}

\begin{figure}
   \centering
   \includegraphics[width=9cm]{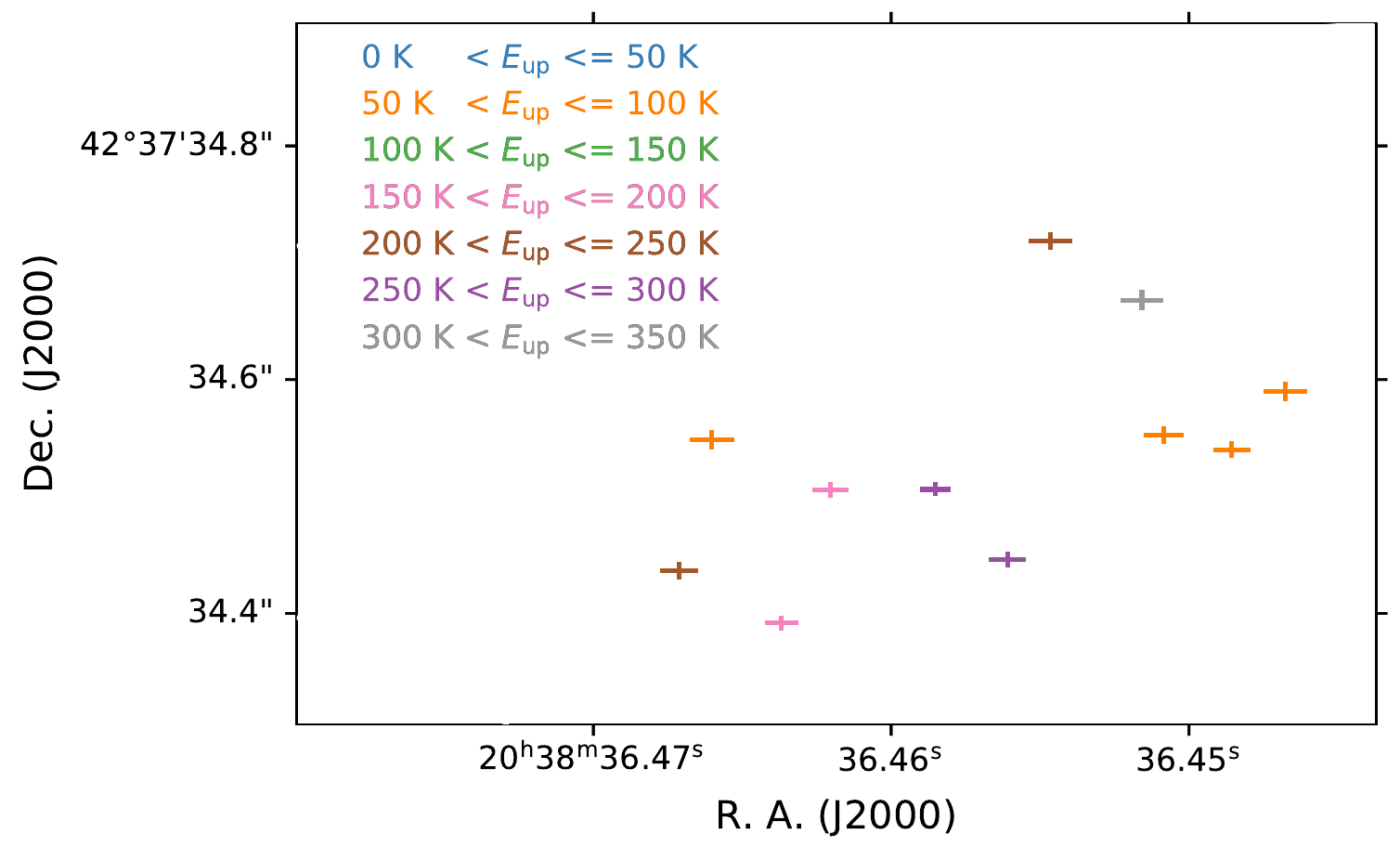}
      \caption{Peak positions of CH$_3$OH line transitions, with the upper-level energy of each transition represented in colour. The data points do not follow a systematic pattern.}
      \label{Fig:N30_CH3OH_Tex}
   \end{figure}

\begin{figure}
   \centering
   \includegraphics[width=9cm]{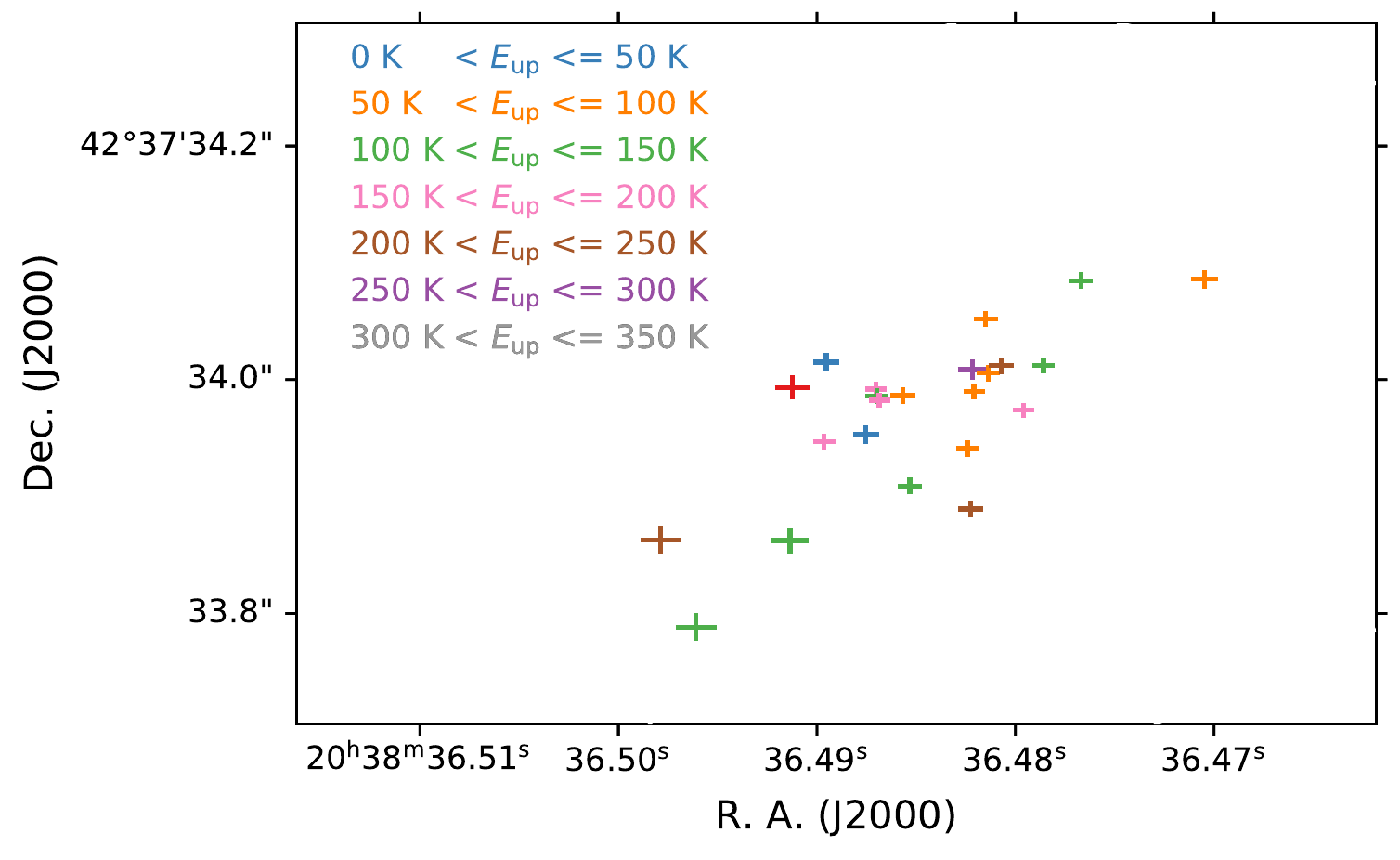}
      \caption{Peak positions of SO$_2$ line transitions, with the upper-level energy of each transition represented in colour. As with the CH$_3$OH line transitions, there does not seem to be any pattern with upper-level energy.}
      \label{Fig:N30_SO2_Tex}
   \end{figure}

\begin{figure}
   \centering
   \includegraphics[width=9cm]{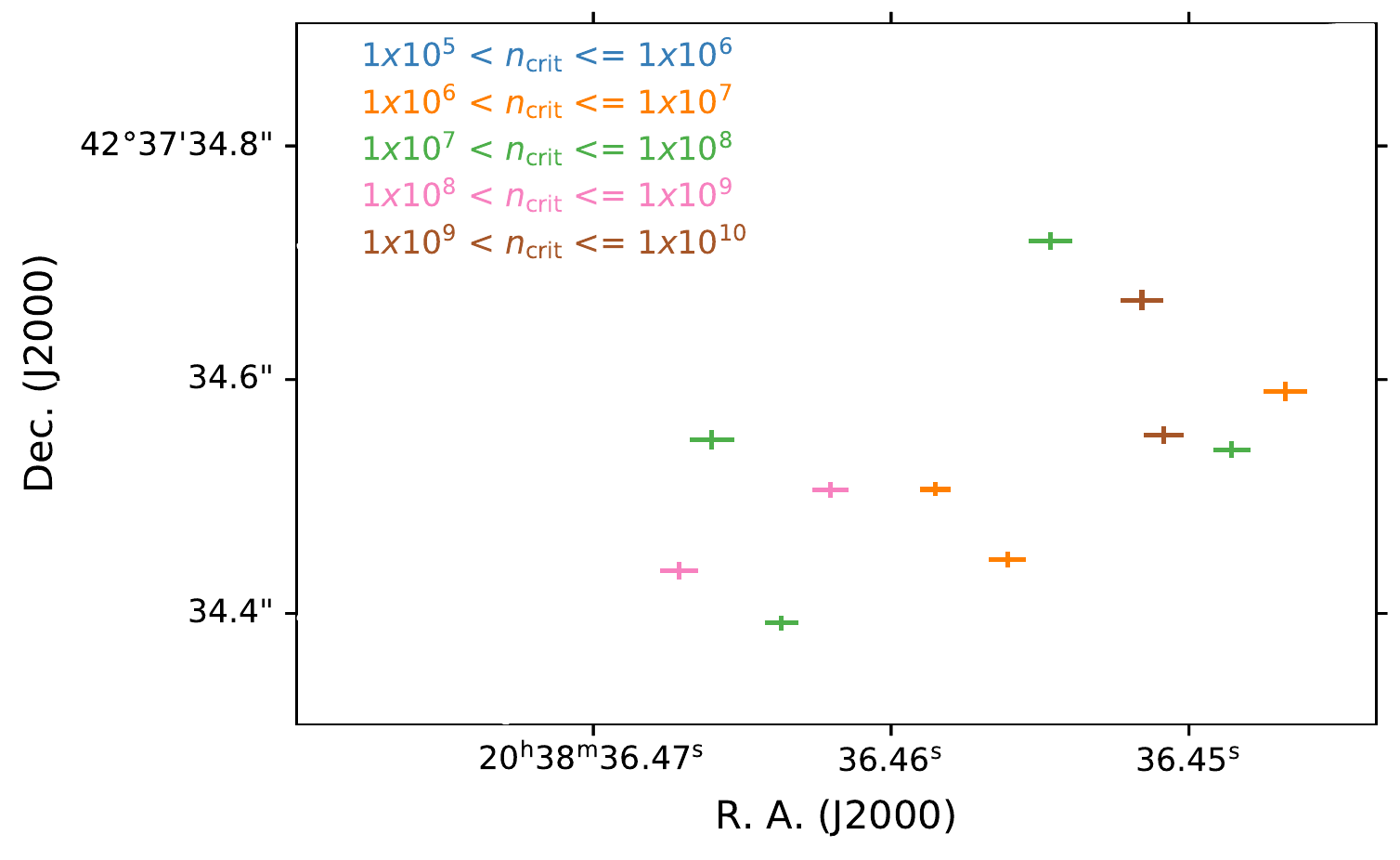}
      \caption{Peak positions of the CH$_3$OH line transitions, with the critical density now represented in colour. As with the upper-level energy, there does not seem to be a pattern depending on critical density.}
      \label{Fig:N30_CH3OH_ncrit}
   \end{figure}

\begin{figure}
   \centering
   \includegraphics[width=9cm]{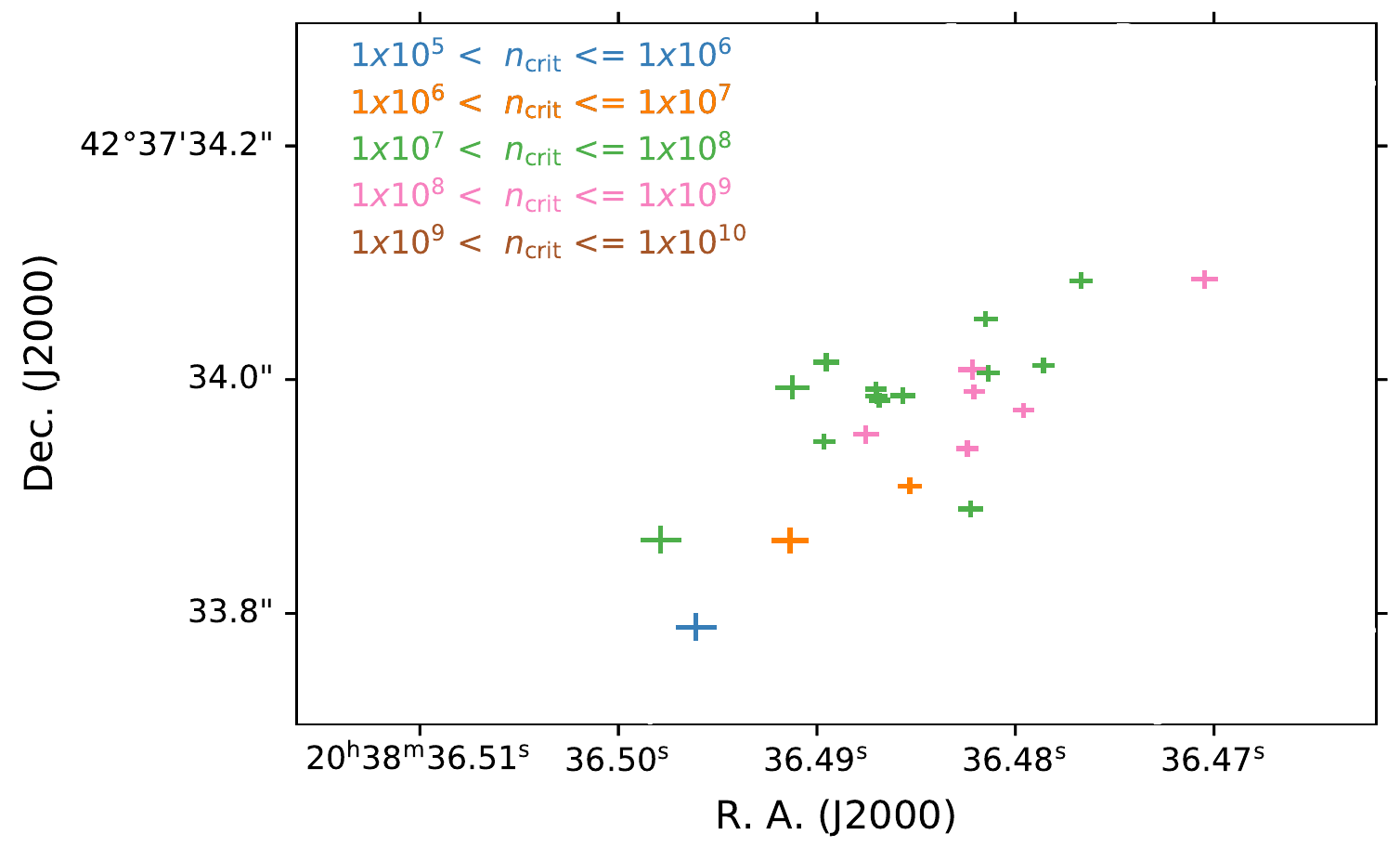}
      \caption{Peak positions of SO$_2$ line transitions, with the critical density represented in colour. Again, no systematic pattern is seen.}
      \label{Fig:N30_SO2_ncrit}
   \end{figure}

We find that all molecules that do not show significant extended emission peak along a linear gradient coinciding with the axis marked by the red- and blueshifted components of the H$_2$CO and CS emission. This axis runs parallel to but is offset from the axis drawn between the continuum peaks of MM1a and MM1b, as shown in Fig. \ref{Fig:N30_Mol_Grad}. To investigate whether or not this observed molecular gradient is the result of a difference in upper-level energy of the molecular transitions, that is, a temperature gradient, the peak positions of different transitions of CH$_3$OH and SO$_2$ are represented in Figs. \ref{Fig:N30_CH3OH_Tex} and \ref{Fig:N30_SO2_Tex}, where the upper-level energies are represented in different colours. As can be seen from the plot, there does not seem to be any pattern, and the positions of the respective lines seem to be distributed randomly. More specifically, the peaks are distributed within less than one-third of the beam, that is, the spread is over $\sim$0$\farcs$4, whereas the beam size is $\sim$1$\farcs$5. A similar exercise was performed for the critical densities ($n_\mathrm{crit}$), where $n_\mathrm{crit} = A_{\mathrm{ij}}/K_{\mathrm{ij}}$, with $A_{\mathrm{ij}}$ the Einstein A coefficient for spontaneous emission, and $K_{\mathrm{ij}}$ the rate coefficient of the transition from the upper level $i$ to the lower level $j$, for temperatures of 100 K \citep[rate coefficients were obtained from the Leiden Atomic and Molecular Database;][]{Schoier-2005-LAMDA}\footnote{https://home.strw.leidenuniv.nl/$\sim$moldata/}. Figures \ref{Fig:N30_CH3OH_ncrit} and \ref{Fig:N30_SO2_ncrit} again show the peak positions of the different transitions of CH$_3$OH and SO$_2$, with the critical densities of different transitions represented in colours. As with the upper-level energies, there does not seem to be any density gradient, and the peaks are localised within a fraction of a beam (0$\farcs$4 -- 0$\farcs$6 vs. the 1\farcs 5 beam), which leads us to conclude that the observed gradient is not an excitation gradient.

%______________________________________________________________
%                                  One column figure
%-------------------------------------------------------------- 
   
Figure \ref{Fig:N30_Mol_Grad} illustrates the observed molecular gradient, where the molecular peak emission is represented in three groups. The O-bearing COMs peak closer to MM1b, between VLA1 and VLA2 (positions 1 and 2), with CH$_3$OH peaking close to VLA2, and the more complex species (C$_2$H$_5$OH, CH$_3$OCHO, and CH$_3$OCH$_3$) peaking closer to VLA1. The exceptions here are the S-bearing species, CS, OCS, and H$_2$CS, which peak between VLA1 and VLA2, while the other S-bearing species peak between VLA2 and VLA3 (position 3). This is also where the N-bearing species peak. Moreover, HCOOH peaks close to the VLA3/MM1a positions, but shows more extended emission in the northwest direction, along the blueshifted outflow axis (see Fig. \ref{Fig:N30_mols}).

\subsection{Other molecules}

\begin{figure}
\centering
\includegraphics[width=9cm]{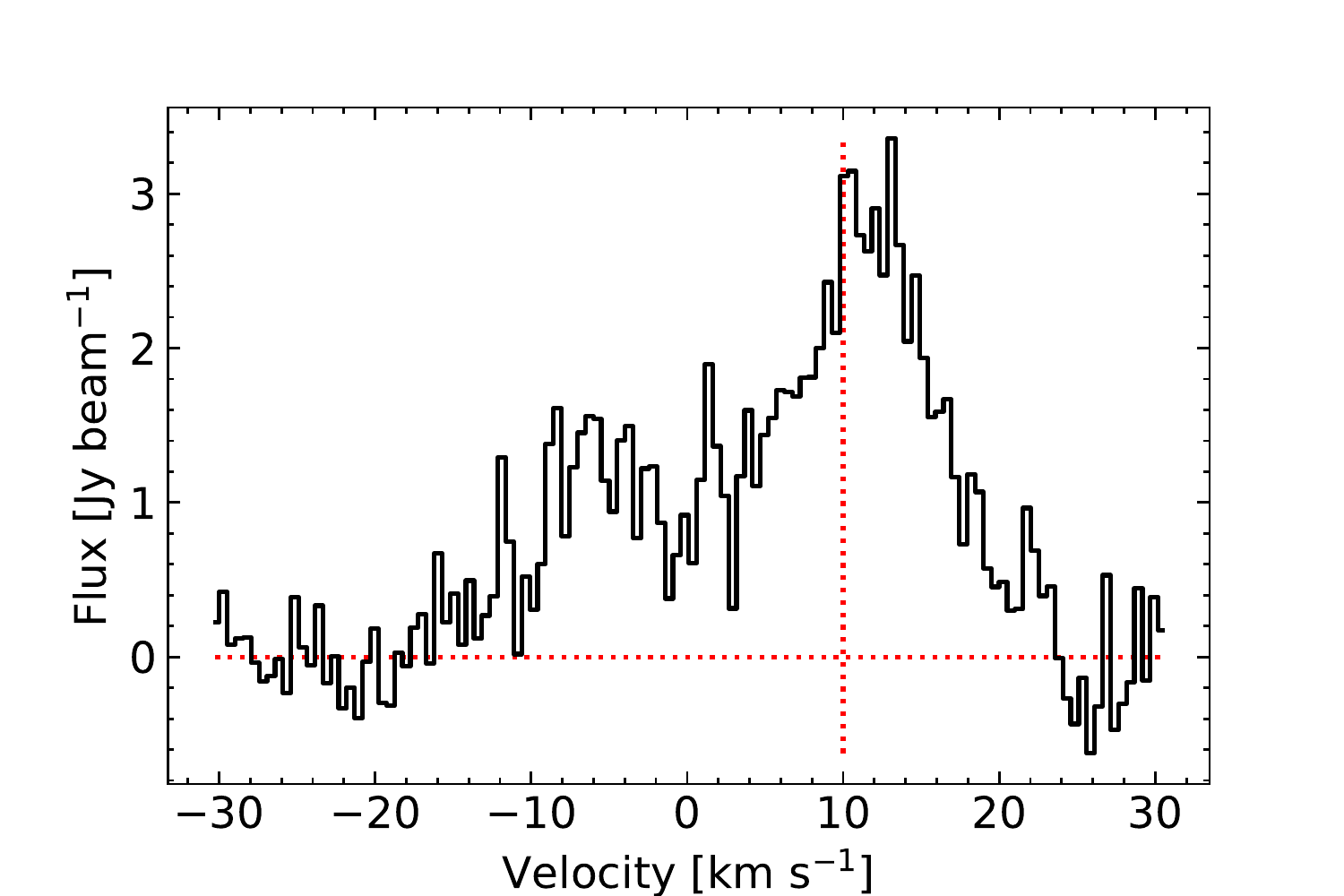}
  \caption{The 347.33 GHz SiO, $J=8-7$ line, showing broad emission, with a weak blueshifted component at $-6 \mathrm{km \ s^{-1}}$. The horizontal red line represents the baseline, while the vertical line is the source velocity derived for MM1a from the line modelling in CASSIS, at 9.5 km s$^{-1}$.}
  \label{Fig:N30_SiO_line}
\end{figure}

The 347.33 GHz SiO, $J=8-7$, transition has a weak blueshifted component at $-6 \mathrm{\ km \ s^{-1}}$ (or perhaps a neighbouring unidentified line, see Fig. \ref{Fig:N30_SiO_line}), which was included in the integrated line map, with the velocity range of the integrated map from $-10 \mathrm{\ km \ s^{-1}}$ to $20\mathrm{\ km \ s^{-1}}$. The source velocity and FWHM listed in Table \ref{table:Molecules_N_SiO_HDO} are for the component centred around $ 10 \mathrm{\ km \ s^{-1}}$. The molecular peak for SiO is at position 3 in Fig. \ref{Fig:N30_Mol_Grad}, together with the S- and N-bearing species.

%______________________________________________________________
%                                  One column figure
%----------------------------------------------------------- 

\begin{figure}
\centering
\includegraphics[width=9cm]{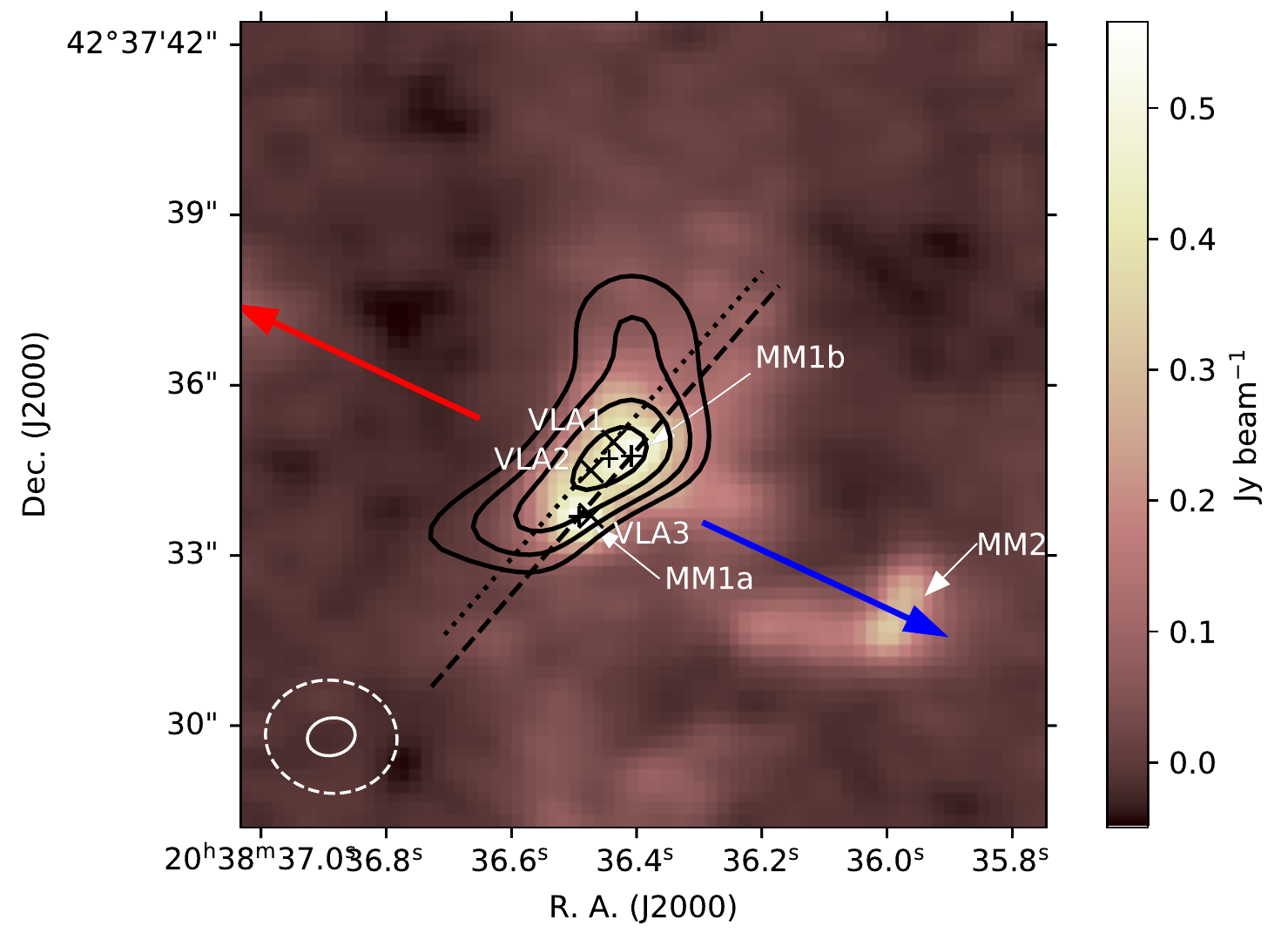}
  \caption{Continuum image of CygX-N30, with contours showing the 335.395 GHz HDO line emission (transition $J=3_{3,1} - 4_{2,2}$). The contour levels are at 3$\sigma$, 4$\sigma$, 5$\sigma$ and 6$\sigma$.}
  \label{Fig:N30_HDO_contours}
\end{figure}

Only one deuterated molecule is detected, the 335.395 GHz HDO, $J=3_{3,1} - 4_{2,2}$ transition. This is shown in contours over the continuum in Fig. \ref{Fig:N30_HDO_contours}. HDO shows extended emission, also following the red- and blueshifted H$_2$CO and CS axes, but with some elongated structure to the north of VLA1 and MM1b. The line peaks at 2.39 Jy beam$^{-1}$, or about 6$\sigma$, and with the peak position located between the VLA1 and VLA2 peaks. 

%______________________________________________________________
%                                  One column figure
%----------------------------------------------------------- 

\subsection{Column densities}
As we observe such a strong chemical gradient in the positions of the molecular peak emission, we derived column densities for the respective molecules at the three positions shown in Fig. \ref{Fig:N30_Mol_Grad}. A synthetic spectrum was constructed with CASSIS using only line emission, assuming local thermodynamic equilibrium (LTE). The inbuilt regular grid function was used to cover a large parameter space and compute the reduced $\chi^2$ minimum in order to determine the best-fit spectral model. Four parameters were used as variables for fitting of the observed spectra. These were the source velocity ($\varv_{\text{source}} $), FWHM, the column density ($N$), and  excitation temperature ($T_\mathrm{ex}$). Estimates for the FWHM and $\varv_{\text{source}} $ were first obtained by looking at a few individual lines and fitting Gaussian profiles to these lines. These estimates were then used with a parameter space of 4 km s$^{-1}$ around the lines, in steps of 0.5 km s$^{-1}$ , which is equivalent to the channel width. The initial parameter space for $N$ was $N_\mathrm{min}=10^{14}$ and $N_\mathrm{max}=10^{19}$, with ten steps in the range, and for $T_\mathrm{ex}$ the range was 90 to 300 K, also with ten steps in this range. A few iterations were then run to decrease the ranges, but keeping the step size larger than the uncertainties. The source size was assumed to be 1.5$''$, or slightly larger than the beam size. Only line data were used in the modelling process. The values obtained for each molecule are shown in Tables \ref{table:N30_column_densities_COMs}, \ref{table:N30_column_densities_Meth}, and \ref{table:N30_column_densities_N-S-}. Table \ref{table:N30_column_densities_COMs} shows the values obtained from fits to the spectrum taken at position 1 (see Fig. \ref{Fig:N30_Mol_Grad}), where the O-bearing species peak around VLA1. Table \ref{table:N30_column_densities_Meth} in turn shows values obtained at peak position 2, between VLA1 and VLA2. Finally, Table \ref{table:N30_column_densities_N-S-} shows the values obtained at position 3, where the N- and S-bearing species peak (between VLA2 and VLA3). The uncertainties for $N$ and $T_\mathrm{ex}$ are mainly a result of how well the spectral model fits the data. This in turn depends on our assumptions of LTE (if the level populations are not in LTE, this will introduce a systematic error), optically thin emission (if emission is optically thick, column densities will be underestimated), and the line shape (we assume Gaussian line profiles). For species with many optically thick lines, this becomes especially problematic because fewer lines can be used in the fit. The optical depth of each line was checked, and only lines with $\tau \lesssim 0.8 $ were used in the modelling. The uncertainties on $N$ and $T_\mathrm{ex}$ were obtained by fixing $N$ (or $T_\mathrm{ex}$), and then changing $T_\mathrm{ex}$ (or $N$) until we were able to observe the impact of the change on the resulting fit, following the method outlined in \citet{calcutt-2018b}. Figures of each of the fitted lines for the molecules at position 2 are shown in Appendix \ref{App:Line_plots}, with the synthetic spectra shown in red and over-plotted on the observed spectra in black. 

In the case of molecules with many optically thick lines, such as CH$_3$OH and SO$_2$, the column densities were obtained from optically thin isotopologue emission. The isotopologue column density was then multiplied by the ISM isotope ratio. As an example, the modelled column density of $^{13}$CH$_3$OH was multiplied by the $^{12}$C/$^{13}$C ratio, which we took to be $\sim 77$ \citep{Wilson&Rood-1994}. 
Similarly, column densities for optically thick sulphur-bearing molecular lines were obtained using the ISM isotopologue ratio $^{32}$S/$^{34}$S $\sim 22$ \citep[][]{Wilson-1999}. In the case of CH$_3$CN, we did not detect its isotopologue $^{13}$CH$_3$CN, and so we used the higher energy transition, CH$_3$CN, v$_8$=1, for the fitting.

Apart from HDO, no other deuterated molecules were detected. Upper limits for the column densities of CH$_2$DOH were found to be 4.6 $\times 10^{16}$ $\mathrm{cm^{-2} }$, 6.8$\times 10^{16}$ $\mathrm{cm^{-2} }$, and 8.9$\times 10^{16}$ $\mathrm{cm^{-2} }$ at positions 1, 2, and 3, respectively. These values were obtained by assuming $T_\mathrm{ex}$, $\varv_{\text{source}}$, and FWHM of $^{13}$CH$_3$OH, and then increasing the column density of the synthetic spectrum until some lines have a strength of $\gtrsim 3 \sigma$ of the observed spectrum. 
The obtained upper limits correspond to a D/H ratio of $\sim$0.1\%. This value includes statistical correction for the three possible symmetries of the CH$_2$DOH molecule, which will give an observed N$_\mathrm{CH2DOH}$/N$_\mathrm{CH3OH}$ ratio a factor of three higher than the actual ratio \citep[see e.g.][]{joergensen-2018,Manigand-2019}. 

A number of more complex species have been detected towards high-mass star-forming regions; for example Sgr(B2) near the Galactic centre. One of the more abundant molecules detected towards this source is acetone (CH$_3$COCH$_3$) with a relative molecular abundance of X, corresponding to $N$(CH$_3$COCH$_3$) / $N$(CH$_3$OH) $\sim$ 0.1 \citep{Belloche-2016}. When adopting the same column-density ratio and physical conditions, we find that the non-detection towards N30 is consistent with the noise level. 

%_____________________________________________________________
%               Two-column Table - Column densities position 1
%_____________________________________________________________
%

\begin{table*}
\caption{Column densities, excitation temperatures, FWHM and $\varv_{\text{peak}}$ derived at the peak positions of H$_2$CS and other COMs, excluding CH$_3$OH (position 1 in Fig. \ref{Fig:N30_Mol_Grad}).} \tablefootmark{a}              
\label{table:N30_column_densities_COMs}      
\centering          
\begin{tabular}{l c c c c  c}    
\hline  \hline 
    Molecule & Column density & $T_\mathrm{ex}$& FWHM & $\varv_{\text{peak}}$  & Peak position\tablefootmark{b} \\ 
     & [$\mathrm{cm^{-2} }$]& [K] & [$\mathrm{km \ s^{-1} }$] & [$\mathrm{km \ s^{-1} }$] & \\
\hline
   \multicolumn{6}{c}{COMs and O-bearing species} \\
\hline   
    CH$_3$OH\tablefootmark{c}    & 1.3 (0.2) $\times 10^{19}$ & \hspace{0.6cm} 140 (25) \hspace{0.6cm} & \hspace{1cm} 5.0 \hspace{1cm}  & \hspace{1cm} 8.5 \hspace{1cm}  & 2 \\
    $^{13}$CH$_3$OH   & 1.7 (0.2) $\times 10^{17}$ & 140 (25) & 5.0 & 8.5  & 1/2\\
    CH$_3$$^{18}$OH\tablefootmark{d} & <3.9 $\times 10^{16}$ & 140  & 5.0 & 8.5  & -\\
    CH$_2$DOH\tablefootmark{d}  & <5.2 $\times 10^{16}$ & 140  & 5.0 & 8.5  &  -\\
    C$_2$H$_5$OH      & 7.2 (2.0) $\times 10^{16}$ & 120 (60) & 4.5 & 9.0  & 1 \\ 
    CH$_3$OCHO        & 1.9 (0.2) $\times 10^{17}$ & 110 (20) & 4.5 & 8.5  & 1 \\ 
    CH$_3$OCH$_3$     & 2.9 (0.2) $\times 10^{17}$ & 100 (20) & 4.0 & 8.5  & 1\\ 
    H$_2$CO\tablefootmark{c}    & 9.2 (1.5) $\times 10^{17}$ & 190 (40) & 7.5 & 8.0 & 1 \\
    H$_{2}$$^{13}$CO  & 1.2 (0.2) $\times 10^{16} $ & 190 (40) & 7.5 & 8.0 & 1 \\
    t-HCOOH            & 2.8 (0.5) $\times 10^{16}$ & 190 (50) & 7.0 & 11.0 & 3\\
    c-HCOOH\tablefootmark{d}     & <2.3 $\times 10^{15}$ & 190 & 7.0 & 11.0 &  3 \\
    HDO\tablefootmark{e}         & 4.6 (2.0) $\times 10^{16}$ & 140 (25) & 4.0 & 10.5  & - \\
\hline 
\multicolumn{6}{c}{S-bearing species} \\
\hline 
    H$_2$CS           &  3.3 (0.5) $\times 10^{16}$ & 170 (50) & 5.5 & 9.0  &  1\\
    $^{33}$SO\tablefootmark{f}   & 4.0 (2.0) $\times 10^{15}$ & 180 (40) & 6.5 & 10.0 & 3 \\
    SO$_{2}$\tablefootmark{c}    & 5.9 (0.2) $\times 10^{17}$ & 180 (40) & 6.5 & 10.0 & 3\\
    $^{33}$SO$_{2}$\tablefootmark{d}   & <6.1 $\times 10^{15}$ & 180 (40) & 6.5 & 10.0 & - \\
    $^{34}$SO$_{2}$   & 2.7 (1.0) $\times 10^{16}$ & 180 (40) & 6.5 & 10.0 & 3\\ 
    OCS\tablefootmark{c}        & 3.1 (0.4) $\times 10^{17}$ & 170 (50) & 5.5 & 9.5 & 2 \\
    OC$^{34}$S        & 1.4 (0.2)$\times 10^{16}$ & 170 (50) & 5.5 & 9.5 & 1 \\

\hline 
\multicolumn{6}{c}{N-bearing species} \\
\hline 
    HNCO                     &  4.0 (1.0)$\times 10^{16}$ & 120 (20) & 6.5 & 9.5   & 3\\
    HC$_3$N                  &  3.0 (1.0)$\times 10^{15}$ & 200 (40) & 7.0 & 9.5 & 3\\
    CH$_3$CN, v$_8$=1\tablefootmark{g}  &  3.6 (2.0)$\times 10^{16}$ & 140 (15) & 7.0 & 10.5 & 2/3 \\ 
    $^{13}$CH$_3$CN\tablefootmark{d}  &  < 6.0 $\times 10^{14}$ & 140 & 7.0 & 10.5 & - \\ 

\hline 

\end{tabular}
\tablefoot{\tablefoottext{a}{Values are derived from synthetic spectrum fitting with CASSIS, with the assumption of LTE.}
\tablefoottext{b}{Positions of peak molecular emission are as represented in Fig. \ref{Fig:N30_Mol_Grad}.}
\tablefoottext{c}{Column densities for optically thick lines derived using ISM ratios from \cite{Wilson&Rood-1994}.}
\tablefoottext{d}{Upper limits were determined by setting $T_\mathrm{ex}$, FWHM, and $\varv_{\text{peak}}$, equal to that of a detected isotopologue.}
\tablefoottext{e}{To estimate a column density for HDO, the excitation temperature of methanol was assumed.}
\tablefoottext{f}{Both SO and $^{34}$SO had optically thick lines, so only the column density of $^{33}$SO could be derived, using $T_\mathrm{ex}$ of $^{34}$SO$_2$.}
\tablefoottext{g}{CH$_3$CN, v$_8$=1, was used for the model fit of CH$_3$CN because it is optically thick, and $^{13}$CH$_3$CN was not detected.}
}
\end{table*}

%_____________________________________________________________
%               Two-column Table - Column densities position 2 
%_____________________________________________________________
%

\begin{table*}
\caption{Column densities, excitation temperatures, FWHM and $\varv_{\text{peak}}$ derived at the peak positions of CH$_3$OH (position 2 in Fig. \ref{Fig:N30_Mol_Grad}) \tablefootmark{a}}.             
\label{table:N30_column_densities_Meth}      
\centering          
\begin{tabular}{l c c c c  c}     % 2 columns 
\hline  \hline 

    Molecule & Column density & $T_\mathrm{ex}$& FWHM & $\varv_{\text{peak}}$  & Peak position\tablefootmark{b}\\ 
     & [$\mathrm{cm^{-2} }$]& [K] & [$\mathrm{km \ s^{-1} }$] & [$\mathrm{km \ s^{-1} }$] & \\
\hline
   \multicolumn{6}{c}{COMs and O-bearing species} \\
\hline  
    CH$_3$OH\tablefootmark{c}    & 1.0 (0.2) $\times 10^{19}$ & \hspace{0.6cm} 120 (25) \hspace{0.6cm} & \hspace{1cm} 4.5 \hspace{1cm}  & \hspace{1cm} 9.0 \hspace{1cm}  & 2 \\
    $^{13}$CH$_3$OH        & 1.3 (0.2) $\times 10^{17}$ & 120 (25) & 4.5 & 9.0  & 1/2\\
    CH$_3$$^{18}$OH\tablefootmark{d} & <3.0 $\times 10^{16}$ & 100 & 4.5 & 9.0  & -\\
    CH$_2$DOH\tablefootmark{d}       & <4.0  $\times 10^{16}$ & 100 & 4.5 & 9.0  & -\\
    C$_2$H$_5$OH           & 6.7 (2.0) $\times 10^{16}$ & 130 (60) & 5.0 & 9.5  & 1\\ 
    CH$_3$OCHO             &  1.4 (0.2) $\times 10^{17}$ & 110 (20) & 4.5 & 9.0 & 1\\ 
    CH$_3$OCH$_3$          &  2.4 (0.2) $\times 10^{17}$ & 110 (20) & 5.0 & 9.0 & 1\\
    H$_2$CO\tablefootmark{c}         &  5.9  (1.5) $\times 10^{17}$ & 160 (40) & 8.0 & 9.0 & 1 \\
    H$_{2}$$^{13}$CO       &  7.6 (2.0) $\times 10^{15} $ & 160 (40) & 8.0 & 9.0 & 1 \\
    t-HCOOH                & 4.0 (0.5) $\times 10^{16}$ & 190 (50) & 7.0 & 11.0 & 3\\
    c-HCOOH\tablefootmark{d}         & <3.3 $\times 10^{15}$ & 190 & 7.0 & 11.0 &  -\\
    HDO$^{ d}$             & 6.8 (2.0) $\times 10^{16}$ & 120 (25) & 4.0 & 10.5  & - \\

\hline 
\multicolumn{6}{c}{S-bearing species} \\
\hline 
    H$_2$CS                &  3.3 (0.5) $\times 10^{16}$ & 140 (50) & 5.5 & 9.0  &  1\\ 
    $^{33}$SO\tablefootmark{f}        & 5.4 (2.0) $\times 10^{15}$ & 130 (40) & 6.5 & 10.0 & 3\\
    SO$_{2}$\tablefootmark{c}         & 7.3 (0.2) $\times 10^{17}$ & 130 (40) & 6.0 & 10.0 & 3\\
    $^{33}$SO$_{2}$\tablefootmark{d} & <8.1 $\times 10^{15}$ & 130 (40) & 6.0 & 10.0 & - \\
    $^{34}$SO$_{2}$        &  3.3 (1.0)$\times 10^{16}$ & 130 (40) & 6.0 & 10.0  & 3 \\ 
    OCS $^{b}$             & 2.4 (0.4) $\times 10^{17}$ & 170 (50) & 5.0 & 10.0 & 2 \\
    OC$^{34}$S             & 1.1 (0.2) $\times 10^{16}$ & 170 (50) & 5.0 & 10.0 & 1 \\

\hline 
\multicolumn{6}{c}{N-bearing species} \\
\hline 
    HNCO                     &  5.0 (1.0)$\times 10^{16}$ & 150 (20) & 6.5 & 10.5   & 3\\
    HC$_3$N                  &  4.0 (1.0)$\times 10^{15}$ & 210 (40) & 7.0 & 10.0 & 3\\
    CH$_3$CN, v$_8$=1\tablefootmark{g}  &  5.5 (2.0)$\times 10^{16}$ & 140 (10) & 7.0 & 10.5 & 2/3 \\ 
    $^{13}$CH$_3$CN\tablefootmark{d}  &  <9.1 $\times 10^{14}$ & 140 & 7.0 & 10.5 & - \\ 

\hline 

\end{tabular}
\tablefoot{\tablefoottext{a}{Values are derived from synthetic spectrum fitting with CASSIS, with the assumption of LTE.}
\tablefoottext{b}{Positions of peak molecular emission are as represented in Fig. \ref{Fig:N30_Mol_Grad}.}
\tablefoottext{c}{Column densities for optically thick lines derived using ISM ratios from \cite{Wilson&Rood-1994}.}
\tablefoottext{d}{Upper limits were determined by setting $T_\mathrm{ex}$, FWHM, and $\varv_{\text{peak}}$, equal to that of a detected isotopologue.}
\tablefoottext{e}{To estimate a column density for HDO, the excitation temperature of methanol was assumed.}
\tablefoottext{f}{Both SO and $^{34}$SO had optically thick lines, so only the column density of $^{33}$SO could be derived, using $T_\mathrm{ex}$ of $^{34}$SO$_2$.}
\tablefoottext{g}{CH$_3$CN, v$_8$=1, was used for the model fit of CH$_3$CN because it is optically thick, and $^{13}$CH$_3$CN was not detected.}
}

\end{table*}

%_____________________________________________________________
%               Two-column Table - Column densities position 3
%_____________________________________________________________
%

\begin{table*}
\caption{Column densities, excitation temperatures, FWHM, and $\varv_{\text{peak}}$ derived at the peak positions of the N- and S- bearing species peak (position 3 in Fig. \ref{Fig:N30_Mol_Grad}).}\tablefootmark{a}             
\label{table:N30_column_densities_N-S-}      
\centering          
\begin{tabular}{l c c c c c}   
\hline  \hline 
    Molecule & Column density & $T_\mathrm{ex}$ & FWHM & $\varv_{\text{peak}}$  & Peak position\tablefootmark{b} \\ 
     & [$\mathrm{cm^{-2} }$]& [K] & [$\mathrm{km \ s^{-1} }$] & [$\mathrm{km \ s^{-1} }$] & \\
\hline
   \multicolumn{6}{c}{COMs and O-bearing species} \\
\hline
    CH$_3$OH\tablefootmark{c}    & 5.7 (1.5) $\times 10^{18}$ & \hspace{0.6cm} 100 (25) \hspace{0.6cm} & \hspace{1cm} 5.0 \hspace{1cm}  & \hspace{1cm} 9.5 \hspace{1cm}  & 2 \\
    $^{13}$CH$_3$OH   & 7.4 (2.0) $\times 10^{16}$ & 100 (25) & 5.0 & 9.5 & 1/2\\
    CH$_3$$^{18}$OH\tablefootmark{d} & <1.7 $\times 10^{16}$ & 100  & 5.0 & 9.5  & -\\
    CH$_2$DOH\tablefootmark{d}  & <2.3 $\times 10^{16}$ & 100  & 5.0 & 9.5 & - \\
    C$_2$H$_5$OH      &  5.5 (2.0) $\times 10^{16}$ & 140 (60) & 5.5 & 10.5  & 1 \\ 
    CH$_3$OCHO        & 9.6 (2.0) $\times 10^{16}$ & 110 (20) & 5.0 & 9.0  & 1 \\ 
    CH$_3$OCH$_3$     & 1.5 (0.2) $\times 10^{17}$ & 100 (20) & 4.5 & 9.5  & 1\\ 
    H$_2$CO\tablefootmark{c}    & 4.3 (1.5) $\times 10^{17}$ & 150 (40) & 8.5 & 9.5 & 1/2 \\
    H$_{2}$$^{13}$CO  & 5.6 (2.0) $\times 10^{15} $ & 150 (40) & 8.5 & 9.5 & 1 \\

    t-HCOOH           & 4.8 (0.5) $\times 10^{16} $ & 170 (50) & 7.0 & 11.0 & 3\\
    c-HCOOH           & <4.0  $\times 10^{15} $ & 170 (50) & 7.0 & 11.0 & 3\\
    HDO\tablefootmark{e}        & 8.9 (2.0) $\times 10^{16} $ & 100 (25) & 4.0 & 11.0 & - \\

\hline 
\multicolumn{6}{c}{S-bearing species} \\
\hline 
    H$_2$CS          & 2.4 (0.5) $\times 10^{16} $ & 140 (50) & 5.5 & 9.5 & 1\\ 

    $^{33}$SO\tablefootmark{f}   & 7.0 (2.0) $\times 10^{15}$ & 130 (40) & 6.5 & 10.0 & 3\\
  
    SO$_{2}$\tablefootmark{c}   & 1.1 (0.2) $\times 10^{18}$ & 130 (40) & 6.5 & 10.0 & 3\\
    $^{33}$SO$_{2}$\tablefootmark{d}   & <1.2 $\times 10^{16}$ & 130 & 6.5 & 10.0 & - \\
    $^{34}$SO$_{2}$   & 4.8 (1.0) $\times 10^{16}$ & 130 (40) & 6.5 & 10.0 & 3\\ 
    OCS$^{b}$        & 1.4 (0.4) $\times 10^{17}$ & 150 (50) & 5.0 & 10.0 & 2 \\
    OC$^{34}$S        & 6.5 (2.0)$\times 10^{15}$ & 170 (50) & 5.0 & 10.0 & 1 \\

\hline 
\multicolumn{6}{c}{N-bearing species} \\
\hline 
    HNCO               &  5.9 (1.0) $\times 10^{16}$ & 160 (40) & 6.5 & 10.5 & 3\\
    HC$_3$N            &  4.0 (1.0) $\times 10^{15}$ & 240 (40) & 7.0 & 10.0 & 3\\
    CH$_3$CN, v$_8$=1\tablefootmark{g}  &  6.0 (2.0) $\times 10^{16}$ & 140 (10) & 7.0 & 10.5 & 2/3 \\ 
    $^{13}$CH$_3$CN\tablefootmark{d}  &  <1.0 $\times 10^{15}$ & 140 & 7.0 & 10.5 & - \\ 

\hline 

\end{tabular}
\tablefoot{\tablefoottext{a}{Values are derived from synthetic spectrum fitting with CASSIS, with the assumption of LTE.}
\tablefoottext{b}{Positions of peak molecular emission are as represented in Fig. \ref{Fig:N30_Mol_Grad}.}
\tablefoottext{c}{Column densities for optically thick lines derived using ISM ratios from \cite{Wilson&Rood-1994}.}
\tablefoottext{d}{Upper limits were determined by setting $T_\mathrm{ex}$, FWHM, and $\varv_{\text{peak}}$, equal to that of a detected isotopologue.}
\tablefoottext{e}{To estimate a column density for HDO, the excitation temperature of methanol was assumed.}
\tablefoottext{f}{Both SO and $^{34}$SO had optically thick lines, so only the column density of $^{33}$SO could be derived, using $T_\mathrm{ex}$ of $^{34}$SO$_2$.}
\tablefoottext{g}{CH$_3$CN, v$_8$=1, was used for the model fit of CH$_3$CN because it is optically thick, and $^{13}$CH$_3$CN was not detected.}
}

\end{table*}

\begin{figure*}
\centering
\includegraphics[width=16cm]{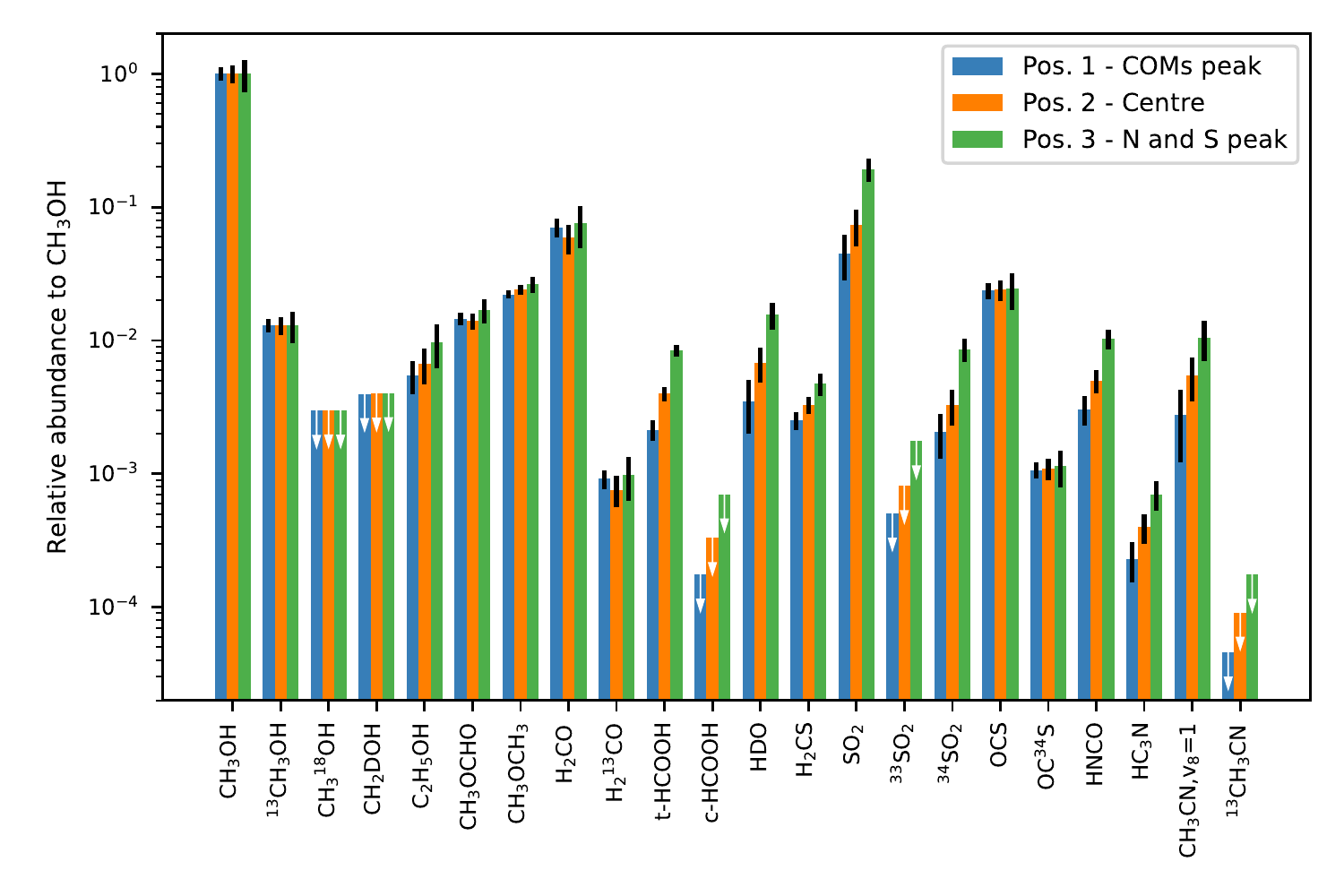}
  \caption{Relative abundances of each molecule relative to the column density of CH$_3$OH at each respective position. The white arrows represent upper limits, whereas the black bars represent the uncertainties (see Tables \ref{table:N30_column_densities_COMs}, \ref{table:N30_column_densities_Meth} and \ref{table:N30_column_densities_N-S-}).}
  \label{Fig:nmol_rel_CH3OH}
\end{figure*}

\begin{figure*}
\centering
\includegraphics[width=16cm]{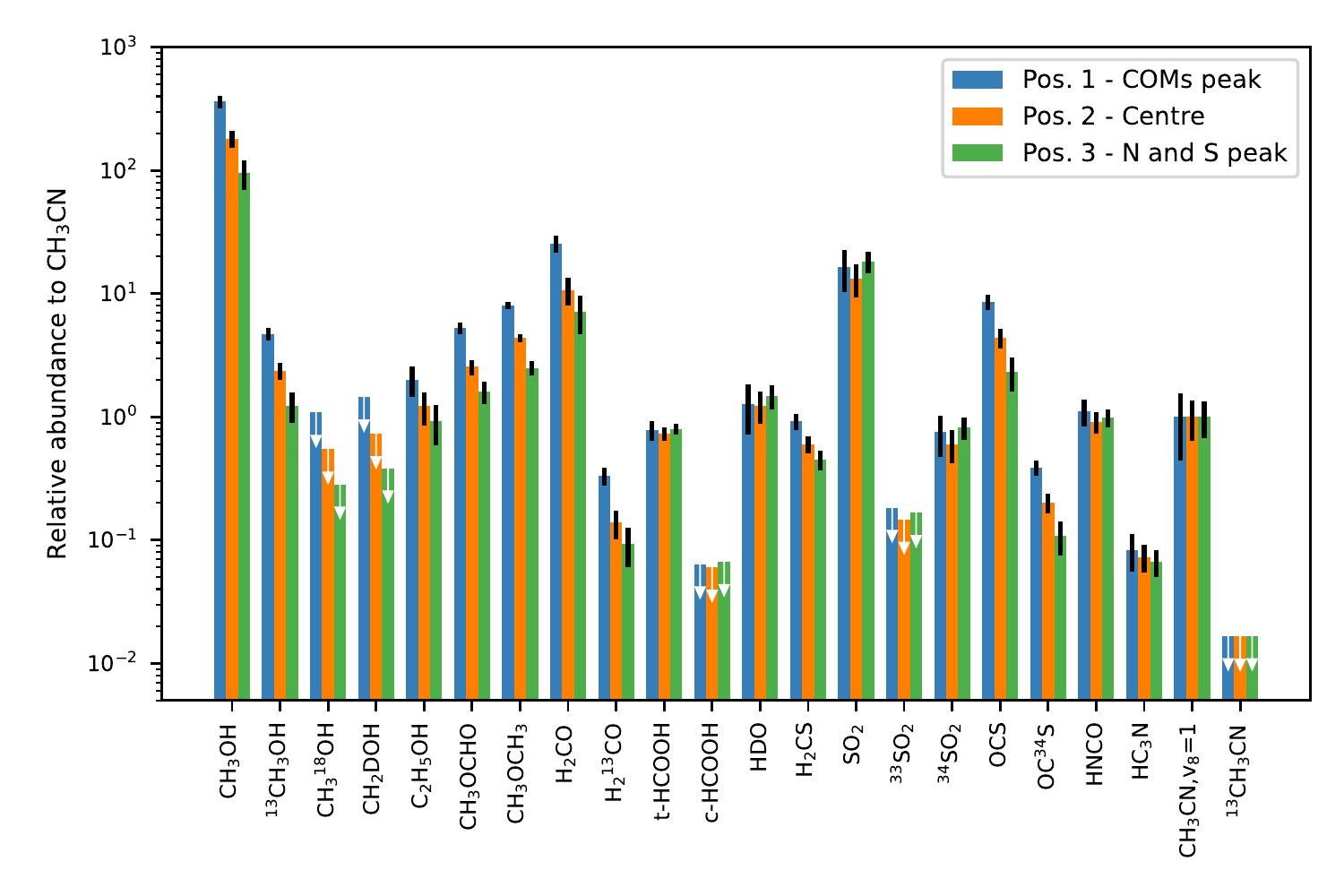}
  \caption{Relative abundances of each molecule relative to the column density of CH$_3$CN at each respective position. The white arrows and black bars are as in Fig. \ref{Fig:nmol_rel_CH3OH}.}
  \label{Fig:nmol_rel_CH3CN}
\end{figure*}

Relative abundances of the modelled molecules are shown in Figs. \ref{Fig:nmol_rel_CH3OH} and \ref{Fig:nmol_rel_CH3CN}, with Fig. \ref{Fig:nmol_rel_CH3OH}, showing the column densities of each molecule with respect to the column density of CH$_3$OH, at each respective position, whereas Fig. \ref{Fig:nmol_rel_CH3CN} shows the column densities with respect to the column density of CH$_3$CN. As with the integrated intensity maps, we see how the O-bearing species peak towards position 1, while the N- and S- bearing species peak towards position 3.

%______________________________________________________________
%                                  One column figure
%----------------------------------------------------------- 

%______________________________________________________________
%                                  One column figure
%----------------------------------------------------------- 

\section{Discussion}
\label{Sec:Discussion}

In this section, we discuss our results in more detail, starting with a comparison of the abundances towards N30 with other sources, followed by discussions on excitation temperatures and the D/H ratio observed towards N30. We end the section with a discussion of the continuum cores, and the observed molecular gradient.

\subsection{Abundances compared to other sources}

To put the inferred abundances into a broader context, we compare them to those observed towards different sources, looking at the differences and similarities. An example of such a comparison is presented by \cite{Drozdovskaya-2019}, in which abundances towards the low-mass protostar IRAS 16293B are compared with those measured towards Comet 67P/C-G, with the cometary data used as a representation of the initial ingredients that formed the Earth and the other Solar System planets. From their results, the authors concluded that the volatile composition of cometesimals and planetesimals is partially inherited from the pre- and protostellar phases of the evolution. \cite{Joergensen-2020} extended this comparison to include sources of different masses and located in different environments. They compared, for example, IRAS 16293B and HH212 (both low-mass sources located relatively close to the Sun), AFGL 4176 (a high-mass sources also in our Galactic neighbourhood), and Sgr B2 (N2), a high-mass source located close to the Galactic centre. They also included the shocked regions L1157 B1 and Orion KL as well as Comet 67P/C-G in their comparison. The result of this comparison was that the abundances towards these sources are in general very similar, which is an indication that the underlying chemistry is relatively independent of the differences in their physical conditions. There are some differences, however. These latter authors found, for example, that abundances observed towards the protostellar sources tend to agree better with one another than with those towards Comet 67P/C-G, which suggests that some chemical processing occurs between the protostellar and cometary stages \citep[see also][]{Drozdovskaya-2019}. In this section, we compare the available column densities observed towards the sources from \cite{Joergensen-2020} to the those we observe towards N30. The column densities taken are relative to CH$_3$OH, CH$_3$CN, and SO$_2$, representing COM, N-, and S-bearing species, respectively. These sources were selected simply because of the amount of data available in the literature for each of them. While there are many sources in the literature with available column densities, it is rare to find sources observed with wide enough frequency coverage so that column densities can be found for a large amount of molecules. We therefore compare N30 to these sources. As N30 shows a strong chemical gradient, it is important to place the column densities measured towards this source in context with those of the sources from these studies, and in this way achieve a better understanding of the physical nature of the observed chemical gradient. 

\begin{figure*}
    \centering
    \includegraphics[width=16.5cm]{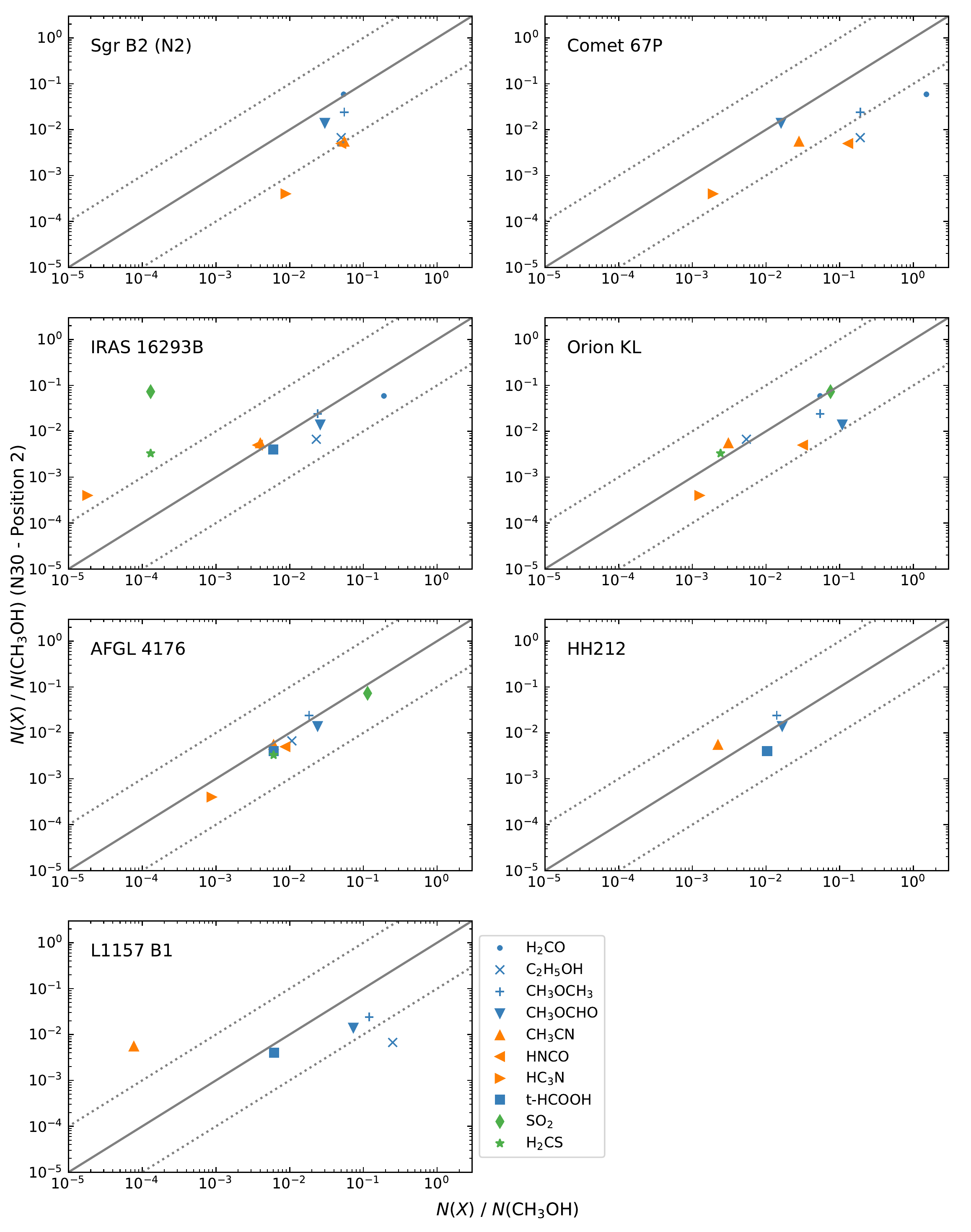}
    \caption{Column densities towards N30, position 2, with respect to CH$_3$OH on the y-axis compared to different sources plotted on the x-axis. The solid line represents equal abundances, whereas the dotted lines represent an order of magnitude difference.}
    \label{Fig:N30_Column_density_compare_sources-wrt_CH3OH_multi}%
\end{figure*}

Figure \ref{Fig:N30_Column_density_compare_sources-wrt_CH3OH_multi} shows the column density ratios observed towards N30 compared with those observed towards the sources discussed above. The figure shows the column density ratios towards position 2 (see Fig. \ref{Fig:N30_Mol_Grad}). Similar figures are provided for positions 1 and 3 in Appendix \ref{App:ColumnDensities_compare}. The panels in the figure show the column densities of each molecule with respect to CH$_3$OH, with N30 represented on the y-axis, and each respective source represented on the x-axis. The upper-left panel shows the Galactic centre source, Sgr B2 (N2), which is part of the Sgr B2 molecular cloud, one of the most active high-mass star forming regions in the Galaxy; it was part of the high-resolution EMoCA survey \citep[Exploring Molecular Complexity with ALMA][]{Belloche-2016, Belloche-2017, Muller-2016a, Bonfand-2019}. For this source, the O-bearing species show similar abundances (differences within one order of magnitude), while the N-bearing species observed towards N30 have lower ratios compared to CH$_3$OH (typically by a factor 10). 

The upper-right panel shows abundances measured by the ROSINA instrument of the Rosetta mission towards Comet 67P/C-G, as presented by \cite{Drozdovskaya-2019} in their comparison to the low-mass source IRAS 16293B which was the target of the PILS program by \cite{joergensen-2016}. Abundances towards Comet 67P/C-G are in general higher than that observed towards N30, with no real differences between O- and N-bearing species. However, for IRAS 16293B, the O-bearing species are very similar to N30, while the N- and S-bearing species are in general higher towards N30. This is in contrast to the Galactic centre source Sgr B2 (N2), which also shows similar O-bearing species, but with higher abundances in N-bearing species. This would suggest that N30 has abundances in N-bearing species somewhere in between the Galactic centre (high-mass) source, and the solar neighbourhood (low-mass) source IRAS 16293B. 

Orion KL, which is not a `traditional' hot core but is believed to be the result of an explosive event where a pre-existing dense region is heated from the outside \citep[][]{Friedel-2008,2011A&A...529A..24Zapata,Crockett-2015,Pagani-2017}, shows, in general, similar abundances to those observed towards N30. The high-mass source AFGL 4176 \citep[][]{Bogelund-2019} shows abundances in even better agreement to N30. HH212 is a low-mass source, with COMs observed in its disc atmosphere, that is, above and below the disc \citep{Lee-2019}. This source also shows similarities to N30, although the available data in the literature for this source are still limited. L1157-B1 \citep[][]{Arce-2008, Sugimura-2011, Lefloch-2017} on the other hand is a shocked region, with the available column densities for this source suggesting that the O-bearing species are a little less abundant in N30, while the only N-bearing species available for comparison, CH$_3$CN, is significantly more abundant in N30. Nevertheless, when all the sources are compared,  the abundances observed towards AFGL 4176 agree best with those towards N30. At a distance of $\sim 3.7$ kpc, this source is more distant than N30 ($d \sim 1.3$ kpc), and more luminous, with an upper limit of $2\times 10^5 $ $L_\odot$ \citep[compared to N30, with a luminosity of $\sim2.04 \times 10^4$ $L_\odot$;][]{Cao-2019}. It also has a large disc-like structure, as was found by \cite{Johnston-2015} using ALMA observations of CH$_3$CN on scales of $\sim 1200$ AU. Furthermore, \cite{Bogelund-2019} found (also from ALMA observations, at a resolution of $\sim 1285$ AU) that the O- and N-bearing species in this source also peak in slightly different locations, with N-bearing species peaking slightly closer to the continuum peak than the O-bearing species. This would suggest that a combination of processes could be causing the sublimation of the COMs from the dust grains rather than the traditional hot core model. The agreement with Orion KL also strengthens the case that it is a combination of processes that result in the COMs we observe.

It therefore seems that, when comparing the abundances observed towards N30, they agree more with other high-mass sources, such as AFGL 4176 (a more traditional hot core, but with some chemical differentiation) and Orion KL (a high-mass, shocked region), than with the abundances observed towards low-mass sources, such as IRAS 16293B, or the outflow region, L1157-B1. The agreement is also better with these high-mass sources than compared to high-mass sources in the Galactic centre. The environment here, which has a cosmic ray ionisation rate of a factor of 50 higher than usually assumed for the Galactic disc \citep[][]{Guzman-2015,Bonfand-2019}, resulting in a higher dust temperature. This presents some challenges for our understanding of the formation of COMs \citep[][]{Joergensen-2020}, which are believed to form primarily on grain surfaces at low temperatures. However, it might explain the high abundance of N-bearing species in the Galactic centre, which evolves over longer timescales in the gas phase \citep[][]{Garrod-2017,Bogelund-2019}. It might also explain the low abundance of S- and N-bearing species in low-mass sources, where the temperatures are lower than in their high-mass counterparts. \cite{Bogelund-2019} suggest that AFGL 4176 is a very young source, where little processing of the chemical inventory by the protostar has occurred, which is why they observe N-bearing species to be low in abundance compared to Sgr B2. For the shocked region L1157-B1, the O-bearing species are sputtered off of the dust grains, but there is not sufficient time for the nitrogen chemistry to evolve in the gas phase \citep[see chemical models by e.g. ][]{Garrod-2013,Barone-2015,Garrod-2017,Codella-2017}. The agreement between the abundances towards N30 MM1 and AFGL 4176 would suggest that the sources are at a similar stage of evolution, although at scales of $\sim 1200$ AU, AFGL 4176 appears to be a single source, as opposed to N30, which is at least a binary source at scales of $\sim 1300$ AU, with the MM1b/VLA1 core probably less evolved than the MM1a/VLA3 core. 

\begin{figure}
\centering
\includegraphics[width=9cm]{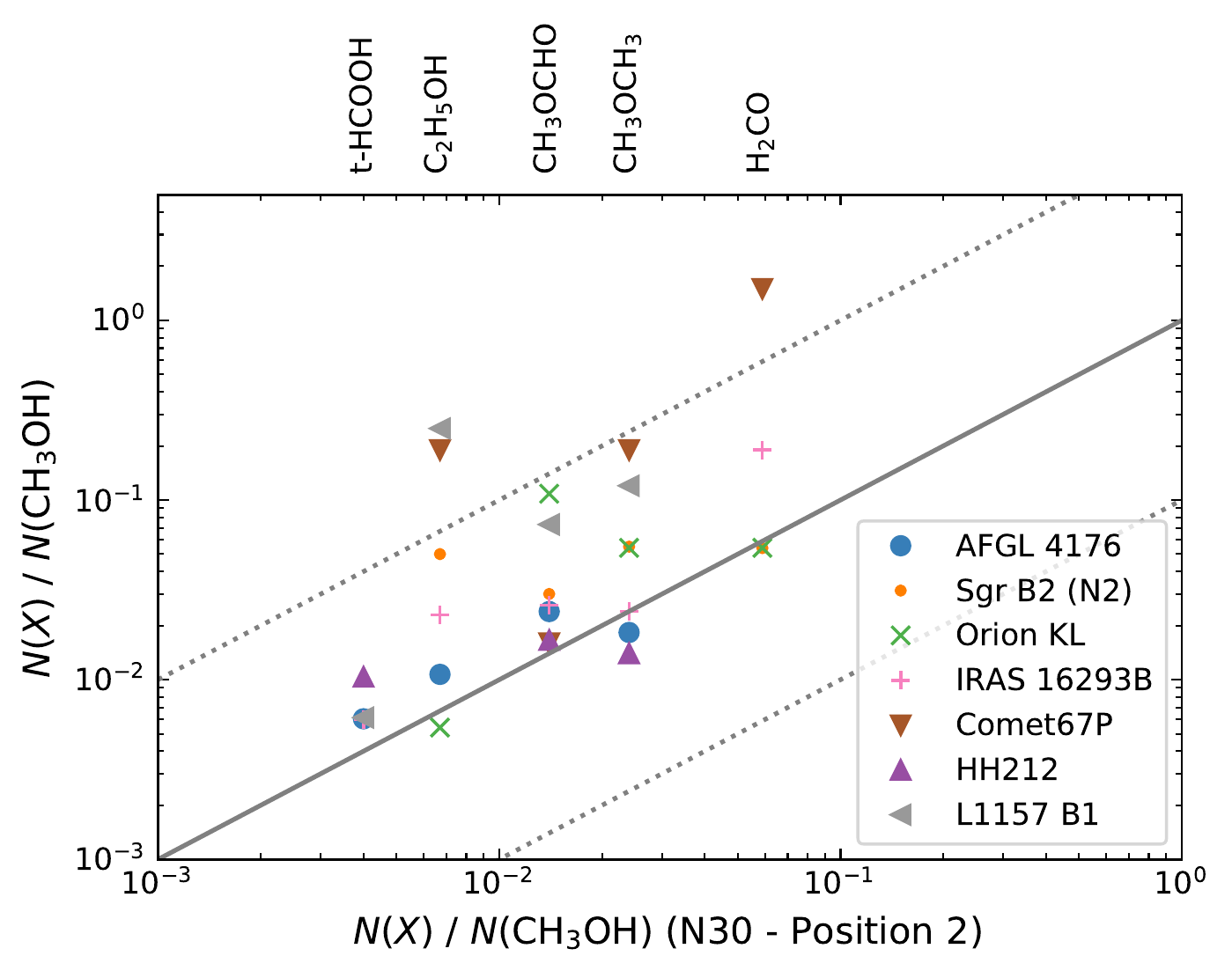}
  \caption{Column densities with respect to CH$_3$OH towards N30, position 2, on the x-axis compared to different sources plotted on the y-axis. Each source is marked with a different marker. The solid and dotted lines are as in Fig. \ref{Fig:N30_Column_density_compare_sources-wrt_CH3OH_multi}}
  \label{Fig:N30_Column_density_compare_sources-wrt_CH3OH}
\end{figure}

Figure \ref{Fig:N30_Column_density_compare_sources-wrt_CH3OH} shows the abundances of O-bearing species normalised to CH$_3$OH, with N30 on the x-axis and the respective sources depicted with different symbols on the y-axis. Figure \ref{Fig:N30_Column_density_compare_sources-wrt_wrt_CH3CN,SO2} shows the N and S-bearing species normalised to CH$_3$CN and SO$_2$, respectively. Some trends that can be seen in Fig. \ref{Fig:N30_Column_density_compare_sources-wrt_CH3OH} are that the abundances of O-bearing species in Comet 67P and L1157-B1 are in general higher than the protostellar sources (around one order of magnitude), while the shocked region Orion KL shows higher relative abundances in CH$_3$OCHO and CH$_3$OCH$_3$ (5--10 times higher), but with H$_2$CO and C$_2$H$_5$OH comparable to N30. IRAS 16293B has abundances in O-bearing species that are a little higher than N30 (around five times), while Sgr B2 (N2) has a higher abundance in C$_2$H$_5$OH and (five to ten times), and with abundances of CH$_3$OCHO, CH$_3$OCH$_3$, and H$_2$CO  similar  to those for N30. As discussed above for Fig. \ref{Fig:N30_Column_density_compare_sources-wrt_CH3OH_multi}, we also see from this figure that the best agreement is with the high-mass source AFGL 4176. This can also be seen in the abundances with respect to CH$_3$CN and SO$_2$ in Fig. \ref{Fig:N30_Column_density_compare_sources-wrt_wrt_CH3CN,SO2}, with AFGL 4176 showing similar abundances, while Orion KL has higher abundances in N-bearing species but a similar abundance of H$_2$CS. However, for the N- and S-bearing species, the number of molecular species available for comparison is limited. 

Therefore, while N30 clearly shows a strong chemical gradient with different species peaking in different locations, the inferred abundances are typical of what is seen towards other intermediate- and high-mass protostars. This suggests that, no matter the physical origin of the chemical gradient, the chemistry does not change significantly between different sources. 

%
%______________________________________________________________
%         Two column figure
%___________________________________________________________________

%______________________________________________________________
%                                  One column figure
%----------------------------------------------------------- 

%______________________________________________________________
%                                  One column figure
%----------------------------------------------------------- 
\begin{figure}
\centering
\includegraphics[width=9cm]{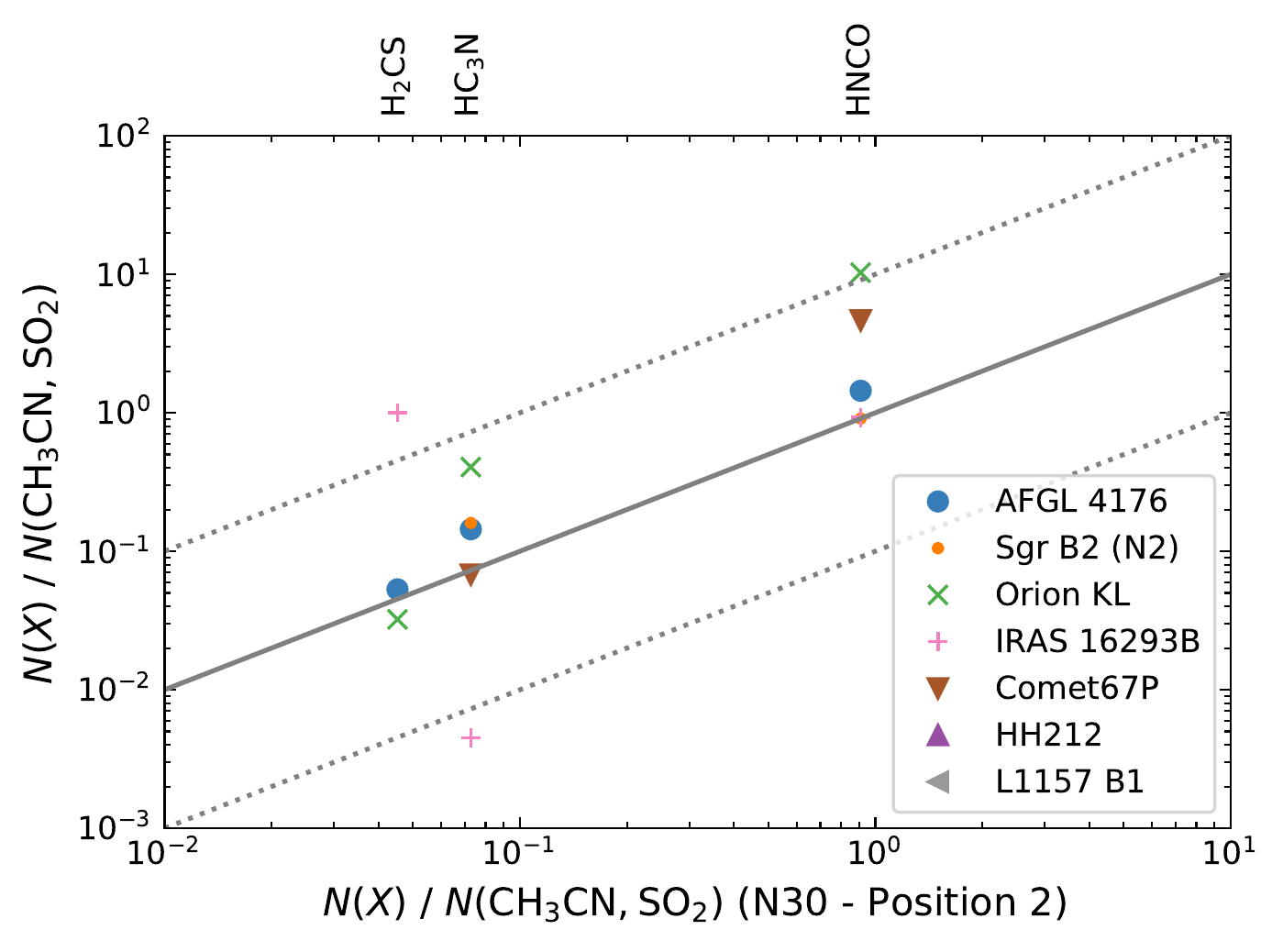}
  \caption{Column densities towards N30, position 2 (on the x-axis), with respect to CH$_3$CN for the N-bearing species and SO$_2$ for H$_2$CS, compared to different sources plotted on the y-axis. Each source is marked with a different marker. The solid and dotted lines are as in Fig. \ref{Fig:N30_Column_density_compare_sources-wrt_CH3OH_multi}}
  \label{Fig:N30_Column_density_compare_sources-wrt_wrt_CH3CN,SO2}
\end{figure}

\subsection{Excitation temperatures}

We find some differences when comparing the excitation temperatures of different molecular species. The N- and S-bearing species seem to be slightly warmer than the O-bearing species, with ranges from $130$ to $200$ K and $100$ to $140$ K, respectively. HC$_3$N seems to be warmer than other molecular species, with a range of $200-240$ K between the three positions in Fig. \ref{Fig:N30_Mol_Grad}. t-HCOOH also seems to be warmer ($170 - 190$ K) than other O-bearing species, but considering that it peaks at the position with the N-bearing species, this might not be strange, because it does not seem to be connected with the other O-bearing molecules. H$_2$$^{13}$CO also has a warmer excitation temperature, in the range of $150 - 190$ K.

Other authors have found more significant differences in the excitation temperatures of N- and O-bearing species towards other sources. An example is Orion KL, in which the O-bearing species trace  gas of lower temperatures ---ranging between $100$ and $150$ K--- than the N-bearing species, which trace gas of $200-300$ K \citep{Bell-2014,Crockett-2015}. \cite{joergensen-2018} also found a range of $80 - 300$ K in excitation temperatures of different molecular species in IRAS 16293, and suggested that the differences observed could be attributed to binding energy. \cite{van'tHoff-2020} argues that the so-called `soot line' located close to the protostar at a temperature of $\sim$ 300 K is marked by an excess of hydrocarbons and nitriles sublimated from of carbon grains inside this line. The O-bearing COMs on the other hand desorb around the water snowline, which is located further out from the protostar at a temperature of $\sim$ 100 K. However, in the case of N30,  we do not observe the COMs concentrated at a central position, but rather spread out in a chemical gradient, which suggest that it is probably not a traditional hot core(s) but rather a combination of factors causing the chemical gradient and the difference in excitation temperature that we observe, including two or more protostars (see Section \ref{Subsection:Continuum_Mol_Grad}).

\subsection{D/H ratio}

We find an upper limit on the D/H ratio derived for N(CH$_2$DOH)/N(CH$_3$OH) of about $0.1\%$. Compared to IRAS 16293B, which has values of $2$\%-3$\%$, this is very low. It is also lower than values reported by \cite{Neill-2013} for Orion KL, in which they found similar D/H ratios for water, formaldehyde, and methanol, $0.2\%$-$0.8\%$. However, for Sgr B2 (N2), \cite{Belloche-2016} reported a XD/XH ratio for CH$_2$DOH of $0.12\%$, which they argue might be because of the higher kinetic temperatures in the Galactic centre region compared to nearby star-forming regions. \cite{Bogelund-2019} do not detect any deuterated species towards AFGL 4176, which further strengthens the agreement between this source and N30, as discussed in the abundance comparison above. Other high-mass sources have also shown low D/H ratios \citep[e.g. ][]{Ratajczak-2011,Bogelund-2018}. \cite{Bogelund-2018} conclude that warm formation temperatures of $\sim 30$ K could account for the low deuteration in the NGC 6334I regions, with the pre-stellar cloud heated by a nearby O-type star and associated HII region. In the case of N30, the presence of the strong radiation field caused by the nearby Cyg--OB2 association \citep[projected distance of $\sim$23 pc at a distance of 1.3 kpc;][]{Reipurth-2008} may have heated the pre-stellar cloud and resulted in the low levels of deuteration we observe. 

An alternative is that the prestellar phase is cold but short, as is expected for higher mass sources. In this scenario, time is the bottleneck in setting the deuteration, as opposed to above where temperature is the limiting factor. Either way, the low degree of deuteration in sources such as this one is not unexpected. 

\subsection{Continuum cores, and the observed molecular gradient}
\label{Subsection:Continuum_Mol_Grad}
We find that the O-bearing species have their peak emission concentrated towards the radio (cm) continuum source, VLA1, close to but to the west of the MM1b submm continuum core. S- and N-bearing species on the other hand peak between VLA2 and VLA3 to the north of MM1a. All molecules peak on a gradient traced by the red- and blueshifted axis of H$_2$CO and CS emission. This axis is nearly perpendicular to the larger scale \cite[$\sim$1 pc;][]{Fischer-1985} bipolar outflow axis seen in CO emission, which might indicate that the observed submm and radio continuum sources are located in a disc-like structure \citep[see][]{Hutawarakorn-2002}. It is important to note here that this disc-like structure is more akin to a pseudo-disc that is not rotation-supported, but just a rotating structure around the protostars. Looking at the H$_2$CO and CS maps in Figs. \ref{Fig:N30_H2CO_contours} and \ref{Fig:N30_CS_contours}, it is also clear that this structure is very large, with a radius of around 3000 AU, and encompass all the continuum sources \citep[see also ][]{Hutawarakorn-2002}. Similar disc-like structures are also observed in other sources, such as G35.20-0.74N, in which \cite{Hutawarakorn-1999} observed a large-scale outflow structure seen in CO emission perpendicular to a disc-like structure containing four hot cores. Chemical differentiation between the different cores was also observed for this source \citep[with separation between the cores of $1000 - 2000$ AU;][]{Allen-2017}, with O- and S-bearing species detected towards the inner two cores (B1 and B2), but few N-bearing species, while the outer two cores (B3 and A) have strong detections of N-, S-, and O-bearing species. Another source is G328.2551$-$0.5321, for which \cite{Csengeri-2018} found N-bearing COMs peaking towards the protostar, while the O-bearing COMs peak towards two spots, offset from the protostar \citep[see also][]{Csengeri-2019}. The authors interpreted the two spots as accretion shocks onto a disc with an estimated distance of $200-800$ AU from the central protostar. These large disc-like structures are therefore not uncommon in high-mass protostars, and chemical differentiation also seems to be a common feature. Furthermore, with the high angular resolution provided by interferometric observations, more and more examples of sources are identified in which the observed COM emission cannot be explained by the canonical hot-core scenario in which COMs sublimate from the heated dust grains close to the protostar. This was also seen by \cite{Belloche-2020} for low-mass protostars in the CALYPSO survey.

For N30, we observe chemical differentiation between S- and N-bearing species associated with the VLA2 and VLA3/MM1a cores 
on the one hand, and COMs and O-bearing species associated
with the VLA1/MM1b core on the other. Molecules that do not follow this differentiation in O- and S-/N-bearing species are t-HCOOH, observed to peak towards position 3 at the N- and S-bearing species position, and CS, OCS, and H$_2$CS that peak towards the O-bearing species position. 
 
Our results therefore suggest that N30 is not a purely traditional hot core in which ice-covered dust grains collapse towards a warm protostar and are heated to temperatures $\gtrsim$ 100 K  as
they get closer, where the ices sublimate and we see a peak of COM emission \citep[see e.g.][ and references therein]{Herbst-2009}. The release of the COMs into the gas phase may then lead to further reactions in the dense warm gas, but this is still centred on the region where the dust temperature exceeds $\sim$ 100 K. Instead, the O-bearing COM emission is likely caused by a combination of processes, including accretion of infalling material onto the disc surface ---which is similar to what was found by \cite{Csengeri-2018} for G328.2551--0.5321---,while the N- and S- bearing species towards VLA3/MM1a might be a slightly more evolved source where the gas-phase chemistry had more time to evolve.

%______________________________________________________________

\section{Summary and Conclusions}
\label{Sec:Conclusion}

We observed CygX-N30 and MM1 with the SMA in the 345 GHz frequency window at a resolution of $\sim$1$''$ and with 32 GHz of continuous frequency coverage. About 400 lines were detected from 29 different molecular species and their isotopologues.

We observe a chemical gradient of molecules along the axis of a disc-like structure perpendicular to the large-scale bi-polar CO emission observed in previous studies. This disc is parallel to but offset from the axis connecting the mm continuum cores (by $\sim$1$''$). The O-bearing molecular emission peaks are close to but offset to the west of the MM1b continuum core and are between the VLA1 and VLA2 radio continuum cores, which fall on the molecular gradient--disc axis. The N- and S-bearing species on the other hand are concentrated closer to the MM1a core, between the VLA2 and VLA3 radio continuum cores. The disc-like structure is observed from the red- blueshifted H$_2$CO and CS emission, which signifies infalling gas onto the surface of the disc-like structure. This implies that the COMs observed towards MM1 are not purely the result of a traditional hot core, but are rather caused by a combination of processes, including accretion. In order to test this hypothesis, it will be useful to re-observe this source in a few years to check for time variability.

Comparing the abundances observed towards N30 MM1 with other sources, we find that it is similar to the high-mass protostar AFGL 4176, and to a lesser extent to the shocked region Orion KL, and to the Galactic centre source Sgr B2 (N2). The O-bearing species show similar abundances with respect to CH$_3$OH towards these sources, whereas for Sgr B2 (N2), the N-bearing species are less abundant in N30, while the low-mass source IRAS 16293B has lower abundances in N- and S-bearing species; the O-bearing species are similar. The agreement with AFGL 4176 seems to suggest that N30 MM1 is at a similar stage of evolution to this source, but with the MM1b/VLA1/VLA2 source at the centre of the disc-like structure probably at an earlier stage than the MM1a/VLA3 core, where the N- and S- bearing species had more time to evolve.

We observe a small difference in excitation temperature between the O-bearing species and the S- and N- bearing species. A larger difference is observed for HC$_3$N, H$_2^{13}$CO, and t-HCOOH, which seem to trace warmer gas than the other species. The low levels of deuteration in N30 MM1 suggest that the grain-surface formation temperature of COMs in the pre-stellar cloud was warm, that is, $\sim$30 K, probably the result of radiation from the nearby Cyg-OB2 association.

In conclusion, these observations highlight the benefit of being able to observe large frequency ranges simultaneously and at high angular resolution in order to shed light on the physical origin of COMs. Particularly, the frequency coverage ensures that a large number of species from different chemical families are covered simultaneously, and the high angular resolution allows spatial correlations between species to be closely examined. This is especially important for the physically complex structure of sources such as N30, which complicates our understanding of the origin of COMs. It is therefore important to study such sources with multiple chemical tracers, and at sufficient angular resolution to be able to disentangle the various physical and chemical processes taking place. Further analysis of the remaining nine sources in the PILS-Cygnus survey will clearly continue to shed light on their origin.

\begin{acknowledgements}
We acknowledge and thank the staff of the SMA for their assistance and continued support. The authors wish to recognise and acknowledge the very significant cultural role and reverence that the summit of Mauna Kea has always had within the indigenous Hawaiian community. We are most fortunate to have had the opportunity to conduct observations from this mountain. The SMA is a joint project between the Smithsonian Astrophysical Observatory and the Academia Sinica Institute of Astronomy and Astrophysics and is funded by the Smithsonian Institution and the Academia Sinica. The research of SJvdW, LEK and MeA is supported by a research grant (19127) from VILLUM FONDEN. JKJ and SM acknowledge support by the European Research Council (ERC) under the European Union’s Horizon 2020 research and innovation programme through ERC Consolidator Grant “S4F” (grant agreement No 646908). Finally, we would like to thank the anonymous referee, who provided constructive comments that helped improve the clarity of the paper. \end{acknowledgements}

%-------------------------------------------------------------------

\bibliographystyle{aa} % style aa.bst
\bibliography{main_arxiv} % 

\begin{thebibliography}{86}
\expandafter\ifx\csname natexlab\endcsname\relax\def\natexlab#1{#1}\fi

\bibitem[{{Allen} {et~al.}(2017){Allen}, {van der Tak}, {S{\'a}nchez-Monge},
  {Cesaroni}, \& {Beltr{\'a}n}}]{Allen-2017}
{Allen}, V., {van der Tak}, F.~F.~S., {S{\'a}nchez-Monge}, {\'A}., {Cesaroni},
  R., \& {Beltr{\'a}n}, M.~T. 2017, \aap, 603, A133

\bibitem[{{Arce} {et~al.}(2008){Arce}, {Santiago-Garc{\'\i}a}, {J{\o}rgensen},
  {Tafalla}, \& {Bachiller}}]{Arce-2008}
{Arce}, H.~G., {Santiago-Garc{\'\i}a}, J., {J{\o}rgensen}, J.~K., {Tafalla},
  M., \& {Bachiller}, R. 2008, \apjl, 681, L21

\bibitem[{{Artur de la Villarmois} {et~al.}(2018){Artur de la Villarmois},
  {Kristensen}, {J{\o}rgensen}, {Bergin}, {Brinch}, {Frimann}, {Harsono},
  {Sakai}, \& {Yamamoto}}]{ArturdelaVillarmois-2018}
{Artur de la Villarmois}, E., {Kristensen}, L.~E., {J{\o}rgensen}, J.~K.,
  {et~al.} 2018, \aap, 614, A26

\bibitem[{{Avery} \& {Chiao}(1996)}]{Avery_and_Chiao-1996}
{Avery}, L.~W. \& {Chiao}, M. 1996, \apj, 463, 642

\bibitem[{{Barone} {et~al.}(2015){Barone}, {Latouche}, {Skouteris}, {Vazart},
  {Balucani}, {Ceccarelli}, \& {Lefloch}}]{Barone-2015}
{Barone}, V., {Latouche}, C., {Skouteris}, D., {et~al.} 2015, \mnras, 453, L31

\bibitem[{{Bell} {et~al.}(2014){Bell}, {Cernicharo}, {Viti}, {Marcelino},
  {Palau}, {Esplugues}, \& {Tercero}}]{Bell-2014}
{Bell}, T.~A., {Cernicharo}, J., {Viti}, S., {et~al.} 2014, \aap, 564, A114

\bibitem[{{Belloche} {et~al.}(2020){Belloche}, {Maury}, {Maret}, {Anderl},
  {Bacmann}, {Andr{\'e}}, {Bontemps}, {Cabrit}, {Codella}, {Gaudel}, {Gueth},
  {Lef{\`e}vre}, {Lefloch}, {Podio}, \& {Testi}}]{Belloche-2020}
{Belloche}, A., {Maury}, A.~J., {Maret}, S., {et~al.} 2020, \aap, 635, A198

\bibitem[{{Belloche} {et~al.}(2017){Belloche}, {Meshcheryakov}, {Garrod},
  {Ilyushin}, {Alekseev}, {Motiyenko}, {Margul{\`e}s}, {M{\"u}ller}, \&
  {Menten}}]{Belloche-2017}
{Belloche}, A., {Meshcheryakov}, A.~A., {Garrod}, R.~T., {et~al.} 2017, \aap,
  601, A49

\bibitem[{{Belloche} {et~al.}(2016){Belloche}, {M{\"u}ller}, {Garrod}, \&
  {Menten}}]{Belloche-2016}
{Belloche}, A., {M{\"u}ller}, H.~S.~P., {Garrod}, R.~T., \& {Menten}, K.~M.
  2016, \aap, 587, A91

\bibitem[{{Birks} {et~al.}(2006){Birks}, {Fuller}, \& {Gibb}}]{Birks-2006}
{Birks}, J.~R., {Fuller}, G.~A., \& {Gibb}, A.~G. 2006, \aap, 458, 181

\bibitem[{{B{\o}gelund} {et~al.}(2019){B{\o}gelund}, {Barr}, {Taquet},
  {Ligterink}, {Persson}, {Hogerheijde}, \& {van Dishoeck}}]{Bogelund-2019}
{B{\o}gelund}, E.~G., {Barr}, A.~G., {Taquet}, V., {et~al.} 2019, \aap, 628, A2

\bibitem[{{B{\o}gelund} {et~al.}(2018){B{\o}gelund}, {McGuire}, {Ligterink},
  {Taquet}, {Brogan}, {Hunter}, {Pearson}, {Hogerheijde}, \& {van
  Dishoeck}}]{Bogelund-2018}
{B{\o}gelund}, E.~G., {McGuire}, B.~A., {Ligterink}, N. F.~W., {et~al.} 2018,
  \aap, 615, A88

\bibitem[{{Bonfand} {et~al.}(2019){Bonfand}, {Belloche}, {Garrod}, {Menten},
  {Willis}, {St{\'e}phan}, \& {M{\"u}ller}}]{Bonfand-2019}
{Bonfand}, M., {Belloche}, A., {Garrod}, R.~T., {et~al.} 2019, \aap, 628, A27

\bibitem[{{Calcutt} {et~al.}(2018){Calcutt}, {J{\o}rgensen}, {M{\"u}ller},
  {Kristensen}, {Coutens}, {Bourke}, {Garrod}, {Persson}, {van der Wiel}, {van
  Dishoeck}, \& {Wampfler}}]{calcutt-2018b}
{Calcutt}, H., {J{\o}rgensen}, J.~K., {M{\"u}ller}, H.~S.~P., {et~al.} 2018,
  \aap, 616, A90

\bibitem[{{Campbell} {et~al.}(1982){Campbell}, {Hoffmann}, {Thronson}, {Niles},
  {Nawfel}, \& {Hawrylycz}}]{Campbell-1982}
{Campbell}, M.~F., {Hoffmann}, W.~F., {Thronson}, H.~A., J., {et~al.} 1982,
  \apj, 261, 550

\bibitem[{{Cao} {et~al.}(2019){Cao}, {Qiu}, {Zhang}, {Wang}, {Hu}, \&
  {Liu}}]{Cao-2019}
{Cao}, Y., {Qiu}, K., {Zhang}, Q., {et~al.} 2019, \apjs, 241, 1

\bibitem[{{Carrasco-Gonz{\'a}lez} {et~al.}(2010){Carrasco-Gonz{\'a}lez},
  {Rodr{\'\i}guez}, {Torrelles}, {Anglada}, \&
  {Gonz{\'a}lez-Mart{\'\i}n}}]{Carrasco-Gonzalez-2010}
{Carrasco-Gonz{\'a}lez}, C., {Rodr{\'\i}guez}, L.~F., {Torrelles}, J.~M.,
  {Anglada}, G., \& {Gonz{\'a}lez-Mart{\'\i}n}, O. 2010, \aj, 139, 2433

\bibitem[{{Caselli} \& {Ceccarelli}(2012)}]{Caselli-2012}
{Caselli}, P. \& {Ceccarelli}, C. 2012, \aapr, 20, 56

\bibitem[{{Ceccarelli} {et~al.}(2014){Ceccarelli}, {Caselli},
  {Bockel{\'e}e-Morvan}, {Mousis}, {Pizzarello}, {Robert}, \&
  {Semenov}}]{Ceccarelli-2014}
{Ceccarelli}, C., {Caselli}, P., {Bockel{\'e}e-Morvan}, D., {et~al.} 2014, in
  Protostars and Planets VI, ed. H.~{Beuther}, R.~S. {Klessen}, C.~P.
  {Dullemond}, \& T.~{Henning}, 859

\bibitem[{{Cesaroni} {et~al.}(2010){Cesaroni}, {Hofner}, {Araya}, \&
  {Kurtz}}]{Cesaroni-2010}
{Cesaroni}, R., {Hofner}, P., {Araya}, E., \& {Kurtz}, S. 2010, \aap, 509, A50

\bibitem[{{Codella} {et~al.}(2017){Codella}, {Ceccarelli}, {Caselli},
  {Balucani}, {Barone}, {Fontani}, {Lefloch}, {Podio}, {Viti}, {Feng},
  {Bachiller}, {Bianchi}, {Dulieu}, {Jim{\'e}nez-Serra}, {Holdship}, {Neri},
  {Pineda}, {Pon}, {Sims}, {Spezzano}, {Vasyunin}, {Alves}, {Bizzocchi},
  {Bottinelli}, {Caux}, {Chac{\'o}n-Tanarro}, {Choudhury}, {Coutens}, {Favre},
  {Hily-Blant}, {Kahane}, {Jaber Al-Edhari}, {Laas}, {L{\'o}pez-Sepulcre},
  {Ospina}, {Oya}, {Punanova}, {Puzzarini}, {Quenard}, {Rimola}, {Sakai},
  {Skouteris}, {Taquet}, {Testi}, {Theul{\'e}}, {Ugliengo}, {Vastel}, {Vazart},
  {Wiesenfeld}, \& {Yamamoto}}]{Codella-2017}
{Codella}, C., {Ceccarelli}, C., {Caselli}, P., {et~al.} 2017, \aap, 605, L3

\bibitem[{{Crockett} {et~al.}(2015){Crockett}, {Bergin}, {Neill}, {Favre},
  {Blake}, {Herbst}, {Anderson}, \& {Hassel}}]{Crockett-2015}
{Crockett}, N.~R., {Bergin}, E.~A., {Neill}, J.~L., {et~al.} 2015, \apj, 806,
  239

\bibitem[{{Csengeri} {et~al.}(2019){Csengeri}, {Belloche}, {Bontemps},
  {Wyrowski}, {Menten}, \& {Bouscasse}}]{Csengeri-2019}
{Csengeri}, T., {Belloche}, A., {Bontemps}, S., {et~al.} 2019, \aap, 632, A57

\bibitem[{{Csengeri} {et~al.}(2018){Csengeri}, {Bontemps}, {Wyrowski},
  {Belloche}, {Menten}, {Leurini}, {Beuther}, {Bronfman}, {Commer{\c{c}}on},
  {Chapillon}, {Longmore}, {Palau}, {Tan}, \& {Urquhart}}]{Csengeri-2018}
{Csengeri}, T., {Bontemps}, S., {Wyrowski}, F., {et~al.} 2018, \aap, 617, A89

\bibitem[{{Drozdovskaya} {et~al.}(2019){Drozdovskaya}, {van Dishoeck}, {Rubin},
  {J{\o}rgensen}, \& {Altwegg}}]{Drozdovskaya-2019}
{Drozdovskaya}, M.~N., {van Dishoeck}, E.~F., {Rubin}, M., {J{\o}rgensen},
  J.~K., \& {Altwegg}, K. 2019, \mnras, 490, 50

\bibitem[{{Drozdovskaya} {et~al.}(2015){Drozdovskaya}, {Walsh}, {Visser},
  {Harsono}, \& {van Dishoeck}}]{Drozdovskaya-2015}
{Drozdovskaya}, M.~N., {Walsh}, C., {Visser}, R., {Harsono}, D., \& {van
  Dishoeck}, E.~F. 2015, \mnras, 451, 3836

\bibitem[{{Endres} {et~al.}(2016){Endres}, {Schlemmer}, {Schilke}, {Stutzki},
  \& {M{\"u}ller}}]{Endres-2016-CDMS}
{Endres}, C.~P., {Schlemmer}, S., {Schilke}, P., {Stutzki}, J., \&
  {M{\"u}ller}, H. S.~P. 2016, Journal of Molecular Spectroscopy, 327, 95

\bibitem[{{Fischer} {et~al.}(1985){Fischer}, {Sanders}, {Simon}, \&
  {Solomon}}]{Fischer-1985}
{Fischer}, J., {Sanders}, D.~B., {Simon}, M., \& {Solomon}, P.~M. 1985, \apj,
  293, 508

\bibitem[{{Fish} {et~al.}(2005){Fish}, {Reid}, {Argon}, \& {Zheng}}]{Fish-2005}
{Fish}, V.~L., {Reid}, M.~J., {Argon}, A.~L., \& {Zheng}, X.-W. 2005, \apjs,
  160, 220

\bibitem[{{Friedel} \& {Snyder}(2008)}]{Friedel-2008}
{Friedel}, D.~N. \& {Snyder}, L.~E. 2008, \apj, 672, 962

\bibitem[{{Garrod}(2013)}]{Garrod-2013}
{Garrod}, R.~T. 2013, \apj, 765, 60

\bibitem[{{Garrod} {et~al.}(2017){Garrod}, {Belloche}, {M{\"u}ller}, \&
  {Menten}}]{Garrod-2017}
{Garrod}, R.~T., {Belloche}, A., {M{\"u}ller}, H.~S.~P., \& {Menten}, K.~M.
  2017, \aap, 601, A48

\bibitem[{{Gibb} {et~al.}(2003){Gibb}, {Hoare}, {Little}, \&
  {Wright}}]{Gibb-2003}
{Gibb}, A.~G., {Hoare}, M.~G., {Little}, L.~T., \& {Wright}, M.~C.~H. 2003,
  \mnras, 339, 1011

\bibitem[{{Guzm{\'a}n} {et~al.}(2015){Guzm{\'a}n}, {Sanhueza}, {Contreras},
  {Smith}, {Jackson}, {Hoq}, \& {Rathborne}}]{Guzman-2015}
{Guzm{\'a}n}, A.~E., {Sanhueza}, P., {Contreras}, Y., {et~al.} 2015, \apj, 815,
  130

\bibitem[{{Haschick} {et~al.}(1981){Haschick}, {Reid}, {Burke}, {Moran}, \&
  {Miller}}]{Haschick-1981}
{Haschick}, A.~D., {Reid}, M.~J., {Burke}, B.~F., {Moran}, J.~M., \& {Miller},
  G. 1981, \apj, 244, 76

\bibitem[{{Herbst} \& {van Dishoeck}(2009)}]{Herbst-2009}
{Herbst}, E. \& {van Dishoeck}, E.~F. 2009, \araa, 47, 427

\bibitem[{{Hunter} {et~al.}(1994){Hunter}, {Taylor}, {Felli}, \&
  {Tofani}}]{hunter-1994}
{Hunter}, T.~R., {Taylor}, G.~B., {Felli}, M., \& {Tofani}, G. 1994, \aap, 284,
  215

\bibitem[{{Hutawarakorn} \& {Cohen}(1999)}]{Hutawarakorn-1999}
{Hutawarakorn}, B. \& {Cohen}, R.~J. 1999, \mnras, 303, 845

\bibitem[{{Hutawarakorn} {et~al.}(2002){Hutawarakorn}, {Cohen}, \&
  {Brebner}}]{Hutawarakorn-2002}
{Hutawarakorn}, B., {Cohen}, R.~J., \& {Brebner}, G.~C. 2002, \mnras, 330, 349

\bibitem[{{Johnston} {et~al.}(2015){Johnston}, {Robitaille}, {Beuther}, {Linz},
  {Boley}, {Kuiper}, {Keto}, {Hoare}, \& {van Boekel}}]{Johnston-2015}
{Johnston}, K.~G., {Robitaille}, T.~P., {Beuther}, H., {et~al.} 2015, \apjl,
  813, L19

\bibitem[{{J{\o}rgensen} {et~al.}(2020){J{\o}rgensen}, {Belloche}, \&
  {Garrod}}]{Joergensen-2020}
{J{\o}rgensen}, J.~K., {Belloche}, A., \& {Garrod}, R.~T. 2020, \araa, 58, 727

\bibitem[{{J{\o}rgensen} {et~al.}(2004){J{\o}rgensen}, {Hogerheijde}, {Blake},
  {van Dishoeck}, {Mundy}, \& {Sch{\"o}ier}}]{Joergensen-2004}
{J{\o}rgensen}, J.~K., {Hogerheijde}, M.~R., {Blake}, G.~A., {et~al.} 2004,
  \aap, 415, 1021

\bibitem[{{J{\o}rgensen} {et~al.}(2018){J{\o}rgensen}, {M{\"u}ller}, {Calcutt},
  {Coutens}, {Drozdovskaya}, {{\"O}berg}, {Persson}, {Taquet}, {van Dishoeck},
  \& {Wampfler}}]{joergensen-2018}
{J{\o}rgensen}, J.~K., {M{\"u}ller}, H.~S.~P., {Calcutt}, H., {et~al.} 2018,
  \aap, 620, A170

\bibitem[{{J{\o}rgensen} {et~al.}(2016){J{\o}rgensen}, {van der Wiel},
  {Coutens}, {Lykke}, {M{\"u}ller}, {van Dishoeck}, {Calcutt}, {Bjerkeli},
  {Bourke}, {Drozdovskaya}, {Favre}, {Fayolle}, {Garrod}, {Jacobsen},
  {{\"O}berg}, {Persson}, \& {Wampfler}}]{joergensen-2016}
{J{\o}rgensen}, J.~K., {van der Wiel}, M.~H.~D., {Coutens}, A., {et~al.} 2016,
  \aap, 595, A117

\bibitem[{{Kauffmann} {et~al.}(2008){Kauffmann}, {Bertoldi}, {Bourke}, {Evans},
  \& {Lee}}]{kauffmann08}
{Kauffmann}, J., {Bertoldi}, F., {Bourke}, T.~L., {Evans}, N.~J., I., \& {Lee},
  C.~W. 2008, \aap, 487, 993

\bibitem[{{Kn{\"o}dlseder}(2000)}]{Knodlseder-2000}
{Kn{\"o}dlseder}, J. 2000, \aap, 360, 539

\bibitem[{{Kurtz} {et~al.}(2000){Kurtz}, {Cesaroni}, {Churchwell}, {Hofner}, \&
  {Walmsley}}]{Kurtz-2000}
{Kurtz}, S., {Cesaroni}, R., {Churchwell}, E., {Hofner}, P., \& {Walmsley},
  C.~M. 2000, in Protostars and Planets IV, ed. V.~{Mannings}, A.~P. {Boss}, \&
  S.~S. {Russell}, 299--326

\bibitem[{{Lee} {et~al.}(2019){Lee}, {Codella}, {Li}, \& {Liu}}]{Lee-2019}
{Lee}, C.-F., {Codella}, C., {Li}, Z.-Y., \& {Liu}, S.-Y. 2019, \apj, 876, 63

\bibitem[{{Lefloch} {et~al.}(2017){Lefloch}, {Ceccarelli}, {Codella}, {Favre},
  {Podio}, {Vastel}, {Viti}, \& {Bachiller}}]{Lefloch-2017}
{Lefloch}, B., {Ceccarelli}, C., {Codella}, C., {et~al.} 2017, \mnras, 469, L73

\bibitem[{{Manigand} {et~al.}(2019){Manigand}, {Calcutt}, {J{\o}rgensen},
  {Taquet}, {M{\"u}ller}, {Coutens}, {Wampfler}, {Ligterink}, {Drozdovskaya},
  {Kristensen}, {van der Wiel}, \& {Bourke}}]{Manigand-2019}
{Manigand}, S., {Calcutt}, H., {J{\o}rgensen}, J.~K., {et~al.} 2019, \aap, 623,
  A69

\bibitem[{{McMullin} {et~al.}(2007){McMullin}, {Waters}, {Schiebel}, {Young},
  \& {Golap}}]{McMullin-2007}
{McMullin}, J.~P., {Waters}, B., {Schiebel}, D., {Young}, W., \& {Golap}, K.
  2007, in Astronomical Society of the Pacific Conference Series, Vol. 376,
  Astronomical Data Analysis Software and Systems XVI, ed. R.~A. {Shaw},
  F.~{Hill}, \& D.~J. {Bell}, 127

\bibitem[{{Minh} {et~al.}(2010){Minh}, {Su}, {Chen}, {Liu}, {Yan}, \&
  {Kim}}]{minh-2010}
{Minh}, Y.~C., {Su}, Y.~N., {Chen}, H.~R., {et~al.} 2010, \apj, 723, 1231

\bibitem[{{Minier} {et~al.}(2001){Minier}, {Conway}, \& {Booth}}]{Minier-2001}
{Minier}, V., {Conway}, J.~E., \& {Booth}, R.~S. 2001, \aap, 369, 278

\bibitem[{{Motte} {et~al.}(2007){Motte}, {Bontemps}, {Schilke}, {Schneider},
  {Menten}, \& {Brogui{\`e}re}}]{Motte-2007}
{Motte}, F., {Bontemps}, S., {Schilke}, P., {et~al.} 2007, \aap, 476, 1243

\bibitem[{{M{\"u}ller} {et~al.}(2016){M{\"u}ller}, {Belloche}, {Xu}, {Lees},
  {Garrod}, {Walters}, {van Wijngaarden}, {Lewen}, {Schlemmer}, \&
  {Menten}}]{Muller-2016a}
{M{\"u}ller}, H. S.~P., {Belloche}, A., {Xu}, L.-H., {et~al.} 2016, \aap, 587,
  A92

\bibitem[{{M{\"u}ller} {et~al.}(2005){M{\"u}ller}, {Schl{\"o}der}, {Stutzki},
  \& {Winnewisser}}]{Muller-2005-CDMS}
{M{\"u}ller}, H. S.~P., {Schl{\"o}der}, F., {Stutzki}, J., \& {Winnewisser}, G.
  2005, Journal of Molecular Structure, 742, 215

\bibitem[{{M{\"u}ller} {et~al.}(2001){M{\"u}ller}, {Thorwirth}, {Roth}, \&
  {Winnewisser}}]{Muller-2001-CDMS}
{M{\"u}ller}, H.~S.~P., {Thorwirth}, S., {Roth}, D.~A., \& {Winnewisser}, G.
  2001, \aap, 370, L49

\bibitem[{{Neill} {et~al.}(2013){Neill}, {Crockett}, {Bergin}, {Pearson}, \&
  {Xu}}]{Neill-2013}
{Neill}, J.~L., {Crockett}, N.~R., {Bergin}, E.~A., {Pearson}, J.~C., \& {Xu},
  L.-H. 2013, \apj, 777, 85

\bibitem[{{\"O}berg(2016)}]{Oberg-2016}
{\"O}berg, K.~I. 2016, Chemical Reviews, 116, 9631, {P}MID: 27099922

\bibitem[{{Odenwald} \& {Schwartz}(1993)}]{Odenwald-1993}
{Odenwald}, S.~F. \& {Schwartz}, P.~R. 1993, \apj, 405, 706

\bibitem[{{Orozco-Aguilera} {et~al.}(2017){Orozco-Aguilera}, {Zapata},
  {Hirota}, {Qin}, \& {Masqu{\'e}}}]{Orozco-Aguilera-2017}
{Orozco-Aguilera}, M.~T., {Zapata}, L.~A., {Hirota}, T., {Qin}, S.-L., \&
  {Masqu{\'e}}, J.~M. 2017, \apj, 847, 66

\bibitem[{{Ossenkopf} \& {Henning}(1994)}]{Ossenkopf-1994}
{Ossenkopf}, V. \& {Henning}, T. 1994, \aap, 291, 943

\bibitem[{{Pagani} {et~al.}(2017){Pagani}, {Favre}, {Goldsmith}, {Bergin},
  {Snell}, \& {Melnick}}]{Pagani-2017}
{Pagani}, L., {Favre}, C., {Goldsmith}, P.~F., {et~al.} 2017, \aap, 604, A32

\bibitem[{{Persi} {et~al.}(2006){Persi}, {Tapia}, \& {Smith}}]{Persi-2006}
{Persi}, P., {Tapia}, M., \& {Smith}, H.~A. 2006, \aap, 445, 971

\bibitem[{{Pickett} {et~al.}(1998){Pickett}, {Poynter}, {Cohen}, {Delitsky},
  {Pearson}, \& {M{\"u}ller}}]{Pickett-1998-JPL}
{Pickett}, H.~M., {Poynter}, R.~L., {Cohen}, E.~A., {et~al.} 1998, \jqsrt, 60,
  883

\bibitem[{{Podio} {et~al.}(2015){Podio}, {Codella}, {Gueth}, {Cabrit},
  {Bachiller}, {Gusdorf}, {Lee}, {Lefloch}, {Leurini}, {Nisini}, \&
  {Tafalla}}]{Podio-2015}
{Podio}, L., {Codella}, C., {Gueth}, F., {et~al.} 2015, \aap, 581, A85

\bibitem[{{Ratajczak} {et~al.}(2011){Ratajczak}, {Taquet}, {Kahane},
  {Ceccarelli}, {Faure}, \& {Quirico}}]{Ratajczak-2011}
{Ratajczak}, A., {Taquet}, V., {Kahane}, C., {et~al.} 2011, \aap, 528, L13

\bibitem[{{Reipurth} \& {Schneider}(2008)}]{Reipurth-2008}
{Reipurth}, B. \& {Schneider}, N. 2008, {Star Formation and Young Clusters in
  Cygnus}, ed. B.~{Reipurth}, Vol.~4, 36

\bibitem[{{Rodr{\'\i}guez-Kamenetzky}
  {et~al.}(2020){Rodr{\'\i}guez-Kamenetzky}, {Carrasco-Gonz{\'a}lez},
  {Torrelles}, {Vlemmings}, {Rodr{\'\i}guez}, {Surcis}, {G{\'o}mez},
  {Cant{\'o}}, {Goddi}, {Kim}, {Kim}, {A{\~n}ez-L{\'o}pez}, {Curiel}, \& {van
  Langevelde}}]{Rodriguez-Kamenetzky-2020}
{Rodr{\'\i}guez-Kamenetzky}, A., {Carrasco-Gonz{\'a}lez}, C., {Torrelles},
  J.~M., {et~al.} 2020, \mnras, 496, 3128

\bibitem[{{Rygl} {et~al.}(2012){Rygl}, {Brunthaler}, {Sanna}, {Menten}, {Reid},
  {van Langevelde}, {Honma}, {Torstensson}, \& {Fujisawa}}]{Rygl-2012}
{Rygl}, K.~L.~J., {Brunthaler}, A., {Sanna}, A., {et~al.} 2012, \aap, 539, A79

\bibitem[{{Sch{\"o}ier} {et~al.}(2002){Sch{\"o}ier}, {J{\o}rgensen}, {van
  Dishoeck}, \& {Blake}}]{Schoier-2002}
{Sch{\"o}ier}, F.~L., {J{\o}rgensen}, J.~K., {van Dishoeck}, E.~F., \& {Blake},
  G.~A. 2002, \aap, 390, 1001

\bibitem[{{Sch{\"o}ier} {et~al.}(2005){Sch{\"o}ier}, {van der Tak}, {van
  Dishoeck}, \& {Black}}]{Schoier-2005-LAMDA}
{Sch{\"o}ier}, F.~L., {van der Tak}, F.~F.~S., {van Dishoeck}, E.~F., \&
  {Black}, J.~H. 2005, \aap, 432, 369

\bibitem[{{Shepherd}(2001)}]{shepherd-2001}
{Shepherd}, D.~S. 2001, \apj, 546, 345

\bibitem[{{Shepherd} {et~al.}(2004){Shepherd}, {Kurtz}, \&
  {Testi}}]{shepherd-2004}
{Shepherd}, D.~S., {Kurtz}, S.~E., \& {Testi}, L. 2004, \apj, 601, 952

\bibitem[{{Shepherd} {et~al.}(2003){Shepherd}, {Testi}, \&
  {Stark}}]{shepherd-2003}
{Shepherd}, D.~S., {Testi}, L., \& {Stark}, D.~P. 2003, \apj, 584, 882

\bibitem[{{Snyder} {et~al.}(2005){Snyder}, {Lovas}, {Hollis}, {Friedel},
  {Jewell}, {Remijan}, {Ilyushin}, {Alekseev}, \& {Dyubko}}]{Snyder-2005}
{Snyder}, L.~E., {Lovas}, F.~J., {Hollis}, J.~M., {et~al.} 2005, \apj, 619, 914

\bibitem[{{Sugimura} {et~al.}(2011){Sugimura}, {Yamaguchi}, {Sakai}, {Umemoto},
  {Sakai}, {Takano}, {Aikawa}, {Hirano}, {Liu}, {Millar}, {Nomura}, {Su},
  {Takakuwa}, \& {Yamamoto}}]{Sugimura-2011}
{Sugimura}, M., {Yamaguchi}, T., {Sakai}, T., {et~al.} 2011, \pasj, 63, 459

\bibitem[{{Surcis} {et~al.}(2011){Surcis}, {Vlemmings}, {Curiel}, {Hutawarakorn
  Kramer}, {Torrelles}, \& {Sarma}}]{Surcis-2011}
{Surcis}, G., {Vlemmings}, W.~H.~T., {Curiel}, S., {et~al.} 2011, \aap, 527,
  A48

\bibitem[{{Surcis} {et~al.}(2009){Surcis}, {Vlemmings}, {Dodson}, \& {van
  Langevelde}}]{Surcis-2009}
{Surcis}, G., {Vlemmings}, W.~H.~T., {Dodson}, R., \& {van Langevelde}, H.~J.
  2009, \aap, 506, 757

\bibitem[{{Torrelles} {et~al.}(1997){Torrelles}, {G{\'o}mez}, {Rodr{\'\i}guez},
  {Ho}, {Curiel}, \& {V{\'a}zquez}}]{1997ApJ...489..744T}
{Torrelles}, J.~M., {G{\'o}mez}, J.~F., {Rodr{\'\i}guez}, L.~F., {et~al.} 1997,
  \apj, 489, 744

\bibitem[{{van 't Hoff} {et~al.}(2020){van 't Hoff}, {Bergin}, {J{\o}rgensen},
  \& {Blake}}]{van'tHoff-2020}
{van 't Hoff}, M. L.~R., {Bergin}, E.~A., {J{\o}rgensen}, J.~K., \& {Blake},
  G.~A. 2020, \apjl, 897, L38

\bibitem[{{Vastel} {et~al.}(2015){Vastel}, {Bottinelli}, {Caux}, {Glorian}, \&
  {Boiziot}}]{Vastel_CASSIS-2015}
{Vastel}, C., {Bottinelli}, S., {Caux}, E., {Glorian}, J.~M., \& {Boiziot}, M.
  2015, in SF2A-2015: Proceedings of the Annual meeting of the French Society
  of Astronomy and Astrophysics, 313--316

\bibitem[{{Wilson}(1999)}]{Wilson-1999}
{Wilson}, T.~L. 1999, Reports on Progress in Physics, 62, 143

\bibitem[{{Wilson} \& {Rood}(1994)}]{Wilson&Rood-1994}
{Wilson}, T.~L. \& {Rood}, R. 1994, \araa, 32, 191

\bibitem[{{Zapata} {et~al.}(2011){Zapata}, {Schmid-Burgk}, \&
  {Menten}}]{2011A&A...529A..24Zapata}
{Zapata}, L.~A., {Schmid-Burgk}, J., \& {Menten}, K.~M. 2011, \aap, 529, A24

\bibitem[{{Zhu} {et~al.}(2011){Zhu}, {Zhao}, \& {Wright}}]{Zhu-2011}
{Zhu}, L., {Zhao}, J.-H., \& {Wright}, M.~C.~H. 2011, \apj, 740, 114

\end{thebibliography}

%\Online

\begin{appendix} %First online appendix
\section{Observing log}
\label{sect:obslog}

The full observing log is provided in Table \ref{table:obslog}. Specifically, this includes the observing dates and calibrators used for each night of the observations. The observing strategy was to observe two science targets between a pair of gain observations, and to loop over all ten targets in each track. This ensured uniform sensitivity. Furthermore, a random source was chosen to start each track, to ensure near uniform $uv$ coverage for all sources, and thus similar image fidelity.

\begin{table*}[h]
\caption{Observing log.}             
\label{table:obslog}      
\centering          
\begin{tabular}{c c c c c c c}     % 8 columns 
\hline
\hline 
Observing date  &       No of antennas  &       Configuration   &       Bandpass        &       Flux    &       Gain & $\tau$(225 GHz)       \\ \hline
21/06/2017      &       7       &       COM     &       3c454.3 &       Titan, Neptune &       mwc349a & 0.05--0.07    \\
22/06/2017      &       7       &       COM     &       3c273, 3c454.3  &       Titan, Neptune &       mwc349a & 0.10  \\
27/06/2017      &       7       &       COM     &       3c454.3 &       Callisto        &       mwc349a & 0.08  \\
10/07/2017      &       7       &       COM     &       3c454.3 &       Callisto        &       mwc349a & 0.05--0.06 \\
07/08/2017      &       6       &       COM     &       3c84    &       Titan, Uranus  &       mwc349a & 0.05  \\
20/10/2017      &       8       &       EXT     &       3c84    &       Uranus  &       mwc349a & 0.02--0.03    \\
22/10/2017      &       7       &       EXT     &       3c84    &       Uranus  &       mwc349a & 0.08--0.10    \\
08/11/2017      &       8       &       EXT     &       3c84    &       Uranus  &       mwc349a & 0.07  \\
09/11/2017      &       7       &       EXT     &       3c84    &       Uranus  &       mwc349a & 0.07  \\
10/11/2017      &       8       &       EXT     &       3c84    &       Uranus  &       mwc349a & 0.07  \\ \hline
\end{tabular}
\end{table*}

\section{Column density comparison}
\label{App:ColumnDensities_compare}

The column density comparisons with other sources at positions 1 and 3 (see Fig. \ref{Fig:N30_Mol_Grad}) are provided here, with the comparison at position 2 shown and discussed in the text. There is a slight difference, as expected from the difference in column densities of the different positions, with a slightly better agreement with Orion KL and AFGL 4176 at position 3. 
%______________________________________________________________
%         Two column figure
%___________________________________________________________________
\begin{figure*}
    \centering
    \includegraphics[width=16.5cm]{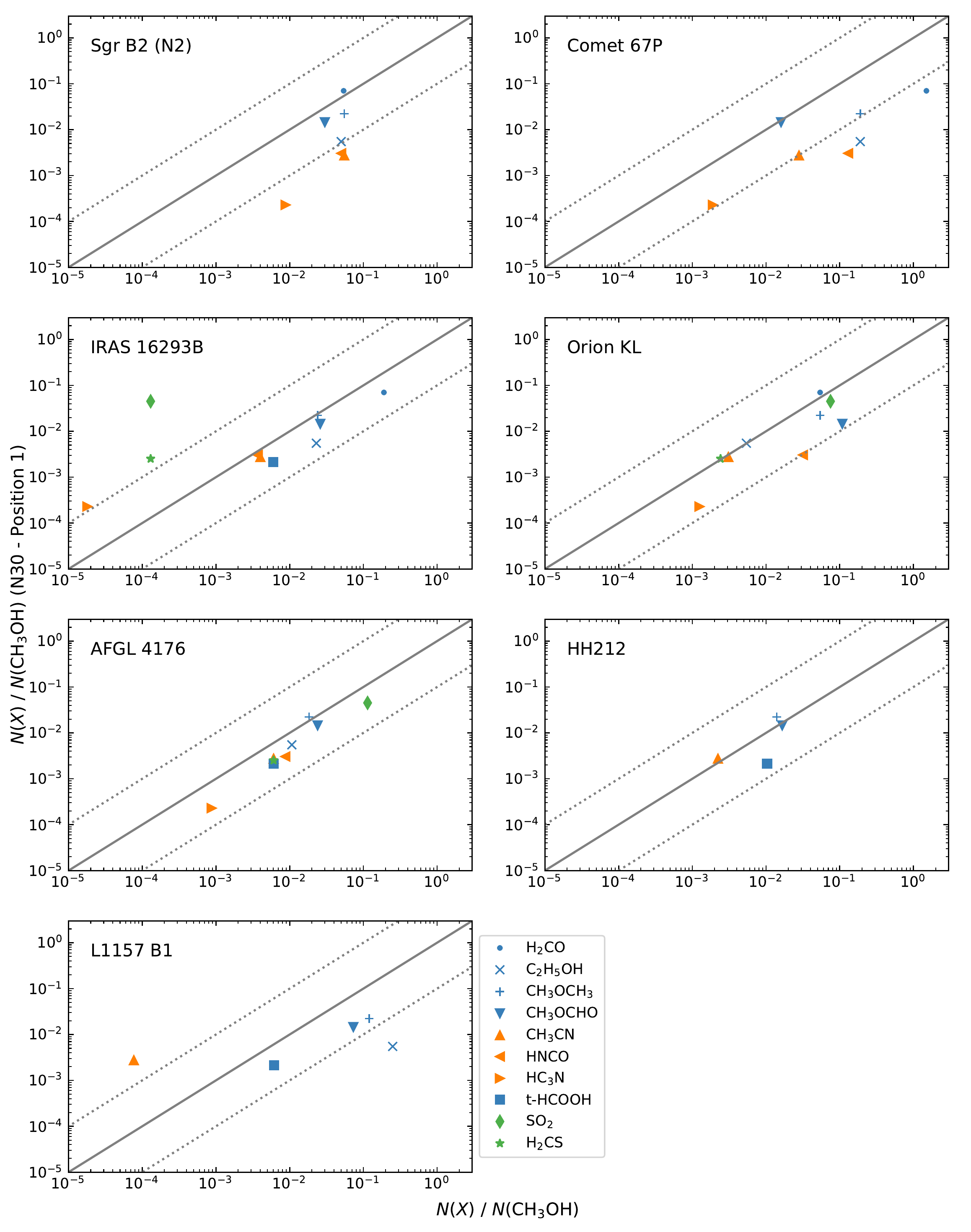}
    \caption{Column densities towards N30, position 1, with respect to CH$_3$OH, on the y-axis, compared to different sources plotted on the x-axis. The solid line represents equal abundances, whereas the dotted lines represent an order of magnitude difference.}
    \label{Fig:N30_pos1_Column_density_compare_sources-wrt_CH3OH_multi}%
\end{figure*}
%______________________________________________________________
%         Two column figure
%___________________________________________________________________
\begin{figure*}
    \centering
    \includegraphics[width=16.5cm]{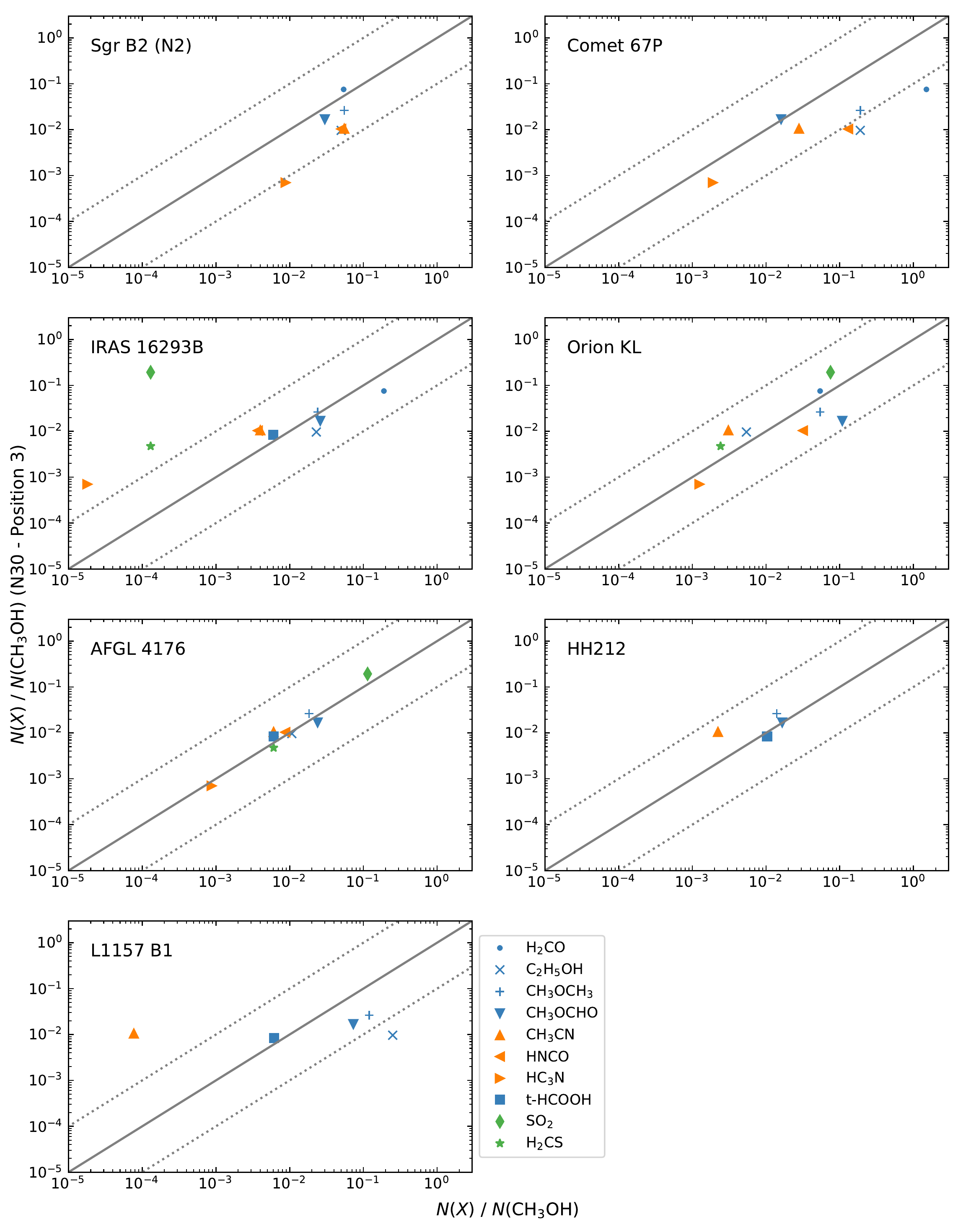}
    \caption{Column densities towards N30, position 3, with respect to CH$_3$OH, on the y-axis, compared to different sources plotted on the x-axis. The solid line represents equal abundances, whereas the dotted lines represent an order of magnitude difference.}
    \label{Fig:N30_pos3_Column_density_compare_sources-wrt_CH3OH_multi}%
\end{figure*}
%______________________________________________________________
%                                  One column figure
%----------------------------------------------------------- 
\begin{figure}
\centering
\includegraphics[width=9cm]{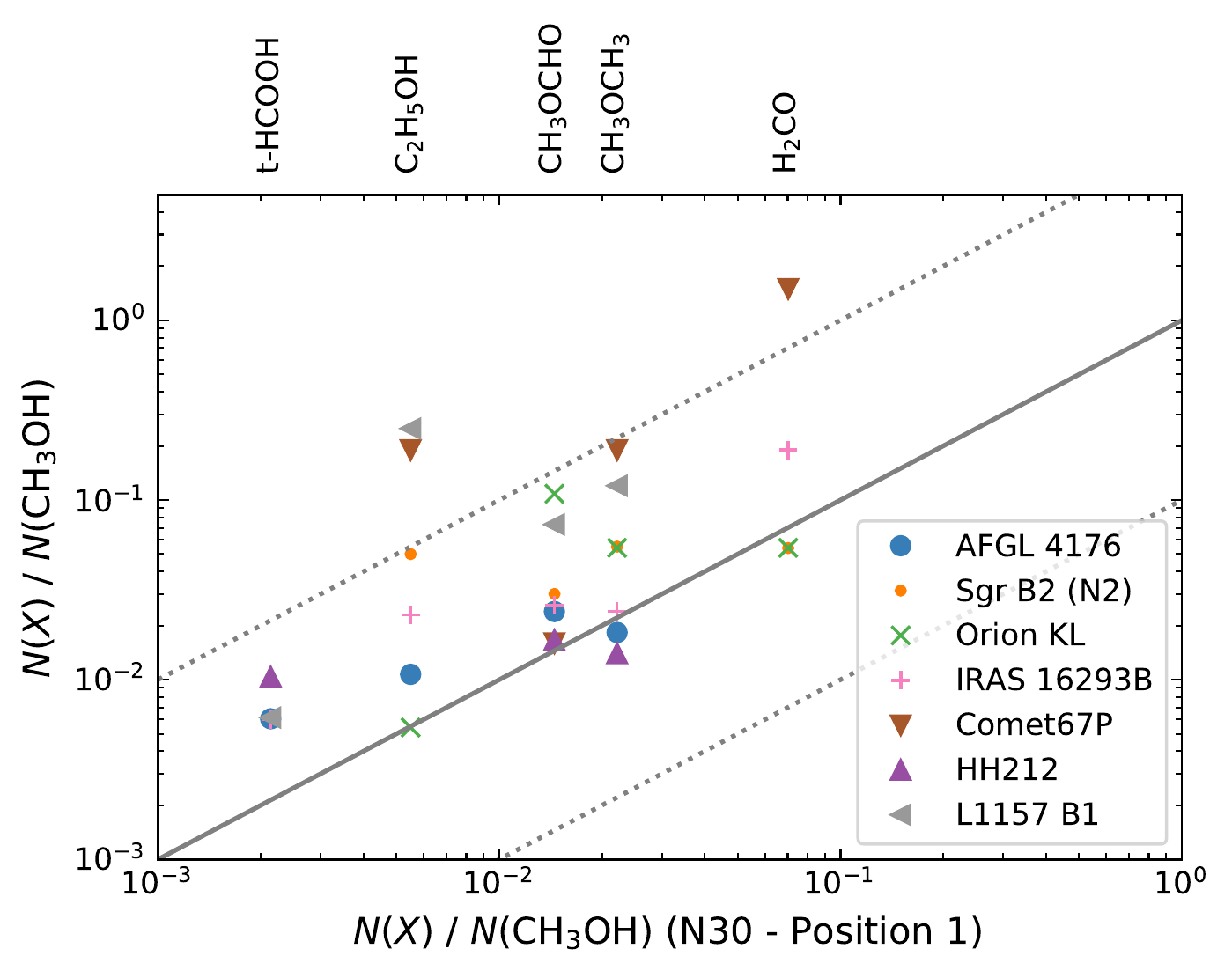}
  \caption{Column densities with respect to CH$_3$OH towards N30, position 1, on the x-axis, compared to different sources plotted on the y-axis. Each source is marked with a different marker. The solid and dotted lines are as in Fig. \ref{Fig:N30_Column_density_compare_sources-wrt_CH3OH_multi}}
  \label{Fig:N30_pos1_Column_density_compare_sources-wrt_CH3OH}
\end{figure}
%______________________________________________________________
%                                  One column figure
%----------------------------------------------------------- 
\begin{figure}
\centering
\includegraphics[width=9cm]{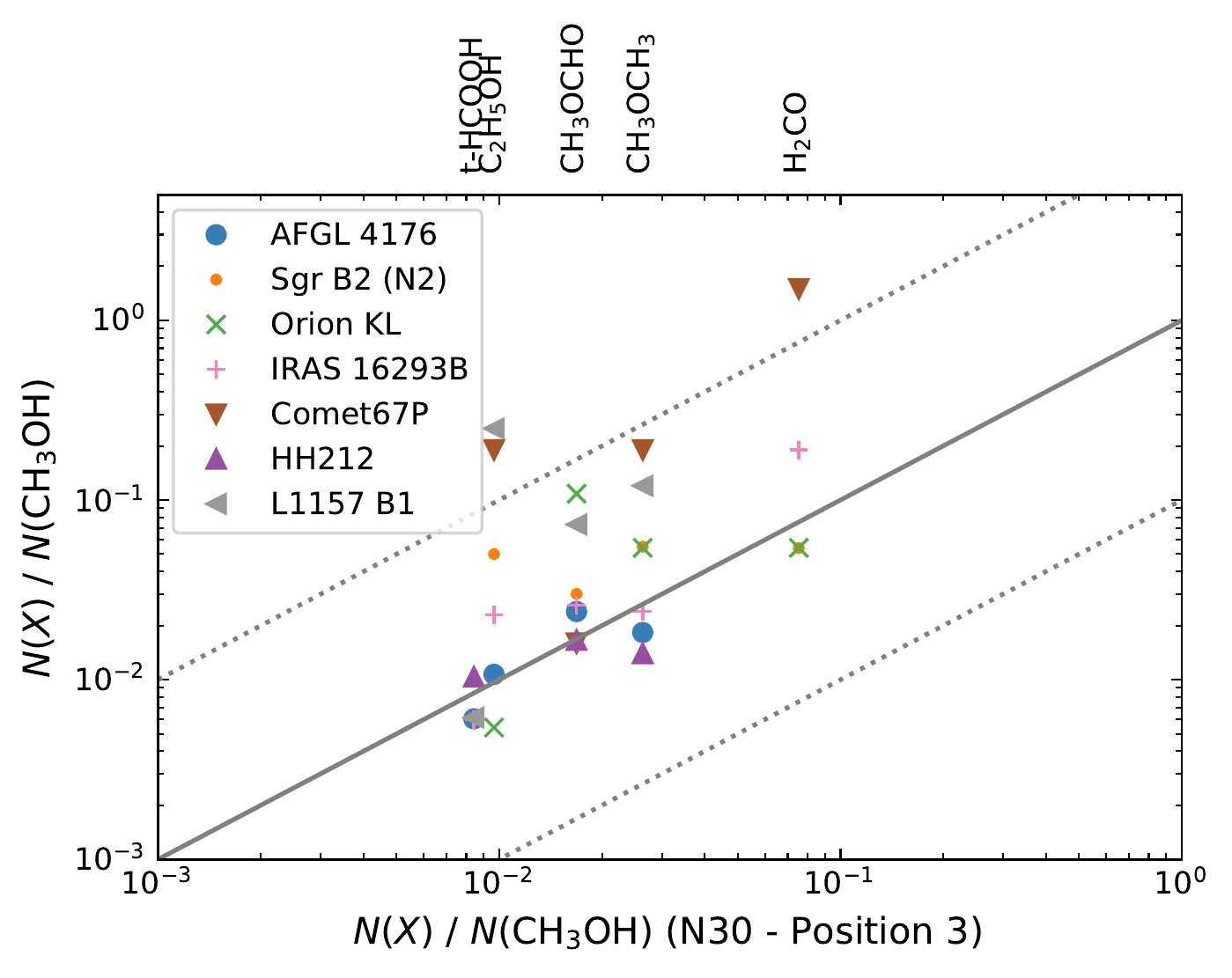}
  \caption{Column densities with respect to CH$_3$OH towards N30, position 3, on the x-axis, compared to different sources plotted on the y-axis. Each source is marked with a different marker. The solid and dotted lines are as in Fig. \ref{Fig:N30_pos1_Column_density_compare_sources-wrt_CH3OH_multi}}
  \label{Fig:N30_pos3_Column_density_compare_sources-wrt_CH3OH}
\end{figure}
%______________________________________________________________
%                                  One column figure
%----------------------------------------------------------- 
\begin{figure}
\centering
\includegraphics[width=9cm]{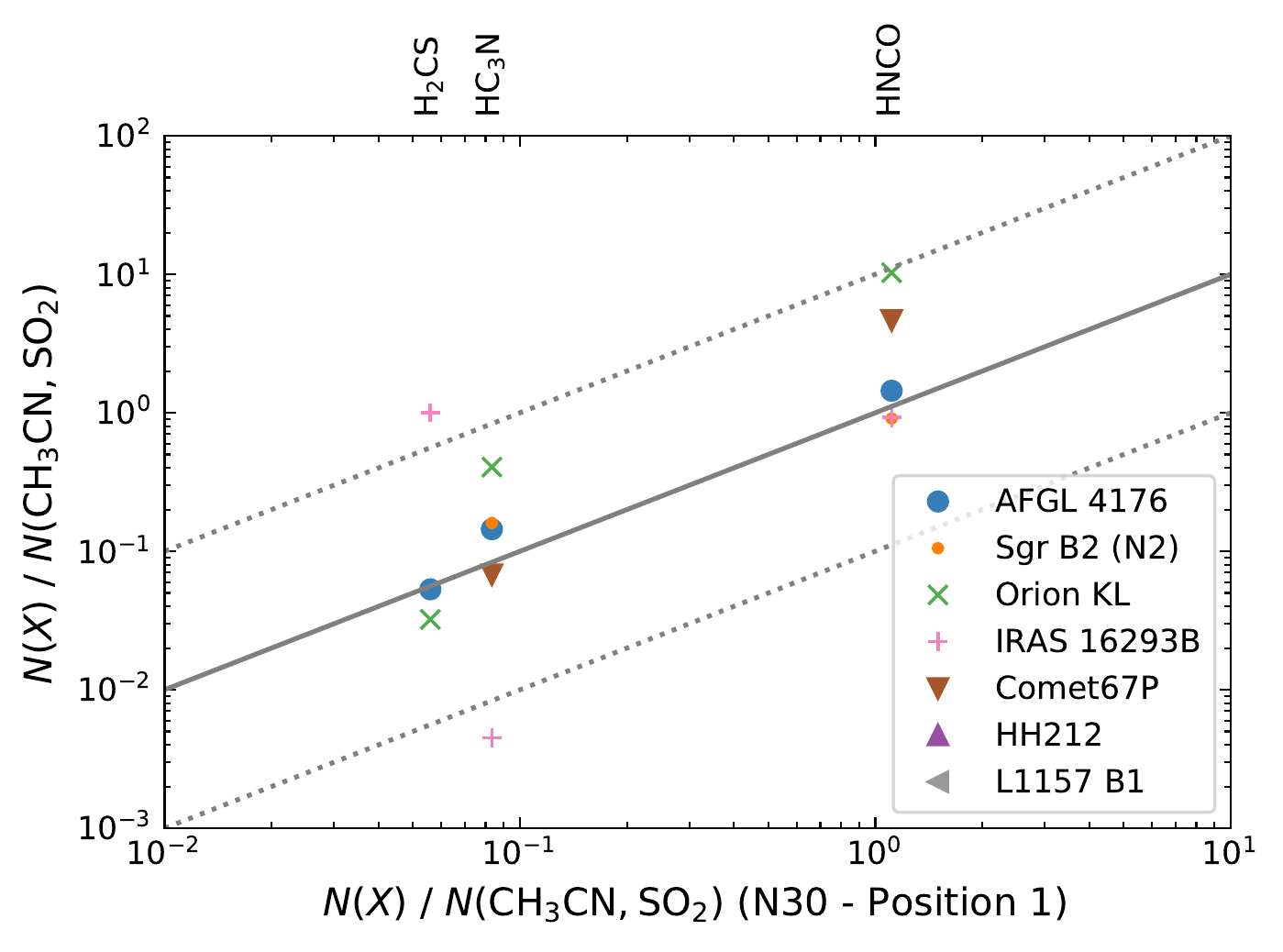}
  \caption{Column densities towards N30 position 1 (on the x-axis), with respect to CH$_3$CN, for the N-bearing species, and SO$_2$ for H$_2$CS, compared to different sources plotted on the y-axis. Each source is marked with a different marker. The solid and dotted lines are as in Fig. \ref{Fig:N30_pos1_Column_density_compare_sources-wrt_CH3OH_multi}}
  \label{Fig:N30_pos1_Column_density_compare_sources-wrt_wrt_CH3CN,SO2}
\end{figure}

%______________________________________________________________
%                                  One column figure
%----------------------------------------------------------- 
\begin{figure}
\centering
\includegraphics[width=9cm]{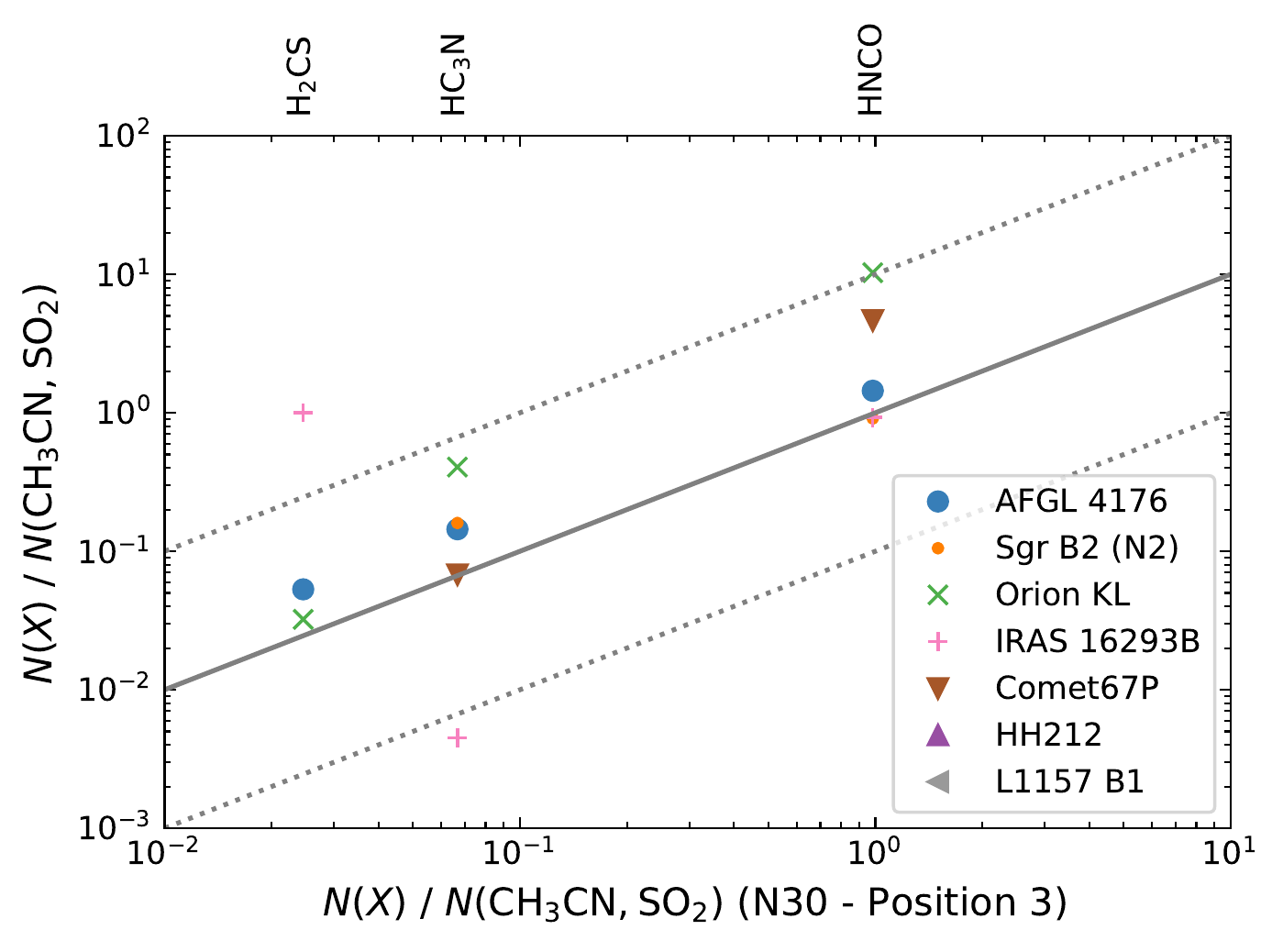}
  \caption{Column densities towards N30 position 3 (on the x-axis), with respect to CH$_3$CN, for the N-bearing species, and SO$_2$ for H$_2$CS, compared to different sources plotted on the y-axis. Each source is marked with a different marker. The solid and dotted lines are as in Fig. \ref{Fig:N30_pos1_Column_density_compare_sources-wrt_CH3OH_multi}}
  \label{Fig:N30_pos3_Column_density_compare_sources-wrt_wrt_CH3CN,SO2}
\end{figure}

\section{Position--velocity maps}
\label{App:PV_plots}
\begin{figure*}
\centering
\includegraphics[width=16cm]{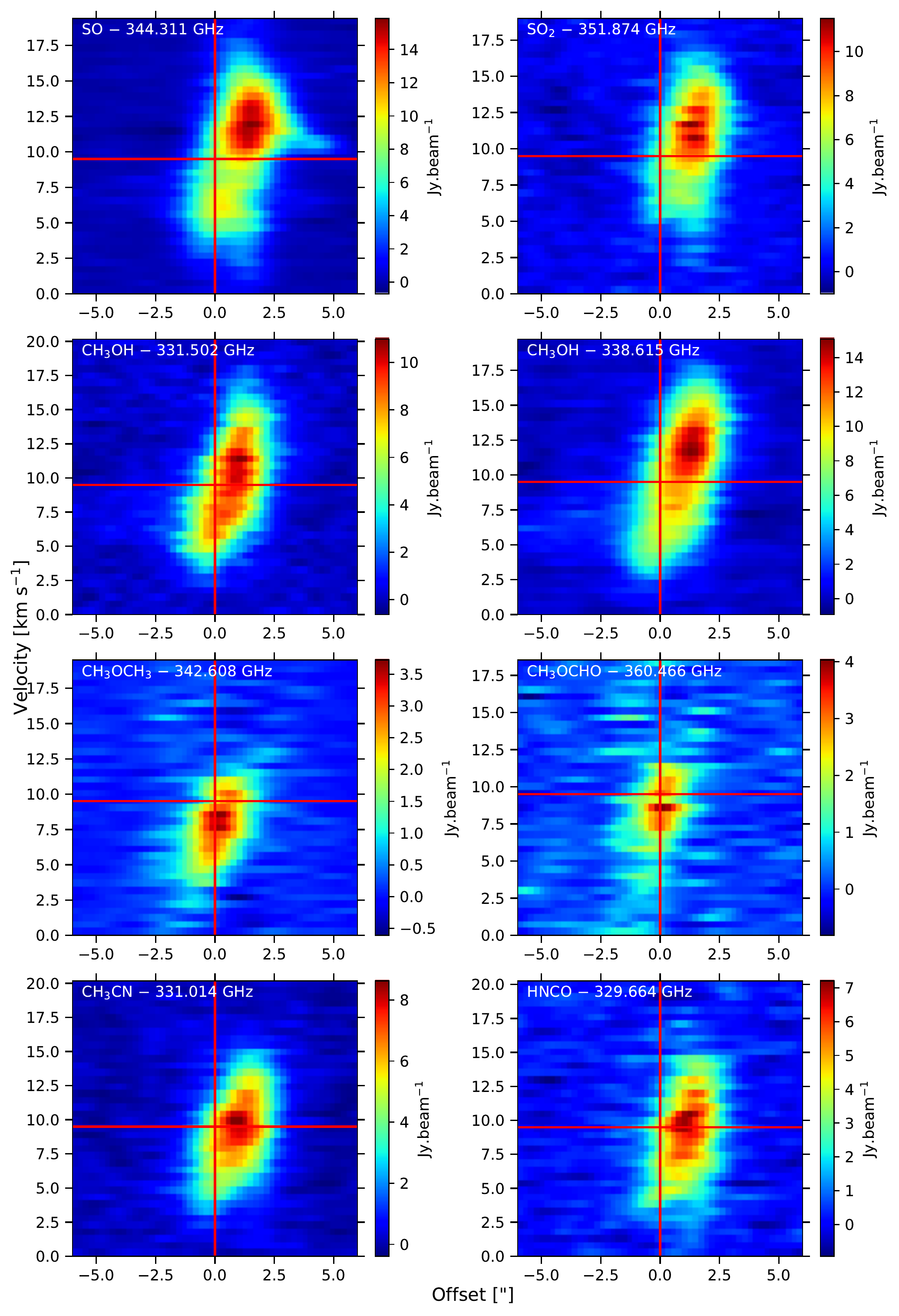}
  \caption{Position--velocity maps of selected molecular lines. The red vertical line represents the position of 0 $''$ offset from the centre position between the peak red- and blueshifted H$_2$CO emission (see \ref{Fig:N30_PV_plot_H2CO}). The horizontal line represents the systemic velocity, also for H$_2$CO.}
  \label{Fig:PV_plots}
\end{figure*}

Position--velocity maps of selected molecular lines are presented in Fig. \ref{Fig:PV_plots}, with the maps of H$_2$CO and CS shown (Figs. \ref{Fig:N30_PV_plot_H2CO} and \ref{Fig:N30_PV_plot_CS} respectively) and discussed in the text. A full kinematics study of the source is beyond the scope of this work, but it is clear from Fig. \ref{Fig:PV_plots} that different molecules trace different regions in the system, which is also seen in the molecular gradient that we observe.

\section{Molecular line plots}
\label{App:Line_plots}

In this Appendix we present plots of all the lines for each of the molecules listed in Table \ref{table:N30_column_densities_Meth}, taken at position 2 in Fig. \ref{Fig:N30_Mol_Grad}, for which we made a model fit. In all the plots, the model is overplotted in red over the observed spectrum in black. Only lines detected above $3\sigma$ $\sim$10 K intensity are shown.

%______________________________________________________________
%                                  Two column figure
%----------------------------------------------------------- 
% 13CH3OH lines
    \begin{figure*}
    \centering
    \includegraphics[width=17.5cm]{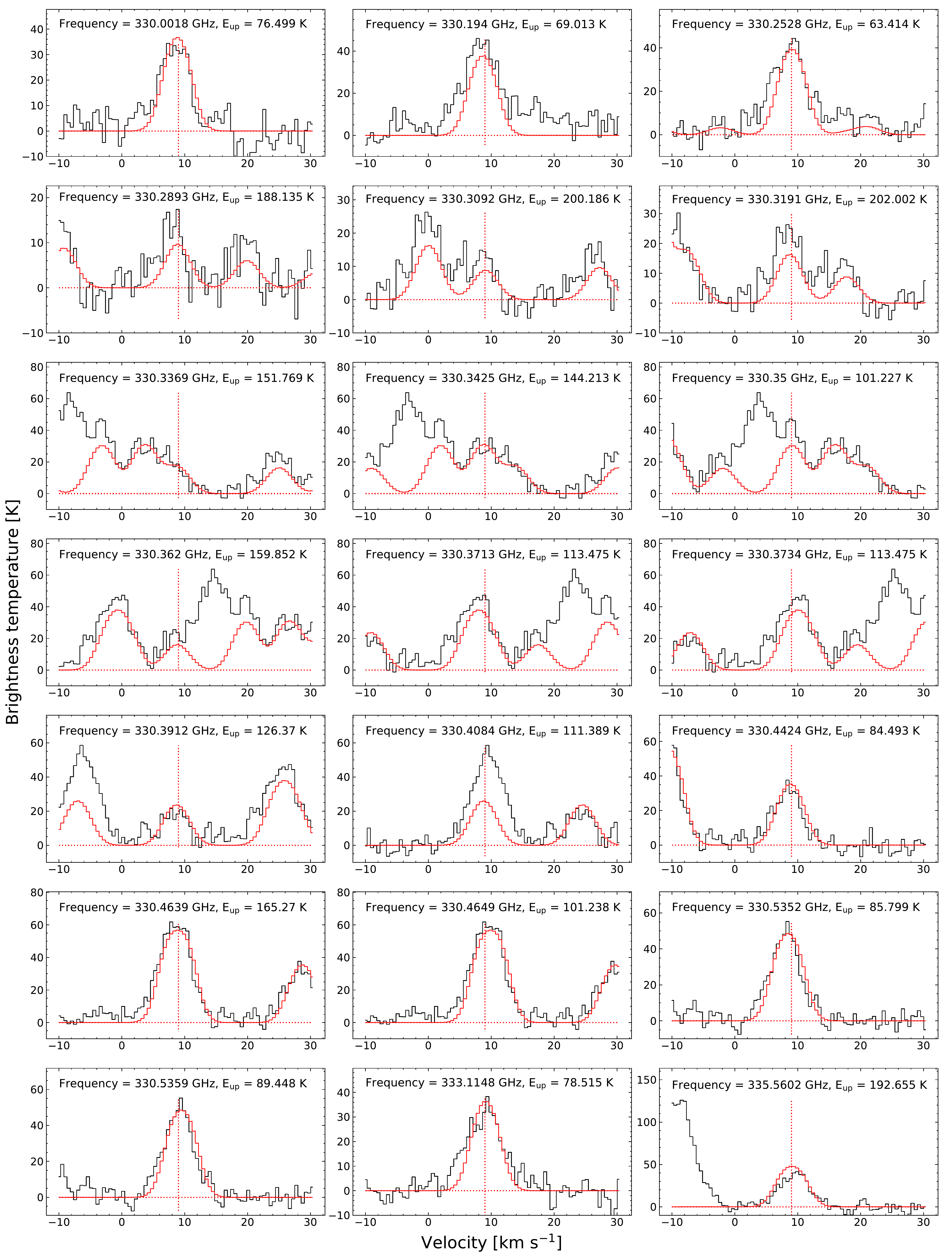}
      \caption{$^{13}$CH$_3$OH synthetic spectrum overplotted on the observed spectrum. The fitted values obtained were column density = 1.3 $\times 10^{16} \mathrm{cm^{-2}}$, excitation temperature (T$_{\mathrm{ex}}$) = 120.0 K, line width (FWHM) = 4.5 $\mathrm{km \ s^{-1}}$, and source velocity ($\varv_{\text{source}}$) = 9.0 $\mathrm{km \ s^{-1}}.$}
      \label{Fig:13CH3OH_lines_1}
    \end{figure*} 
    
    \begin{figure*}
    \centering
    \includegraphics[width=17.5cm]{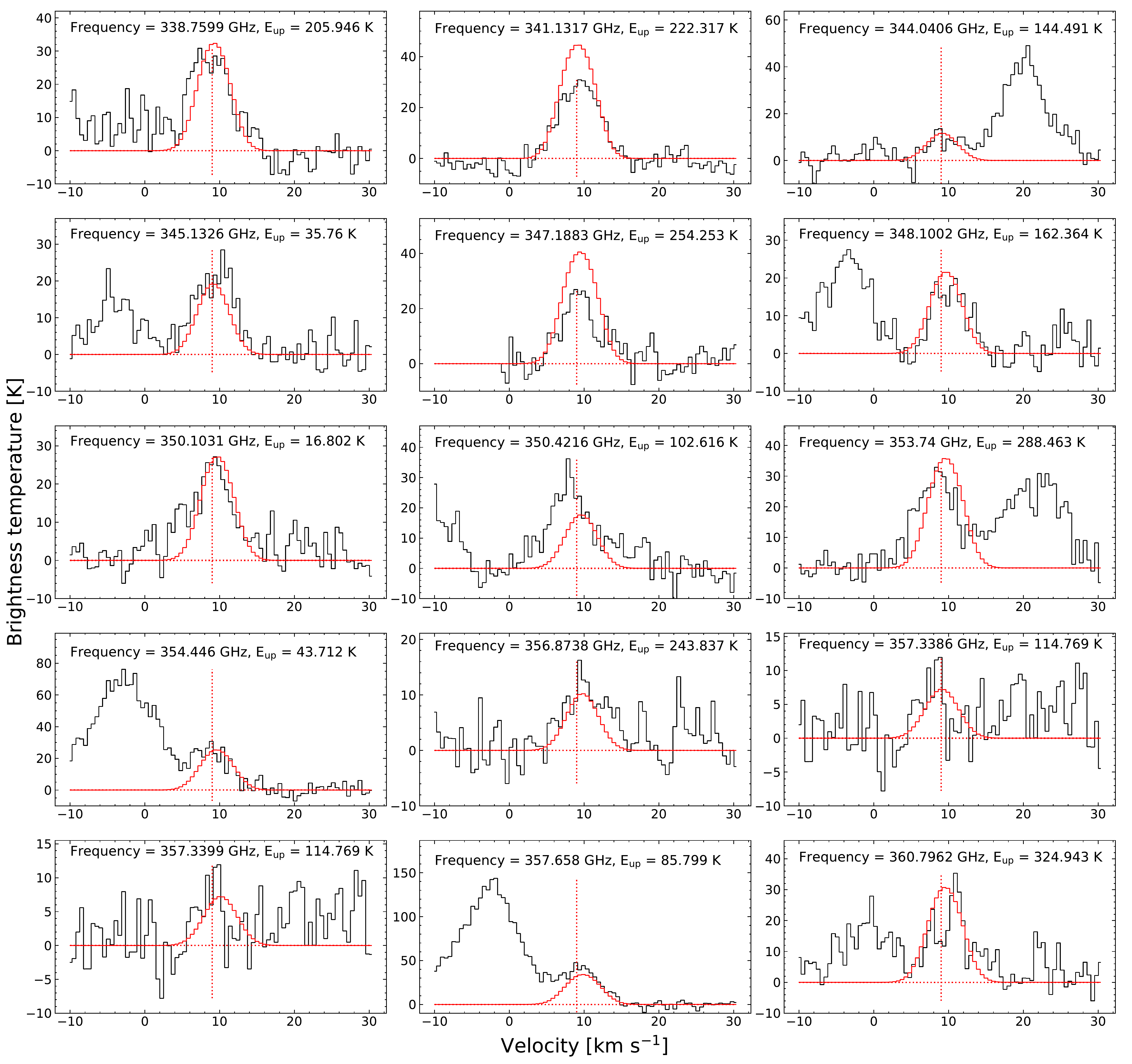}
      \caption{$^{13}$CH$_3$OH synthetic spectrum overplotted on the observed spectrum. The fitted values obtained were column density = 1.3$\times 10^{16} \mathrm{cm^{-2}}$, excitation temperature (T$_{\mathrm{ex}}$) = 120.0 K, line width (FWHM) = 4.5 $\mathrm{km \ s^{-1}}$, and source velocity ($\varv_{\text{source}}$) = 9.00 $\mathrm{km \ s^{-1}}.$}
      \label{Fig:13CH3OH_lines_2}
    \end{figure*} 
    
% CH3OH lines
    \begin{figure*}
    \centering
    \includegraphics[width=17.5cm]{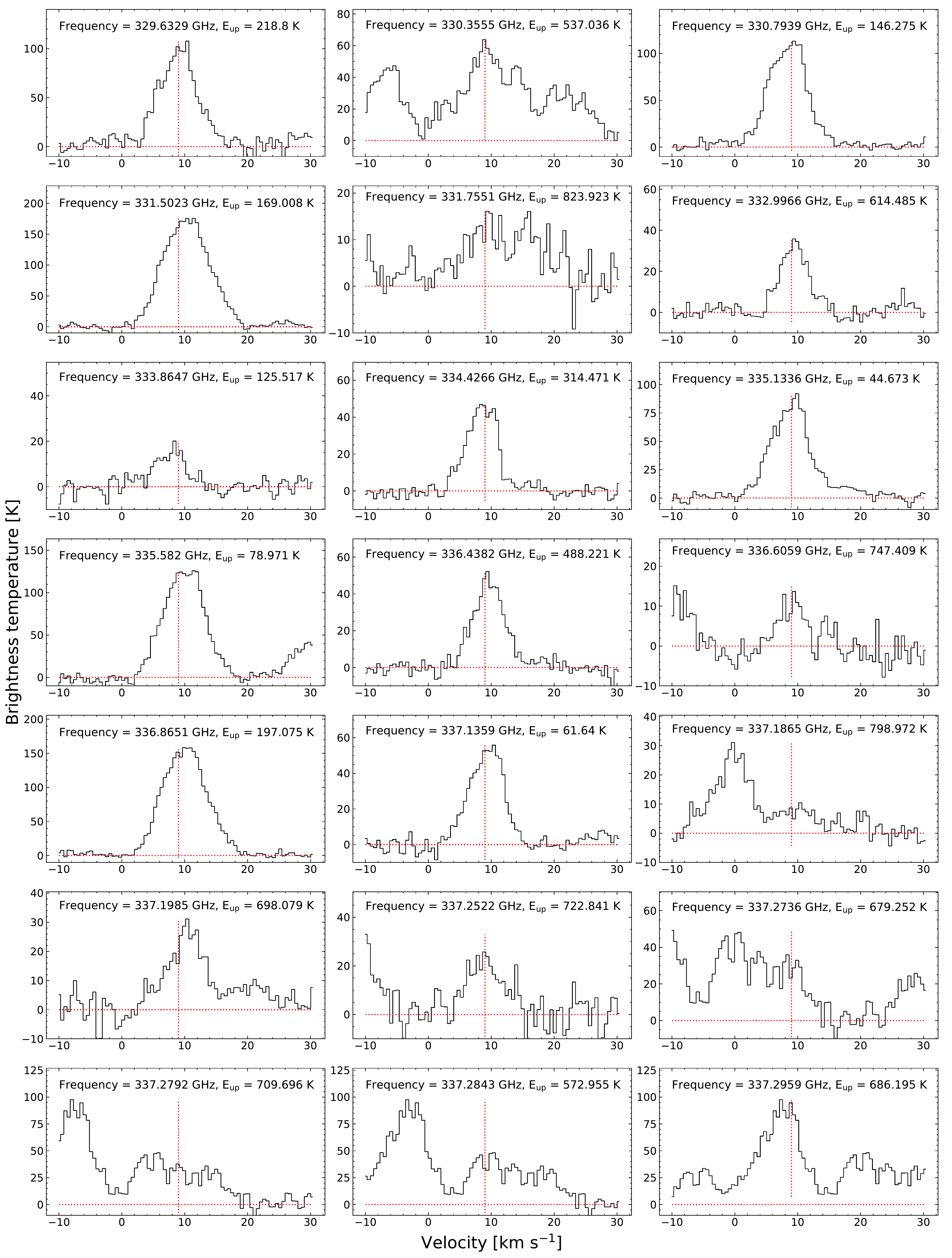}
      \caption{CH$_3$OH observed spectral lines. In this case the column density was obtained by multiplying the fitted column density of $^{13}$CH$_3$OH with the ISM isotopologue ratio of $^{12}$C/$^{13}$C = 77. The obtained column density was 1.0$\times 10^{19} \mathrm{cm^{-2}}$.}
      \label{Fig:CH3OH_lines_1}
    \end{figure*} 

    \begin{figure*}
    \centering
    \includegraphics[width=17.5cm]{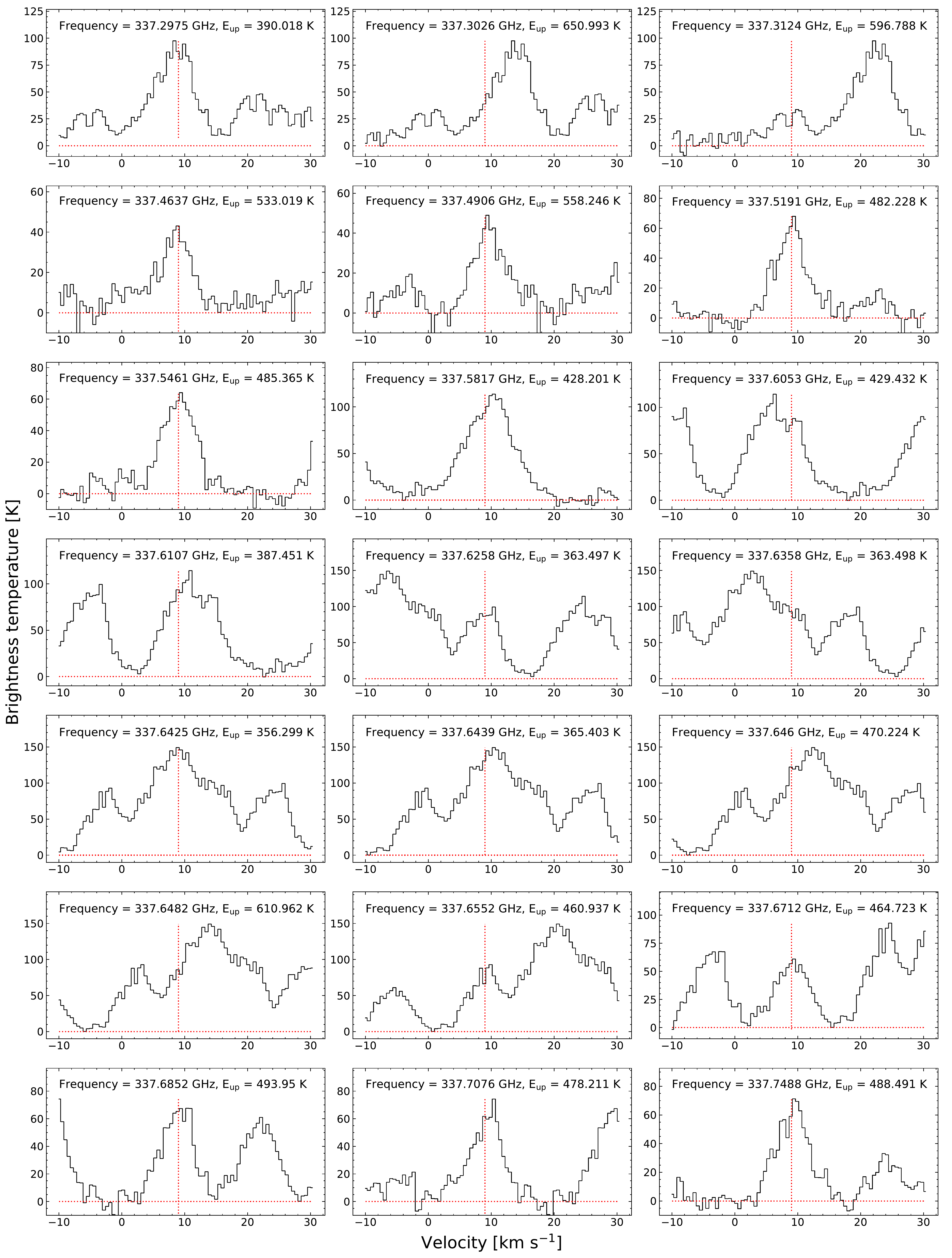}
      \caption{CH$_3$OH observed spectral lines. In this case the column density was obtained by multiplying the fitted column density of $^{13}$CH$_3$OH with the ISM isotopologue ratio of $^{12}$C/$^{13}$C = 77. The obtained column density was 1.0$\times 10^{19} \mathrm{cm^{-2}}$.}
      \label{Fig:CH3OH_lines_2}
    \end{figure*}
    
    \begin{figure*}
    \centering
    \includegraphics[width=17.5cm]{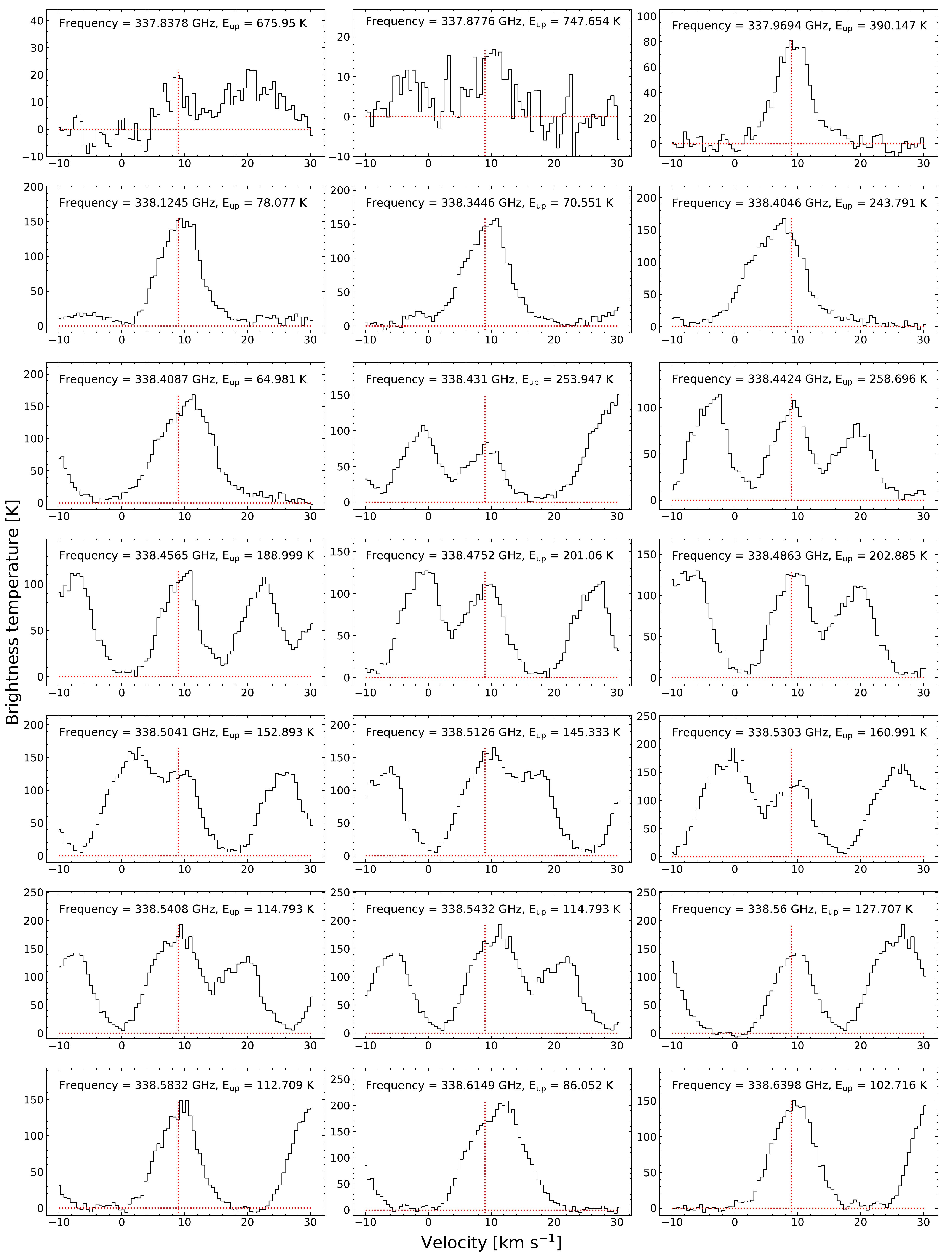}
      \caption{CH$_3$OH observed spectral lines. In this case the column density was obtained by multiplying the fitted column density of $^{13}$CH$_3$OH with the ISM isotopologue ratio of $^{12}$C/$^{13}$C = 77. The obtained column density was 1.0$\times 10^{19} \mathrm{cm^{-2}}$.}
      \label{Fig:CH3OH_lines_3}
    \end{figure*} 

    \begin{figure*}
    \centering
    \includegraphics[width=17.5cm]{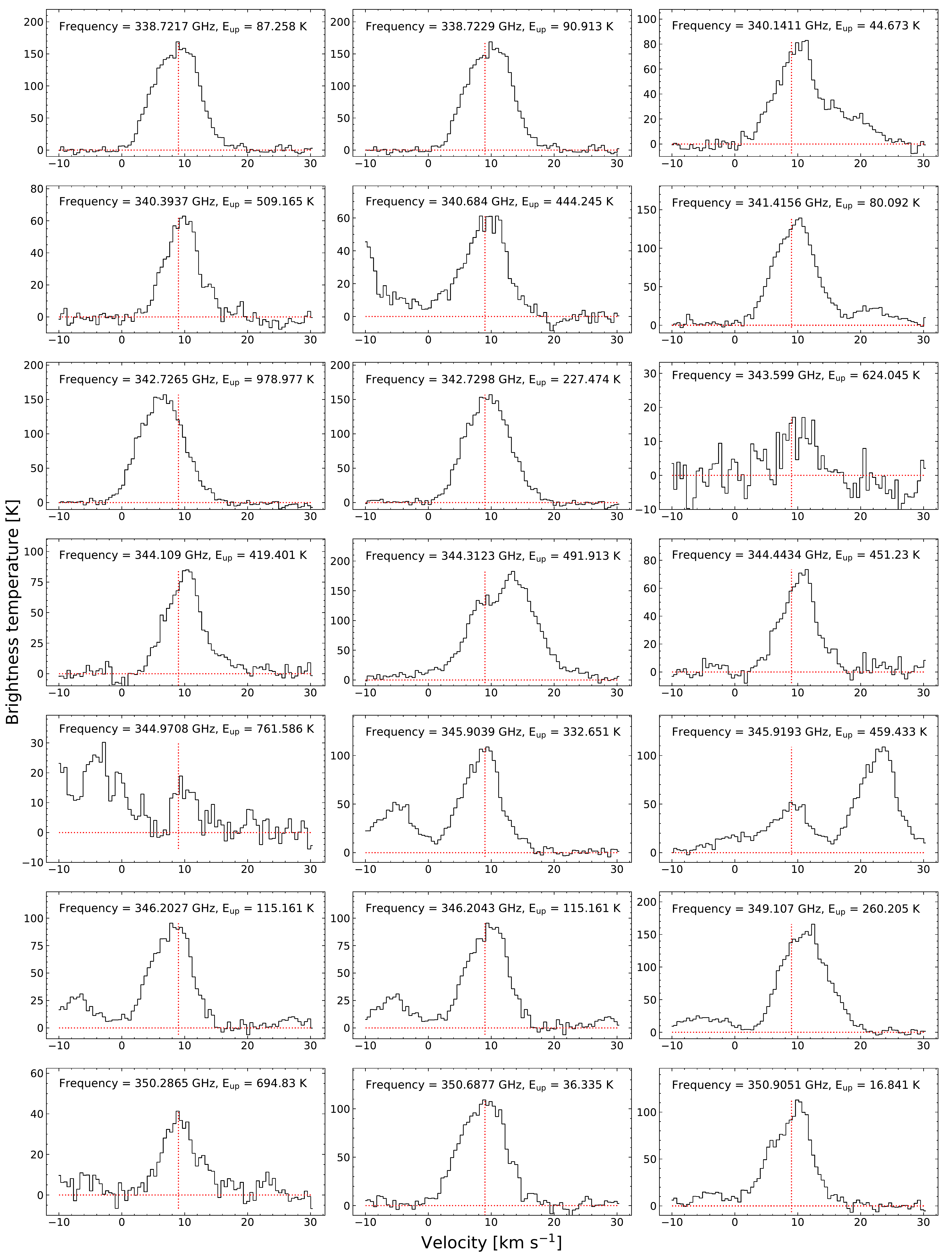}
      \caption{CH$_3$OH observed spectral lines. In this case the column density was obtained by multiplying the fitted column density of $^{13}$CH$_3$OH with the ISM isotopologue ratio of $^{12}$C/$^{13}$C = 77. The obtained column density was 1.0$\times 10^{19} \mathrm{cm^{-2}}$.}
      \label{Fig:CH3OH_lines_4}
    \end{figure*} 
    
    \begin{figure*}
    \centering
    \includegraphics[width=17.5cm]{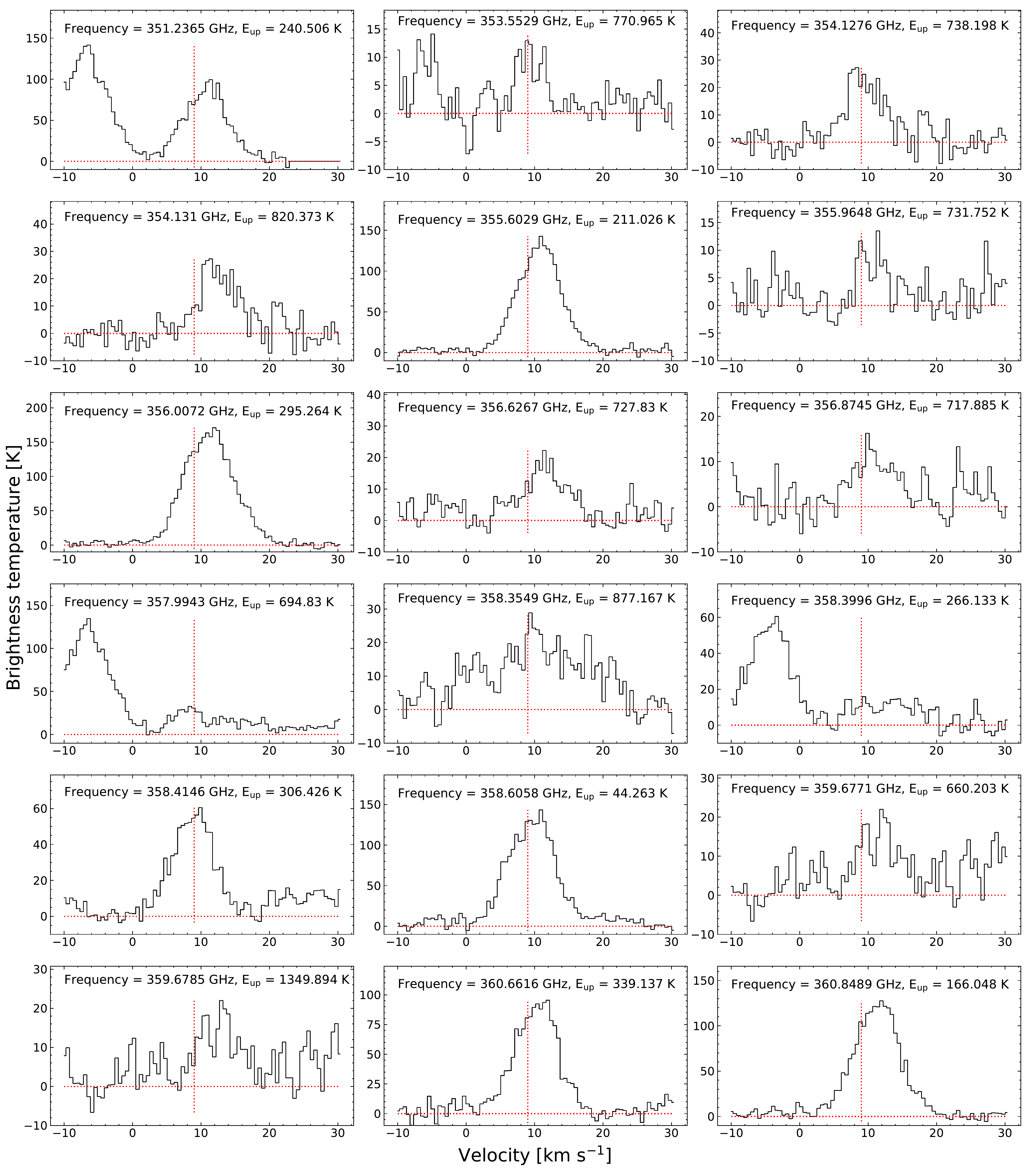}
      \caption{CH$_3$OH observed spectral lines. In this case the column density was obtained by multiplying the fitted column density of $^{13}$CH$_3$OH with the ISM isotopologue ratio of $^{12}$C/$^{13}$C = 77. The obtained column density was 1.0$\times 10^{19} \mathrm{cm^{-2}}$.}
      \label{Fig:CH3OH_lines_5}
    \end{figure*}
        
% C2H5OH lines
    \begin{figure*}
    \centering
    \includegraphics[width=17.5cm]{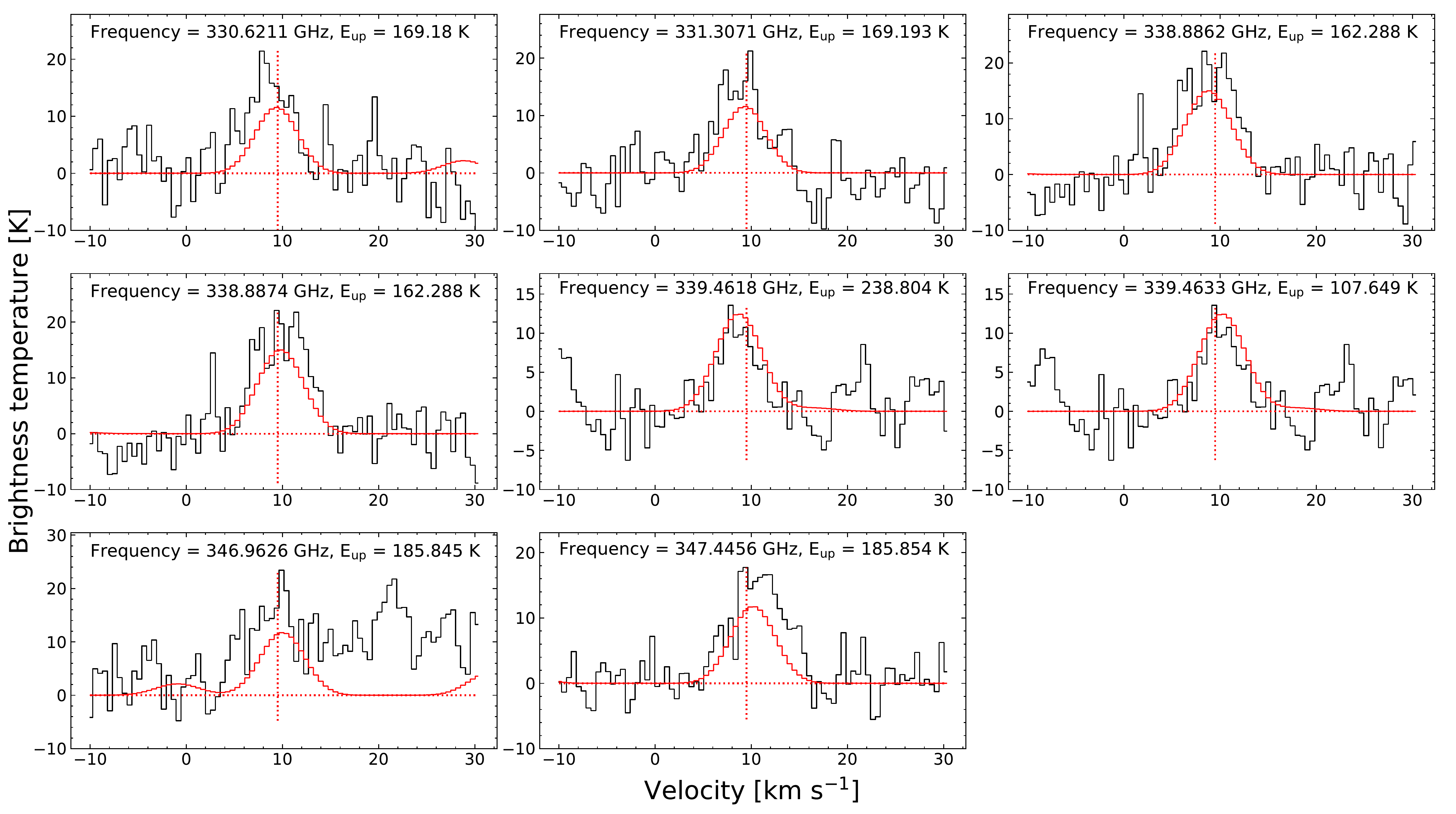}
      \caption{C$_2$H$_5$OH model lines and synthetic spectrum. The model fitted values obtained were column density = 6.7$\times 10^{16} \mathrm{cm^{-2}}$, excitation temperature (T$_{\mathrm{ex}}$) = 130.0 K, line width (FWHM) = 5.0 $\mathrm{km \ s^{-1}}$, and source velocity ($\varv_{\text{source}}$) = 9.5.}
      \label{Fig:C2H5OH_lines_1}
    \end{figure*} 
    
% CH3OCHO lines
    \begin{figure*}
    \centering
    \includegraphics[width=17.5cm]{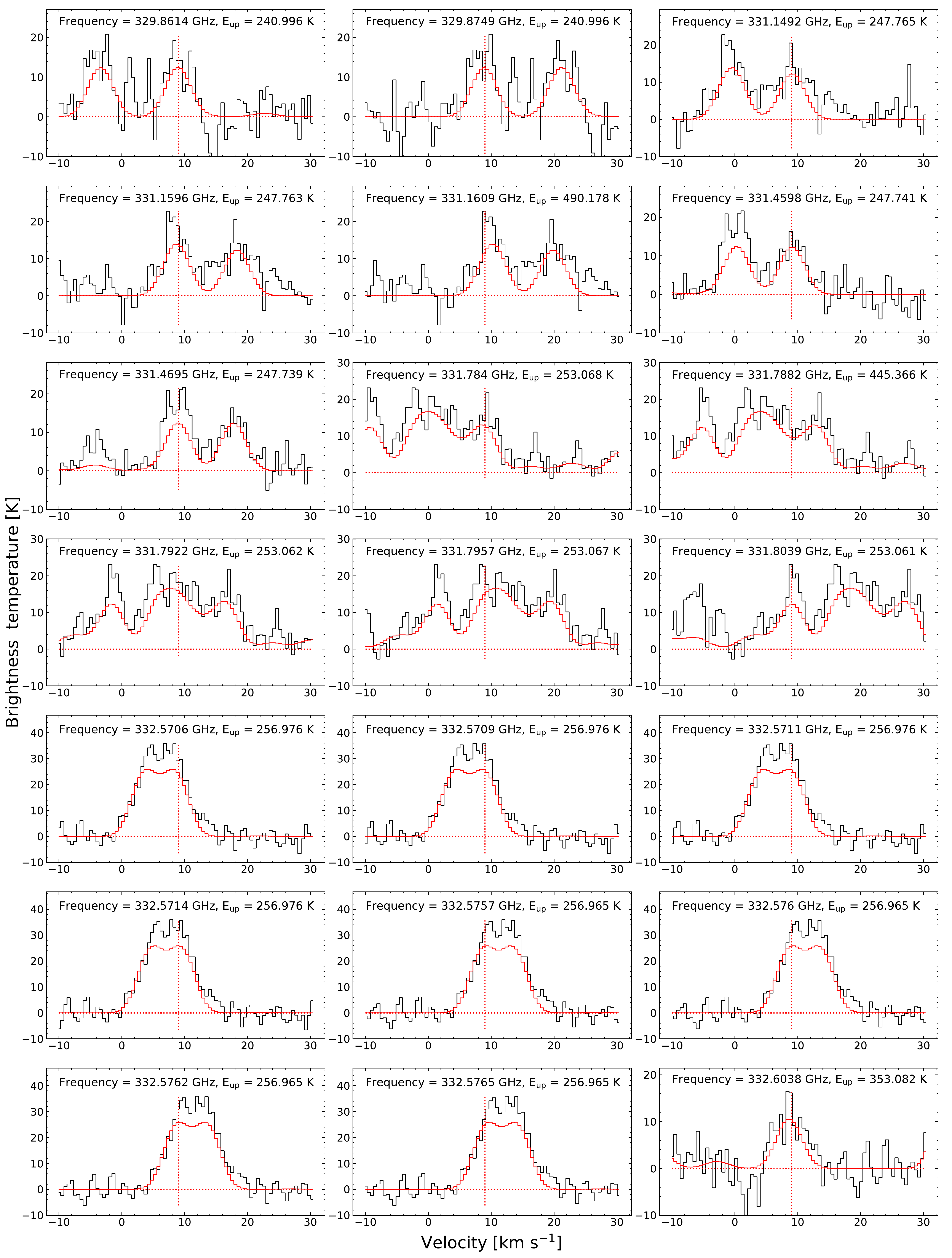}
      \caption{CH$_3$OCHO model lines and synthetic spectrum. The model fitted values obtained were column density = 1.4$\times 10^{17} \mathrm{cm^{-2}}$, excitation temperature (T$_{\mathrm{ex}}$) = 110.0 K, line width (FWHM) = 4.5 $\mathrm{km \ s^{-1}}$, and source velocity ($\varv_{\text{source}}$) = 9.0 $\mathrm{km \ s^{-1}}.$}
      \label{Fig:CH3OCHO_lines_1}
    \end{figure*} 
    
    \begin{figure*}
    \centering
    \includegraphics[width=17.5cm]{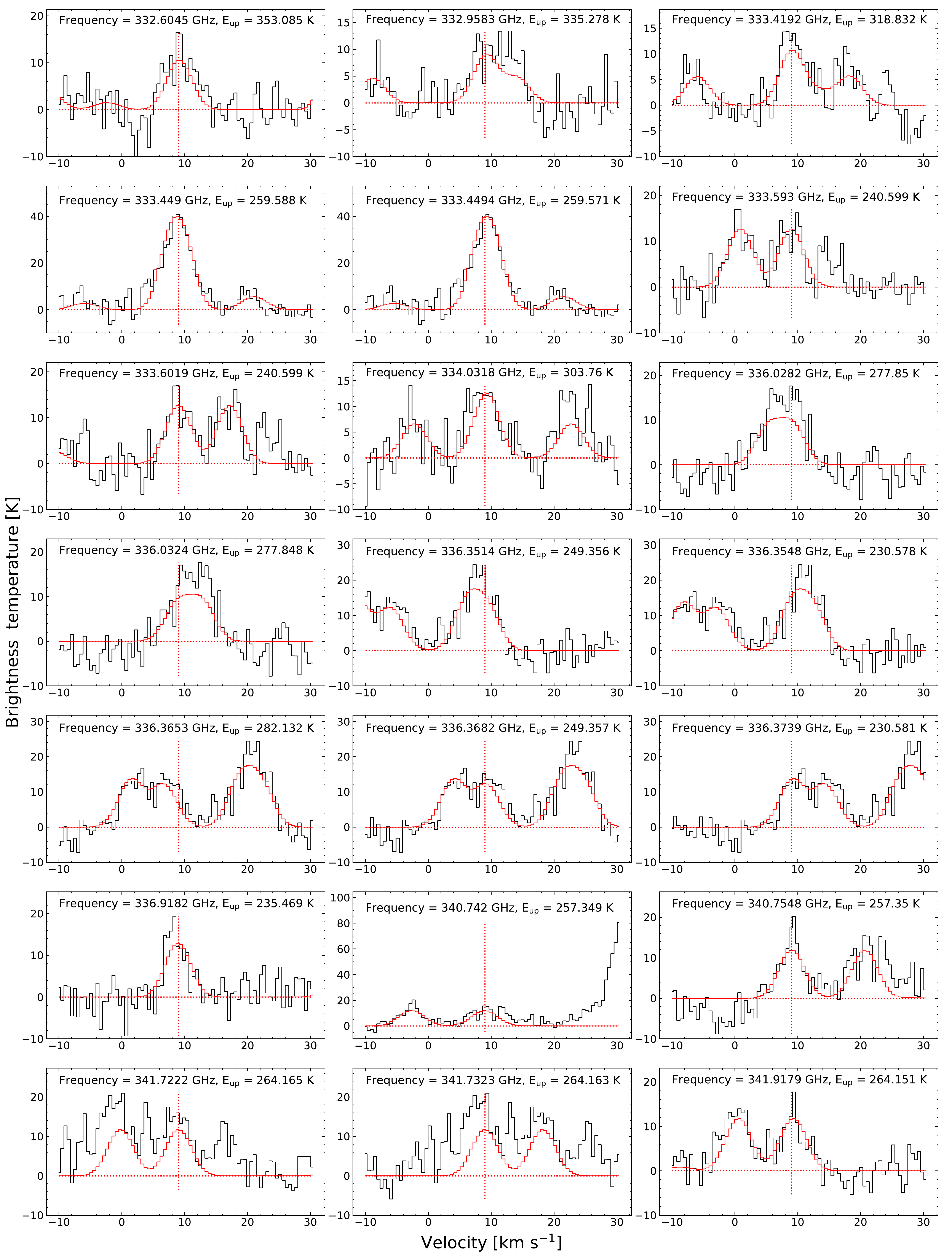}
      \caption{CH$_3$OCHO model lines and synthetic spectrum. The model fitted values obtained were column density = 1.4$\times 10^{17} \mathrm{cm^{-2}}$, excitation temperature (T$_{\mathrm{ex}}$) = 110.0 K, line width (FWHM) = 4.5 $\mathrm{km \ s^{-1}}$, and source velocity ($\varv_{\text{source}}$) = 9.0 $\mathrm{km \ s^{-1}}.$}
      \label{Fig:CH3OCHO_lines_2}
    \end{figure*} 
    
    \begin{figure*}
    \centering
    \includegraphics[width=17.5cm]{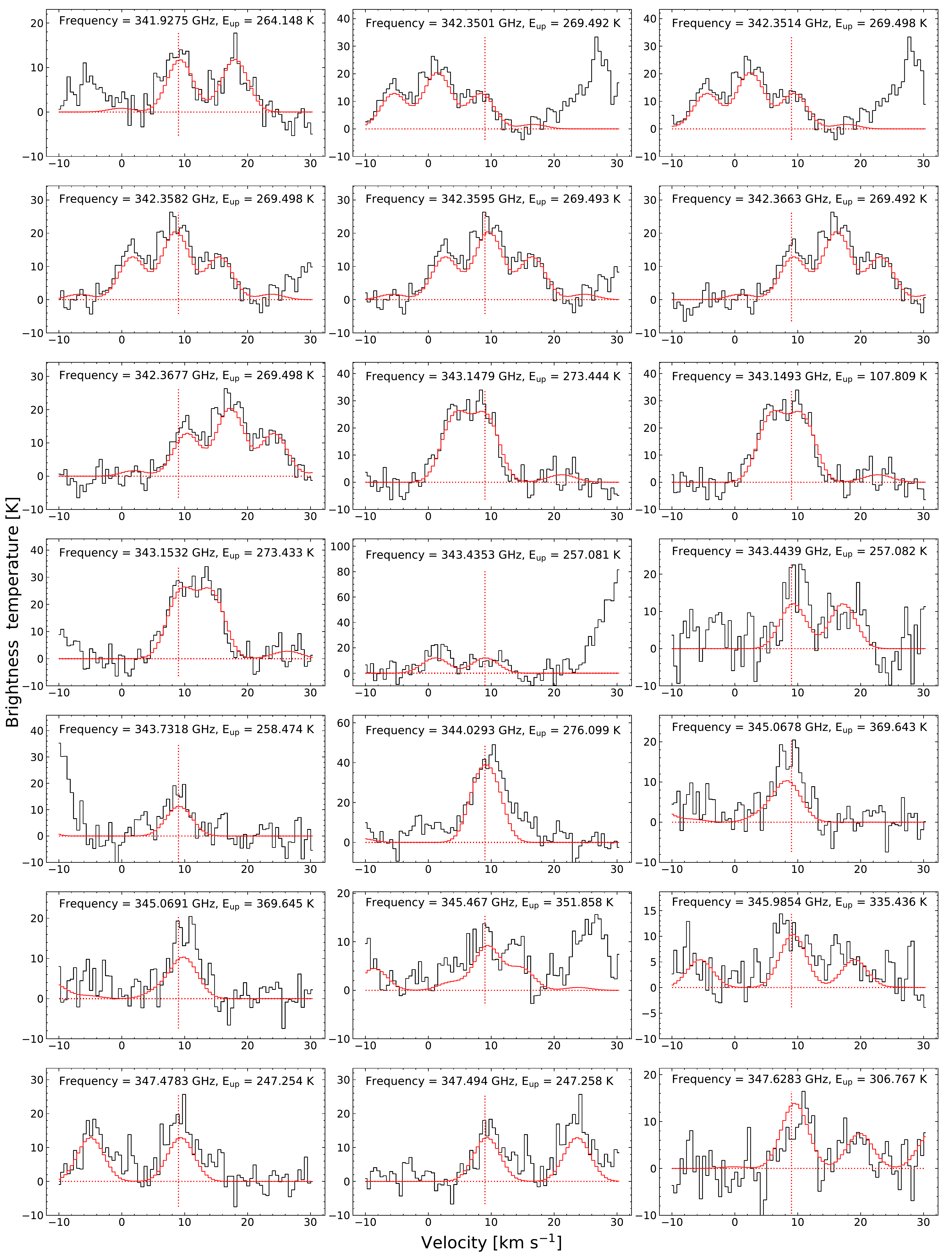}
      \caption{CH$_3$OCHO model lines and synthetic spectrum. The model fitted values obtained were column density = 1.4$\times 10^{17} \mathrm{cm^{-2}}$, excitation temperature (T$_{\mathrm{ex}}$) = 110.0 K, line width (FWHM) = 4.5 $\mathrm{km \ s^{-1}}$, and source velocity ($\varv_{\text{source}}$) = 9.0 $\mathrm{km \ s^{-1}}.$}
      \label{Fig:CH3OCHO_lines_3}
    \end{figure*} 
    
    \begin{figure*}
    \centering
    \includegraphics[width=17.5cm]{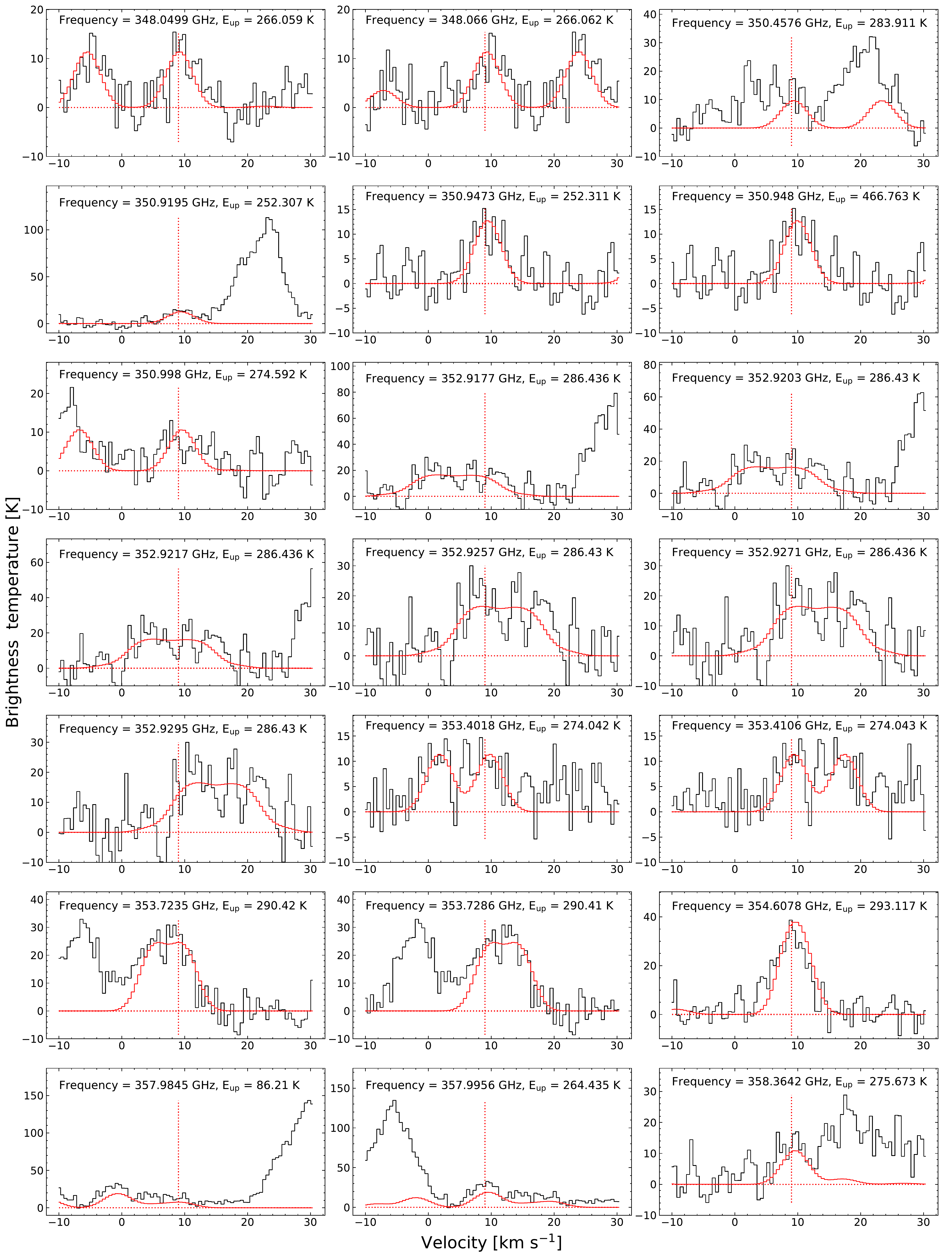}
      \caption{CH$_3$OCHO model lines and synthetic spectrum. The model fitted values obtained were column density = 1.4$\times 10^{17} \mathrm{cm^{-2}}$, excitation temperature (T$_{\mathrm{ex}}$) = 110.0 K, line width (FWHM) = 4.5 $\mathrm{km \ s^{-1}}$, and source velocity ($\varv_{\text{source}}$) = 9.0 $\mathrm{km \ s^{-1}}.$}
      \label{Fig:CH3OCHO_lines_4}
    \end{figure*} 
    
    \begin{figure*}
    \centering
    \includegraphics[width=17.5cm]{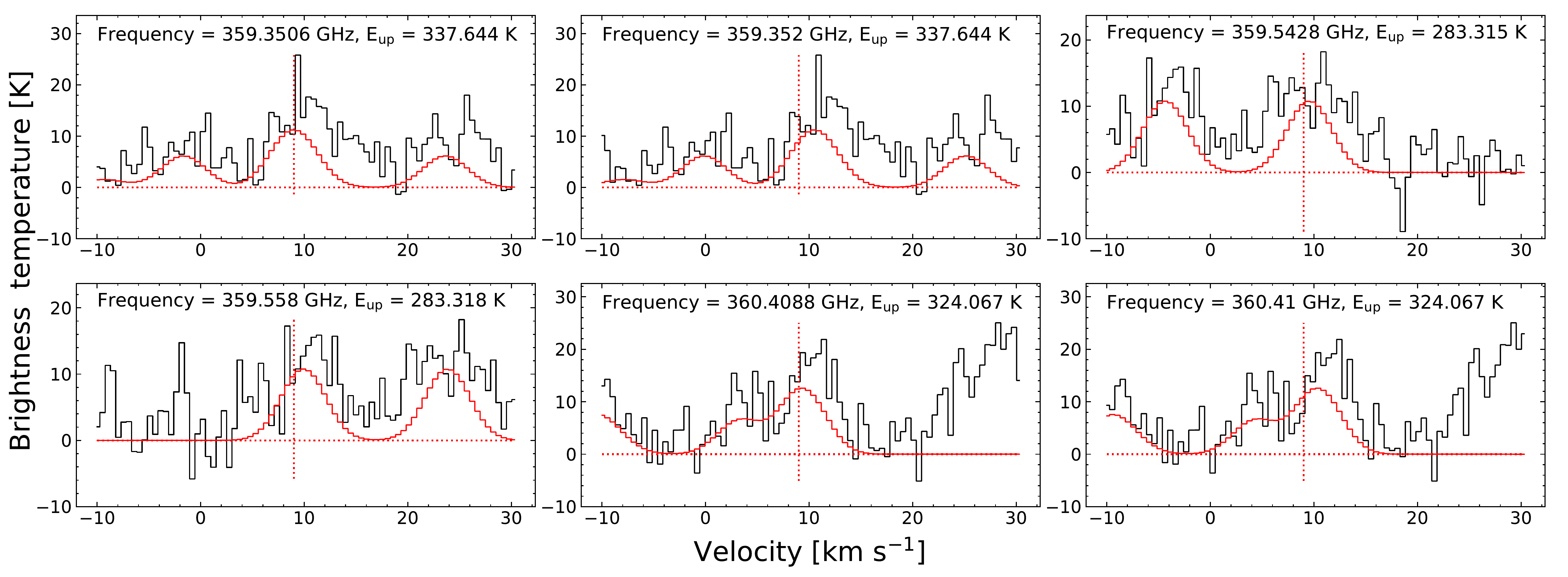}
      \caption{CH$_3$OCHO model lines and synthetic spectrum. The model fitted values obtained were column density = 1.4$\times 10^{17} \mathrm{cm^{-2}}$, excitation temperature (T$_{\mathrm{ex}}$) = 110.0 K, line width (FWHM) = 4.5 $\mathrm{km \ s^{-1}}$, and source velocity ($\varv_{\text{source}}$) = 9.0 $\mathrm{km \ s^{-1}}.$}
      \label{Fig:CH3OCHO_lines_5}
    \end{figure*} 
        
% CH3OCH3 lines
    \begin{figure*}
    \centering
    \includegraphics[width=17.5cm]{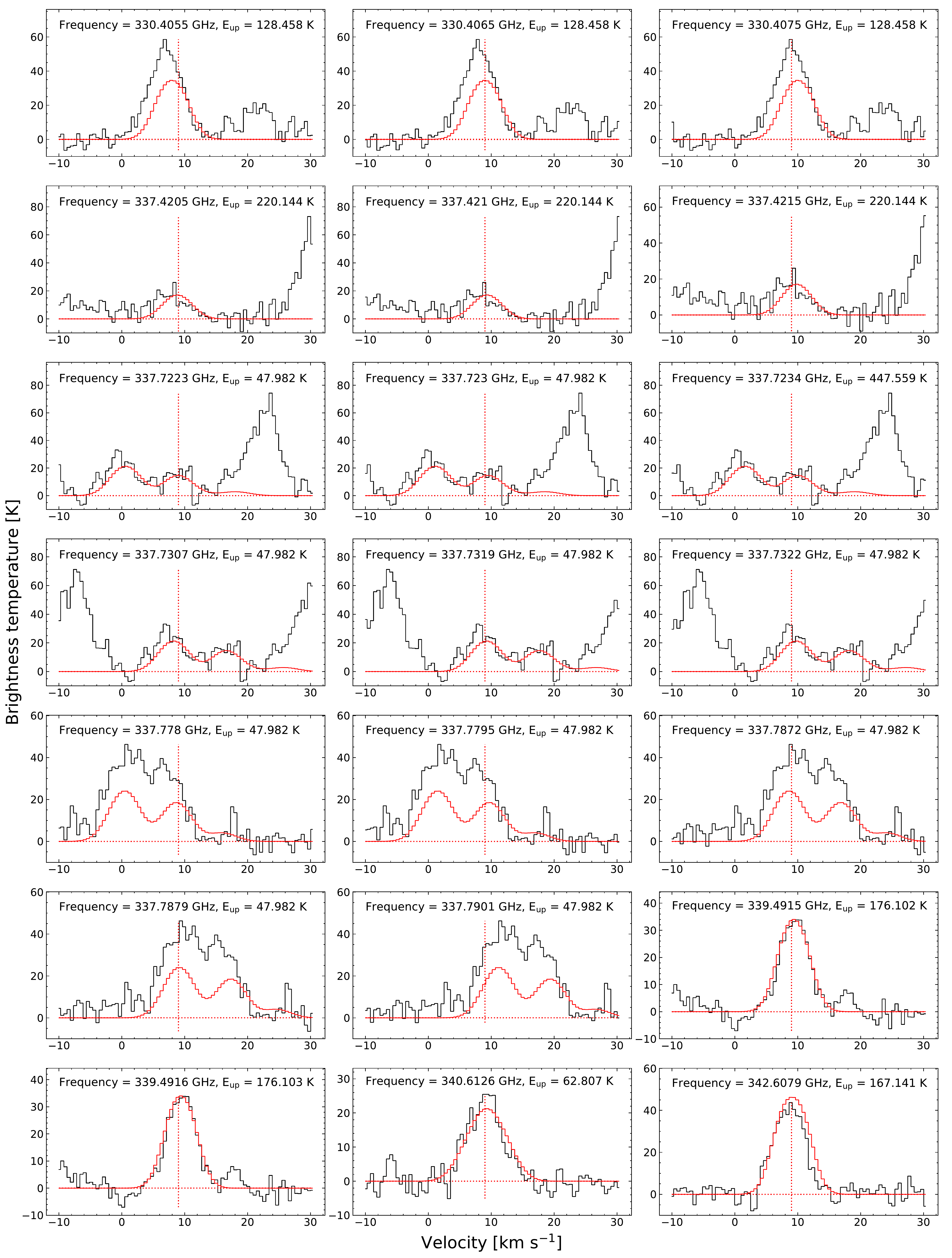}
      \caption{CH$_3$OCH$_3$ model lines and synthetic spectrum. The model fitted values obtained were column density = 2.4$\times 10^{17} \mathrm{cm^{-2}}$, excitation temperature (T$_{\mathrm{ex}}$) = 110.0 K, line width (FWHM) = 5.0 $\mathrm{km \ s^{-1}}$, and source velocity ($\varv_{\text{source}}$) = 9.0 $\mathrm{km \ s^{-1}}.$}
      \label{Fig:CH3OCH3_lines_1}
    \end{figure*} 
    
    \begin{figure*}
    \centering
    \includegraphics[width=17.5cm]{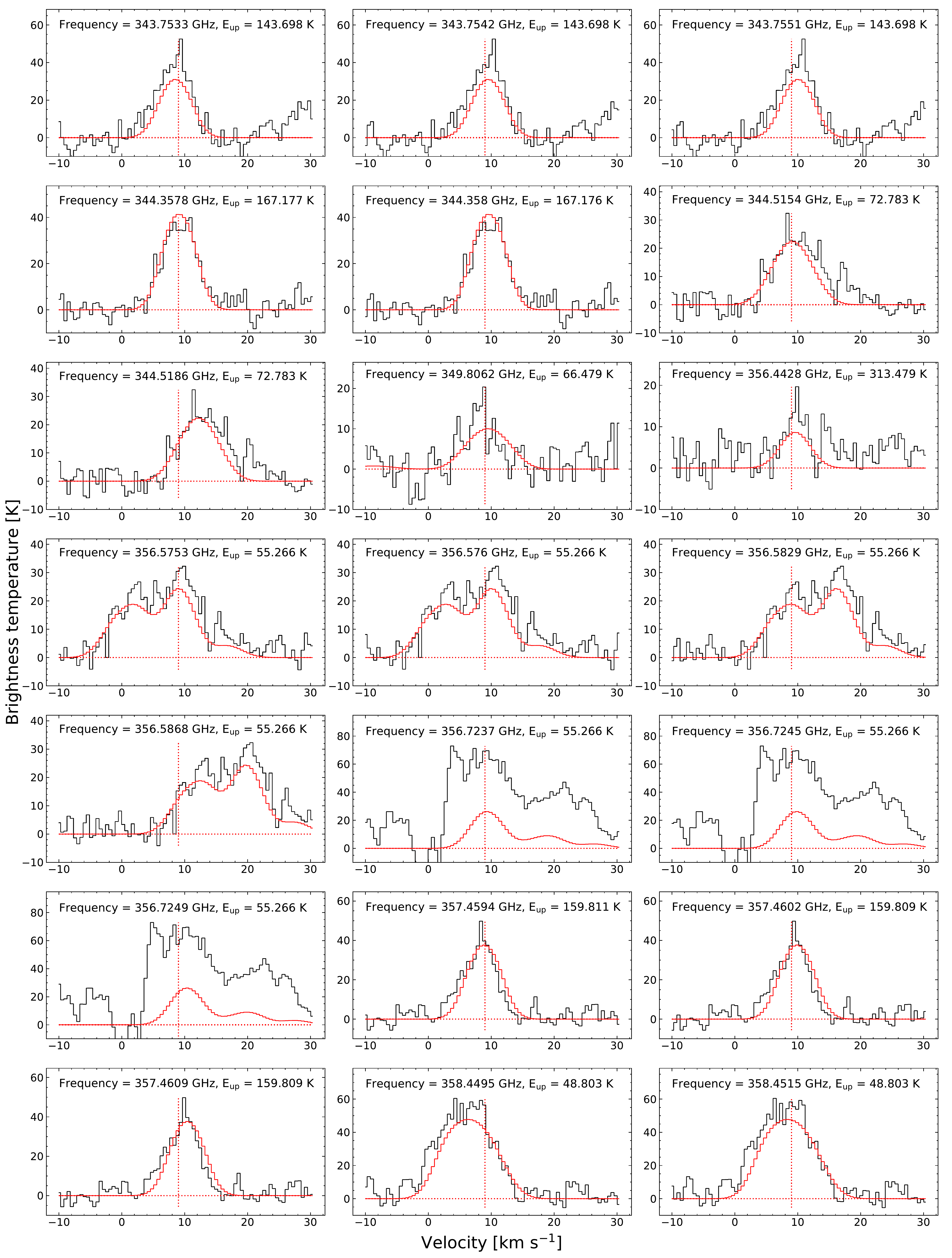}
      \caption{CH$_3$OCH$_3$ model lines and synthetic spectrum. The model fitted values obtained were column density = 2.4$\times 10^{17} \mathrm{cm^{-2}}$, excitation temperature (T$_{\mathrm{ex}}$) = 110.0 K, line width (FWHM) = 5.0 $\mathrm{km \ s^{-1}}$, and source velocity ($\varv_{\text{source}}$) = 9.0 $\mathrm{km \ s^{-1}}.$}
      \label{Fig:CH3OCH3_lines_2}
    \end{figure*}   
    
    \begin{figure*}
    \centering
    \includegraphics[width=17.5cm]{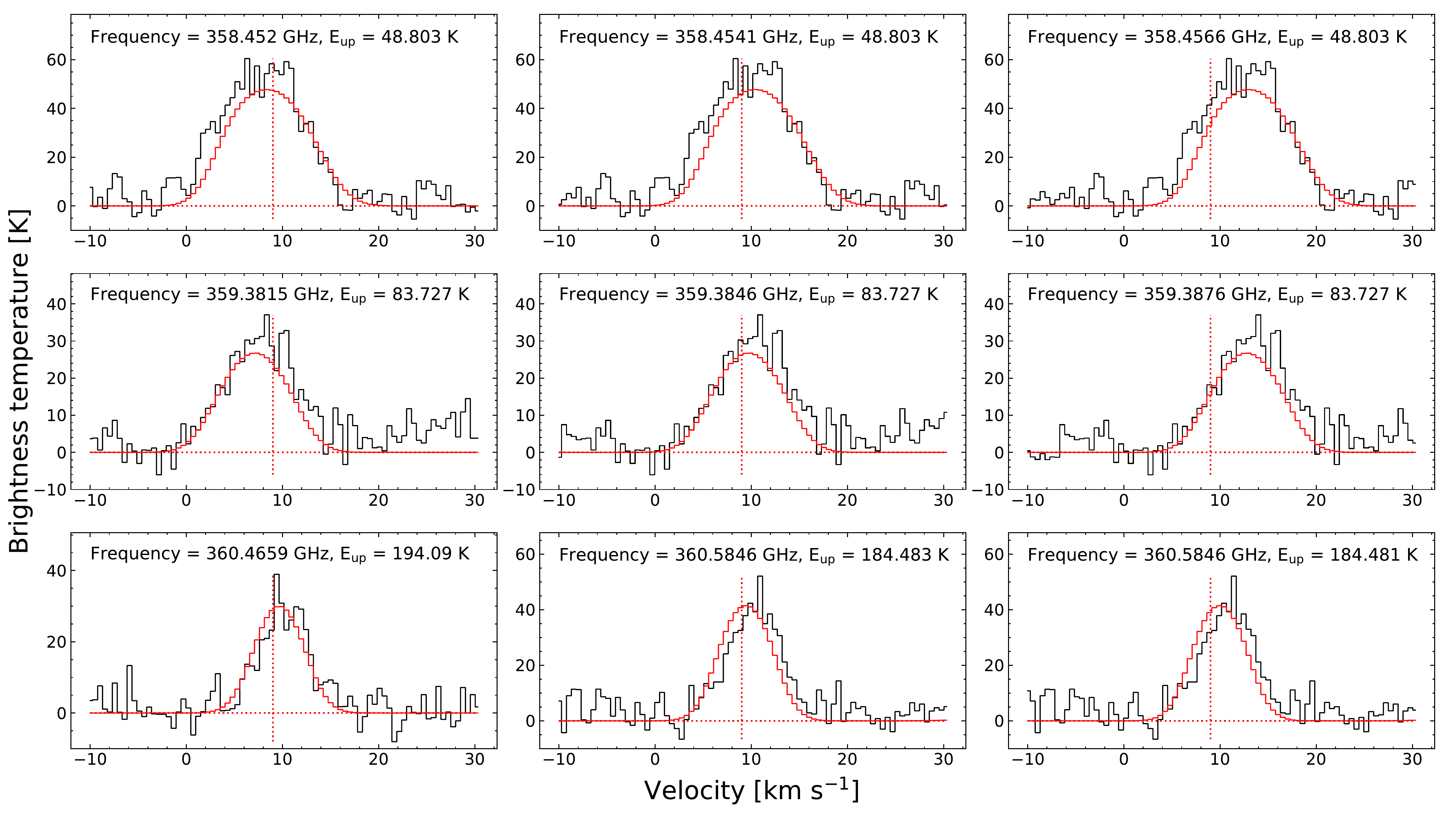}
      \caption{CH$_3$OCH$_3$ model lines and synthetic spectrum. The model fitted values obtained were column density = 2.4$\times 10^{17} \mathrm{cm^{-2}}$, excitation temperature (T$_{\mathrm{ex}}$) = 110.0 K, line width (FWHM) = 5.0 $\mathrm{km \ s^{-1}}$, and source velocity ($\varv_{\text{source}}$) = 9.0 $\mathrm{km \ s^{-1}}.$}
      \label{Fig:CH3OCH3_lines_3}
    \end{figure*}    
    
% H2C-13-O lines
    \begin{figure*}
    \centering
    \includegraphics[width=17.5cm]{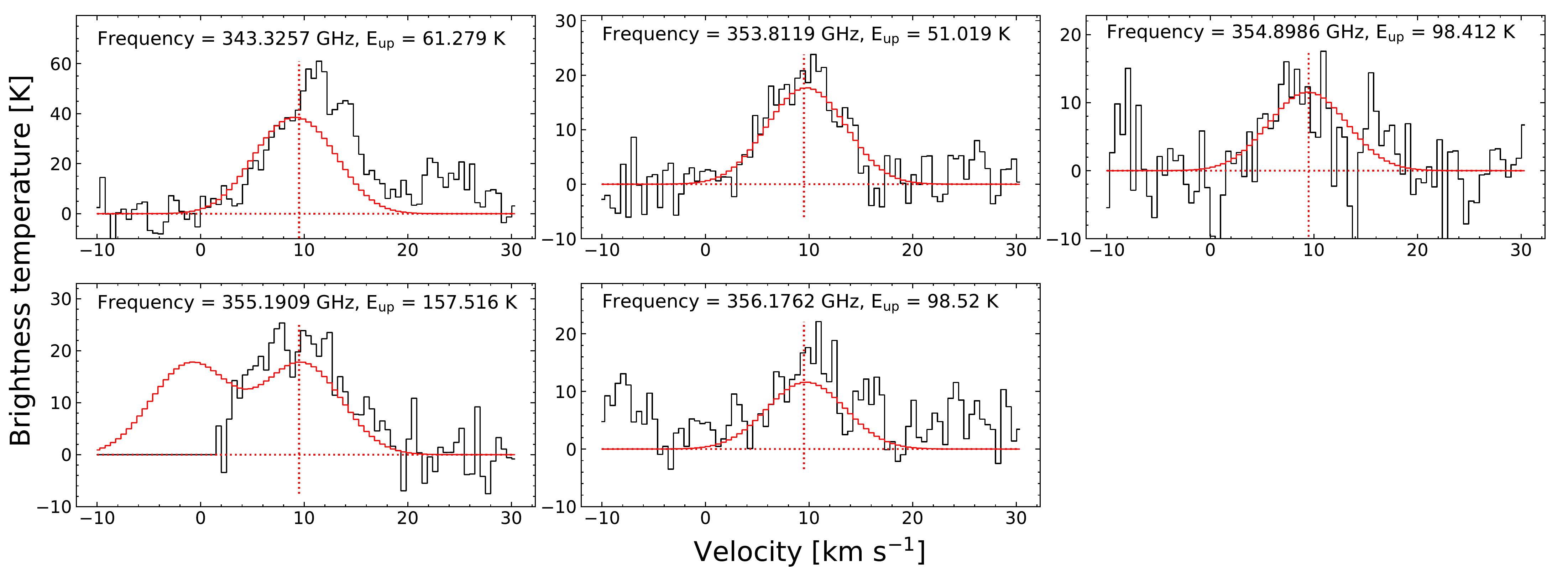}
      \caption{H$_2$$^{13}$CO model lines and synthetic spectrum. The model fitted values obtained were column density = 7.6$\times 10^{15} \mathrm{cm^{-2}}$, excitation temperature (T$_{\mathrm{ex}}$) = 160.0 K, line width (FWHM) = 8.0 $\mathrm{km \ s^{-1}}$, and source velocity ($\varv_{\text{source}}$) = 9.0 $\mathrm{km \ s^{-1}}.$}
      \label{Fig:H2C-13-O_lines_1}
    \end{figure*} 
    
% t-HCOOH lines
    \begin{figure*}
    \centering
    \includegraphics[width=17.5cm]{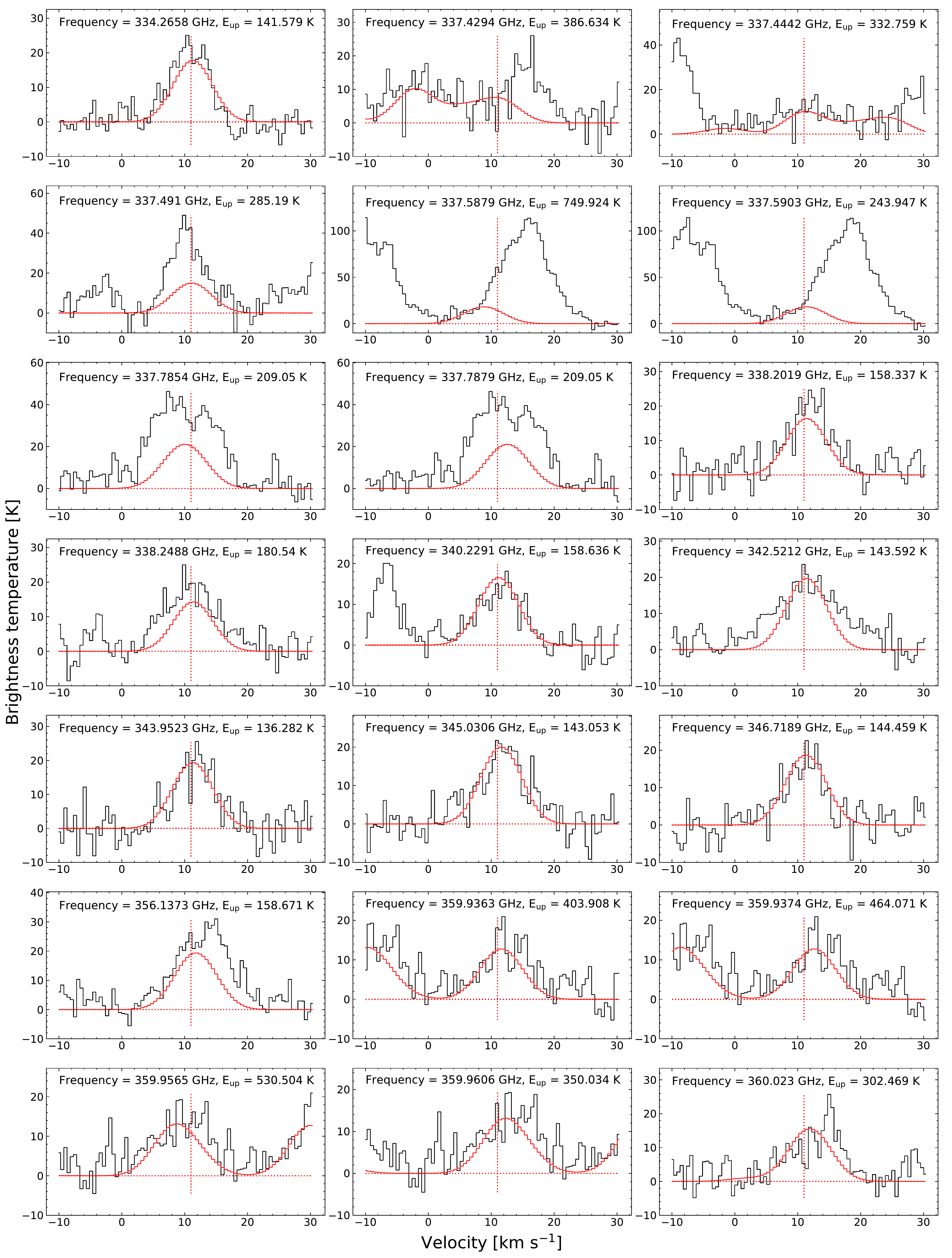}
      \caption{t-HCOOH model lines and synthetic spectrum. The model fitted values obtained were column density = 4.0$\times 10^{16} \mathrm{cm^{-2}}$, excitation temperature (T$_{\mathrm{ex}}$) = 190.0 K, line width (FWHM) = 7.0 $\mathrm{km \ s^{-1}}$, and source velocity ($\varv_{\text{source}}$) = 11.0 $\mathrm{km \ s^{-1}}.$}
      \label{Fig:t-HCOOH_lines_1}
    \end{figure*} 
    
   \begin{figure*}
    \centering
    \includegraphics[width=17.5cm]{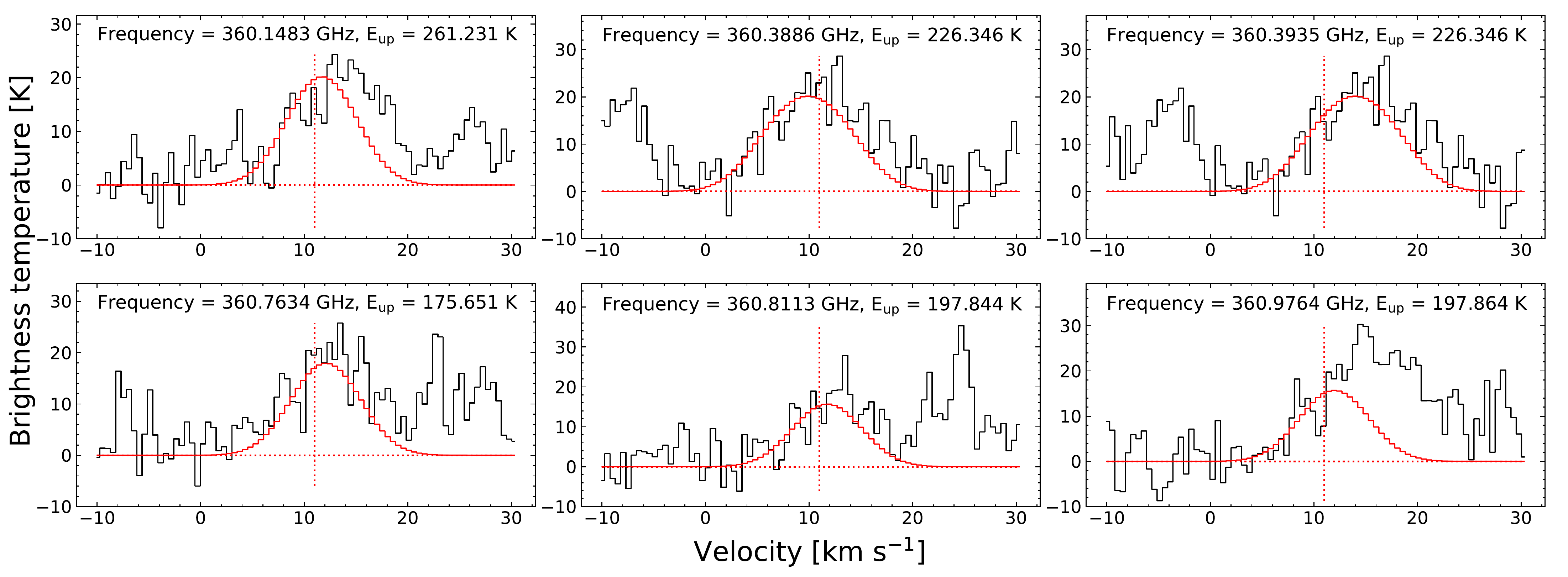}
      \caption{t-HCOOH model lines and synthetic spectrum. The model fitted values obtained were column density = 4.0$\times 10^{16} \mathrm{cm^{-2}}$, excitation temperature (T$_{\mathrm{ex}}$) = 190.0 K, line width (FWHM) = 7.0 $\mathrm{km \ s^{-1}}$, and source velocity ($\varv_{\text{source}}$) = 11.0 $\mathrm{km \ s^{-1}}.$}
      \label{Fig:t-HCOOH_lines_2}
    \end{figure*}

% 33SO lines
    \begin{figure*}
    \centering
    \includegraphics[width=17.5cm]{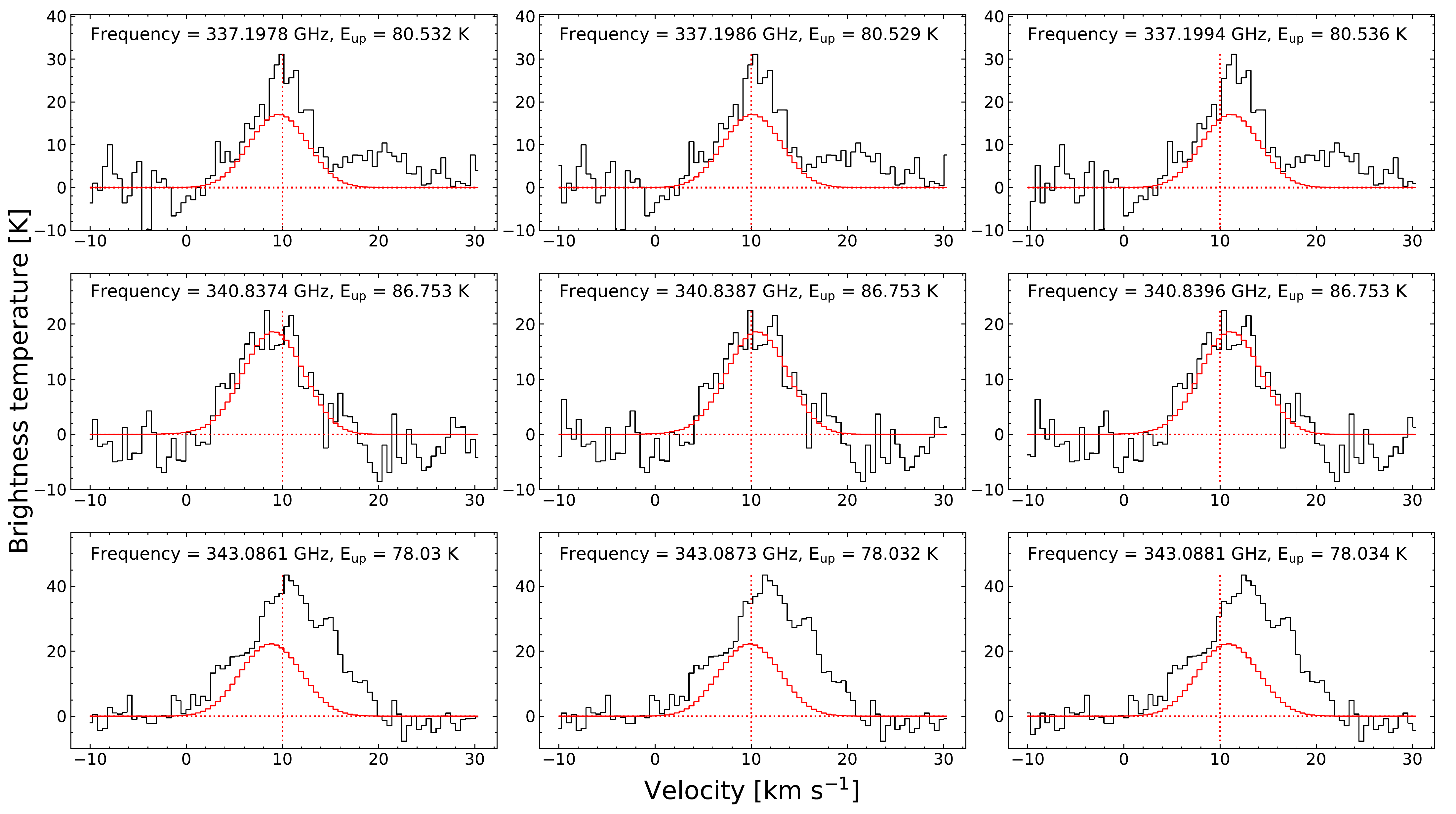}
      \caption{$^{33}$SO model lines and synthetic spectrum. The model fitted values obtained were column density = 5.0$\times 10^{15} \mathrm{cm^{-2}}$, excitation temperature (T$_{\mathrm{ex}}$) = 140.0 K, line width (FWHM) = 6.5 $\mathrm{km \ s^{-1}}$, and source velocity ($\varv_{\text{source}}$) = 10.0 $\mathrm{km \ s^{-1}}.$}
      \label{Fig:33SO_lines_1}
    \end{figure*} 
    
% 34SO2 lines
    \begin{figure*}
    \centering
    \includegraphics[width=17.5cm]{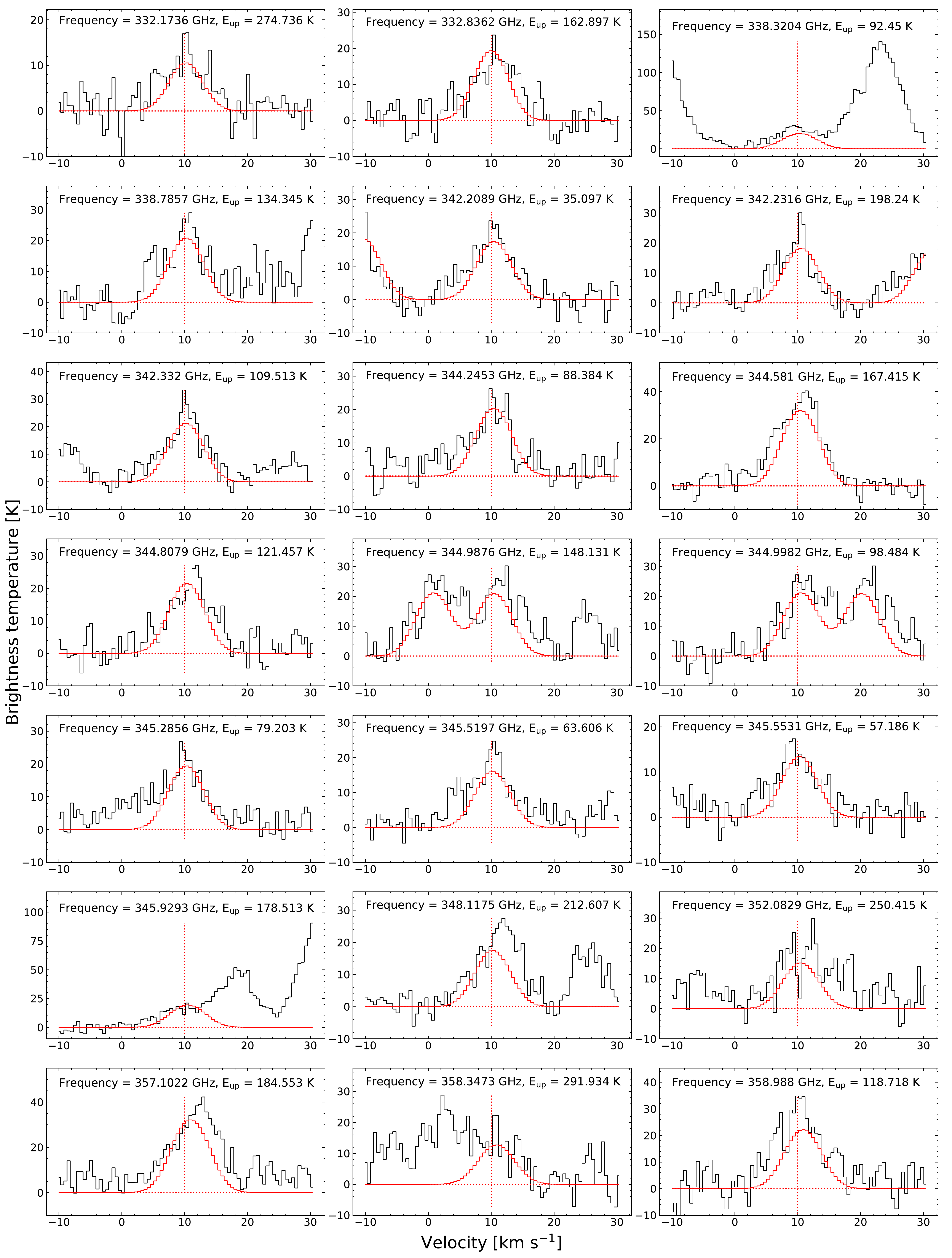}
      \caption{$^{34}$SO$_2$ model lines and synthetic spectrum. The model fitted values obtained were column density = 3.3$\times 10^{16} \mathrm{cm^{-2}}$, excitation temperature (T$_{\mathrm{ex}}$) = 130.0 K, line width (FWHM) = 6.0 $\mathrm{km \ s^{-1}}$, and source velocity ($\varv_{\text{source}}$) = 10.0 $\mathrm{km \ s^{-1}}.$}
      \label{Fig:34SO2_lines_1}
    \end{figure*} 

% SO2 lines
    \begin{figure*}
    \centering
    \includegraphics[width=17.5cm]{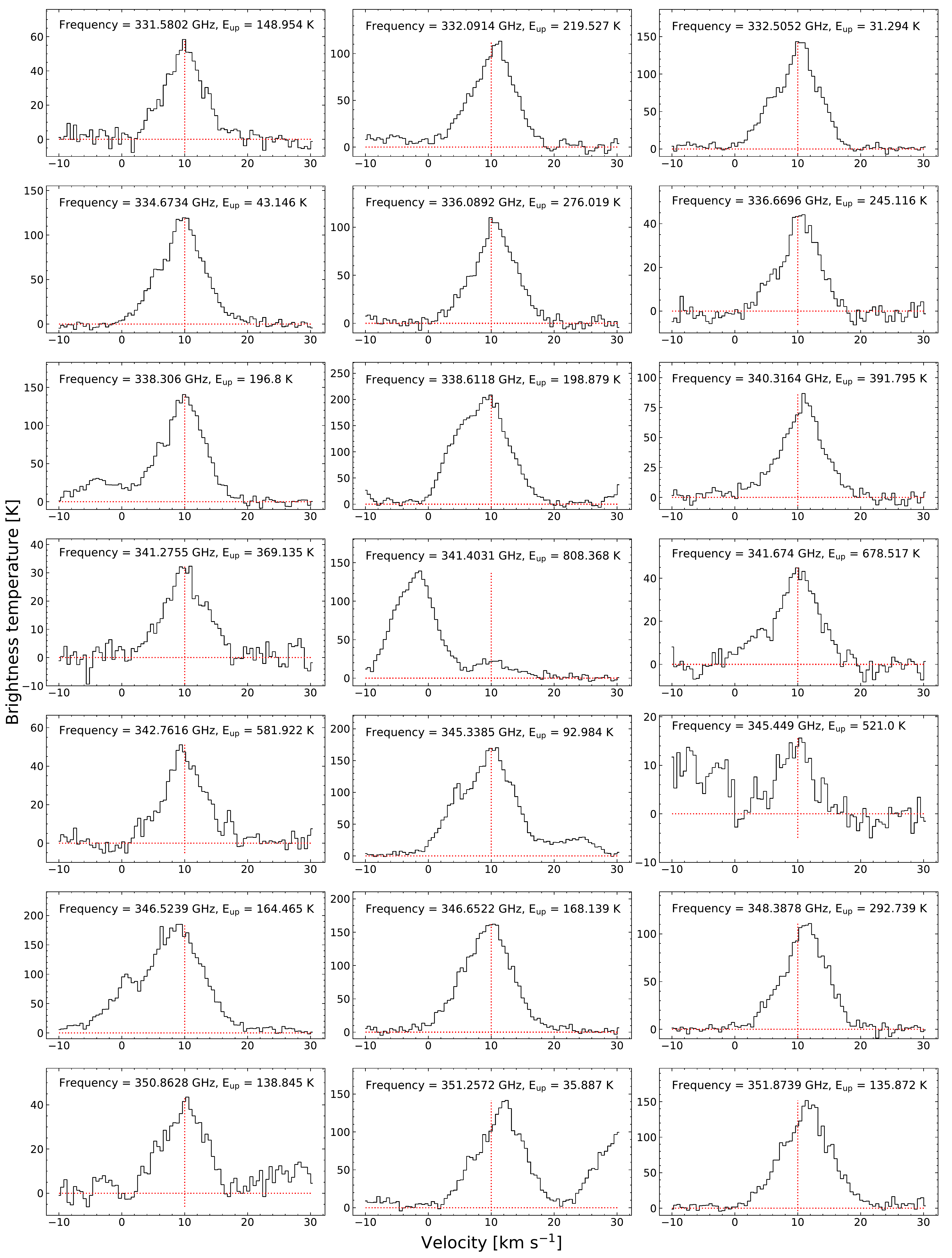}
      \caption{SO$_2$ observed spectral lines. In this case the column density was obtained by multiplying the fitted column density of $^{34}$SO$_2$ with the ISM isotopologue ratio of $^{34}$S/$^{32}$S = 22. The obtained column density was 7.3$\times 10^{17} \mathrm{cm^{-2}}$.}
      \label{Fig:SO2_lines_1}
    \end{figure*} 

    \begin{figure*}
    \centering
    \includegraphics[width=17.5cm]{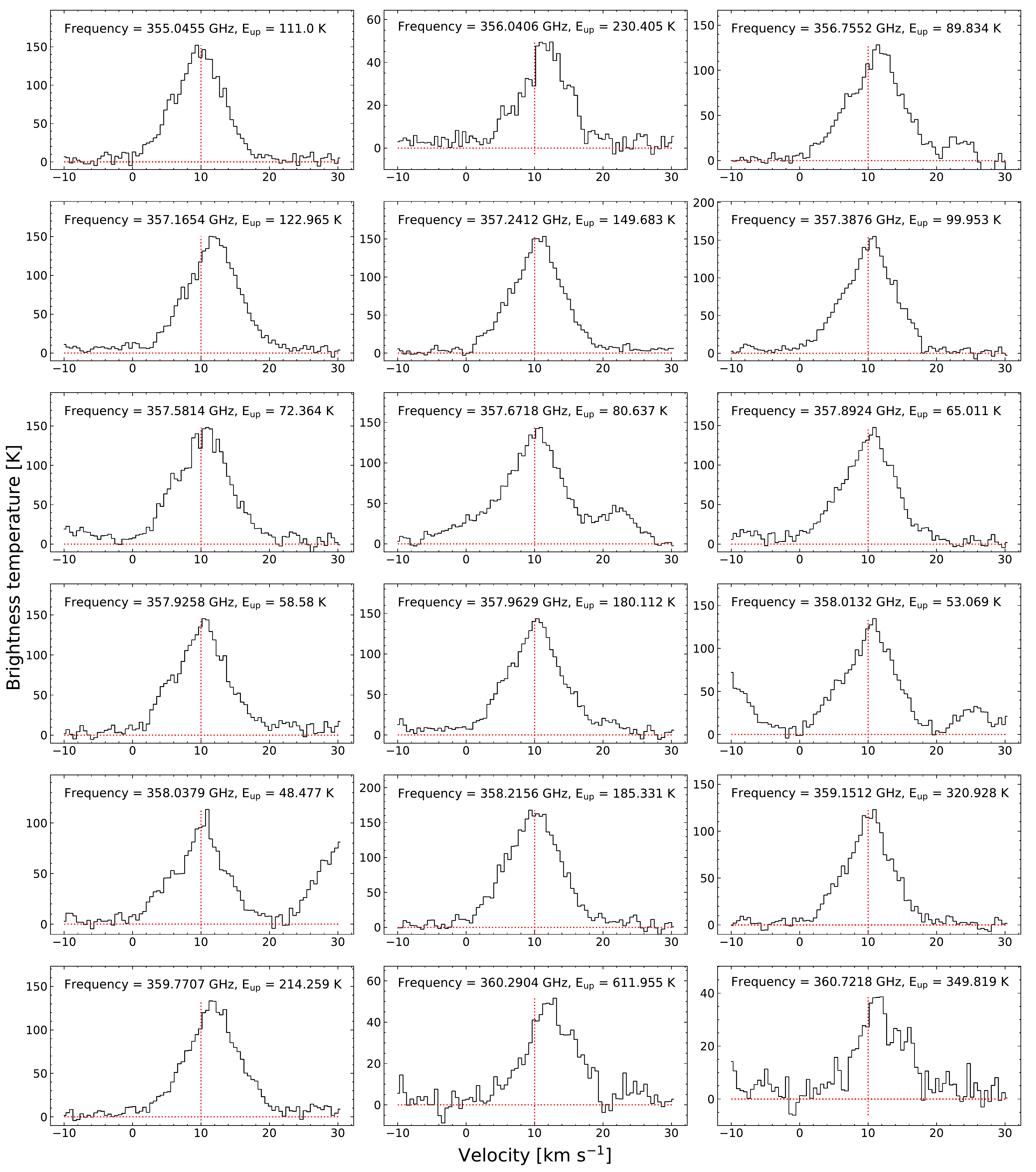}
      \caption{SO$_2$ observed spectral lines. In this case the column density was obtained by multiplying the fitted column density of $^{34}$SO$_2$ with the ISM isotopologue ratio of $^{34}$S/$^{32}$S = 22. The obtained column density was 7.3$\times 10^{17} \mathrm{cm^{-2}}$.}
      \label{Fig:SO2_lines_2}
    \end{figure*} 

% H2CS lines
    \begin{figure*}
    \centering
    \includegraphics[width=17.5cm]{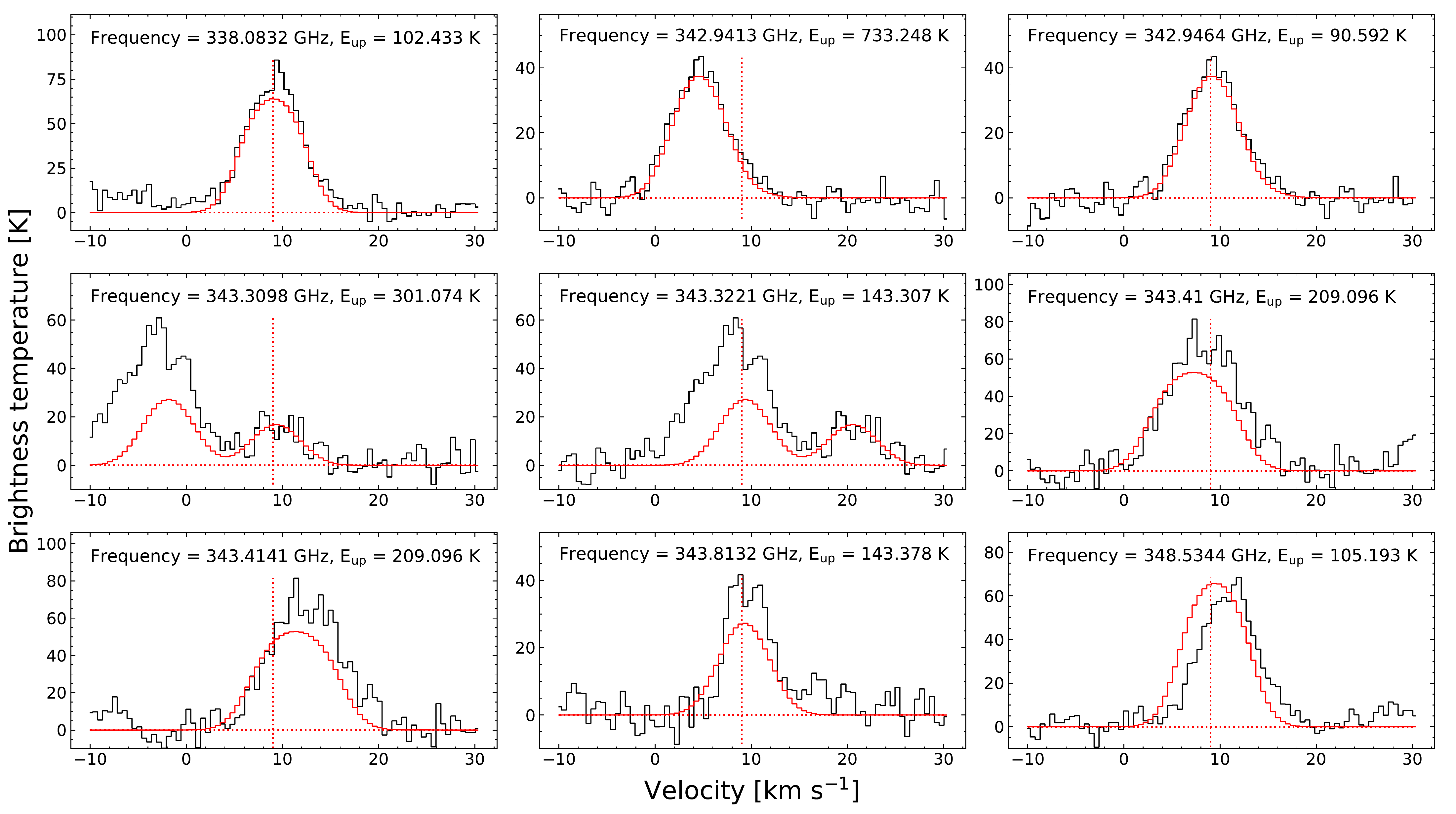}
      \caption{H$_2$CS model lines and synthetic spectrum. The model fitted values obtained were column density = 3.3$\times 10^{16} \mathrm{cm^{-2}}$, excitation temperature (T$_{\mathrm{ex}}$) = 140.0 K, line width (FWHM) = 5.5 $\mathrm{km \ s^{-1}}$, and source velocity ($\varv_{\text{source}}$) = 9.0 $\mathrm{km \ s^{-1}}.$}
      \label{Fig:H2CS_lines_1}
    \end{figure*} 

% OC34S lines
    \begin{figure*}
    \centering
    \includegraphics[width=17.5cm]{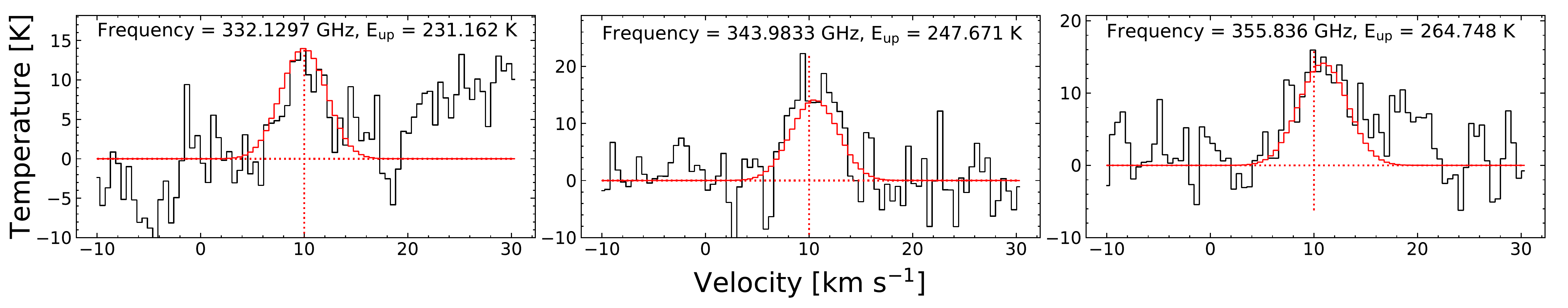}
      \caption{OC$^{34}$S model lines and synthetic spectrum. The model fitted values obtained were column density = 1.1$\times 10^{16} \mathrm{cm^{-2}}$, excitation temperature (T$_{\mathrm{ex}}$) = 170.0 K, line width (FWHM) = 5.0 $\mathrm{km \ s^{-1}}$, and source velocity ($\varv_{\text{source}}$) = 10.0 $\mathrm{km \ s^{-1}}$}
      \label{Fig:OC34S_lines_1}
    \end{figure*} 
    
% OCS lines
    \begin{figure*}
    \centering
    \includegraphics[width=17.5cm]{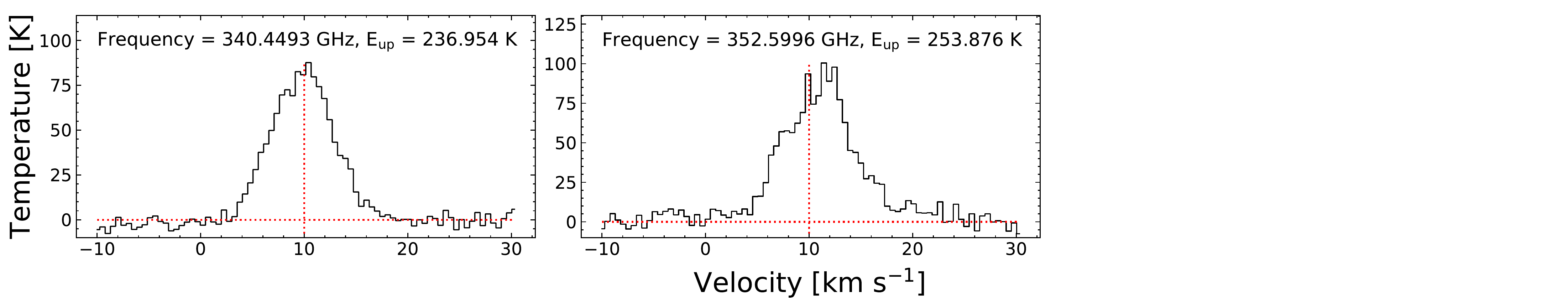}
      \caption{OCS observed spectral lines. In this case the column density was obtained by multiplying the fitted column density of OC$^{34}$S with the ISM isotopologue ratio of $^{34}$S/$^{32}$S = 22. The obtained column density was 2.4$\times 10^{17} \mathrm{cm^{-2}}$.}
      \label{Fig:OCS_lines_1}
    \end{figure*} 
    
% HNCO lines
    \begin{figure*}
    \centering
    \includegraphics[width=17.5cm]{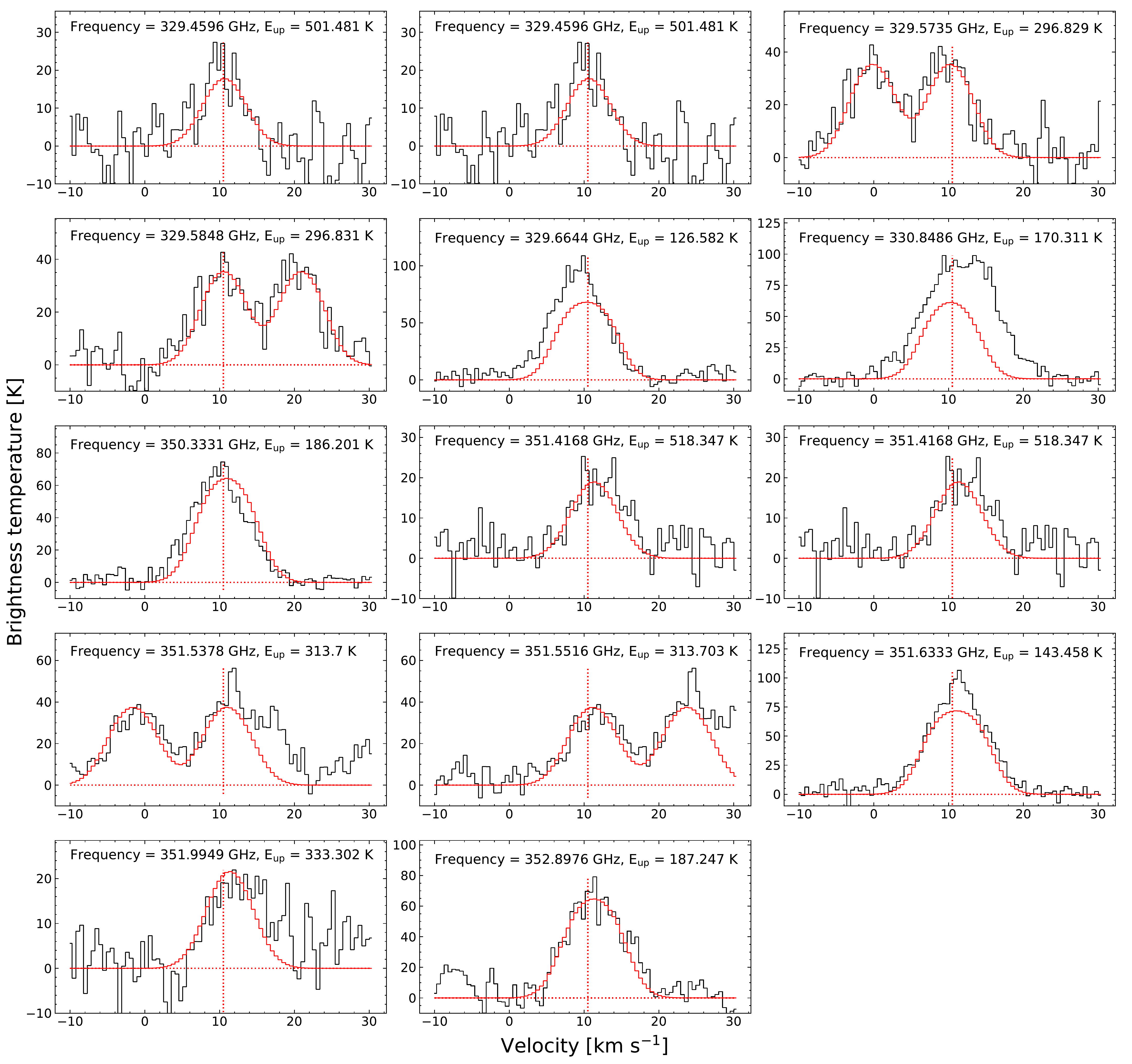}
      \caption{HNCO model lines and synthetic spectrum. The model fitted values obtained were column density = 5.0$\times 10^{16} \mathrm{cm^{-2}}$, excitation temperature (T$_{\mathrm{ex}}$) = 150.0 K, line width (FWHM) = 6.5 $\mathrm{km \ s^{-1}}$, and source velocity ($\varv_{\text{source}}$) = 10.5 $\mathrm{km \ s^{-1}}.$}
      \label{Fig:HNCO_lines_1}
    \end{figure*} 
 
% HC3N lines
    \begin{figure*}
    \centering
    \includegraphics[width=17.5cm]{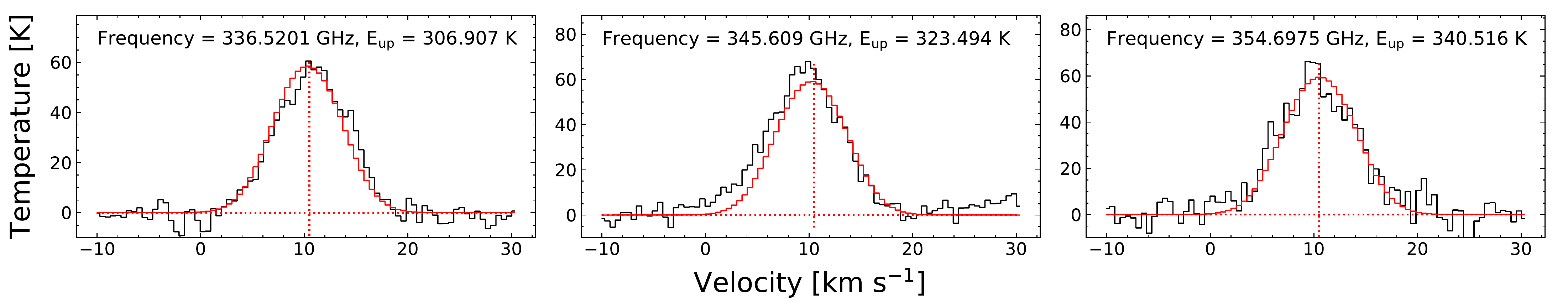}
      \caption{HC$_3$N model lines and synthetic spectrum. The model fitted values obtained were column density = 4.0$\times 10^{16} \mathrm{cm^{-2}}$, excitation temperature (T$_{\mathrm{ex}}$) = 210.0 K, line width (FWHM) = 7.0 $\mathrm{km \ s^{-1}}$, and source velocity ($\varv_{\text{source}}$) = 10.0 $\mathrm{km \ s^{-1}}.$}
      \label{Fig:HC3N_lines_1}
    \end{figure*}

% CH3CN,v8=1 lines
    \begin{figure*}
    \centering
    \includegraphics[width=17.5cm]{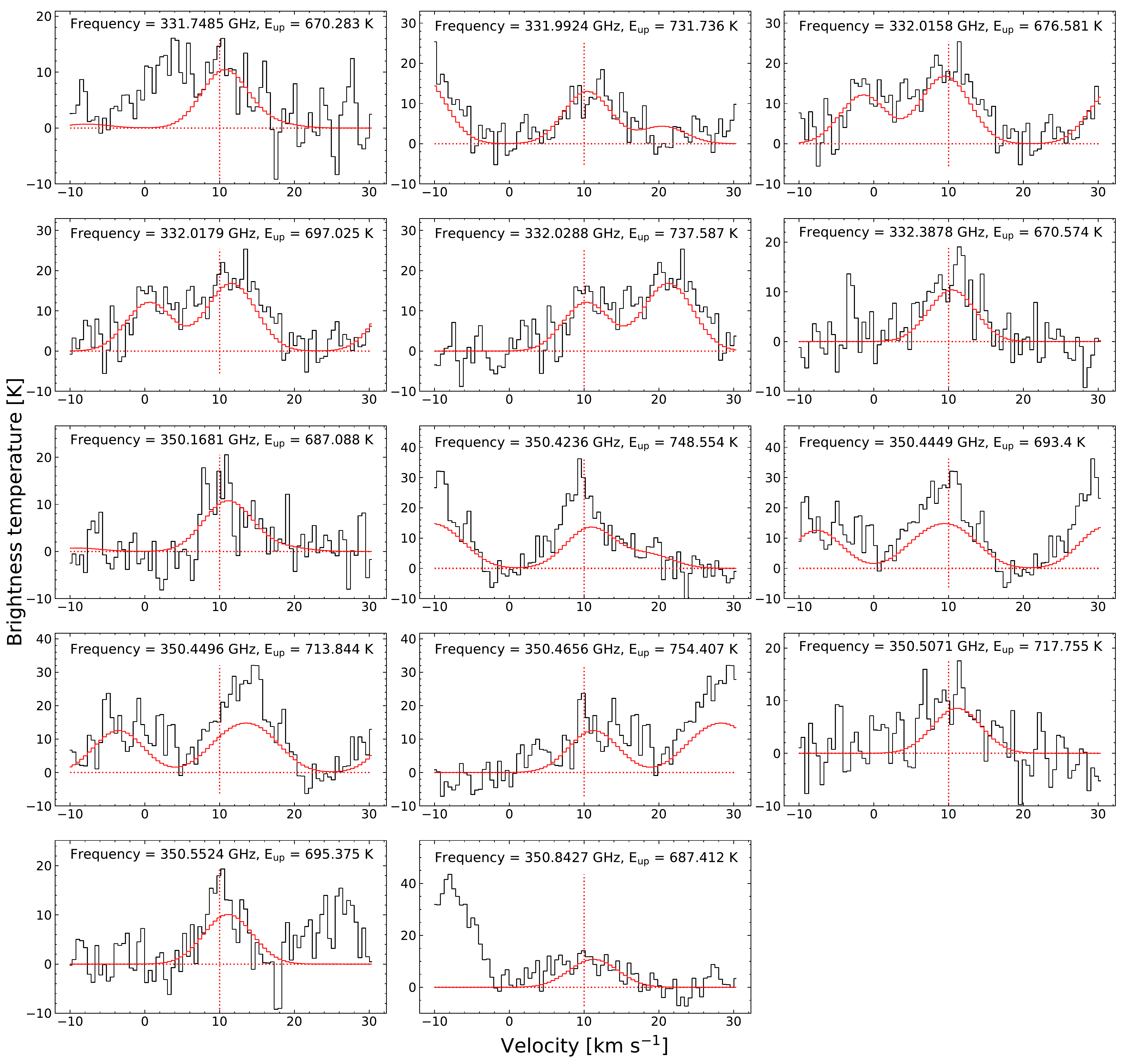}
      \caption{CH$_3$CN, v8=1 model lines and synthetic spectrum. The model fitted values obtained were column density = 5.5$\times 10^{16} \mathrm{cm^{-2}}$, excitation temperature (T$_{\mathrm{ex}}$) = 140.0 K, line width (FWHM) = 7.0 $\mathrm{km \ s^{-1}}$, and source velocity ($\varv_{\text{source}}$) = 10.5 $\mathrm{km \ s^{-1}}.$}
      \label{Fig:CH3CN,v8=1_lines_1}
    \end{figure*} 

% CH3CN,v=0 lines
    \begin{figure*}
    \centering
    \includegraphics[width=17.5cm]{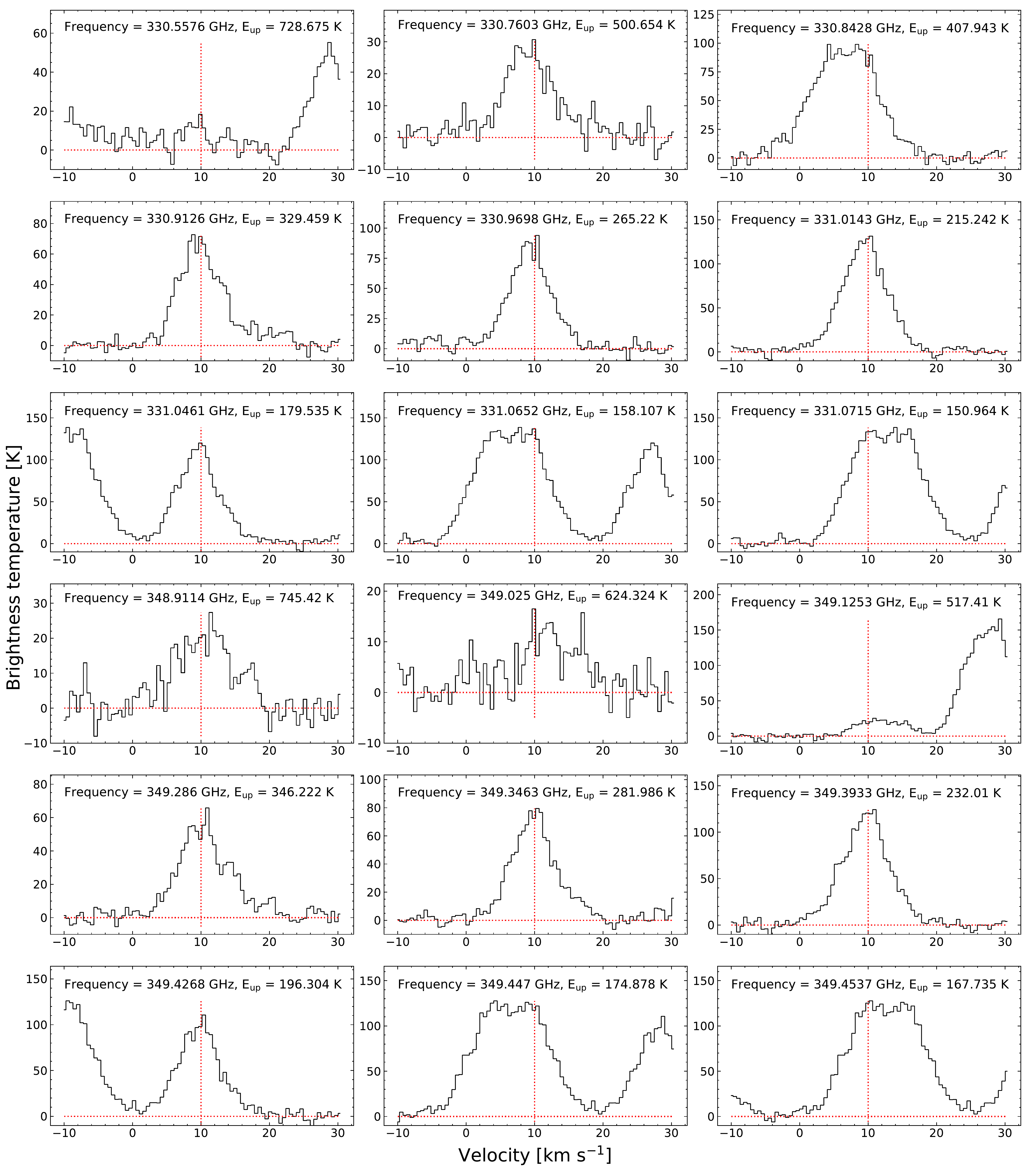}
      \caption{CH$_3$CN, v=0 observed spectral lines. In this case the column density was obtained using the higher energy $v=1$ transition. The obtained column density was 5.5$\times 10^{16} \mathrm{cm^{-2}}$.}
      \label{Fig:CH3CN,v=0_lines_1}
    \end{figure*} 
\end{appendix}

\clearpage
\onecolumn

\setcounter{table}{7}

%_____________________________________________________________
%                                       Two column long Table 
%_____________________________________________________________

\begin{longtable}{l c c c c r}     % 8 columns 
\caption{\label{table:Molecules_all} Detected molecular transitions. The listed line transitions are from best fit line models obtained using the software package CASSIS, on the spectrum taken at the position of peak emission of CH$_3$OH (position 2 in Fig. \ref{Fig:N30_Mol_Grad}). The model parameters are listed in Table \ref{table:N30_column_densities_Meth}. Only transitions above 10 K are listed ($\sim 3 \sigma$), corresponding to about 1.0 $\mathrm{Jy \ beam^{-1}}$.} \\
\hline\hline 
     Frequency & Molecule & Transition & $E_{\mathrm{up}}$/$k_{\mathrm{B}}$ & $A_{\mathrm{ij}}$  & Opacity ($\tau$)  \\ %$T_{\text{ex}}$ &
      $[\mathrm{GHz}] $  &  & &  $[\mathrm{K}]$ & $\times 10^{-4} [\mathrm{s^{-1}}]$  &   \\ 
\hline
\endfirsthead
\caption{continued.}\\
\hline\hline
     Frequency & Molecule & Transition & $E_{\mathrm{up}}$/$k_{\mathrm{B}}$ & $A_{\mathrm{ij}}$  & Opacity ($\tau$)    \\ %$T_{\text{ex}}$ &
      $[\mathrm{GHz}] $  &  & &  $[\mathrm{K}]$ & $\times 10^{-4} [\mathrm{s^{-1}}]$  &  \\ 
\hline
\endhead
\hline
\endfoot
%%%
  329.330552 & C$^{18}$O & $3 - 2$ &      31.61 &       0.02 &       1.39 \\
  329.385477 & SO & $2_{1} - 1_{0}$ &      15.81 &       0.14 &       0.64 \\
  329.459589 & HNCO & $15_{3,13} - 14_{3,12}$ &     501.48 &       4.34 &       0.13 \\
  329.459599 & HNCO & $15_{3,12} - 14_{3,11}$ &     501.48 &       4.34 &       0.13 \\
  329.573452 & HNCO & $15_{2,14} - 14_{2,13}$ &     296.83 &       4.72 &       0.62 \\
  329.584800 & HNCO & $15_{2,13} - 14_{2,12}$ &     296.83 &       4.72 &       0.62 \\
  329.632881 & CH$_3$OH & $12_{2,10,0} - 11_{3,9,0}$ &     218.80 &       0.60 &      61.09 \\
  329.664367 & HNCO & $15_{0,15} - 14_{0,14}$ &     126.58 &       5.04 &       2.24 \\
  329.861407 & CH$_3$OCHO & $27_{5,23,1} - 26_{5,22,1}$ &     241.00 &       5.23 &       0.24 \\
  329.874892 & CH$_3$OCHO & $27_{5,23,0} - 26_{5,22,0}$ &     241.00 &       5.23 &       0.24 \\
  330.001752 & $^{13}$CH$_3$OH & $7_{0,7,0} - 6_{0,6,0}$ &      76.50 &       1.58 &       0.84 \\
  330.194042 & $^{13}$CH$_3$OH & $7_{-1,7,0} - 6_{-1,6,0}$ &      69.01 &       1.55 &       0.88 \\
  330.252798 & $^{13}$CH$_3$OH & $7_{0,7,+0} - 6_{0,6,+0}$ &      63.41 &       1.58 &       0.94 \\
  330.289345 & $^{13}$CH$_3$OH & $7_{-5,2,0} - 6_{-5,1,0}$ &     188.13 &       0.77 &       0.16 \\
  330.309191 & $^{13}$CH$_3$OH & $7_{5,3,0} - 6_{5,2,0}$ &     200.19 &       0.77 &       0.15 \\
  330.319110 & $^{13}$CH$_3$OH & $7_{5,2,-0} - 6_{5,1,-0}$ &     202.00 &       0.78 &       0.14 \\
  330.336917 & $^{13}$CH$_3$OH & $7_{-4,4,0} - 6_{-4,3,0}$ &     151.77 &       1.06 &       0.30 \\
  330.342534 & $^{13}$CH$_3$OH & $7_{4,4,-0} - 6_{4,3,-0}$ &     144.21 &       1.07 &       0.32 \\
  330.349987 & $^{13}$CH$_3$OH & $7_{2,6,-0} - 6_{2,5,-0}$ &     101.23 &       1.46 &       0.63 \\
  330.355512 & CH$_3$OH & $20_{3,17,0} - 19_{4,16,0}$ &     537.04 &       0.64 &       4.41 \\
  330.362042 & $^{13}$CH$_3$OH & $7_{4,3,0} - 6_{4,2,0}$ &     159.85 &       1.07 &       0.28 \\
  330.371293 & $^{13}$CH$_3$OH & $7_{3,5,+0} - 6_{3,4,+0}$ &     113.47 &       1.29 &       0.50 \\
  330.373443 & $^{13}$CH$_3$OH & $7_{3,4,-0} - 6_{3,3,-0}$ &     113.47 &       1.29 &       0.50 \\
  330.391188 & $^{13}$CH$_3$OH & $7_{-3,5,0} - 6_{-3,4,0}$ &     126.37 &       1.30 &       0.45 \\
  330.405456 & CH$_3$OCH$_3$ & $16_{2,15,3} - 15_{1,14,3}$ &     128.46 &       1.65 &       0.10 \\
  330.406503 & CH$_3$OCH$_3$ & $16_{2,15,1} - 15_{1,14,1}$ &     128.46 &       1.65 &       0.41 \\
  330.407549 & CH$_3$OCH$_3$ & $16_{2,15,0} - 15_{1,14,0}$ &     128.46 &       1.65 &       0.26 \\
  330.408395 & $^{13}$CH$_3$OH & $7_{3,4,0} - 6_{3,3,0}$ &     111.39 &       1.29 &       0.51 \\
  330.442421 & $^{13}$CH$_3$OH & $7_{1,6,0} - 6_{1,5,0}$ &      84.49 &       1.59 &       0.79 \\
  330.463873 & $^{13}$CH$_3$OH & $11_{1,10,-0} - 11_{0,11,+0}$ &     165.27 &       3.90 &       1.51 \\
  330.464947 & $^{13}$CH$_3$OH & $7_{2,5,+0} - 6_{2,4,+0}$ &     101.24 &       1.46 &       0.63 \\
  330.535222 & $^{13}$CH$_3$OH & $7_{2,5,0} - 6_{2,4,0}$ &      85.80 &       1.44 &       0.71 \\
  330.535890 & $^{13}$CH$_3$OH & $7_{-2,6,0} - 6_{-2,5,0}$ &      89.45 &       1.46 &       0.69 \\
  330.557569 & CH$_3$CN, v=0 & $18_{9,0} - 17_{-9,0}$ &     728.67 &      23.60 &       0.08 \\
  330.587965 & $^{13}$CO & $3 - 2$ &      31.73 &       0.02 &       1.45 \\
  330.760284 & CH$_3$CN, v=0 & $18_{7,0} - 17_{7,0}$ &     500.65 &      26.80 &       0.41 \\
  330.793887 & CH$_3$OH & $8_{3,5,2} - 9_{2,7,2}$ &     146.28 &       0.54 &      76.08 \\
  330.842762 & CH$_3$CN, v=0 & $18_{6,0} - 17_{-6,0}$ &     407.94 &      28.00 &       0.80 \\
  330.848569 & HNCO & $15_{1,14} - 14_{1,13}$ &     170.31 &       5.01 &       1.62 \\
  330.912608 & CH$_3$CN, v=0 & $18_{5,0} - 17_{5,0}$ &     329.46 &      29.10 &       1.41 \\
  330.969794 & CH$_3$CN, v=0 & $18_{4,0} - 17_{4,0}$ &     265.22 &      30.00 &       2.22 \\
  331.014296 & CH$_3$CN, v=0 & $18_{3,0} - 17_{-3,0}$ &     215.24 &      30.70 &       3.17 \\
  331.046096 & CH$_3$CN, v=0 & $18_{2,0} - 17_{2,0}$ &     179.53 &      31.20 &       4.09 \\
  331.065182 & CH$_3$CN, v=0 & $18_{1,0} - 17_{1,0}$ &     158.11 &      31.50 &       4.76 \\
  331.071544 & CH$_3$CN, v=0 & $18_{0,0} - 17_{0,0}$ &     150.96 &      31.60 &       5.01 \\
  331.149215 & CH$_3$OCHO & $28_{4,25,1} - 27_{4,24,1}$ &     247.76 &       5.34 &       0.23 \\
  331.159562 & CH$_3$OCHO & $28_{4,25,0} - 27_{4,24,0}$ &     247.76 &       5.34 &       0.23 \\
  331.160928 & CH$_3$OCHO & $27_{11,17,3} - 26_{11,16,3}$ &     490.18 &       4.62 &       0.02 \\
  331.459814 & CH$_3$OCHO & $28_{3,25,2} - 27_{3,24,2}$ &     247.74 &       5.36 &       0.23 \\
  331.469465 & CH$_3$OCHO & $28_{3,25,0} - 27_{3,24,0}$ &     247.74 &       5.36 &       0.23 \\
  331.469664 & CH$_3$OCHO & $9_{6,4,0} - 8_{5,4,1}$ &     138.16 &       1.61 &       0.04 \\
  331.502319 & CH$_3$OH & $11_{1,10,0} - 11_{0,11,0}$ &     169.01 &       3.93 &     595.50 \\
  331.580244 & SO$_{2}$ & $11_{6,6} - 12_{5,7}$ &     148.95 &       0.43 &       0.80 \\
  331.748518 & CH$_3$CN, v$_8$=1 & $18_{-1,3} - 17_{1,3}$ &     670.28 &      31.40 &       0.15 \\
  331.755099 & CH$_3$OH & $15_{5,10,5} - 16_{6,11,5}$ &     823.92 &       1.27 &       0.37 \\
  331.784009 & CH$_3$OCHO & $29_{3,27,1} - 28_{3,26,1}$ &     253.07 &       5.43 &       0.23 \\
  331.788199 & CH$_3$OCHO & $31_{1,31,4} - 30_{1,30,4}$ &     445.37 &       6.26 &       0.05 \\
  331.792159 & CH$_3$OCHO & $29_{3,27,0} - 28_{3,26,0}$ &     253.06 &       5.43 &       0.23 \\
  331.795710 & CH$_3$OCHO & $29_{2,27,2} - 28_{2,26,2}$ &     253.07 &       5.43 &       0.23 \\
  331.803922 & CH$_3$OCHO & $29_{2,27,0} - 28_{2,26,0}$ &     253.06 &       5.43 &       0.23 \\
  331.992354 & CH$_3$CN, v$_8$=1 & $18_{2,2} - 17_{-2,2}$ &     731.74 &      31.20 &       0.09 \\
  332.015818 & CH$_3$CN, v$_8$=1 & $18_{0,2} - 17_{0,2}$ &     676.58 &      31.60 &       0.14 \\
  332.017856 & CH$_3$CN, v$_8$=1 & $18_{1,2} - 17_{1,2}$ &     697.02 &      31.50 &       0.12 \\
  332.028758 & CH$_3$CN, v$_8$=1 & $18_{4,3} - 17_{-4,3}$ &     737.59 &      30.00 &       0.09 \\
  332.091431 & SO$_{2}$ & $21_{2,20} - 21_{1,21}$ &     219.53 &       1.51 &       3.00 \\
  332.120135 & CH$_3$OCHO & $23_{1,22,1} - 22_{2,20,0}$ &     292.14 &       0.34 &       0.01 \\
  332.129700 & OC$^{34}$S & $28 - 27$ &     231.16 &       1.07 &       0.15 \\
  332.173573 & $^{34}$SO$_{2}$ & $23_{3,21} - 23_{2,22}$ &     274.74 &       2.54 &       0.16 \\
  332.387753 & CH$_3$CN, v$_8$=1 & $18_{1,3} - 17_{-1,3}$ &     670.57 &      31.60 &       0.15 \\
  332.505242 & SO$_{2}$ & $4_{3,1} - 3_{2,2}$ &      31.29 &       3.29 &       5.82 \\
  332.570617 & CH$_3$OCHO & $30_{1,29,2} - 29_{2,28,1}$ &     256.98 &       0.80 &       0.03 \\
  332.570888 & CH$_3$OCHO & $30_{2,29,1} - 29_{2,28,1}$ &     256.98 &       5.53 &       0.23 \\
  332.571107 & CH$_3$OCHO & $30_{1,29,2} - 29_{1,28,2}$ &     256.98 &       5.53 &       0.23 \\
  332.571378 & CH$_3$OCHO & $30_{2,29,1} - 29_{1,28,2}$ &     256.98 &       0.80 &       0.03 \\
  332.575715 & CH$_3$OCHO & $30_{1,29,0} - 29_{2,28,0}$ &     256.96 &       0.80 &       0.03 \\
  332.575985 & CH$_3$OCHO & $30_{2,29,0} - 29_{2,28,0}$ &     256.96 &       5.53 &       0.23 \\
  332.576202 & CH$_3$OCHO & $30_{1,29,0} - 29_{1,28,0}$ &     256.96 &       5.53 &       0.23 \\
  332.576472 & CH$_3$OCHO & $30_{2,29,0} - 29_{1,28,0}$ &     256.96 &       0.80 &       0.03 \\
  332.603808 & CH$_3$OCHO & $27_{14,13,2} - 26_{14,12,2}$ &     353.08 &       4.11 &       0.07 \\
  332.604456 & CH$_3$OCHO & $27_{14,13,0} - 26_{14,12,0}$ &     353.08 &       4.11 &       0.07 \\
  332.836225 & $^{34}$SO$_{2}$ & $16_{4,12} - 16_{3,13}$ &     162.90 &       3.02 &       0.32 \\
  332.958301 & CH$_3$OCHO & $27_{13,15,0} - 26_{13,14,0}$ &     335.28 &       4.33 &       0.08 \\
  332.996563 & CH$_3$OH & $22_{2,20,2} - 21_{3,18,2}$ &     614.49 &       0.63 &       2.16 \\
  333.114779 & $^{13}$CH$_3$OH & $7_{1,6,-0} - 6_{1,5,-0}$ &      78.52 &       1.59 &       0.82 \\
  333.419204 & CH$_3$OCHO & $27_{12,16,0} - 26_{12,15,0}$ &     318.83 &       4.54 &       0.10 \\
  333.448976 & CH$_3$OCHO & $31_{1,31,1} - 30_{1,30,1}$ &     259.59 &       5.73 &       0.24 \\
  333.449419 & CH$_3$OCHO & $31_{1,31,0} - 30_{1,30,0}$ &     259.57 &       5.64 &       0.24 \\
  333.593050 & CH$_3$OCHO & $27_{4,23,2} - 26_{4,22,2}$ &     240.60 &       5.42 &       0.24 \\
  333.601947 & CH$_3$OCHO & $27_{4,23,0} - 26_{4,22,0}$ &     240.60 &       5.42 &       0.24 \\
  333.864722 & CH$_3$OH & $9_{1,8,1} - 8_{2,6,2}$ &     125.52 &       0.01 &       1.53 \\
  333.900983 & $^{34}$SO & $8_{7} - 7_{6}$ &      79.86 &       4.69 &       1.81 \\
  334.031781 & CH$_3$OCHO & $27_{11,17,0} - 26_{11,16,0}$ &     303.76 &       4.74 &       0.12 \\
  334.265833 & t-HCOOH & $15_{2,14} - 14_{2,13}$ &     141.58 &       4.17 &       0.18 \\
  334.426571 & CH$_3$OH & $3_{0,3,4} - 2_{1,2,4}$ &     314.47 &       0.56 &       5.88 \\
  334.673353 & SO$_{2}$ & $8_{2,6} - 7_{1,7}$ &      43.15 &       1.27 &       3.83 \\
  335.133570 & CH$_3$OH & $2_{2,1,0} - 3_{1,2,0}$ &      44.67 &       0.27 &      30.12 \\
  335.395500 & HDO & $3_{3,1} - 4_{2,2}$ &     335.27 &       0.26 &       0.19 \\
  335.560207 & $^{13}$CH$_3$OH & $12_{1,11,-0} - 12_{0,12,+0}$ &     192.66 &       4.04 &       1.32 \\
  335.582017 & CH$_3$OH & $7_{1,7,0} - 6_{1,6,0}$ &      78.97 &       1.63 &     386.60 \\
  336.028165 & CH$_3$OCHO & $27_{9,19,0} - 26_{9,18,0}$ &     277.85 &       5.14 &       0.16 \\
  336.032357 & CH$_3$OCHO & $27_{9,19,1} - 26_{9,18,1}$ &     277.85 &       4.90 &       0.15 \\
  336.089228 & SO$_{2}$ & $23_{3,21} - 23_{2,22}$ &     276.02 &       2.67 &       3.68 \\
  336.351390 & CH$_3$OCHO & $27_{6,22,1} - 26_{6,21,1}$ &     249.36 &       5.49 &       0.22 \\
  336.354829 & CH$_3$OCHO & $26_{5,21,2} - 25_{5,20,2}$ &     230.58 &       5.57 &       0.26 \\
  336.365317 & CH$_3$OCHO & $15_{6,9,5} - 14_{5,9,5}$ &     282.13 &       0.37 &       0.01 \\
  336.368224 & CH$_3$OCHO & $27_{6,22,0} - 26_{6,21,0}$ &     249.36 &       5.49 &       0.22 \\
  336.373878 & CH$_3$OCHO & $26_{5,21,0} - 25_{5,20,0}$ &     230.58 &       5.57 &       0.26 \\
  336.438224 & CH$_3$OH & $14_{7,7,0} - 15_{6,9,0}$ &     488.22 &       0.36 &       2.76 \\
  336.520084 & HC$_3$N & $37 - 36$ &     306.91 &      30.50 &       0.56 \\
  336.553811 & SO & $11_{10} - 10_{10}$ &     142.88 &       0.06 &       0.70 \\
  336.605889 & CH$_3$OH & $7_{1,7,6} - 6_{1,6,6}$ &     747.41 &       1.64 &       0.48 \\
  336.669581 & SO$_{2}$ & $16_{7,9} - 17_{6,12}$ &     245.12 &       0.58 &       0.71 \\
  336.865149 & CH$_3$OH & $12_{1,11,0} - 12_{0,12,0}$ &     197.07 &       4.07 &     491.30 \\
  336.918184 & CH$_3$OCHO & $26_{6,20,0} - 25_{6,19,0}$ &     235.47 &       5.53 &       0.24 \\
  337.060513 & C$^{17}$O & $3_{1.5} - 2_{2.5}$ &      32.35 &       0.00 &       0.00 \\
  337.060709 & C$^{17}$O & $3_{0.5} - 2_{1.5}$ &      32.35 &       0.01 &       0.00 \\
  337.060831 & C$^{17}$O & $3_{2.5} - 2_{3.5}$ &      32.35 &       0.00 &       0.00 \\
  337.060936 & C$^{17}$O & $3_{2.5} - 2_{2.5}$ &      32.35 &       0.01 &       0.01 \\
  337.060988 & C$^{17}$O & $3_{4.5} - 2_{3.5}$ &      32.35 &       0.02 &       0.04 \\
  337.060988 & C$^{17}$O & $3_{5.5} - 2_{4.5}$ &      32.35 &       0.02 &       0.06 \\
  337.061050 & C$^{17}$O & $3_{1.5} - 2_{1.5}$ &      32.35 &       0.01 &       0.01 \\
  337.061123 & C$^{17}$O & $3_{3.5} - 2_{3.5}$ &      32.35 &       0.01 &       0.01 \\
  337.061214 & C$^{17}$O & $3_{0.5} - 2_{0.5}$ &      32.35 &       0.02 &       0.01 \\
  337.061226 & C$^{17}$O & $3_{3.5} - 2_{2.5}$ &      32.35 &       0.01 &       0.02 \\
  337.061471 & C$^{17}$O & $3_{2.5} - 2_{1.5}$ &      32.35 &       0.01 &       0.01 \\
  337.061553 & C$^{17}$O & $3_{1.5} - 2_{0.5}$ &      32.35 &       0.01 &       0.01 \\
  337.061951 & C$^{17}$O & $3_{4.5} - 2_{4.5}$ &      32.35 &       0.00 &       0.01 \\
  337.062093 & C$^{17}$O & $3_{3.5} - 2_{4.5}$ &      32.35 &       0.00 &       0.00 \\
  337.135853 & CH$_3$OH & $3_{3,1,1} - 4_{2,3,1}$ &      61.64 &       0.16 &      20.65 \\
  337.186488 & CH$_3$OH & $7_{0,7,7} - 6_{0,6,7}$ &     798.97 &       1.68 &       0.30 \\
  337.197845 & $^{33}$SO & $8_{7,7.5} - 7_{6,6.5}$ &      80.53 &       4.70 &       0.07 \\
  337.198488 & CH$_3$OH & $7_{5,3,8} - 6_{5,2,8}$ &     698.08 &       0.83 &       0.40 \\
  337.198620 & $^{33}$SO & $8_{7,8.5} - 7_{6,7.5}$ &      80.53 &       4.83 &       0.08 \\
  337.199371 & $^{33}$SO & $8_{7,5.5} - 7_{6,4.5}$ &      80.54 &       4.65 &       0.05 \\
  337.252172 & CH$_3$OH & $7_{3,5,6} - 6_{3,4,6}$ &     722.84 &       1.39 &       0.52 \\
  337.273561 & CH$_3$OH & $7_{4,4,6} - 6_{4,3,6}$ &     679.25 &       1.13 &       0.66 \\
  337.279180 & CH$_3$OH & $7_{2,5,8} - 6_{2,4,8}$ &     709.70 &       1.54 &       0.66 \\
  337.284320 & CH$_3$OH & $7_{0,7,6} - 6_{0,6,6}$ &     572.96 &       1.68 &       2.84 \\
  337.295913 & CH$_3$OH & $7_{3,5,7} - 6_{3,4,7}$ &     686.20 &       1.37 &       0.74 \\
  337.297484 & CH$_3$OH & $7_{1,7,3} - 6_{1,6,3}$ &     390.02 &       1.65 &      17.36 \\
  337.302644 & CH$_3$OH & $7_{2,6,7} - 6_{2,5,7}$ &     650.99 &       1.55 &       1.20 \\
  337.312360 & CH$_3$OH & $7_{1,7,8} - 6_{1,6,8}$ &     596.79 &       1.65 &       2.19 \\
  337.337010 & CH$_3$OH & $23_{4,20,4} - 22_{2,20,4}$ &    1001.45 &       0.00 &       0.00 \\
  337.396459 & C$^{34}$S & $7_{0} - 6_{0}$ &      64.77 &       8.00 &       0.48 \\
  337.420459 & CH$_3$OCH$_3$ & $21_{2,19,0} - 20_{3,18,0}$ &     220.14 &       1.11 &       0.09 \\
  337.421003 & CH$_3$OCH$_3$ & $21_{2,19,1} - 20_{3,18,1}$ &     220.14 &       1.11 &       0.15 \\
  337.421547 & CH$_3$OCH$_3$ & $21_{2,19,5} - 20_{3,18,5}$ &     220.14 &       1.11 &       0.06 \\
  337.429366 & t-HCOOH & $15_{9,6} - 14_{9,5}$ &     386.63 &       2.80 &       0.03 \\
  337.444214 & t-HCOOH & $15_{8,7} - 14_{8,6}$ &     332.76 &       3.13 &       0.05 \\
  337.463703 & CH$_3$OH & $7_{6,1,3} - 6_{6,0,3}$ &     533.02 &       0.45 &       1.13 \\
  337.490562 & CH$_3$OH & $7_{6,2,5} - 6_{6,1,5}$ &     558.25 &       0.44 &       0.86 \\
  337.491036 & t-HCOOH & $15_{7,9} - 14_{7,8}$ &     285.19 &       3.42 &       0.07 \\
  337.519138 & CH$_3$OH & $7_{3,4,4} - 6_{3,3,4}$ &     482.23 &       1.38 &       5.77 \\
  337.546116 & CH$_3$OH & $7_{5,2,3} - 6_{5,1,3}$ &     485.37 &       0.82 &       3.31 \\
  337.580147 & $^{34}$SO & $8_{8} - 7_{7}$ &      86.07 &       4.89 &       1.99 \\
  337.581680 & CH$_3$OH & $7_{4,4,4} - 6_{4,3,4}$ &     428.20 &       1.13 &       8.09 \\
  337.587936 & t-HCOOH & $15_{14,1} - 14_{14,0}$ &     749.92 &       0.56 &       0.00 \\
  337.590319 & t-HCOOH & $15_{6,9} - 14_{6,8}$ &     243.95 &       3.68 &       0.09 \\
  337.605288 & CH$_3$OH & $7_{2,6,5} - 6_{2,5,5}$ &     429.43 &       1.56 &      11.02 \\
  337.610661 & CH$_3$OH & $7_{3,5,5} - 6_{3,4,5}$ &     387.45 &       1.37 &      14.77 \\
  337.625753 & CH$_3$OH & $7_{2,6,3} - 6_{2,5,3}$ &     363.50 &       1.55 &      21.13 \\
  337.635754 & CH$_3$OH & $7_{2,5,3} - 6_{2,4,3}$ &     363.50 &       1.55 &      21.13 \\
  337.642478 & CH$_3$OH & $7_{1,7,4} - 6_{1,6,4}$ &     356.30 &       1.65 &      24.26 \\
  337.643915 & CH$_3$OH & $7_{0,7,4} - 6_{0,6,4}$ &     365.40 &       1.69 &      22.64 \\
  337.646042 & CH$_3$OH & $7_{4,3,5} - 6_{4,2,5}$ &     470.22 &       1.14 &       5.33 \\
  337.648209 & CH$_3$OH & $7_{5,2,5} - 6_{5,1,5}$ &     610.96 &       0.83 &       0.95 \\
  337.655199 & CH$_3$OH & $7_{3,5,3} - 6_{3,4,3}$ &     460.94 &       1.38 &       7.09 \\
  337.671238 & CH$_3$OH & $7_{2,5,4} - 6_{2,4,4}$ &     464.72 &       1.55 &       7.70 \\
  337.685248 & CH$_3$OH & $7_{5,3,4} - 6_{5,2,4}$ &     493.95 &       0.83 &       3.07 \\
  337.707568 & CH$_3$OH & $7_{1,6,5} - 6_{1,5,5}$ &     478.21 &       1.65 &       7.16 \\
  337.722348 & CH$_3$OCH$_3$ & $7_{4,4,1} - 6_{3,3,1}$ &      47.98 &       0.96 &       0.22 \\
  337.723002 & CH$_3$OCH$_3$ & $7_{4,4,5} - 6_{3,3,5}$ &      47.98 &       1.94 &       0.05 \\
  337.723390 & CH$_3$OCH$_3$ & $31_{2,30,0} - 31_{1,31,0}$ &     447.56 &       0.61 &       0.01 \\
  337.730739 & CH$_3$OCH$_3$ & $7_{4,4,0} - 6_{3,3,0}$ &      47.98 &       1.94 &       0.16 \\
  337.731869 & CH$_3$OCH$_3$ & $7_{4,4,3} - 6_{3,3,3}$ &      47.98 &       1.22 &       0.07 \\
  337.732193 & CH$_3$OCH$_3$ & $7_{4,3,1} - 6_{3,3,1}$ &      47.98 &       0.98 &       0.22 \\
  337.748830 & CH$_3$OH & $7_{0,7,3} - 6_{0,6,3}$ &     488.49 &       1.69 &       6.61 \\
  337.778023 & CH$_3$OCH$_3$ & $7_{4,4,1} - 6_{3,4,1}$ &      47.98 &       0.98 &       0.22 \\
  337.779477 & CH$_3$OCH$_3$ & $7_{4,3,5} - 6_{3,4,5}$ &      47.98 &       1.94 &       0.16 \\
  337.785430 & t-HCOOH & $15_{5,11} - 14_{5,10}$ &     209.05 &       3.90 &       0.12 \\
  337.787213 & CH$_3$OCH$_3$ & $7_{4,3,0} - 6_{3,4,0}$ &      47.98 &       1.94 &       0.27 \\
  337.787868 & CH$_3$OCH$_3$ & $7_{4,3,1} - 6_{3,4,1}$ &      47.98 &       0.96 &       0.22 \\
  337.787949 & t-HCOOH & $15_{5,10} - 14_{5,9}$ &     209.05 &       3.90 &       0.12 \\
  337.790087 & CH$_3$OCH$_3$ & $7_{4,4,3} - 6_{3,4,3}$ &      47.98 &       0.72 &       0.04 \\
  337.837801 & CH$_3$OH & $20_{6,14,2} - 21_{5,17,2}$ &     675.95 &       0.60 &       0.99 \\
  337.877550 & CH$_3$OH & $7_{1,6,6} - 6_{1,5,6}$ &     747.65 &       1.65 &       0.48 \\
  337.969438 & CH$_3$OH & $7_{1,6,3} - 6_{1,5,3}$ &     390.15 &       1.66 &      17.37 \\
  338.083195 & H$_2$CS & $10_{1,10} - 9_{1,9}$ &     102.43 &       5.77 &       1.73 \\
  338.124488 & CH$_3$OH & $7_{0,7,1} - 6_{0,6,1}$ &      78.08 &       1.70 &     400.70 \\
  338.201860 & t-HCOOH & $15_{3,13} - 14_{3,12}$ &     158.34 &       4.23 &       0.17 \\
  338.248816 & t-HCOOH & $15_{4,11} - 14_{4,10}$ &     180.54 &       4.09 &       0.14 \\
  338.305993 & SO$_{2}$ & $18_{4,14} - 18_{3,15}$ &     196.80 &       3.27 &       6.43 \\
  338.320356 & $^{34}$SO$_{2}$ & $13_{2,12} - 12_{1,11}$ &      92.45 &       2.27 &       0.33 \\
  338.344588 & CH$_3$OH & $7_{1,7,2} - 6_{1,6,2}$ &      70.55 &       1.67 &     424.10 \\
  338.404610 & CH$_3$OH & $7_{6,2,1} - 6_{6,1,1}$ &     243.79 &       0.45 &      20.32 \\
  338.408698 & CH$_3$OH & $7_{0,7,0} - 6_{0,6,0}$ &      64.98 &       1.70 &     457.60 \\
  338.430975 & CH$_3$OH & $7_{6,1,2} - 6_{6,0,2}$ &     253.95 &       0.45 &      18.46 \\
  338.442367 & CH$_3$OH & $7_{6,1,0} - 6_{6,0,0}$ &     258.70 &       0.45 &      17.56 \\
  338.456536 & CH$_3$OH & $7_{5,3,2} - 6_{5,2,2}$ &     189.00 &       0.83 &      64.94 \\
  338.475226 & CH$_3$OH & $7_{5,2,1} - 6_{5,1,1}$ &     201.06 &       0.83 &      57.59 \\
  338.486322 & CH$_3$OH & $7_{5,2,0} - 6_{5,1,0}$ &     202.88 &       0.84 &      56.78 \\
  338.504065 & CH$_3$OH & $7_{4,4,2} - 6_{4,3,2}$ &     152.89 &       1.15 &     128.00 \\
  338.512632 & CH$_3$OH & $7_{4,4,0} - 6_{4,3,0}$ &     145.33 &       1.15 &     138.20 \\
  338.530257 & CH$_3$OH & $7_{4,3,1} - 6_{4,2,1}$ &     160.99 &       1.15 &     118.70 \\
  338.540826 & CH$_3$OH & $7_{3,5,0} - 6_{3,4,0}$ &     114.79 &       1.39 &     226.80 \\
  338.543152 & CH$_3$OH & $7_{3,4,0} - 6_{3,3,0}$ &     114.79 &       1.39 &     226.80 \\
  338.559963 & CH$_3$OH & $7_{3,4,2} - 6_{3,3,2}$ &     127.71 &       1.40 &     200.70 \\
  338.583216 & CH$_3$OH & $7_{3,5,1} - 6_{3,4,1}$ &     112.71 &       1.39 &     232.30 \\
  338.611810 & SO$_{2}$ & $20_{1,19} - 19_{2,18}$ &     198.88 &       2.87 &       6.15 \\
  338.614936 & CH$_3$OH & $7_{1,6,1} - 6_{1,5,1}$ &      86.05 &       1.71 &     372.40 \\
  338.639802 & CH$_3$OH & $7_{2,5,0} - 6_{2,4,0}$ &     102.72 &       1.58 &     290.30 \\
  338.721693 & CH$_3$OH & $7_{2,6,1} - 6_{2,5,1}$ &      87.26 &       1.55 &     333.20 \\
  338.722898 & CH$_3$OH & $7_{2,5,2} - 6_{2,4,2}$ &      90.91 &       1.57 &     325.00 \\
  338.759948 & $^{13}$CH$_3$OH & $13_{0,13,+0} - 12_{1,12,+0}$ &     205.95 &       2.18 &       0.67 \\
  338.785687 & $^{34}$SO$_{2}$ & $14_{4,10} - 14_{3,11}$ &     134.34 &       3.08 &       0.34 \\
  339.341459 & SO & $3_{3} - 2_{3}$ &      25.51 &       0.14 &       1.34 \\
  339.398442 & CH$_3$OCHO & $11_{7,4,2} - 11_{6,5,2}$ &     116.89 &       1.71 &       0.06 \\
  339.398465 & CH$_3$OCHO & $11_{7,5,2} - 11_{6,6,2}$ &     116.89 &       1.71 &       0.06 \\
  339.491473 & CH$_3$OCH$_3$ & $19_{1,18,0} - 18_{2,17,0}$ &     176.10 &       2.03 &       0.23 \\
  339.491586 & CH$_3$OCH$_3$ & $19_{1,18,5} - 18_{2,17,5}$ &     176.10 &       2.03 &       0.14 \\
  339.857269 & $^{34}$SO & $8_{9} - 7_{8}$ &      77.34 &       5.08 &       2.44 \\
  339.978923 & CH$_3$OCHO & $9_{4,6,2} - 8_{3,5,2}$ &      57.90 &       2.11 &       0.10 \\
  340.052575 & C$^{33}$S & $7_{0} - 6_{0}$ &      65.28 &       8.19 &       0.27 \\
  340.141143 & CH$_3$OH & $2_{2,0,0} - 3_{1,3,0}$ &      44.67 &       0.28 &      30.23 \\
  340.189247 & CH$_3$OCHO & $6_{5,2,2} - 5_{4,1,2}$ &      49.00 &       3.42 &       0.12 \\
  340.229102 & t-HCOOH & $15_{3,12} - 14_{3,11}$ &     158.64 &       4.30 &       0.17 \\
  340.316406 & SO$_{2}$ & $28_{2,26} - 28_{1,27}$ &     391.80 &       2.58 &       1.73 \\
  340.393659 & CH$_3$OH & $16_{6,11,0} - 17_{5,12,0}$ &     509.17 &       0.50 &       3.41 \\
  340.449273 & OCS & $28 - 27$ &     236.95 &       1.15 &       1.79 \\
  340.612619 & CH$_3$OCH$_3$ & $10_{3,7,1} - 9_{2,8,1}$ &      62.81 &       1.23 &       0.34 \\
  340.683969 & CH$_3$OH & $11_{1,11,4} - 10_{0,10,4}$ &     444.25 &       1.10 &      10.10 \\
  340.714155 & SO & $8_{7} - 7_{6}$ &      81.25 &       4.99 &      63.85 \\
  340.741990 & CH$_3$OCHO & $28_{5,24,1} - 27_{5,23,1}$ &     257.35 &       5.78 &       0.22 \\
  340.754756 & CH$_3$OCHO & $28_{5,24,0} - 27_{5,23,0}$ &     257.35 &       5.78 &       0.22 \\
  340.837357 & $^{33}$SO & $8_{8,6.5} - 7_{7,5.5}$ &      86.75 &       4.89 &       0.06 \\
  340.838679 & $^{33}$SO & $8_{8,8.5} - 7_{7,7.5}$ &      86.75 &       4.92 &       0.07 \\
  340.839639 & $^{33}$SO & $8_{8,9.5} - 7_{7,8.5}$ &      86.75 &       5.03 &       0.08 \\
  341.131665 & $^{13}$CH$_3$OH & $13_{1,12,-0} - 13_{0,13,+0}$ &     222.32 &       4.19 &       1.12 \\
  341.275524 & SO$_{2}$ & $21_{8,14} - 22_{7,15}$ &     369.13 &       0.69 &       0.41 \\
  341.403068 & SO$_{2}$ & $40_{4,36} - 40_{3,37}$ &     808.37 &       4.11 &       0.16 \\
  341.415615 & CH$_3$OH & $7_{1,6,0} - 6_{1,5,0}$ &      80.09 &       1.71 &     389.60 \\
  341.673961 & SO$_{2}$ & $36_{5,31} - 36_{4,32}$ &     678.52 &       4.34 &       0.41 \\
  341.722187 & CH$_3$OCHO & $29_{4,26,1} - 28_{4,25,1}$ &     264.17 &       5.88 &       0.21 \\
  341.732284 & CH$_3$OCHO & $29_{4,26,0} - 28_{4,25,0}$ &     264.16 &       5.88 &       0.21 \\
  341.917897 & CH$_3$OCHO & $29_{3,26,2} - 28_{3,25,2}$ &     264.15 &       5.89 &       0.22 \\
  341.927501 & CH$_3$OCHO & $29_{3,26,0} - 28_{3,25,0}$ &     264.15 &       5.89 &       0.22 \\
  342.208857 & $^{34}$SO$_{2}$ & $5_{3,3} - 4_{2,2}$ &      35.10 &       3.10 &       0.28 \\
  342.231633 & $^{34}$SO$_{2}$ & $20_{1,19} - 19_{2,18}$ &     198.24 &       3.06 &       0.29 \\
  342.332013 & $^{34}$SO$_{2}$ & $12_{4,8} - 12_{3,9}$ &     109.51 &       3.06 &       0.35 \\
  342.350119 & CH$_3$OCHO & $30_{2,28,0} - 29_{3,27,0}$ &     269.49 &       0.77 &       0.03 \\
  342.351420 & CH$_3$OCHO & $30_{3,28,1} - 29_{3,27,1}$ &     269.50 &       5.98 &       0.21 \\
  342.358225 & CH$_3$OCHO & $30_{2,28,2} - 29_{2,27,2}$ &     269.50 &       5.98 &       0.21 \\
  342.359508 & CH$_3$OCHO & $30_{3,28,0} - 29_{3,27,0}$ &     269.49 &       5.98 &       0.21 \\
  342.366296 & CH$_3$OCHO & $30_{2,28,0} - 29_{2,27,0}$ &     269.49 &       5.98 &       0.21 \\
  342.367680 & CH$_3$OCHO & $30_{3,28,1} - 29_{2,27,2}$ &     269.50 &       0.77 &       0.03 \\
  342.521194 & t-HCOOH & $16_{1,16} - 15_{1,15}$ &     143.59 &       4.56 &       0.20 \\
  342.607898 & CH$_3$OCH$_3$ & $19_{0,19,5} - 18_{1,18,5}$ &     167.14 &       3.30 &       0.24 \\
  342.726474 & CH$_3$OH & $20_{6,15,5} - 19_{7,13,5}$ &     978.98 &       0.68 &       0.05 \\
  342.729796 & CH$_3$OH & $13_{1,12,0} - 13_{0,13,0}$ &     227.47 &       4.23 &     393.60 \\
  342.761625 & SO$_{2}$ & $34_{3,31} - 34_{2,32}$ &     581.92 &       3.45 &       0.64 \\
  342.882850 & CS & $7_{0} - 6_{0}$ &      65.83 &       8.40 &       7.64 \\
  342.941346 & H$_2$CS & $10_{7,3} - 9_{7,2}$ &     733.25 &       3.10 &       0.01 \\
  342.946424 & H$_2$CS & $10_{0,10} - 9_{0,9}$ &      90.59 &       6.08 &       0.65 \\
  343.086102 & $^{33}$SO & $8_{9,7.5} - 7_{8,6.5}$ &      78.03 &       5.11 &       0.07 \\
  343.087298 & $^{33}$SO & $8_{9,8.5} - 7_{8,7.5}$ &      78.03 &       5.09 &       0.08 \\
  343.088078 & $^{33}$SO & $8_{9,9.5} - 7_{8,8.5}$ &      78.03 &       5.13 &       0.09 \\
  343.147898 & CH$_3$OCHO & $31_{1,30,2} - 30_{2,29,1}$ &     273.44 &       0.89 &       0.03 \\
  343.149303 & CH$_3$OCHO & $17_{5,12,0} - 16_{4,13,0}$ &     107.81 &       0.25 &       0.02 \\
  343.153227 & CH$_3$OCHO & $31_{1,30,0} - 30_{1,29,0}$ &     273.43 &       6.08 &       0.22 \\
  343.309830 & H$_2$CS & $10_{4,7} - 9_{4,6}$ &     301.07 &       5.12 &       0.12 \\
  343.322082 & H$_2$CS & $10_{2,9} - 9_{2,8}$ &     143.31 &       5.86 &       0.43 \\
  343.325713 & H$_{2}$ $^{13}$CO & $5_{1,5} - 4_{1,4}$ &      61.28 &      11.20 &       0.55 \\
  343.409963 & H$_2$CS & $10_{3,8} - 9_{3,7}$ &     209.10 &       5.56 &       0.76 \\
  343.414146 & H$_2$CS & $10_{3,7} - 9_{3,6}$ &     209.10 &       5.56 &       0.76 \\
  343.435260 & CH$_3$OCHO & $28_{4,24,2} - 27_{4,23,2}$ &     257.08 &       5.92 &       0.22 \\
  343.443944 & CH$_3$OCHO & $28_{4,24,0} - 27_{4,23,0}$ &     257.08 &       5.92 &       0.22 \\
  343.599019 & CH$_3$OH & $13_{1,12,5} - 14_{2,13,5}$ &     624.04 &       0.36 &       0.63 \\
  343.731783 & CH$_3$OCHO & $27_{7,20,2} - 26_{7,19,2}$ &     258.47 &       5.77 &       0.20 \\
  343.753320 & CH$_3$OCH$_3$ & $17_{2,16,3} - 16_{1,15,3}$ &     143.70 &       1.96 &       0.10 \\
  343.754216 & CH$_3$OCH$_3$ & $17_{2,16,1} - 16_{1,15,1}$ &     143.70 &       1.96 &       0.42 \\
  343.755112 & CH$_3$OCH$_3$ & $17_{2,16,0} - 16_{1,15,0}$ &     143.70 &       1.96 &       0.16 \\
  343.813168 & H$_2$CS & $10_{2,8} - 9_{2,7}$ &     143.38 &       5.88 &       0.43 \\
  343.952343 & t-HCOOH & $15_{1,14} - 14_{1,13}$ &     136.28 &       4.60 &       0.20 \\
  343.983268 & OC$^{34}$S & $29 - 28$ &     247.67 &       1.19 &       0.15 \\
  344.029259 & CH$_3$OCHO & $32_{0,32,2} - 31_{1,31,1}$ &     276.10 &       0.77 &       0.03 \\
  344.040629 & $^{13}$CH$_3$OH & $8_{-3,6,0} - 9_{-2,8,0}$ &     144.49 &       0.61 &       0.19 \\
  344.109039 & CH$_3$OH & $18_{2,17,1} - 17_{3,15,1}$ &     419.40 &       0.68 &      12.63 \\
  344.200109 & HC$^{15}$N & $4 - 3$ &      41.30 &      18.80 &       1.34 \\
  344.245346 & $^{34}$SO$_{2}$ & $10_{4,6} - 10_{3,7}$ &      88.38 &       2.96 &       0.33 \\
  344.310612 & SO & $8_{8} - 7_{7}$ &      87.48 &       5.19 &      70.30 \\
  344.312267 & CH$_3$OH & $10_{2,9,5} - 11_{3,9,5}$ &     491.91 &       1.77 &       9.01 \\
  344.357816 & CH$_3$OCH$_3$ & $19_{1,19,3} - 18_{0,18,3}$ &     167.18 &       3.36 &       0.16 \\
  344.358041 & CH$_3$OCH$_3$ & $19_{1,19,0} - 18_{0,18,0}$ &     167.18 &       3.36 &       0.24 \\
  344.443433 & CH$_3$OH & $19_{1,19,0} - 18_{2,16,0}$ &     451.23 &       0.73 &      10.38 \\
  344.515385 & CH$_3$OCH$_3$ & $11_{3,9,1} - 10_{2,8,1}$ &      72.78 &       1.30 &       0.35 \\
  344.518572 & CH$_3$OCH$_3$ & $11_{3,9,0} - 10_{2,8,0}$ &      72.78 &       1.30 &       0.13 \\
  344.581045 & $^{34}$SO$_{2}$ & $19_{1,19} - 18_{0,18}$ &     167.41 &       5.16 &       0.58 \\
  344.807915 & $^{34}$SO$_{2}$ & $13_{4,10} - 13_{3,11}$ &     121.46 &       3.17 &       0.35 \\
  344.970808 & CH$_3$OH & $12_{7,5,4} - 11_{6,5,4}$ &     761.59 &       0.89 &       0.36 \\
  344.987585 & $^{34}$SO$_{2}$ & $15_{4,12} - 15_{3,13}$ &     148.13 &       3.27 &       0.34 \\
  344.998160 & $^{34}$SO$_{2}$ & $11_{4,8} - 11_{3,9}$ &      98.48 &       3.05 &       0.34 \\
  345.030561 & t-HCOOH & $16_{0,16} - 15_{0,15}$ &     143.05 &       4.66 &       0.20 \\
  345.067795 & CH$_3$OCHO & $28_{14,14,2} - 27_{14,13,2}$ &     369.64 &       4.71 &       0.06 \\
  345.069059 & CH$_3$OCHO & $28_{14,15,0} - 27_{14,14,0}$ &     369.64 &       4.71 &       0.06 \\
  345.132599 & $^{13}$CH$_3$OH & $4_{0,4,0} - 3_{-1,3,0}$ &      35.76 &       0.82 &       0.34 \\
  345.285620 & $^{34}$SO$_{2}$ & $9_{4,6} - 9_{3,7}$ &      79.20 &       2.88 &       0.31 \\
  345.338538 & SO$_{2}$ & $13_{2,12} - 12_{1,11}$ &      92.98 &       2.38 &       7.31 \\
  345.448984 & SO$_{2}$ & $26_{9,17} - 27_{8,20}$ &     521.00 &       0.76 &       0.17 \\
  345.466962 & CH$_3$OCHO & $28_{13,16,0} - 27_{13,15,0}$ &     351.86 &       4.94 &       0.08 \\
  345.519656 & $^{34}$SO$_{2}$ & $7_{4,4} - 7_{3,5}$ &      63.61 &       2.59 &       0.25 \\
  345.553093 & $^{34}$SO$_{2}$ & $6_{4,2} - 6_{3,3}$ &      57.19 &       2.35 &       0.20 \\
  345.609010 & HC$_3$N & $38 - 37$ &     323.49 &      33.00 &       0.55 \\
  345.795990 & CO & $3 - 2$ &      33.19 &       0.03 &       - \\
  345.903916 & CH$_3$OH & $16_{1,15,0} - 15_{2,14,0}$ &     332.65 &       1.04 &      40.61 \\
  345.919260 & CH$_3$OH & $18_{3,15,2} - 17_{4,14,2}$ &     459.43 &       0.73 &       8.98 \\
  345.929349 & $^{34}$SO$_{2}$ & $17_{4,14} - 17_{3,15}$ &     178.51 &       3.37 &       0.31 \\
  345.985381 & CH$_3$OCHO & $28_{12,17,0} - 27_{12,16,0}$ &     335.44 &       5.16 &       0.09 \\
  346.202719 & CH$_3$OH & $5_{4,2,0} - 6_{3,3,0}$ &     115.16 &       0.22 &      24.87 \\
  346.204271 & CH$_3$OH & $5_{4,1,0} - 6_{3,4,0}$ &     115.16 &       0.22 &      24.87 \\
  346.523878 & SO$_{2}$ & $16_{4,12} - 16_{3,13}$ &     164.47 &       3.39 &       7.29 \\
  346.528481 & SO & $8_{9} - 7_{8}$ &      78.78 &       5.38 &      86.10 \\
  346.652169 & SO$_{2}$ & $19_{1,19} - 18_{0,18}$ &     168.14 &       5.22 &      12.89 \\
  346.718858 & t-HCOOH & $15_{2,13} - 14_{2,12}$ &     144.46 &       4.66 &       0.19 \\
  346.998344 & H$^{13}$CO$^{+}$ & $4 - 3$ &      41.63 &      32.90 &       0.12 \\
  347.188283 & $^{13}$CH$_3$OH & $14_{1,13,-0} - 14_{0,14,+0}$ &     254.25 &       4.36 &       0.92 \\
  347.330581 & SiO & $8_{0} - 7_{0}$ &      75.02 &      22.00 &       0.15 \\
  347.478251 & CH$_3$OCHO & $27_{5,22,2} - 26_{5,21,2}$ &     247.25 &       6.14 &       0.24 \\
  347.493965 & CH$_3$OCHO & $27_{5,22,0} - 26_{5,21,0}$ &     247.26 &       6.14 &       0.24 \\
  347.628340 & CH$_3$OCHO & $28_{10,19,1} - 27_{10,18,1}$ &     306.77 &       5.49 &       0.13 \\
  347.887461 & CH$_3$OCHO & $20_{5,15,0} - 19_{5,14,0}$ &     262.21 &       3.68 &       0.08 \\
  348.049886 & CH$_3$OCHO & $28_{6,23,1} - 27_{6,22,1}$ &     266.06 &       6.10 &       0.20 \\
  348.065967 & CH$_3$OCHO & $28_{6,23,0} - 27_{6,22,0}$ &     266.06 &       6.10 &       0.20 \\
  348.100194 & $^{13}$CH$_3$OH & $11_{0,11,0} - 10_{1,9,0}$ &     162.36 &       1.08 &       0.39 \\
  348.117469 & $^{34}$SO$_{2}$ & $19_{4,16} - 19_{3,17}$ &     212.61 &       3.50 &       0.27 \\
  348.387800 & SO$_{2}$ & $24_{2,22} - 23_{3,21}$ &     292.74 &       1.91 &       2.25 \\
  348.534365 & H$_2$CS & $10_{1,9} - 9_{1,8}$ &     105.19 &       6.32 &       1.76 \\
  348.786868 & CH$_3$OCHO & $7_{3,5,1} - 6_{2,5,0}$ &      95.90 &       1.29 &       0.03 \\
  348.911401 & CH$_3$CN, v=0 & $19_{9,0} - 18_{-9,0}$ &     745.42 &      28.70 &       0.08 \\
  349.024971 & CH$_3$CN, v=0 & $19_{8,0} - 18_{8,0}$ &     624.32 &      30.50 &       0.20 \\
  349.106997 & CH$_3$OH & $14_{1,13,0} - 14_{0,14,0}$ &     260.20 &       4.41 &     306.30 \\
  349.125287 & CH$_3$CN, v=0 & $19_{7,0} - 18_{7,0}$ &     517.41 &      32.10 &       0.42 \\
  349.286006 & CH$_3$CN, v=0 & $19_{5,0} - 18_{5,0}$ &     346.22 &      34.60 &       1.42 \\
  349.346343 & CH$_3$CN, v=0 & $19_{4,0} - 18_{4,0}$ &     281.99 &      35.50 &       2.23 \\
  349.393297 & CH$_3$CN, v=0 & $19_{3,0} - 18_{-3,0}$ &     232.01 &      36.30 &       3.18 \\
  349.426850 & CH$_3$CN, v=0 & $19_{2,0} - 18_{2,0}$ &     196.30 &      36.80 &       4.09 \\
  349.446987 & CH$_3$CN, v=0 & $19_{1,0} - 18_{1,0}$ &     174.88 &      37.10 &       4.76 \\
  349.453700 & CH$_3$CN, v=0 & $19_{0,0} - 18_{0,0}$ &     167.74 &      37.20 &       5.01 \\
  349.806179 & CH$_3$OCH$_3$ & $11_{2,9,1} - 10_{1,10,1}$ &      66.48 &       0.42 &       0.11 \\
  350.103118 & $^{13}$CH$_3$OH & $1_{1,1,+0} - 0_{0,0,+0}$ &      16.80 &       3.29 &       0.51 \\
  350.168100 & CH$_3$CN, v$_8$=1 & $19_{1,3} - 18_{-1,3}$ &     687.09 &      37.00 &       0.14 \\
  350.286493 & CH$_3$OH & $15_{3,12,4} - 16_{4,13,4}$ &     694.83 &       2.09 &       2.00 \\
  350.333059 & HNCO & $16_{1,16} - 15_{1,15}$ &     186.20 &       5.97 &       1.64 \\
  350.421585 & $^{13}$CH$_3$OH & $8_{1,7,0} - 7_{2,5,0}$ &     102.62 &       0.70 &       0.30 \\
  350.423619 & CH$_3$CN, v$_8$=1 & $19_{-2,2} - 18_{2,2}$ &     748.55 &      36.80 &       0.09 \\
  350.444906 & CH$_3$CN, v$_8$=1 & $19_{0,2} - 18_{0,2}$ &     693.40 &      37.20 &       0.14 \\
  350.449631 & CH$_3$CN, v$_8$=1 & $19_{1,2} - 18_{1,2}$ &     713.84 &      37.10 &       0.12 \\
  350.457580 & CH$_3$OCHO & $28_{8,21,0} - 27_{8,20,0}$ &     283.91 &       6.03 &       0.17 \\
  350.465645 & CH$_3$CN, v$_8$=1 & $19_{4,3} - 18_{-4,3}$ &     754.41 &      35.60 &       0.09 \\
  350.507127 & CH$_3$CN, v$_8$=1 & $19_{3,3} - 18_{3,3}$ &     717.75 &      36.30 &       0.11 \\
  350.552359 & CH$_3$CN, v$_8$=1 & $19_{2,3} - 18_{2,3}$ &     695.38 &      36.80 &       0.14 \\
  350.687662 & CH$_3$OH & $4_{0,4,1} - 3_{1,3,2}$ &      36.34 &       0.87 &     174.00 \\
  350.842670 & CH$_3$CN, v$_8$=1 & $19_{-1,3} - 18_{1,3}$ &     687.41 &      37.20 &       0.14 \\
  350.862756 & SO$_{2}$ & $10_{6,4} - 11_{5,7}$ &     138.84 &       0.44 &       0.72 \\
  350.905100 & CH$_3$OH & $1_{1,1,0} - 0_{0,0,0}$ &      16.84 &       3.32 &     269.20 \\
  350.919517 & CH$_3$OCHO & $27_{6,21,2} - 26_{6,20,2}$ &     252.31 &       6.29 &       0.23 \\
  350.947331 & CH$_3$OCHO & $27_{6,21,0} - 26_{6,20,0}$ &     252.31 &       6.29 &       0.23 \\
  350.947970 & CH$_3$OCHO & $30_{4,27,4} - 29_{3,26,5}$ &     466.76 &       0.69 &       0.00 \\
  350.998044 & CH$_3$OCHO & $28_{7,22,1} - 27_{7,21,1}$ &     274.59 &       6.17 &       0.19 \\
  351.236479 & CH$_3$OH & $9_{5,4,1} - 10_{4,6,1}$ &     240.51 &       0.36 &      20.00 \\
  351.257223 & SO$_{2}$ & $5_{3,3} - 4_{2,2}$ &      35.89 &       3.36 &       6.30 \\
  351.416798 & HNCO & $16_{3,14} - 15_{3,13}$ &     518.35 &       5.31 &       0.14 \\
  351.416812 & HNCO & $16_{3,13} - 15_{3,12}$ &     518.35 &       5.31 &       0.14 \\
  351.537795 & HNCO & $16_{2,15} - 15_{2,14}$ &     313.70 &       5.75 &       0.63 \\
  351.551573 & HNCO & $16_{2,14} - 15_{2,13}$ &     313.70 &       5.75 &       0.63 \\
  351.633257 & HNCO & $16_{0,16} - 15_{0,15}$ &     143.46 &       6.13 &       2.27 \\
  351.768645 & H$_2$CO & $5_{1,5} - 4_{1,4}$ &      62.45 &      12.00 &    - \\
  351.873873 & SO$_{2}$ & $14_{4,10} - 14_{3,11}$ &     135.87 &       3.43 &       7.85 \\
  351.994870 & HNCO & $23_{1,23} - 24_{0,24}$ &     333.30 &       2.31 &       0.31 \\
  352.082921 & $^{34}$SO$_{2}$ & $21_{4,18} - 21_{3,19}$ &     250.41 &       3.65 &       0.23 \\
  352.599570 & OCS & $29 - 28$ &     253.88 &       1.28 &       1.75 \\
  352.897581 & HNCO & $16_{1,15} - 15_{1,14}$ &     187.25 &       6.10 &       1.64 \\
  352.917739 & CH$_3$OCHO & $31_{3,29,1} - 30_{3,28,1}$ &     286.44 &       6.56 &       0.20 \\
  352.920275 & CH$_3$OCHO & $31_{2,29,0} - 30_{3,28,0}$ &     286.43 &       0.85 &       0.03 \\
  352.921697 & CH$_3$OCHO & $31_{2,29,2} - 30_{2,28,2}$ &     286.44 &       6.56 &       0.20 \\
  352.925693 & CH$_3$OCHO & $31_{3,29,0} - 30_{3,28,0}$ &     286.43 &       6.56 &       0.20 \\
  352.927094 & CH$_3$OCHO & $31_{3,29,1} - 30_{2,28,2}$ &     286.44 &       0.85 &       0.03 \\
  352.929508 & CH$_3$OCHO & $31_{2,29,0} - 30_{2,28,0}$ &     286.43 &       6.56 &       0.20 \\
  353.401848 & CH$_3$OCHO & $29_{4,25,2} - 28_{4,24,2}$ &     274.04 &       6.46 &       0.20 \\
  353.410588 & CH$_3$OCHO & $29_{4,25,0} - 28_{4,24,0}$ &     274.04 &       6.46 &       0.20 \\
  353.552866 & CH$_3$OH & $17_{3,14,4} - 17_{2,15,4}$ &     770.97 &       0.43 &       0.21 \\
  353.723522 & CH$_3$OCHO & $32_{1,31,2} - 31_{2,30,1}$ &     290.42 &       0.97 &       0.03 \\
  353.728572 & CH$_3$OCHO & $32_{1,31,0} - 31_{2,30,0}$ &     290.41 &       0.98 &       0.03 \\
  353.739964 & $^{13}$CH$_3$OH & $15_{1,14,-0} - 15_{0,15,+0}$ &     288.46 &       4.54 &       0.75 \\
  353.811872 & H$_{2}^{13}$CO & $5_{0,5} - 4_{0,4}$ &      51.02 &      12.70 &       0.21 \\
  354.127629 & CH$_3$OH & $19_{2,17,3} - 18_{1,17,3}$ &     738.20 &       1.43 &       1.09 \\
  354.131005 & CH$_3$OH & $26_{1,26,0} - 25_{3,23,0}$ &     820.37 &       0.00 &       0.00 \\
  354.445952 & $^{13}$CH$_3$OH & $4_{1,3,0} - 3_{0,3,0}$ &      43.71 &       1.27 &       0.47 \\
  354.505477 & HCN & $4 - 3$ &      42.53 &      20.50 &     - \\
  354.607764 & CH$_3$OCHO & $33_{0,33,2} - 32_{1,32,1}$ &     293.12 &       0.70 &       0.02 \\
  354.697463 & HC$_3$N & $39 - 38$ &     340.52 &      35.70 &       0.54 \\
  354.898595 & H$_{2}^{13}$CO & $5_{2,4} - 4_{2,3}$ &      98.41 &      10.80 &       0.13 \\
  355.045517 & SO$_{2}$ & $12_{4,8} - 12_{3,9}$ &     111.00 &       3.40 &       7.96 \\
  355.190900 & H$_{2}$$^{13}$CO & $5_{3,3} - 4_{3,2}$ &     157.52 &       8.25 &       0.21 \\
  355.602945 & CH$_3$OH & $13_{0,13,0} - 12_{1,12,0}$ &     211.03 &       2.53 &     258.70 \\
  355.835971 & OC$^{34}$S & $30 - 29$ &     264.75 &       1.32 &       0.14 \\
  355.964812 & CH$_3$OH & $16_{3,13,4} - 16_{2,14,4}$ &     731.75 &       0.44 &       0.30 \\
  356.007235 & CH$_3$OH & $15_{1,14,0} - 15_{0,15,0}$ &     295.26 &       4.60 &     231.90 \\
  356.040644 & SO$_{2}$ & $15_{7,9} - 16_{6,10}$ &     230.41 &       0.64 &       0.74 \\
  356.137265 & t-HCOOH & $16_{2,15} - 15_{2,14}$ &     158.67 &       5.07 &       0.19 \\
  356.176243 & H$_{2}$$^{13}$CO & $5_{2,3} - 4_{2,2}$ &      98.52 &      10.90 &       0.13 \\
  356.442778 & CH$_3$OCH$_3$ & $25_{3,22,1} - 24_{4,21,1}$ &     313.48 &       0.91 &       0.06 \\
  356.575254 & CH$_3$OCH$_3$ & $8_{4,5,5} - 7_{3,4,5}$ &      55.27 &       2.09 &       0.17 \\
  356.576016 & CH$_3$OCH$_3$ & $8_{4,5,1} - 7_{3,4,1}$ &      55.27 &       1.47 &       0.32 \\
  356.582852 & CH$_3$OCH$_3$ & $8_{4,5,0} - 7_{3,4,0}$ &      55.27 &       2.09 &       0.28 \\
  356.586761 & CH$_3$OCH$_3$ & $8_{4,4,1} - 7_{3,4,1}$ &      55.27 &       0.62 &       0.14 \\
  356.626667 & CH$_3$OH & $23_{4,20,2} - 22_{5,18,2}$ &     727.83 &       0.80 &       0.81 \\
  356.723697 & CH$_3$OCH$_3$ & $8_{4,4,1} - 7_{3,5,1}$ &      55.27 &       1.47 &       0.32 \\
  356.724457 & CH$_3$OCH$_3$ & $8_{4,4,0} - 7_{3,5,0}$ &      55.27 &       2.10 &       0.17 \\
  356.724864 & CH$_3$OCH$_3$ & $8_{4,5,3} - 7_{3,5,3}$ &      55.27 &       1.17 &       0.06 \\
  356.734223 & HCO$^{+}$ & $4 - 3$ &      42.80 &      35.70 &       1.50 \\
  356.755190 & SO$_{2}$ & $10_{4,6} - 10_{3,7}$ &      89.83 &       3.28 &       7.54 \\
  356.873814 & $^{13}$CH$_3$OH & $13_{2,12,-0} - 12_{3,9,-0}$ &     243.84 &       0.79 &       0.16 \\
  356.874537 & CH$_3$OH & $18_{8,11,2} - 19_{7,12,2}$ &     717.88 &       0.49 &       0.43 \\
  357.102182 & $^{34}$SO$_{2}$ & $20_{0,20} - 19_{1,19}$ &     184.55 &       5.81 &       0.56 \\
  357.165390 & SO$_{2}$ & $13_{4,10} - 13_{3,11}$ &     122.97 &       3.51 &       8.02 \\
  357.241193 & SO$_{2}$ & $15_{4,12} - 15_{3,13}$ &     149.68 &       3.62 &       7.72 \\
  357.338598 & $^{13}$CH$_3$OH & $5_{4,2,-0} - 6_{3,3,-0}$ &     114.77 &       0.24 &       0.06 \\
  357.339947 & $^{13}$CH$_3$OH & $5_{4,1,+0} - 6_{3,4,+0}$ &     114.77 &       0.24 &       0.06 \\
  357.387579 & SO$_{2}$ & $11_{4,8} - 11_{3,9}$ &      99.95 &       3.38 &       7.84 \\
  357.459408 & CH$_3$OCH$_3$ & $18_{2,17,3} - 17_{1,16,3}$ &     159.81 &       2.31 &       0.10 \\
  357.460164 & CH$_3$OCH$_3$ & $18_{2,17,1} - 17_{1,16,1}$ &     159.81 &       2.31 &       0.42 \\
  357.460920 & CH$_3$OCH$_3$ & $18_{2,17,0} - 17_{1,16,0}$ &     159.81 &       2.31 &       0.26 \\
  357.581449 & SO$_{2}$ & $8_{4,4} - 8_{3,5}$ &      72.36 &       3.06 &       6.47 \\
  357.657954 & $^{13}$CH$_3$OH & $7_{2,5,0} - 6_{1,5,0}$ &      85.80 &       1.63 &       0.69 \\
  357.671821 & SO$_{2}$ & $9_{4,6} - 9_{3,7}$ &      80.64 &       3.20 &       7.09 \\
  357.892442 & SO$_{2}$ & $7_{4,4} - 7_{3,5}$ &      65.01 &       2.87 &       5.67 \\
  357.925848 & SO$_{2}$ & $6_{4,2} - 6_{3,3}$ &      58.58 &       2.60 &       4.68 \\
  357.962905 & SO$_{2}$ & $17_{4,14} - 17_{3,15}$ &     180.11 &       3.73 &       7.08 \\
  357.984520 & CH$_3$OCHO & $10_{9,1,0} - 9_{8,2,0}$ &      86.21 &       0.99 &       0.06 \\
  357.994262 & CH$_3$OH & $15_{3,12,4} - 15_{2,13,4}$ &     694.83 &       0.44 &       0.41 \\
  357.995629 & CH$_3$OCHO & $28_{5,23,2} - 27_{5,22,2}$ &     264.44 &       6.71 &       0.22 \\
  358.013154 & SO$_{2}$ & $5_{4,2} - 5_{3,3}$ &      53.07 &       2.18 &       3.46 \\
  358.037887 & SO$_{2}$ & $4_{4,0} - 4_{3,1}$ &      48.48 &       1.45 &       1.95 \\
  358.215633 & SO$_{2}$ & $20_{0,20} - 19_{1,19}$ &     185.33 &       5.83 &      12.44 \\
  358.347316 & $^{34}$SO$_{2}$ & $23_{4,20} - 23_{3,21}$ &     291.93 &       3.86 &       0.19 \\
  358.354940 & CH$_3$OH & $18_{4,15,3} - 19_{5,14,3}$ &     877.17 &       1.44 &       0.25 \\
  358.364215 & CH$_3$OCHO & $28_{7,21,2} - 27_{7,20,2}$ &     275.67 &       6.59 &       0.19 \\
  358.399594 & CH$_3$OH & $14_{2,13,1} - 14_{1,14,2}$ &     266.13 &       0.01 &       0.60 \\
  358.414648 & CH$_3$OH & $10_{6,5,1} - 11_{5,6,1}$ &     306.43 &       0.34 &      10.37 \\
  358.449468 & CH$_3$OCH$_3$ & $5_{5,1,5} - 4_{4,0,5}$ &      48.80 &       3.71 &       0.07 \\
  358.451541 & CH$_3$OCH$_3$ & $5_{5,0,3} - 4_{4,0,3}$ &      48.80 &       3.71 &       0.14 \\
  358.452015 & CH$_3$OCH$_3$ & $5_{5,0,1} - 4_{4,0,1}$ &      48.80 &       3.71 &       0.55 \\
  358.454082 & CH$_3$OCH$_3$ & $5_{5,1,1} - 4_{4,1,1}$ &      48.80 &       3.71 &       0.55 \\
  358.456618 & CH$_3$OCH$_3$ & $5_{5,1,0} - 4_{4,0,0}$ &      48.80 &       3.71 &       0.20 \\
  358.605799 & CH$_3$OH & $4_{1,3,1} - 3_{0,3,1}$ &      44.26 &       1.32 &     234.40 \\
  358.987971 & $^{34}$SO$_{2}$ & $15_{2,14} - 14_{1,13}$ &     118.72 &       2.92 &       0.35 \\
  359.151158 & SO$_{2}$ & $25_{3,23} - 25_{2,24}$ &     320.93 &       3.10 &       2.89 \\
  359.350585 & CH$_3$OCHO & $29_{11,19,0} - 28_{11,18,0}$ &     337.64 &       6.07 &       0.10 \\
  359.351992 & CH$_3$OCHO & $29_{11,18,0} - 28_{11,17,0}$ &     337.64 &       6.07 &       0.10 \\
  359.381524 & CH$_3$OCH$_3$ & $12_{3,10,3} - 11_{2,9,3}$ &      83.73 &       1.43 &       0.09 \\
  359.384589 & CH$_3$OCH$_3$ & $12_{3,10,1} - 11_{2,9,1}$ &      83.73 &       1.43 &       0.35 \\
  359.387642 & CH$_3$OCH$_3$ & $12_{3,10,0} - 11_{2,9,0}$ &      83.73 &       1.43 &       0.22 \\
  359.542805 & CH$_3$OCHO & $29_{6,24,1} - 28_{6,23,1}$ &     283.31 &       6.75 &       0.19 \\
  359.558047 & CH$_3$OCHO & $29_{6,24,0} - 28_{6,23,0}$ &     283.32 &       6.75 &       0.19 \\
  359.677119 & CH$_3$OH & $14_{3,11,4} - 14_{2,12,4}$ &     660.20 &       0.45 &       0.54 \\
  359.678517 & CH$_3$OH & $33_{3,31,1} - 33_{2,31,2}$ &    1349.89 &       1.37 &       0.00 \\
  359.770685 & SO$_{2}$ & $19_{4,16} - 19_{3,17}$ &     214.26 &       3.85 &       6.20 \\
  359.936339 & t-HCOOH & $16_{9,7} - 15_{9,6}$ &     403.91 &       3.64 &       0.04 \\
  359.937371 & t-HCOOH & $16_{10,6} - 15_{10,5}$ &     464.07 &       3.24 &       0.02 \\
  359.956533 & t-HCOOH & $16_{11,5} - 15_{11,4}$ &     530.50 &       2.80 &       0.01 \\
  359.960609 & t-HCOOH & $16_{8,8} - 15_{8,7}$ &     350.03 &       3.99 &       0.05 \\
  360.022974 & t-HCOOH & $16_{7,10} - 15_{7,9}$ &     302.47 &       4.30 &       0.07 \\
  360.148273 & t-HCOOH & $16_{6,11} - 15_{6,10}$ &     261.23 &       4.58 &       0.10 \\
  360.290404 & SO$_{2}$ & $34_{5,29} - 34_{4,30}$ &     611.96 &       4.80 &       0.64 \\
  360.388616 & t-HCOOH & $16_{5,12} - 15_{5,11}$ &     226.35 &       4.82 &       0.12 \\
  360.393461 & t-HCOOH & $16_{5,11} - 15_{5,10}$ &     226.35 &       4.82 &       0.12 \\
  360.408817 & CH$_3$OCHO & $29_{10,20,0} - 28_{10,19,0}$ &     324.07 &       6.30 &       0.12 \\
  360.409986 & CH$_3$OCHO & $29_{10,19,2} - 28_{10,18,2}$ &     324.07 &       5.65 &       0.11 \\
  360.465944 & CH$_3$OCH$_3$ & $20_{1,19,3} - 19_{2,18,3}$ &     194.09 &       2.55 &       0.09 \\
  360.584554 & CH$_3$OCH$_3$ & $20_{0,20,5} - 19_{1,19,5}$ &     184.48 &       3.89 &       0.08 \\
  360.584633 & CH$_3$OCH$_3$ & $20_{0,20,1} - 19_{1,19,1}$ &     184.48 &       3.89 &       0.61 \\
  360.661633 & CH$_3$OH & $3_{1,3,3} - 4_{2,3,3}$ &     339.14 &       2.64 &      18.91 \\
  360.721829 & SO$_{2}$ & $20_{8,12} - 21_{7,15}$ &     349.82 &       0.77 &       0.46 \\
  360.763437 & t-HCOOH & $16_{3,14} - 15_{3,13}$ &     175.65 &       5.16 &       0.17 \\
  360.796220 & $^{13}$CH$_3$OH & $16_{1,15,-0} - 16_{0,16,+0}$ &     324.94 &       4.74 &       0.59 \\
  360.811283 & t-HCOOH & $16_{4,13} - 15_{4,12}$ &     197.84 &       5.02 &       0.15 \\
  360.848946 & CH$_3$OH & $11_{0,11,1} - 10_{1,9,1}$ &     166.05 &       1.21 &     160.20 \\
  360.976377 & t-HCOOH & $16_{4,12} - 15_{4,11}$ &     197.86 &       5.03 &       0.15 \\
\hline
\end{longtable}

\end{document}